%% file: thesis.tex
\documentclass[10pt,letterpaper]{report}
\usepackage{amssym,mod-aaspp4,thesis,mymacros,graphics}

\renewcommand{\arraystretch}{0.7}  
\title{DYNAMICS AND OBSERVATIONAL APPEARANCE OF CIRCUMSTELLAR DISKS}
\author{Andrew Frederick Nelson}
\theyear{1~9~9~9}
\acks{\input{acknowledgements}}     
\ded{\input{ded}}                   
\abstractfile{\input{abstract}}     
\reffile{\input{references}}        

\begin{document}

\doublespace
%
%
\beforepreface
%
%
%
\include{intro}
\include{isodiskchap}

\include{table-sph}
\include{table-ppm}
\include{table-cmp}
\include{isodiskfigs}
\include{interlude1}

\include{cooldiskchap}

\include{table-cool}
\include{cooldiskfigs}
\include{interlude2}

\include{RVchap}

\include{range-tab}
\include{startab}
\include{signif-tab}
\include{RVfigs}
\include{conclusions}

%
%
\listofrefs                   

\end{document}

%% file: acknowledgements.tex
\singlespace
{\small
When someone submits a PhD thesis, there are always lots of people who have 
contributed directly and indirectly to its completion. On this page I'll 
say a few words about some of the more important ones who've helped in
one way or another. Most of the people who might ever read this page 
will know some of these people. The rest of you ought to. 

It goes without saying that a thesis committee is necessary for getting
`out'. Beyond the necessary, mine was just all around great. As my thesis
advisor, Willy Benz provided direction, funding, humor and ideas to me 
on many occasions over the past years. It was a real privilege to work
with Roger Angel and to see how a good idea evolves over time into something
that becomes an important contribution to What We Know. Dave Arnett provided
a wry perspective about how things work in astronomy and many sage comments 
over the years. Adam Burrows, with his voracious appetite for knowledge, 
provided advice and intelligent criticism of my work. Fulvio Melia was 
always ready with a friendly smile and a kind word. Mike Shupe and 
John Rutherfoord, while not directly involved in my thesis work, provided
me an opportunity to explore other fields before deciding on astrophysics
and an outsider's perspective once I did decide on astrophysics. 

Over the years, my office mates have been great sounding boards all 
sorts of crack pot ideas, bad jokes, puns and silly songs with the words
rearranged to fit the circumstances. They also proved willing (or at least
gullible enough) to take all the pennies that filled up my pockets over 
the years. Every time I had too many I'd give some to them. The original 
members of the old Parker room office are: Crystal Martin, Chad Engelbracht, 
Chris Fryer, Paul Harding and me. Later additions from next door include
Aimee and Audra and Dave and Jennifer. I'm very glad to have been in such 
good company for so many years. 

No one can be happy just doing one thing. I'm no exception.
Three important areas in my life outside of astrophysics are music,
karate and family. I'll note people in the first two areas here but 
acknowledge my family in the dedication which follows. 

For several years now, karate has provided me an outlet for
all the frustrations that built up when things weren't going
right or when I just needed someplace else to go. My teachers
and fellow students at Ko Sho Karate have provided learning and 
friendship and more emotional support than I ever expected when I 
first signed up. Johnny Linebarger, who I can't think of as 
anyone but `Sensei', Kim Linebarger, Elaine, Ralph, Norma and
all the rest of the gang: you'll always have a place in my heart 
and memory.

Finally, there are several people in music who influenced me in fundamentally
important ways. John Grossman and Dorothy Walker taught me a quiet 
dedication to and surpassing love of music. Crispin Campbell taught
me how to learn: not just how to learn how to play the cello but
how to learn anything. All the wonderful vocalists at St. Olaf
taught me everything I know about choral music and, here in Arizona,
Grayson Hirst taught me most of what I know about how to sing.

}

%% file: ded.tex
\begin{center}

\vspace{1in}
Probably the most important part of my life is my family. \\
My parents, Jim and Judy Nelson. \\
My sister Margot and her husband Jeff. \\
And my brother, Charlie. \\
Hopefully someday soon there will be other additions. \\  
\vspace{1in}
I want to sincerely thank them all for past, present \\
and future support in every endeavor I've been a part of.
\end{center}

%% file: abstract.tex
\specdouble

In my thesis I present a study of the dynamics and observational
characteristics of massive circumstellar disks in two dimensions
($r$, $\phi$) using two complimentary hydrodynamic codes: a `Smoothed
Particle Hydrodynamic' (SPH) code and a `Piecewise Parabolic Method' (PPM)
code. I also study the detection limits available to radial velocity 
searches for low mass companions to main sequence stars. This thesis is 
organized as a series of published or submitted papers, connected by 
introductory and concluding material. I strongly recommend that 
readers of this abstract obtain the published versions of each of these 
papers.

I first outline the progress which has been made in the modeling of the 
structure and origins of the solar system, then in chapter \ref{isodisk}
({\it The Astrophysical Journal} v502, p342, 
with W. Benz, F. Adams and D. Arnett), I proceed with numerical simulations
of circumstellar disks using both hydrodynamic codes assuming a `locally 
isothermal' equation of state. The disks studied range in mass from 
$0.05 M_*$ to $1.0 M_*$ and in initial minimum Toomre $Q$ value from 
$1.1$ to $3.0$. Massive disks ($M_D>0.2M_*$) tend to form grand design 
spiral structure with 1--3 arms, while low mass disks ($M_D\leq0.2M_*$)
tend to form filamentary, $>$4 armed spiral structures. Disks with minimum
Toomre stability $Q\lesssim 2.0$ are dynamically active and structures
within each disk become distorted, break apart and reform on orbital 
time scales.  Spiral arms in disks with $Q\lesssim 1.5$ frequently 
collapse into clumps. I perform a detailed comparison of the two numerical
techniques employed and conclude that SPH is unable to follow the linear 
instability regime of the disk systems due to noise inherent in the method.  

In chapter \ref{cooldisk} (submitted to {\it The
Astrophysical Journal} with W. Benz and T. Ruzmaikina), I relax the
assumption the locally isothermal evolution assumption and
instead include simple heating and cooling prescriptions for the 
system. Under these physical conditions, the spiral arm
growth is suppressed in the inner 1/3 of the disks relative to the
isothermal evolution and in the remainder, changes character to
more diffuse spiral structures. I synthesize spectral energy distributions
(SEDs) from the simulations and compare them to fiducial SEDs derived
from observed systems. The size distribution of grains in the inner 
disk can have marked consequences on the near infrared portion of the 
SED. After being vaporized in a hot midplane region, the grains
do not reform quickly into the size distribution on which most opacity
calculations are based. With the original opacities, near infrared 
emission suppressed relative to observations, with a plausible
modification to the opacity, a more realistic SED is obtained. At long
wavelengths, insufficient flux is produced and we conclude that the
internal heating processes included in our model (due e.g. to gravitational
torques) do not provide a large fraction of the thermal energy present in
the outer portions of accretion disks.

In chapter \ref{RVchap} ({\it The Astrophysical Journal} v500,
p940 with Roger Angel), I examine the limits which may be placed upon the
detection of planets, brown dwarfs and low mass stellar companions using
radial velocity measurements. I derive an analytic expression describing
the amplitude limits for periodic signals which may be obtained from a
set of data of known duration, number of measurements and precision. I
have verified the formalism with Monte Carlo experiments and outlined the
regions of its validity. I have used the technique to suggest a strategy 
for continuing large radial velocity searches for low mass companions.

In chapter \ref{conclusions}, I outline several problems
which may be profitably addressed by building on this work.

\doublespace

%% file: intro.tex
\chapter{How We Got to This Point and Where We're Going}

Understanding the character and origin of our solar system and, more 
generally, of other stars and planetary systems is a problem which has 
been under study for a large portion of human history. In this
introductory chapter I will outline some of the important steps in the
development of our understanding of our solar system and of the origin
of solar systems in general. I hope that you as a reader finds this as 
amusing to read as it was amusing for me to write.

Modern records show Greek, Babylonian and Chinese scholars as some of 
the earliest to make a systematic study of the subject. Due to limited
access to their work, similar studies in other parts of the world (e.g.
Central America) have contributed to a lesser extent to the current paradigms.
\footnote{In the spirit of referencing original source material, I have done
so where possible throughout much of this introduction. In the spirit of 
obtaining a good historical background of ancient astronomy, I'd recommend
Dreyer's {\it History of Astronomy} or Heath's {\it Aristarchus of Samos} 
as excellent introductions to the subject.}

Some of the earliest to put forth a model for the solar system were the
Pythagoreans, a school founded by the man whose name it bears. The work of
this school is unfortunately largely available only through secondhand 
sources, since they did not generally widely publish their work. They 
hypothesized a model of a central fire with a spherical earth, the sun,
the moon and the stars orbiting around it on circular trajectories. 
Later Pythagoreans theorized that the earth instead rotated in space
(Heraclides, ca. 340 BC).  The circumference of the earth was later 
determined from observations of shadow lengths at different latitudes by
Eratosthenes (Eratosthenes ca. 220BC). Also following up the heliocentric
theory of the heavens, Aristarchus (ca. 260 BC) determined the size of 
and distance to the sun and moon.

Eudoxus of Cnidus (ca. 365 BC) proposed what, after a number of modifications,
became one of two standard models of the motion of stars through the heavens. 
In this model, celestial objects orbited the earth on spheres of greater 
or lesser distance from the earth. Most objects (`the fixed stars') were 
fixed to an imaginary celestial sphere which rotated around the earth once 
a day. Five `stars' which behaved peculiarly were given the special 
designation `planets' and were fixed to separate imaginary spheres. The
sun and moon were similarly distinguished and given their own spheres.

Due primarily to the efforts of Ptolemy in his landmark treatise on the
subject (Ptolemy ca.~150AD), the geocentric model gained favor over the 
heliocentric models and remained the standard for fifteen centuries. He 
summarizes and extends the work of many previous astronomers (notably 
Hipparchus ca. 160 BC, whose work and that of most of his predecessors is
largely unavailable), and describes a cosmology in which the earth occupied
the center of the universe, celestial objects moved in uniform circular 
motion around the center, each at a different distance. Smaller circles
(`epicycles') were invoked to fix discrepancies between the model and
the observations. 

A formalism called `algebra' (al-Khowarizmi ca. 820 AD), suitable for rigorous
mathematical models of astronomical observations, was developed and put
on a level footing with the geometric mathematics of the Greeks. 
Many of the geometric constructs of earlier workers (e.g. the 
Tables of Chords, calculated by various authors) were transformed into 
the new algebraic form as tables of sines and cosines. In spite of the 
concurrent studies of astronomy, by the same workers, algebra was not
incorporated into the physical models for many more centuries.

Precise observations of the heavens and the redevelopment of the 
heliocentric model of Aristarchus by Copernicus (Copernicus ca. 1530, 1543)
began to overturn the geocentric universe with a heliocentric model in 
which the earth was one member of a system of objects traveling in circular 
orbits around the sun.  Circular epicycles were still required to reproduce 
the observed motion but were of smaller size and fewer were required to 
reproduce the observations.

Developments in instrumentation (Brahe 1598), and observational techniques 
(Brahe \& Kepler 1602) soon resulted in published catalogues (Kepler,
Brahe \& Eckebrecht 1627) of new, precise astronomical observations of the
motions of celestial objects in the solar system over the course of several 
decades of time. Building on this voluminous database, Kepler (1609, 1619) 
fit the observations into an empirical framework of laws governing the motion 
of the planets around the sun. This framework in fact forms the basis of
most orbital motion problems today and the `Keplerian' motion noted
throughout this thesis (minus the many astrological implications)
is identical to that outlined in Kepler's original works.

Parallel developments in instrumentation (Digges 1571, Lippershey 1608) 
enabled, by means of a `telescope', many revolutionary discoveries and
observations to be made of the Jovian moons, phases of Venus and Mercury,
mountains on the moon (Galilei 1610) and sunspots (Fabricius 1611) which
rebutted nearly all of the original criticisms of the Copernican model. Only
three criticisms (absence of stellar parallax measurements, the possible 
presence of an `aether' and the precession of the orbit of the planet 
Mercury) remained much longer, being laid to rest only in 1838 (Bessel
1838,1840), 1887 (Michelson and Morley 1887) and 1917 (Einstein 1917),
respectively. A review article (Galilei 1632) summarized and compared the
so called `Ptolemaic' and `Copernican' models and many of the outstanding
problems with each and, over the next several decades the Copernican
model became the standard.

In the years which followed, Newton (Newton 1687) provided a confirmation 
of Kepler's empirically derived framework on an entirely theoretical basis.
The epicyclic frequency discussed at many points in the following
chapters (see esp. ch. \ref{isodisk} and \ref{cooldisk}) is based upon a 
solution to orbital motion of celestial bodies made within Newton's 
theoretical framework. Based upon this same solution, we may now note that 
had Ptolemaic system allowed for the possibility of elliptical epicycles 
(specifically with an axis ratio of 2:1) the discrepancies between 
observations and the Ptolemaic theory would have been much less pronounced 
and the Copernican model may therefore have been much less readily accepted.

Most of the scientific efforts made until the late 1700's and early
1800's went into describing the physical inventory of the solar system
and the character of the motion of the bodies within it.  Within
the past 200 years, observations have been both precise enough
and of a sufficiently diverse nature that a reasonable effort to understand 
the origin of the solar system and of stars like the sun has been made.
With this mathematical and cosmological framework, the development of a
theory of the origin of the solar system began to take the shape it holds
til this day. Kant (1755) proposed a `solar nebula hypothesis' wherein 
the sun and the planets in the solar system began as a cloud of gas which then 
proceeded to collapse. Laplace (1796) placed the study of the solar system
as a whole onto a solid mathematical framework and outlined a possible theory  
of how the collapse might take place.  Physical conditions under which collapse 
could occur were outlined by Jeans (Jeans 1902) in which he relates the balance 
between the self gravity (characterized by the mass density) and internal 
pressure (characterized by the sound speed) of a cloud.  

Observations of stars outside the solar system (Hertzsprung 1911 and 
independently Russell 1914) showed that a number of differences exist
in the apparent color and luminosity of between various stars in the heavens.
These differences could be attributed on the one hand, to a `main series'
of stars with similar radii but widely differing colors 
and luminosities, and on the other hand, to other sequences of giant 
stars with widely ranging radii. Once the sizes and luminosities of
stars became known it was apparent that they were emitting prodigious
amounts of energy, but the source of the energy was unclear (see e.g.
the review of Jeans 1929). Perrin (1919) and later Eddington (1920) proposed 
that the source of the energy radiating from stars was derived from 
transmutation of the elements, a theory which over the next several
decades was shown to be correct (see e.g. Bethe \& Critchfield 1938, 
Bethe 1939, and the review of Burbidge \etal 1957). 

Regardless of their source of energy, stars had to evolve from some other
form of matter before they started to undergo any nuclear fusion, and 
therefore we might hope to observe some stars which are quite young if
the evolutionary process is ongoing. Joy (1945, 1949) discovered a new class 
of stars, which he named `T~Tauri' stars after one of the brightest objects 
in the new class, which were variable in their brightness over time but which 
were not at all periodic. They also did not fit well into the 
Hertzsprung/Russell picture of dwarf and giant stars, since they lay some 
1--3 magnitudes above the main sequence (Herbig 1952, Walker 1956), which 
is consistent with a star undergoing gravitational collapse prior to ignition
of nuclear fusion.  The stars in the class were nearly always associated 
with dense nebulosity and molecular clouds, leading to the belief that 
these stars were very young. Further evidence in favor of this hypothesis
includes the discoveries that on the one hand the photospheres of this
class of stars are very active photospheres (Herbig 1970) and on the other 
hand that they emit energy far in excess of stellar blackbody emission
in the infrared (Mendoza 1966, 1968). Taken together these observations
suggest a model of a cool disk slowly losing its orbital energy (the 
IR emission) and accreting onto the central star (the photospheric 
activity).

The solar nebula hypothesis and the development of mathematical models
of the collapse itself provided a strong impetus for the study of accretion
disks, since with the collapse of any realistic cloud of gas there will be 
some component of angular momentum. Among the first to make a systematic
study of disks were the Lindblads, who studied mainly galactic disk systems 
and developed the theory of spiral structure in a nearly annual series of
papers (e.g. Lindblad 1960, 1963) and showed that spiral structure could be
explained as a quasi-static phenomenon by the fact that the quantity, 
$\Omega(r) - \kappa(r)/2$, where $\Omega(r)$ and $\kappa(r)$ are the angular
and epicyclic frequencies of matter in the disk, is approximately 
constant over a large range of radii. This implies that resonant effects
may build up the disk into a stable pattern as matter in part of the disk 
orbiting the star at one radius acts upon matter orbiting at a much
different location.

If in fact such accretion disks exist around some stars (and it is now
quite evident both from direct and indirect observations that they do--see
discussion throughout this work), they must be transient objects, since 
we see few or no signs of disks around most stars. Lynden-Bell \& Pringle
(1973) showed that a viscous disk will transport mass radially inward 
through the disk while angular momentum will be transported radially 
outward, ultimately ending with nearly all of the disk mass transported
into the star and all of the angular momentum transported to infinity.
Molecular viscosity however had long since been shown to be far too
small to evolve the disk in the available 10$^6$--10$^7$~yr (von~Weizacker 
1943).  

Work by Shakura and Sunyaev (1974) 
used dimensional arguments to parameterize the magnitude of the viscosity 
as a function of the sound speed and the scale height of the disk, but 
this parameterization has little physical basis and has primarily been
used as a black box dissipation source. A free parameter, `$\alpha$', 
combines all of the unknown physical dissipation into a single dimensionless
number. In practice, and in the absence of a better model, tuning $\alpha$ 
to the value which best fits the model to the observations has become the 
method of choice for many disk models.

For some disks a viscous model is inadequate for a correct understanding
of the morphology of the system. Galactic disks, for example, show large 
spiral shaped density variations. On smaller scales, a similar physical
picture may apply to the collapse of the molecular clouds during the 
star formation process, though direct evidence for such structure is not 
yet available. Lin \& Shu (1964, 1966) theorized that spiral structure
could in fact be characterized as density waves and so the 
formidable mathematical apparatus developed to describe waves
could be applied. The development of this density wave theory led at
the same time (Toomre 1964) to an extension of the Jean's criterion
for cloud collapse to centrifugally supported disk systems. This 
condition defines the relationship between the local gas or particle
pressure (characterized as a local velocity dispersion of particles
or as a sound speed), rotational shear and local self gravity 
(characterized by the local surface density of mass) under which a disk
will become unstable to ring formation and eventual local collapse. 
In chapter's \ref{isodisk} and \ref{cooldisk}, we will study disk systems
which are marginally stable according to this criterion and which develop
spiral structure which we will compare to theoretical predictions made based
on developments of the spiral density wave model of accretion disks.

Without advances in computational hardware (Atanasoff 1940, Eckert \etal
1945, von~Neumann 1945) and algorithms for the numerical solution of 
differential equations (Kutta 1901, Courant, Friedrichs \& 
Lewy 1928, for hydrodynamic equations in particular see e.g.  von~Neumann \& 
Richtmyer 1950, Godunov 1959, Lucy 1977, Gingold \& Monaghan 1977) 
the loss of one or two dimensional symmetry (ie. planar, spherical or 
cylindrical symmetry) may have proven insurmountable in attempts to model 
the behavior of astrophysical systems in general and forming stars or the 
solar system in particular. The data were becoming precise enough that 
physical models had to account for more phenomena than could be included in
purely analytic (mathematical) methods. However, with these new algorithms
and hardware tools and the rate of increase in the density and
complexity of electronic components (Moore 1965), the consequent increase 
in the speed, size and complexity of computational models has been able to
keep pace with observed systems (at least somewhat!). The work 
presented everywhere in this thesis could not have proceeded
without the development of such computational resources.

The state of the study of star formation as of about ten years ago is 
summarized in the review article of Shu, Adams \& Lizano (1987). They 
describe four stages during the formation of stars. The first stage
is characterized by the formation of dense cores of gas and dust within
molecular clouds. At some point, a core collapses to form a protostar with 
a disk surrounding it. Later, the star/disk system begins to eject matter via
bipolar outflows and continues to accrete matter from the cloud. Finally,
accretion and outflow cease, the disk decays and the star evolves onto the
main sequence. While this paradigm provides a qualitative picture of the
star formation process, many important questions remain.

One example is that this paradigm only models the formation of single
stars. Unfortunately, we know that there are lots of stars which are observed
to be in binary or even higher order multiples.  A survey of one class of 
such objects (`spectroscopic binaries', or binaries whose observed spectral
lines are observed to vary periodically and whose motions can be fit to 
Keplerian orbits) by Duquennoy and Mayor (1991) showed that in fact most
stars near the sun are in multiples. Somehow the theory has to account for
these stars as well. More recent observations (Simon \etal 1995, Ghez \etal
1993, Leinert \etal 1993, Reipurth \& Zinnecker 1993) have also 
shown that very young stars are often found in multiple systems. In fact,
those observations showed that a higher fraction of young stars are in 
multiple systems than exist in older stars like those studied by Duquennoy
and Mayor.

Another area where understanding is limited is in the evolution of
circumstellar disks. How do they form? What is their morphology at
different times during the evolution? Do they form spiral density wave
structures similar to those seen in galaxies, and do such structures
collapse to form binary companions? What mass do such companions have:
are they low mass objects like planets or brown dwarfs, or higher mass
objects like binary star? The predictions of the $\alpha$ model
for the temperature structure of disks during their evolution do not
agree with observations. It is also completely inapplicable to a massive
disk where spiral structures may develop. Better models must be developed
in order to understand these properties of disks.

Still another area of limited understanding is in the formation and
evolution of low mass companions like the terrestrial planets and
gas giants in the solar system, or of the low mass companions and
brown dwarfs recently discovered around other nearby stars. How do 
these low mass companions form and/or move within the star/disk system?

This thesis continues and builds upon work outlined above (``standing upon
the shoulders of giants''---Newton 1675/1676). It is primarily an attempt to
better understand the physical processes important during the origin and
evolution of solar type stars and of the solar system, the observational 
consequences of those processes and the limits which can be placed upon
detecting certain signatures of already formed systems. This work is 
organized as a series of published or publishable papers connected 
by short interludes which describe a little of the background behind
the work which follows. The final chapter is an attempt to outline
of few of the directions which will be profitable avenues for further
research. In the paragraphs below I outline, in a short abstract form,
chapter \ref{isodisk} in terms of the improvements beyond previous 
work and the results which were obtained from this study. 

Previous studies of circumstellar disks have provided analyses based on
two strategies: (1) analytic or perturbative techniques and (2) numerical 
simulations of either cloud collapse resulting in a coarsely resolved disk 
or of already formed, but spatially narrow, tori. Working in collaboration 
with my advisors Willy Benz and Dave Arnett, as well as Fred Adams at the
University of Michigan, I have studied the dynamical evolution of massive
circumstellar disks that may form early in the history of young stellar
systems using numerical simulations. 

In chapter \ref{isodisk}, we present a series of two dimensional 
hydrodynamic simulations of massive disks around protostars. We have used
two complementary numerical hydrodynamic methods (the `Piecewise Parabolic
Method' and the `Smoothed Particle Hydrodynamic' method) to simulate the 
growth and evolution of spiral arms within the disks. The simulations 
using each code are compared to discover differences due to error in
the methods used. For this problem, the strengths of the codes overlap 
only in a limited fashion, but similarities exist in their predictions,
including spiral arm pattern speeds and morphological features.

In these calculations, we have studied the evolution of massive 
circumstellar disks with larger radial extent and over a wider range of 
parameter space than has been possible before. From the earliest times,
their evolution is a strongly dynamic process rather than a smooth 
progression toward eventual nonlinear behavior. Processes that occur in
both the extreme inner and outer radial regions affect the growth of 
instabilities over the entire disk. Effects important for the global 
morphology of the system can originate at quite small distances from the
star. Therefore, analyses which neglect the inner disk will not model 
the evolution of the system correctly. 

The disks studied here range in mass from $0.05 M_*$ to
$1.0 M_*$ and in initial minimum Toomre $Q$ value from $1.1$ to
$3.0$. We adopt simple power laws for the initial density and
temperature in the disk with an isothermal ($\gamma=1$) equation
of state. The disks are locally isothermal. We allow the central
star to move freely in response to growing perturbations. The
simulations using each code are compared to discover differences
due to limitations in the methods used. For this problem, the strengths 
of the codes overlap only in a limited fashion, but similarities exist
in their predictions, including spiral arm pattern speeds and
morphological features. Our results represent limiting cases (i.e.
systems evolved isothermally) rather than true physical systems.

We show that disks with minimum Toomre stability $Q\lesssim 1.5$ (recall 
that $Q<1$ implies instability to the growth of ring-like structures) are
dynamically active and spiral structures growing within each disk become
distorted, break apart and reform on orbital time scales. A marked change
in the character of spiral structure occurs in simulations with differing
disk mass. Low mass disks ($M_D\leq0.2M_*$) form filamentary spiral 
structures with $\gtrsim$5 arms. High mass disks form grand design
spiral structures with 1--3 arms. Eventual collapse of such structures 
in more physically realistic models may be responsible for producing 
some stellar or brown dwarf companions. A detailed picture of their
evolution is therefore required to understand the formation of such
companions.

In our SPH simulations, disks with initial minimum $Q=1.5$ or lower
break up into proto-binary or proto-planetary clumps.  However, these
simulations cannot follow the physics important for the flow and must
be terminated before the system has completely evolved. At their
termination, PPM simulations with similar initial conditions show uneven
mass distributions within spiral arms, suggesting that clumping behavior
might result if they were carried further. Simulations of tori, for which
SPH and PPM are directly comparable, do show clumping in both codes.
Concern that the point-like nature of SPH exaggerates clumping, that our
representation of the gravitational potential in PPM is too coarse, and
that our physics assumptions are too simple, suggest caution in
interpretation of the clumping in both the disk and torus simulations.

We calculate approximate growth rates for the spiral patterns and
compared the results of our simulations to the predictions of
linearized analyses. We examine in particular the SLING mechanism
proposed by Adams, Ruden and Shu (1989). They show that a resonance
between the motion of the star and a one armed spiral pattern may 
stimulate growth of the one armed pattern in the outer disk even if 
other spiral patterns are suppressed. Our simulations show that the
one-armed ($m=1$) spiral arm is not the fastest growing pattern of
most disks. Also, due to the dynamic nature of the growth, a resonant
growth mechanism such as SLING may be of limited value because pattern
speeds and amplitudes display wide, short term variations. 

Several qualitative features of the SLING mechanism are reproduced in 
our simulations. The changeover in behavior between filamentary spiral
structures and grand design structures occurs at the disk mass for 
which the SLING instability is predicted to become active. Approximate 
growth rates fitted to the spiral patterns present show that the one 
armed pattern growth rates are similar to those predicted by Adams \etal,
though they are neither the largest amplitude nor the fastest growing 
patterns for most systems.

%% file: isodiskchap.tex
\def\etal{\emph{et~al.\ }}
\def\qmin{$Q_{\rm min}$}
\def\mrat{$M_D/M_*$}

\chapter{Dynamics of Circumstellar Disks\label{isodisk}}

Over the past several years a broad paradigm of star formation has
been developed (see Shu, Adams \& Lizano 1987). First, a cloud of gas 
and dust collapses and forms a protostar with a surrounding disk.  
Later the star/disk system ejects matter in outflows as well as
continuing to accrete matter from the cloud.  Finally, accretion and
outflow cease and the star gradually loses its disk and evolves onto
the main sequence. While this paradigm provides for a good qualitative
picture of the star formation process, many important issues 
require further work. For example, observations by several groups
(Simon \etal 1995, Ghez \etal 1993, Leinert \etal 1993, Reipurth \&
Zinnecker 1993) show that young stars in many different star forming
regions are commonly found in binary or higher order multiple systems,
with a broad peak in separation distance at around 30 AU.  In
addition, many of the higher order multiples show hierarchical
characteristics: a distant companion orbiting a close binary for
example. In what manner are multiple systems such as these formed?

A variety of studies have been undertaken to model the processes
leading to the observed systems. One class of models begins with the
collapse of a cloud of matter. These results (Bate \etal 1995, Foster
\& Boss 1996, Boss 1995, Burkert \& Bodenheimer 1993, Bonnell \&
Bastien 1992, Myhill \& Kaula 1992) show that both single stars and
multiple systems can be formed from the collapse and subsequent
fragmentation of rotating, spherical or elongated molecular cloud
cores.  This class of simulations focus on the collapse phase, but do
not follow in detail the dynamics of disks formed from the material with
initially higher angular momentum.

In addition, a number of models extended beyond the initial collapse
(Bonnell 1994, Pickett \etal 1996, Woodward \etal 1994) have shown
that post-collapse objects can be driven into fragmentation, or into
spiral arm and bar formation prior to the development of a Keplerian 
disk.  Laughlin \& Bodenheimer (1994) have simulated the evolution of a
collapsing cloud in 2D and then followed its late time behavior with a
3D disk simulation. They have found that such a collapse leads to a
core plus a long lived, broad torus. The development of $m=1$ and
$m=2$ spiral patterns may lead to late time fragmentation of the torus
($m$ is the number of arms in the spiral pattern).

As a star-disk or multiple-star-disk system evolves, the dynamics of
the disk itself as well as its interaction with the star or binary becomes
important in determining the final configuration of the system. Depending
on its mass and temperature, a disk may develop spiral density waves 
and viscous phenomena of varying importance. Each may be capable of
processing matter through the disk as well as influencing how the
disk eventually decays away as the star evolves onto the main sequence.

A variety of mechanisms for producing of spiral instabilities in
disks around single stars have been suggested. An incomplete list 
includes the linear perturbation results of Adams, Ruden \& Shu 
(1989; hereafter ARS) who suggest a mechanism (`SLING'-- see Shu
\etal 1990; hereafter STAR) by which a resonance between the 
star and a one armed ($m$=1) spiral mode may become globally unstable.  
Both perturbation theory (Papaloizou \& Lin 1989) and numerical
calculation (Papaloizou \& Savonjie 1991, Heemskirk \etal 1992) have
shown another instability mechanism based on the distribution of
specific vorticity (termed ``vortensity'') which can influence
evolution in disks and tori. It is driven primarily by wave
interactions at corotation and can act either to suppress or amplify
spiral waves in the disk, depending on the vortensity gradient there.
Another family of instabilities is based upon vortensity
gradients at the boundaries of the disk or torus. The SWING amplifier 
(Goldreich \& Lynden-Bell 1965, Julian \& Toomre 1966, Goldreich \&
Tremaine 1978) provides an instability channel whereby low
amplitude leading spiral arms unwind and are transformed
into much larger amplitude trailing waves. A feedback cycle then 
creates additional leading waves and the instability grows. 

This paper is a continuation of work by two of us (Adams \& Benz 1992,
hereafter AB92), who began modeling of disks of mass $M_D\gtrsim 0.5 M_*$ 
and observed formation of spiral arms and clumps. We present a series of
two dimensional numerical simulations of circumstellar disks
with masses between $0.05 M_*$ and $1.0 M_*$. We 
attempt to characterize the growth of instabilities and pay
particular attention to the existence and effect of the SLING 
instability.  In section \ref{codes}, we outline the 
numerical methods used and discuss the limitations of each code
and their effects on our simulations. In section \ref{phyasmpt}, we
outline the initial conditions adopted for the disks studied and in
section \ref{results}, we first describe qualitatively the results of
our simulations and then begin a quantitative analysis
of the pattern growth, the correspondence between
two hydrodynamic codes, and the correspondence between linear analyses
and hydrodynamic simulations. In Section \ref{summary}, we
summarize the results and their significance in the evolution of stars
and star systems.

\section{The Codes\label{codes}}
\subsection{Solving the Hydrodynamic Equations}

In order to understand the properties of protostellar disks
we have adapted two complementary hydrodynamic codes to the task
of simulating such evolution: the Smoothed Particle Hydrodynamic
(SPH) method and the Piecewise Parabolic Method (PPM).  These codes
use very different techniques for solving the equations of
hydrodynamics, and it is hoped that, by the use of such widely different
techniques, numerical artifacts can be sorted out from true physical
evolution. Each code has unique features that allow the
simulation of these systems in some regimes not accessible to the other.

The SPH method (see reviews by Benz 1990, Monaghan 1992) uses a
procedure by which hydrodynamic quantities and their derivatives 
are calculated from an
interpolation technique over neighboring particles. The interpolation
kernel used in our simulations is the standard B-spline kernel with
compact support.  The smoothing length $h$ is varied over time in a manner
such that the number of neighbors is approximately conserved, subject
to the condition that a minimum value of $h\sim R_D/1700$ (where $R_D$ 
is the disk radius) is set to ensure time steps do not become too small.
A second order Runge-Kutta-Fehlberg integrator which includes time step
control is used to evolve the system in time. Being gridless, the
main advantage of the SPH method in our context lies in its ability to
follow structure formation anywhere in the disk without the
limitations associated with a prescribed grid. The two main disadvantages
of the SPH technique are  (1) the inherent random noise level associated 
with the discrete representation of the fluid and 
(2) the high shear component of the dissipation connected with the 
mathematical formulation of the artificial viscosity.

We also have adapted the PROMETHEUS hydrodynamic code (Fryxell,
M\"uller \& Arnett 1989, 1991) to the problem of evolving disks
around protostars. PROMETHEUS is based on the `Piecewise Parabolic
Method' (PPM) of Colella \& Woodward (1984) in which a high order
polynomial interpolation is used to determine cell edge values used in
calculating a second order solution to the Riemann shock tube problem at 
each cell boundary.  The interpolation is modified in regions of sharp
discontinuities to track shocks and contact discontinuities more closely
and retain their sharpness, while a monotonizing condition smoothes out
unphysical oscillations. The solution to the one-dimensional Riemann 
problem is then used to calculate fluxes and advance the solution in
time.  This code was selected because of its low numerical dissipation
and its excellent resolution of discontinuities and shocks.

Both codes incorporate self-gravity using modified versions of the
binary tree described in Benz \etal (1990) which approximates the
gravity of groups of distant particles in a multipole expansion while
calculating interactions of nearby particles explicitly. Gravitational
forces due to neighbor particles are softened to avoid divergences as
particles pass near each other. Due to the organization of the grid,
the tree construction can be considerably simplified in the PPM
version by substituting a procedure by which adjacent grid cells 
(modeled as point masses for the purpose of the gravity calculation)
or groups of grid cells become progressively higher nodes in the
tree. Two simulations run at higher resolution (simulations {\it pch2}
and {\it pch6} in table \ref{ppm-tabl} below) implemented an FFT
based solution to Poisson's equation (Binney \& Tremaine
1987, pp. 96ff). Results for a disk simulation at identical resolution
showed that the tree and the FFT solutions gave identical
dynamical results. However the FFT version proved to be substantially
faster. The torus simulations of section \ref{sphppm}, which are 
more sensitive to resolution, are also more sensitive to the implementation 
of the Poisson solver. In these cases the simulations using the tree code 
gave slower pattern growth rates than simulations using the FFT.

It is important to make a distinction between the resolution of the
hydrodynamics and that of the representation of the gravitational
potential. Just as PPM is well adapted for discontinuities, SPH is
well adapted for gravitational clumping.
The density reconstruction procedure utilized by PPM
contains more structure
than is available from the $N$ grid point algorithm used here.
Better resolution of the gravitational potential may be
possible using densities defined at both the cell centers and at cell
interfaces. Further, this grid effectively implies a gravitational
softening which is about one cell in size. This algorithm uses only the
cell center information, and references below to grid resolution in PPM
simulations will imply this fact.

\subsection{Viscosity in the Codes\label{visc-sec}}

Because disk evolution is partially driven by viscosity
in the disk, we must look carefully at issues related
to numerical viscosity. Except for codes based on a
local solution of the Riemann shock problem such as PPM, most methods
require implementation of an artificial viscosity to enforce stability
and/or improve the shock treatment by the code. In this regard SPH is
no exception and our version of the code implements the standard form
discussed in Benz (1990).  We use bulk and
von~Neumann-Richtmyer (so called `$\bar\alpha$' and `$\beta$')
viscosities to simulate viscous pressures which are linear and
quadratic in the velocity divergence. We use the standard values for
each of the coefficients, with $\bar\alpha=1$ and $\beta=2$.  We
incorporate a switch (see Balsara 1995) which acts to reduce substantially 
the large undesirable shear component associated with the standard form.

The bulk component of the artificial viscosity $\bar\alpha$ in the SPH
code can be identified with a kinematic viscosity $\nu$ (see Murray
1995) using the relation
\begin{equation}
\nu = {{ \bar\alpha c_s h}\over 8},
\end{equation}
where $c_s$ is the sound speed and $h$ is the smoothing length of a
particle. It is possible to  relate the coefficient of bulk artificial
viscosity $\bar\alpha$ to the $\alpha$-parameter of the standard viscous
prescription of accretion disks.  We equate the artificial viscosity to the
Shakura \& Sunyaev (1973) viscosity (defined by $\nu = \alpha c_s H$ and 
the scale height, $H$, is defined as $H=c_s/\Omega$, the local sound speed 
over the angular rotation rate. Solving for $\alpha$ yields 
\begin{equation}   \label{sph-alpha}
\alpha = {{ f\bar\alpha h \Omega}\over {8c_s} }, 
\end{equation}
where $f$ is the shear reduction factor discussed in Benz 1990 suitably
averaged over particles and time. 
We caution the reader that the identification of the SPH form of the
viscosity is not necessarily equivalent to that of the Shakura and
Sunyaev form, especially because of the approximate manner in which 
the Balsara switch must be taken into account. We estimate equation 
[\ref{sph-alpha}] may be valid to a factor of a few but should not be
taken as exact.

For a nearly Keplerian disk with a temperature $T \propto r^{-1/2}$
and a roughly linear variation of the smoothing length, $h$, with radius,
we obtain $\alpha \propto r^{-1/4}$. Depending on the temperature 
constant describing each disk ($T_0$, see section \ref{init}), $\alpha$ 
is of order $\sim10^{-2}$.  Only at small radii ($r\lesssim 2$AU) and low 
disk mass (for which $T_0$ becomes correspondingly small for a specified
value of \qmin) does $\alpha$ rise to values in the range
$\alpha\sim0.1-1$.  These values of $\alpha$ imply that the viscous
time scale, $\tau= r^2/\nu$, remains significantly longer than the few
orbital time scales we simulate. For most of
the disk, the SPH viscosity is small enough not to affect the
evolution of the disk significantly.  The von~Neumann ($\beta$) term
in the viscosity does not mirror the alpha prescription as the bulk
term does. Derived from the assumption that the viscosity is
proportional to square of the velocity divergence, its effect is limited
to portions of the flow in which shocks occur.

The numerical viscosity inherent to the PPM code is 
difficult to quantify. The nonlinear nature of the Riemann solver
(with the associated PPM `switches' to sharpen discontinuities and
enforce monotonicity) renders an artificial viscosity term
unnecessary.  However, a small numerical viscosity still appears
in the code. Porter \& Woodward (1994) derive fits for numerical
dissipation proportional to the third and fourth powers of ${\delta
x}/\lambda$ where ${\delta x}$ is a cell dimension and $\lambda$ is
the wavelength of a disturbance. Thus, large scale disturbances like
the spiral arms will experience little
dissipation, but small scale motions will be damped more.

\section{Physical Assumptions and Constraints\label{phyasmpt}}

Because our simulations involve dimensionless quantities such as the 
disk/star mass ratio and the Toomre stability parameter $Q$, the physics
itself is scalable to systems of different size. We shall express 
all quantities in units with values typical of the early stages of
protostellar evolution. These units are also comparable (for the
most massive disk simulations) to the final dimensions of our own
solar system.  The star mass will be assumed $M_* = 0.5 M_\odot$
and the disk radius $R_D$ = 50 AU. Time units are given in either
years or the disk orbit period defined by $T_D=2\pi\sqrt{R_D^3/GM_*}$ 
which, with the mass and radius given above, is equal to 500 years.

\subsection{Circumstellar Disk Initial Conditions \label{init}}

The initial conditions for prototype low and high mass disks are 
summarized graphically in figures \ref{dinit-ppm} and \ref{dinit-sph} 
for our PPM and SPH simulations respectively.  We assume that the
disks are vertically thin so that two dimensional ($r$,$\phi$) 
simulations are justified. The variables of interest (density,
pressure, etc.) are taken to be vertically integrated quantities.  
Magnetic fields are neglected in our simulations.

In functional form, the disk mass is initially distributed according 
to a density power law
\begin{equation}
\Sigma(r) = \Sigma_0 \left[ 1 + \left({r\over r_c}\right)^2\right]^
{-{p\over{2}}},   \label{denslaw}
\end{equation}
where the power law exponent $p$ is set to 3/2. 
As we shall discuss in the following section we found that our PPM
simulations implementing the initial density profile of eq. [\ref{denslaw}]
became violently dynamic near the inner grid edge and we could not 
simulate the evolution of the system. Instead, we have chosen to remove 
matter completely at small radii in our PPM runs by adopting an initial 
density law which ensures that little matter remains at small radii or
interacts with the boundary. This density law takes the form
\begin{equation}\label{denslaw-a}
\Sigma(r) = {\Sigma_0{\left\{[{1-e^{-\left({{r-R_0}\over{R_c}}\right)^2}]} 
		 \over{r}\right\}^p}},
\end{equation}
where $R_0$ is set to the radius of the innermost boundary cell and
$R_c$ is set arbitrarily to 6 AU. With this choice, the surface
density is substantially reduced near the inner boundary while
retaining a nearly pure $r^{-3/2}$ distribution for radii greater than
about 10 AU. The temperature is given by a similar power law as
\begin{equation}
T(r) = T_0\left[1 + \left({r\over r_c}\right)^2\right]^{-{q\over{2}}},
    \label{templaw}
\end{equation}
with the exponent $q$ set to 1/2.  The softening radius $r_c$ for
both power laws is set to $r_c=R_D/50$(=1~AU).  

We choose the value of the temperature power law index based on
observed temperature profiles in T~Tauri disks (see Beckwith \etal 
1990; Adams \etal 1990).  The density power law is much less well
constrained, and our choice of $p=3/2$ is roughly consistent with
the infall collapse calculations of Cassen \& Moosman (1981). 
As an additional motivation, this choice of exponents
matches the one adopted by ARS and allows a direct comparison with
their work.

$\Sigma_0$ and $T_0$ are determined from the disk mass and a choice 
of the minimum value over the disk of the Toomre stability parameter 
$Q$, defined as
\begin{equation} \label{Qdef}
Q = {{\kappa c_s}\over{ \pi G \Sigma}},
\end{equation}
where $\kappa$ is the local epicyclic frequency. For an ideal gas with
an isothermal equation of state (see section \ref{eos-sec}), the sound 
speed is defined as
\begin{equation}
c_s^2 = {{kT}\over{\mu m_p}},
\end{equation}
where the mean molecular weight is $\mu$ and we assume the gas is
of solar composition.  

By definition, the Toomre $Q$ parameter divides the region of phase 
space for which axisymmetric disturbances (rings) grow exponentially 
from that in which they are damped. It is derived from the dispersion
relation for waves in a disk given by (see Binney \& Tremaine 1987)
\begin{equation}\label{dispersion-rel}
\omega^2 = c_s^2 k^2 - 2\pi G \Sigma |k| + \kappa^2
\end{equation}
where $\omega$ is the complex pattern frequency and $k$ is
the wavenumber. Mathematically speaking, if $\omega^2>0$ the
wave equation from which the dispersion relation is derived will
have exponentially decaying solutions, while if $\omega^2<0$ 
it will have exponentially growing solutions. The line of 
neutral stability at $\omega=0$ defines $Q$. Equation
\ref{dispersion-rel} then becomes quadratic in the wavenumber 
$k$ and whose solution yields the condition defined by eq.
\ref{Qdef}.

The temperature and density laws above produce a profile for the 
instability parameter $Q$ that is nearly flat over the largest
portion of the disk, with a steep rise at small radii and a 
shallow increase towards the outer edge of the disk.  The minimum 
value of $Q$ in the disk is therefore located at $\sim 30-40$ AU,
depending upon the mass and temperature of a specific disk.

Another common criterion of instability in disks is the $X$
parameter, which is important for so called SWING amplification
and is defined by
\begin{equation}
X = {{r\kappa^2}\over{2\pi mG\Sigma}},
\end{equation}
with $m$ the number of spiral arms (the azimuthal wave number). 
In this instability `leading' spiral waves (i.e. those for which
the spiral winds up in the same direction as the orbital motion)
unwind and become `trailing' spiral waves, which then are reflected
through the origin and the process repeats. Ordinarily, in
Keplerian systems where an inner Lindblad resonance exists,
the SWING instability is suppressed because the trailing waves
reflection instead off of the Lindblad resonance. Nevertheless,
we shall examine the value of the $X$ parameter to make certain
that contributions to the instabliity growth from SWING may be 
neglected. In order for a system to be unstable to SWING 
amplification, the value of $X$ must be $\lesssim$3 in the 
region of interest.

The $X$ parameter shows a similar pattern to that seen for $Q$, but with 
a steeper increase at large radii.  For most of the disks we study, 
$X$ is larger than that required to keep the disk stable for the 
lowest order spiral modes, so that we expect SWING not to contribute
to the growth of instabilities. Like the $Q$ and $X$ profiles, the 
vortensity profile shows a steep increase at small radii. In this 
case, such an increase may serve to stimulate growth due to 
the family of instabilities discussed by Papaloizou and Lin (1989).
We will discuss this possibility in more detail below.

The star is represented as a point mass, free to move in response
to gravitational forces from the surrounding disk. Initially, disk matter 
is placed on circular orbits around the star, with 
rotational equilibrium in the disk and radial velocities set to 
zero. Gravitational and pressure forces are balanced with centrifugal 
forces such that the rotation curve is given by
\begin{equation}
\Omega^2(r) = { {GM_*\over{r^3}} + {1\over{r}}{
       {\partial\Psi_D}\over{\partial{r}}} + {1\over{r}}{
       {{\bf\nabla}{P}}\over{\Sigma} } },    \label{rotlaw}
\end{equation}
where the symbols have their usual meanings and $\Psi_D$, the gravitational
potential of the disk, is calculated numerically with the same potential
solver utilized in the full hydrodynamic code.  The magnitudes 
of the pressure and gravitational forces are small compared to the stellar
term, therefore the disk is nearly Keplerian in character.

\subsection{The Construction of Circumstellar Accretion Disks,
Boundary Conditions and Numerical Resolution \label{bound}}

To complete the specification of the initial state of the systems,
we must define the conditions at the boundaries of each simulation.
The linearized analyses of ARS suggest that the dynamics of an accretion
disk will be relatively insensitive to the implementation
of the inner boundary condition, becoming active only at distances far
from the star. On the other hand, the shape of outer edge of the disk
is predicted to be critical for the eventual growth of the SLING 
instability. In order to search for evidence of the SLING instability
we shall implement boundary conditions which may be favorable to its 
growth.

To ease time step constraints, we set the inner boundary at a greater
distance than that which is physically the case for a star/disk boundary.
With a grid code, we can define the inner boundary by modeling the inner
regions in some steady state approximation or by modifying the density 
law at small radii (in effect modeling tori) to reduce interactions with
the inner boundary. Since ARS predict that the inner regions of the disk
will be relatively stable, instabilities are not expected to grow there,
given a disk initially in rotational equilibrium. Any boundary condition
which does not perturb this equilibrium should be sufficient to describe
the inner disk.  Since by assumption, the inner disk begins in rotational
equilibrium (i.e. with $v_r=0$), no matter will cross the boundary and a 
simple reflecting boundary condition will suffice. The reflecting boundary
will also serve a second function. The four wave cycle (STAR) important for
the amplification of SLING requires a corotation or $Q$-barrier from which
waves can be reflected or refracted during part of the cycle. Until such
resonances may develop on their own further out in the disk, the reflecting
boundary serves as a surrogate for the actual resonances. 

Our PPM simulations showed that for a pure 
power law for the density (omitting the core radius of eq. 
[\ref{denslaw}]), the inner regions of the disk are quite dynamic and
unstable. After a few orbits, matter in the inner
disk moved off its initial circular orbit and began interacting with
the boundary. The effect of these interactions is to give a
``kick'' to the system center of mass as matter reflects off the boundary.
In the worst cases, serious computational problems occurred after 
20-50 orbits of the inner disk edge and the calculations had to be 
stopped.

Several prescriptions for avoiding this behavior were attempted without
real success. These prescriptions included allowing matter to
accrete through the boundary onto the star, attaching the inner disk
matter to the star itself, treating the inner disk as a softened point 
mass at the origin with varying degrees of softening or by treating the 
inner disk matter as an additional point mass free to move in response 
to the star and the rest of the disk. In each case results obtained were
strongly dependent on the prescription followed. We conclude
that the dynamics important for the global
behavior of the physical system extend to quite small radii.

With this degree of activity in the inner disk it becomes reasonable
to assume that a portion of the inner disk matter becomes depleted by
accretion onto the star or ejected in an outflow on short time 
scales. The inner disk may expand in the $z$
direction and become truly three dimensional as the dynamical effects
create dissipation and heating. In light of these ideas,
and in order to concentrate our efforts on the large scale features,
we have chosen to implement the density law of eq. [\ref{denslaw-a}]
and study a system for which little mass exists close to the star
but which retains a power law profile further out. Due to the already
artificial nature of the mass distribution at small radii, little physical
meaning can be attached to mass accretion rates through the inner 
boundary, therefore for simplicity we implement reflecting boundary
conditions to complete the specification of the inner grid edge.

For our SPH simulations, we define the inner boundary by establishing 
an accretion radius at a distance from the current position of the star 
of $R_D$/110(=0.4~AU). This distance is set to be slightly smaller
than the initial position of the innermost ring of particles in the
disk. The gravitational softening radius for the star is set to the same
value. As a particle's trajectory takes it inside the 
accretion/softening radius, its mass and momentum are added to the star
and it is removed from the calculation. This inner boundary 
condition does not prove to be as difficult to manage as in our PPM
simulations. Even though a great deal of activity occurs in the 
inner portion of the disk, no particular computational difficulties were 
experienced. We believe this activity is largely due to crude boundary 
conditions which obscure the true physical behavior of the system.
Particles near the boundary have no neighbor particles further
inward to provide pressure support, while accretion of a particle 
through the boundary implies a sudden loss of pressure support to 
its neighbors further out. Also, the stellar gravitational softening
reduces the effect of the star on the orbit of each SPH particle 
there. A small number of particles near the boundary
are strongly affected.

Because of our interest in characterizing disk instabilities,
especially SLING, we have experimented with several outer boundary
conditions as well. In the PPM simulations we have implemented both 
a reflecting boundary and a boundary condition  in which matter falls 
onto the outer edge of the disk (an ``infall'' boundary). With the pure
reflecting conditions, we imitate the boundary conditions implemented
by ARS which have been identified as critical for the SLING instability. 
With the infall boundary condition, we relax this assumption slightly to 
allow the disk edge to begin outward expansion or begin to break up if
conditions require.

With the infall boundary, the outer disk edge is defined to be 
initially located at a cell interface several cells inward from the
outermost computational cell. We define the disk boundary assuming 
an isothermal shock, so that the density and radial component of the 
velocity are determined directly from the shock jump conditions. Since by
definition a shock implies that the tangential velocity across the 
shock is continuous, we know that at the disk edge, $R_D$, the 
$\phi$ component of the infall velocity is the same as the orbit 
velocity, $R_D\Omega(R_D)$. If we then specify the temperature
of the infalling gas as $T=10$ K, conservation laws for mass, momentum
and energy determine the flow from the shock to the outer grid radius. 
The infall is kept constant throughout the simulation at the 
values which initially define it. We note in passing that a flow
constructed in this manner is quite artificial and may have little
relation to flows in real systems. For our simulations,
infall provides a mechanism by which the outer edge of 
the disk can be reasonably well defined.

In our SPH simulations, we adopt a free outer boundary. This choice
has the advantage of simplicity in implementation, but suffers because
quantities such as the density or pressure are less well defined within
about two smoothing lengths of the boundary (see fig \ref{dinit-sph}).
The result is that over time the surface density at the disk edge spreads
radially to a width of $\sim$5 AU. The disk edge is no longer defined by
a sharp discontinuity, but does remain distinct except for very high
\qmin~simulations, for which the mass at the outer disk
edge is nearly unbound. The sharp outer boundary condition required
for SLING to become active is satisfied under these conditions.

At time zero in our SPH simulations we set approximately 8000 equal
mass particles on a series of concentric rings with the innermost ring
at a radius of $R_D$/100. For our PPM simulations we use an inner to
outer radius ratio of 50 and several grid resolutions. Our main series
of simulations, with reflecting outer boundary conditions, have a 
$64\times102$ cell cylindrical polar ($r,\phi$) grid. Two higher
resolution simulations are performed with a $100\times152$ grid, and
we have explored the use of an infall boundary at two resolutions
of $44\times64$ and $64\times96$. Grid cells are defined to be `squares'
in the sense that $\delta r = Cr\delta\phi$ over the entire grid, with
$C$ a constant $\sim$1. With the resolution used for our simulations,
SPH particle smoothing lengths are less than a few tenths of one AU
in the inner portion of the disk up to $\sim$1 AU in the outer disk.
Grid resolution in the PPM simulations is of order 0.1 AU at the inner
grid edge and $\sim$2 AU at the outer edge.

The relatively low resolution of our simulations results in part
from the large radial extent of the systems we study. Many of the 
important dynamical processes in a disk occur on orbital time scales 
in the outer regions of the disk, but the size of a time step
(the Courant condition) is derived from the cell size at the inner
grid edge, where the cells are the smallest and the velocities are
largest.  Assuming nearly Keplerian rotation around the star, an 
inner grid radius at 1 AU, and a moderate resolution of order 150
azimuth cells, the time step is a few days, while the dynamical
time scale of the disk is a few $T_D=$500 yr. In order to evolve a
given simulation to completion, we must integrate over a half a 
million or more time steps. For `high' resolution simulations of say, 
300 or more azimuthal cells, the number is correspondingly increased. 
With the workstations available, it is not computationally feasible 
to run a large number of models to explore the relevant parameter space.
A similar problem exists for our SPH simulations.

\subsection{The Equation of State and Energy Considerations\label{eos-sec}}

In each code, a vertically integrated gas pressure is implemented
using a single component, `isothermal' ($\gamma=1$) gas equation of
state given by
\begin{equation}\label{eos}
P=\Sigma c_s^2.
\end{equation}
In PROMETHEUS (our version of the PPM
algorithm), a truly isothermal equation of state with $\gamma = 1$ is not 
easily obtained, therefore we use an `almost isothermal' ideal gas with
$\gamma = 1.01$ for these simulations. 

Each simulation is evolved isothermally, by which we mean that the temperature
of each cell, once defined at time zero (by eq. [\ref{templaw}] and an input 
value of \qmin), is fixed thereafter. Loss processes such as radiative cooling 
are assumed to balance local heating processes in the disk. Under this 
assumption, a packet of matter which moves radially inward or outward, 
heats or cools according to the prescribed temperature law. Matter which 
expands or is compressed is heated or cooled according to the same law.

With SPH comes the ability to choose the manner in which one
incorporates the isothermal evolution. We may set the temperature 
of each particle as a function of its distance to the star (Eulerian
implementation), or we may keep its temperature fixed no matter
where the particle moves (Lagrangian implementation). In most of our
simulations, we have chosen to use the Eulerian version.  This choice
is dictated by consistency, since the isothermal assumption implies
that the star must contribute the bulk of the heating, and by the
desire to match as closely as possible the PPM calculations.

\section{Results of Simulations\label{results}}

With the initial conditions outlined above, we have run a series of
simulations with both codes in which we vary disk mass but keep a
constant minimum Toomre parameter \qmin~$=1.5$. A free outer boundary
condition was implemented for each SPH simulation, while one series
of PPM simulations was run with a reflecting outer boundary. A second
series of PPM simulations used an infall through the outer few cells
onto the outer disk edge, which was assumed to be an initially 
stable isothermal shock.  To explore varying stability, we also ran
two SPH and one PPM series varying \qmin~between $1.1$ and a
maximum value defined by the condition that the outer edge of the 
disk remained bound (for disks with high \qmin, pressure forces
begin to dominate over gravity). Each simulation was evolved for 
periods ranging between a small fraction of an orbital period $T_D$ (in 
the case of very low \qmin~runs in which rapid clumping was seen), to 
several complete orbital periods for runs in which clumping was not observed. 

Unless otherwise specified, no explicit initial perturbations have been 
assumed beyond computational roundoff error. Due to the discrete 
representation of the fluid variables, this perturbation translates to
a noise level of order $10^{-3}$ in the hydrodynamic quantities
for the SPH calculations. The relatively large amplitude of the noise
originates from the fact that the hydrodynamic quantities are smoothed
using a fixed number of neighbors (see Herant \& Woosley 1994). An increase
in the number of particles does not necessarily decrease the noise
unless the smoothing extends over a larger number of neighbors. Because 
of its similarity to Monte Carlo methods, the decrease in noise goes as
the square root of the number of neighbors, and so decreases slowly 
with a large increase in computational cost.

For PPM, the noise level can be made as small as machine precision 
(while  double precision is used internal to the code, single precision is 
used in initialization and dumps, so we obtain $\sim 10^{-7}$).  The PPM
simulations are terminated at a perturbation amplitude of 
$\delta\Sigma/\Sigma \sim20$\% because matter on elliptical orbits begins to
interact strongly with the inner and outer boundaries. SPH simulations on 
the other hand, are carried out until clumps begin to form
(clumping causes the time step to drop drastically and halt the evolution).
Highly stable disks, for which clumps do not form, are terminated when
no significant additional evolution is anticipated. Each of the SPH 
simulations run for much of their duration with high amplitude
($\delta\Sigma/\Sigma \sim100$\%) perturbations. Comparison simulations 
on a simple test problem (see section \ref{sphppm} below) were run to 
high perturbation amplitude using both SPH and PPM in order to confirm
the late time behavior of the SPH simulations.

We did not formally introduce a perturbation in our
initial conditions, however two conditions provide indirect seeds
for perturbations. First, the disk is cut off at an inner
radius which, while small, is nonetheless large compared to the
stellar radius. This cut off creates a gravitational potential hump at the
center, and is equivalent to a strong seed for the $m=1$ disturbances.
As the star moves away from the origin, it is further
accelerated by the hump, effectively sliding down
the incline.  We show in figure \ref{gravpot} the gravitational potential 
near the origin for the disks with the characteristics described above 
as well as the tori used in our comparison calculations below (section 
\ref{sphppm}).  By following the procedure of Heemskirk \etal (1992),
who derive an equation of motion for the star including the zeroth
order hump term plus first order perturbations, we note that initially
the growth rate for a $m=1$ pattern will be
\begin{equation}
\gamma_1 = \left(\sqrt{ {{d^2\Psi_D}\over{dr^2}} }\right)_{r=0}.
\end{equation}
Computing numerically the curvature of the hump, we derive a $m=1$ 
growth rate due to the hump of $\gamma_1/\Omega_D\sim 5$. Indeed during
the very earliest stages of our simulations ($t\lesssim 0.1T_D$), we
find a growth rate of this (quite large) magnitude. After the
initial transient, growth rates quickly fall to more sedate levels.
The contribution to the long term global growth of instability due to
this initial perturbation is thought to be a small component of the
total. 

A second indirect seed of dynamical instability is connected to the
fact that the density law has been softened (eq. [\ref{denslaw}]) or
modified (eq. [\ref{denslaw-a}]) in the innermost regions of the disk
in order to avoid a singularity at small radii. This density decrease
creates a region of high vortensity gradient which excites wave
growth (see Papaloizou and Lin 1989). This instability channel also 
requires a seed, but its proximity to the inner edge, where orbit 
times are small, coupled with the hump perturbations, make the time 
scale for its initial excitation quite short.

Features of our simulations are tabulated in Tables \ref{sph-tabl} and 
\ref{ppm-tabl}.  The first column of each table represents the name of 
the simulation for identification. The second column defines the 
resolution (in number of particles or grid size). Initial disk/star
mass ratio and minimum $Q$ are given in columns 3 and 4, while total
simulation time and spiral features of each simulation fill out the
remaining columns.

We illustrate the phenomena seen in our simulations by presenting two
representative cases: mass ratios \mrat~$=1.0$ and \mrat~$=0.2$. Both use
initial values of \qmin~$=1.5$. These disks represent points near either
end of a spectrum of behavior. In section \ref{tempvar}, we show additional
models which vary \qmin, demonstrating behavior along another axis in 
parameter space. We first examine the qualitative nature of the simulations,
then examine in detail the structures which form.  A comparison of
the results of SPH and PPM and limitations imposed by numerical features
is discussed in section \ref{sphppm}.

\subsection{General Observations and Morphology\label{genobs}}

Spiral arm growth occurs with varying rates and amplitudes. Growth is not 
smooth or continuous. Frequently arms change shape, stretch, or break off 
and drift until hit by another passing disturbance. Well developed spiral
arms, while subject to irregular change on short time scales, do survive.  
In figure \ref{sph-himas}, we show a series of snapshots of particle positions
in simulation {\it scv6}, characterized by \mrat~$=1.0$ and an initial minimum
\qmin~$=1.5$. Instability first develops in the central regions of the disk,
and propagates outward in radius. Even early ($0.5 T_D$) in the simulation,
variations of density $\delta\Sigma/\bar\Sigma$ approach 10-50\%; at late 
times they reach unity.

The dominant patterns are two and three-armed spirals with significant
components having other symmetries. At late times we see multiple tails on 
a single arm, arms unevenly spaced in azimuth, and patterns which 
have one arm which is significantly stronger than its counterparts.  
Often such spacing and asymmetry is preceded by the breakup of an arm 
at its base, and subsequent drift through the disk or capture by another 
arm. For example, between the $0.94 T_D$ and the 
$1.41 T_D$ images, an $m=2$ structure breaks up, and reforms as an
asymmetric $m=3$ spiral pattern. It then returns to its previous two
armed structure by $1.73T_D$.

A comparable series of plots for a PPM simulation ({\it pch6}) with
analogous initial conditions is shown in figure \ref{ppm-himas}. The 
variable plotted is density variation defined by
\begin{equation}\label{densvar}
{\Sigma_{ij}-\bar\Sigma_i}\over{\bar\Sigma_i},
\end{equation}
where $i$ and $j$ refer to the grid indices of the $r$ and $\phi$
coordinates respectively, and $\bar\Sigma_i$ is the azimuth average 
of the surface density at radial grid index $i$.  Only positive 
variation contours are shown. The linear spacing between
one contour and the next higher contour is noted in the upper right
hand corner of each plot. The dotted line denotes the disk edge at
50 AU. As in the SPH simulation, instability begins in the inner 
regions of the disk. Complex structures follow at midtime epochs. 
Later behavior shows well defined regular spiral patterns,
with a mix of several patterns dominated by $m=2$ and $m=3$ which 
dynamically reorganize themselves with time.

The simulations above display a number of similar characteristics, though 
on a different spatial scale and mass distribution, to the protostellar 
core/inner disk simulations presented in Pickett \etal (1998). In each case, 
large scale spiral structures grow from marginally stable systems. The 
instabilities begin their growth in the innermost regions of the system and 
proceed to involve the entire disk as the simulation proceeds. At late times
in both sets of simulations, the spiral arms become azimuthally condensed. 
A notable difference between our results and theirs, which will be discussed 
in section \ref{disk-grw} below, is the fact that our simulations exhibit a 
pattern speed which increases toward the center of the disk. In contrast, 
Pickett \etal report constant pattern speeds.

Initial behavior of our low disk mass runs is similar to those of high mass,
with instabilities first becoming apparent in the inner regions of the disk. 
Evolution at later times differs from that for high mass disks. We see the 
rapid development of patterns with large numbers of spiral arms, which 
display a tenuous, filamentary structure not present in higher mass disks. 
The disk shown in figure \ref{sph-lomas} (simulation {\it scv2}) has a five 
armed pattern which predominates, and at late times fragments into multiple 
clumps from each arm. A region of apparent stability against spiral arm
formation  becomes apparent in the extreme innermost regions (see also
section \ref{tempvar}).  Such regions are present to some extent in 
all of our SPH disks except those which form clumps immediately and are
defined by a value of Toomre's $Q$ parameter greater than $\sim$2.

In a low disk mass PPM simulation ({\it pch2}), shown in figure 
\ref{ppm-lomas}, we also find a change in character and an increased
number of spiral arms. As in figure \ref{ppm-himas}, the instability 
begins to form its first spiral structures at amplitudes of 0.01-0.1\%.  
Although the precise number of arms seen does not correspond to that 
in the SPH run (showing instead the 2-4 armed patterns dominant), the
degree of small scale fragmentation in the region around 5-25 AU is
similar. We believe that the partial suppression of the high $m$-number 
patterns can be attributed to the wavelengths of those patterns 
approaching the gravitational softening length implied by the grid. This
statement is supported by the fact that for the low mass disks ({\it pch2}
and {\it pcm2}), the amplitude of the perturbations $\delta\Sigma/\bar\Sigma$, 
is larger in the higher resolution simulation. These simulations do not
resolve the small scale structure. Note that the PPM run with \mrat~$=0.1$
($pcm1$) developed only minimal spiral structure after nearly six full 
disk orbit periods.

Structures observed in the moderate resolution PPM simulations (runs 
{\it pcm1-pcm6}) were qualitatively similar to those observed for
our highest resolution runs ({\it pch2, pch6}), although the growth of
the low mass/low resolution structures was slower. Growth rates are 
similar for the low and moderate resolution high mass disks.  
The simulations may have reached a level of convergence sufficient 
to resolve the large scale features of the evolution, but further 
improvement is desirable.

\subsection{The Effects of Temperature \label{tempvar}}

Two series of SPH simulations were run varying the minimum stability
parameter $Q$, with mass ratios \mrat~$=0.8$ and $0.4$. Other things
being equal, high $Q$ implies high temperature in the disk (eq.~[\ref{Qdef}]). 
We vary $Q$ for different simulations between a minimum value of $Q=1.1$, 
at which the disk is only marginally stable to ring formation, and a
maximum value such that the outer edge of the disk remains bound. 
For the high mass series, this limit was found at \qmin~$=2.3$, while for
the lower mass disks up to \qmin~$=3.0$ were available.

In figure \ref{sph-qvar} we show `late time' behavior of each of the disks 
in the \mrat~$=0.8$ series ({\it sqh1-sqh6}). Below an initial value of
\qmin~$\sim1.4$, strong instability and clump formation occurs during a few 
orbit periods of the inner portion of the disk. The outer disk remains 
largely unaffected during the simulation (which suffers drastic decreases 
in time step size once clumps form). At moderate \qmin~(1.4 to somewhat 
less than 1.7), instability in the inner regions is slowed to the extent 
that spiral instabilities involving the entire disk have time to grow. 
These spiral arms then become filamentary and clump on time scales of one
or two $T_D$.  The last few frames in figures \ref{sph-himas} and 
\ref{sph-lomas} show such behavior for a disk with \qmin~$=1.5$
and \mrat~$=1.0$ and $0.2$. The portions of the spiral arms at large
distances from the central star remain thicker and more diffuse,
while the inner regions evolve toward more sharply defined features.
As \qmin~increases the character of the spiral instabilities changes
from narrow, filamentary structures and clumps in the inner disk 
to thicker arms which develop on disk orbital time scales at higher
initial \qmin.

Above initial \qmin~$\sim2.0$, we see only limited asymmetry and spiral
structure. However, there is a strong transient epoch in which the 
centers of mass of the star and the disk orbit each other at large 
distances (relative to their late time behavior or to other, less stable
(lower \qmin) simulations). Simulations have been carried out to more 
than $4 T_D$ for these cases. This resonance gains in strength with 
increasing \qmin~up to the maximum values simulated. Accretion of disk
matter onto the star occurs at higher rates in these runs as well.
The star makes a hole in which little disk matter remains. 

Figure \ref{hiq-trans} shows an example of this transient for simulation 
{\it sqh6}.  The orbit of the star begins with a slow transient to
relatively large distances from the system center of mass (as large as
$\sim0.05 R_D$ in the disk shown). In the first $\sim 2 T_D$,
the star accretes a large fraction initially located in the inner part
of the disk. After this time the star settles to a smaller orbit, with 
occasional fluctuations  as it moves in response to disk perturbations, 
and continues to accrete from the inner disk. We believe this  
transient is largely due to the high mass accretion rates with
nonaxisymmetric flow.  With such fast accretion, the flow of mass
onto the star is rapid enough that appreciable angular momentum is
swept along as well. A comparable simulation, with the star fixed at
the origin, shows nearly as large an accretion rate. We conclude that
high accretion drives the stellar migration, rather than the reverse
process where the star moves by some other means (caused for example
by a torque from the outer disk) into a region of the disk in which 
high accretion may take place.

Although we find that the accretion rates seen in the most $Q$-stable
disks are higher than low stability disks, it is not clear whether the
magnitude of the accretion rates are correct. In SPH the accretion 
of a particle implies a sudden unphysical loss of pressure support for 
the neighbors of the accreted particle. As the disk reorganizes itself, 
additional particles move inward towards the accretion radius. If the 
mass accretion for all of our disks were to be scaled up or down by a 
common factor, the transient in figure \ref{hiq-trans} might increase or 
decrease in magnitude or even disappear. What we can say with certainty is 
that if a star can accrete matter from the inner disk quickly enough that 
it loses its pressure support further out, accretion of disk material which 
has not lost all of its orbital angular momentum can occur, driving the 
star away from the system center of mass. In the simulations we study
here, such a condition occurs when the accretion rate is above
$\sim 6-8\times 10^{-5} M_\odot$/yr for the high mass series and 
$\sim 2\times 10^{-5} M_\odot$/yr for the lower mass series.

Behavior of the lower disk mass series of SPH simulations with varied 
\qmin~is similar. The overall characteristics of the evolution mimic that
of the higher mass runs but are `stretched' along the $Q$ axis to higher
values of \qmin. Azimuthal condensation of spiral arms is again seen up 
to initial \qmin~$=1.5$, but the \qmin~$=1.7$ run at this mass ratio 
appears to be just beyond the critical stability for clumping: many
preliminary characteristics of clumping such as well resolved spiral
arms and short duration over-density spikes (see below) were evident
but no actual formation occurred at the time we stopped the run at
$T=5 T_D$. Production of thick arms continues as high as 
\qmin~$=2.3$ and global star/disk resonances again manifest themselves
all the way up to the maximum \qmin~values studied.  Distinct
one armed spiral waves form at \qmin~$=2.0$ for short periods, then lose
coherency and fade back into a smooth, global pattern.

One series of PPM simulations was run with varying \qmin. The
late time density variation contours for the series are shown in figure 
\ref{ppm-qvar}.  Because of the low amplitude of the initial noise, these
simulations were continued to $\sim 2 T_D$ even for the lowest 
minimum $Q$ values. In the highest stability (\qmin~$=2$) simulation, 
we find that the strength of the instabilities near the inner boundary 
dominate the instabilities over the disk as a whole. This instability 
does not seem to be the same as the transient seen in the SPH
runs: it is limited to small radii inside the 
density maximum, and does not enter the outer disk at all. Because
of the boundary behavior noted above, simulation of the disks into epochs
having large amplitude variations was possible for only short times.
We could not determine if a large transient in the orbits of the
centers of mass of the star and the disk developed at late times for 
these simulations.

At low and moderate \qmin~($\leq1.7$), there is a great deal of 
correspondence between the qualitative results of our SPH and PPM runs. 
For simulations with moderate initial \qmin~($\sim1.4-1.7$) multiple
spiral arm structures develop with the $m=2$ and $m=3$ patterns 
most prominent. The $m=1$ pattern is present at varying levels
as an asymmetric component of the dominant $m=2$ or 3 patterns.

For the lowest stability simulation, run at \qmin~$=1.1$, density 
variations up to $\sim$40\% are present in the disk and variations
within a single spiral arm produce local density maxima within
that arm. Continued collapse from large amplitude spiral structure into
one or more clumps is not observed, probably because 
we have not resolved the gravitational potential 
or the rotational motion of the matter about a collapsing core to 
the necessary scale.  The evolution of these lowest stability disks
(i.e. simulation {\it pqm1} and {\it pqm2}) at early times in the
simulations are dominated by the growth of the $m=1$ pattern which,
unlike their more stable cousins, is distinct even at the 
$10^{-5}-10^{-6}$ level. Later, these patterns tend to break up and 
reform as $m=2$ and $m=3$ patterns.

\subsection{Spiral Pattern Growth\label{patterns}}

An important connection of numerical simulations to linear perturbation
analyses is to define, if possible, the linearly growing spiral patterns
of a system. To do so requires a specification of the growth rates
and pattern speeds of the dominant spiral patterns in each system.

We compute the growth rates by first computing the amplitude of spiral
patterns by Fourier transforming a set of annuli spanning the disk in
the azimuthal coordinate. The amplitude of each Fourier component is
then defined as $|A_m|=\ln (|\Sigma_m|/|\Sigma_0|)$, where $\Sigma_m$ is
\begin{equation} \label{mode-amp}
\Sigma_m = {{1}\over{\pi}}\int_{R_i}^{R_o}\int_0^{2\pi}
		e^{im\phi}\Sigma(r,\phi)rd\phi dr,
\end{equation}
for $m>0$ and the inner and outer radii of the disk are defined by $R_i$ 
and $R_o$. The $m=0$ term is defined with a normalization of $1/{2\pi}$. 
With this normalization, the $\Sigma_0$ term is the mass of the disk and 
the amplitudes, $A_m$, are dimensionless quantities. The phase angle 
is then defined from the real and imaginary components of the amplitude
\begin{equation}
\phi_m = \tan^{-1}\left[{{Im({\Sigma_m})}\over{Re(\Sigma_m)}}\right].
\end{equation}
Local amplitudes for each component can also be derived for annuli by
neglecting the integration over radius in eq. [\ref{mode-amp}]. Each
Fourier component is computed about the center of mass of the system.

Assuming strictly linear growth for each Fourier component, we can use least 
squares techniques to fit a growth rate, $\gamma_m$, to each amplitude as
a function of time with the equation
\begin{equation}
A _m = \gamma_m t + C_m,  \label{grwtheq}
\end{equation}
where $C_m$ is an constant defining the initial amplitude of the component.
If we keep track of the number of times, $N$, a pattern has wound past 
a phase angle of 2$\pi$ and add $2\pi N$ to the derived phase at each
time, we can derive a pattern speed by a similar fit as
\begin{equation}\label{phidot}
\phi_m = {{\dot\phi_m}\over{m}} t + \phi_{m,0}.
\end{equation}
This definition effectively averages over all short term variations (if
any) in the pattern speed. A periodogram analysis gives similar results to
this fit technique.  The frequency with which we produce dumps of the 
simulation is sufficient to produce accurate pattern speeds over all but
the inner $\sim 3-5$ AU of our disks (limited by aliasing), and over 
the full radial extent of the tori (see section \ref{sphppm}).

We may independently derive an additional global growth rate for the
$m=1$ component by noting that it is the only component which can
contribute to the motion of the star. All higher order components are
symmetric under a rotation smaller than 2$\pi$ radians (i.e. $2\pi/m$
respectively for each Fourier component) and therefore do not contribute 
to the motion of the disk center of mass. By fitting the distance between
the centers of mass of the star and disk as a function of time, we find
a growth rate independent of the precise geometry of the spiral arms in 
the disk. In general we find good agreement between this growth rate
and the value derived from the above procedure.

The analysis of the pattern growth in disks and tori can proceed
at either a local or a global level by either including or excluding
the integration over radius in eq. [\ref{mode-amp}]. If we derive a 
growth rate and pattern speed in a succession of narrow rings in the 
system and compare the values over the entire system, we can readily 
identify structures which are coherently growing and moving over large 
temporal and spatial scales. This feature is limited in a global analysis
because the integration effectively averages the amplitude and speed of
a given pattern over the entire system.

On the other hand, a local analysis can be quite misleading. If we consider
a series of concentric narrow rings making up a disk, we must account 
not only for the growth of instability within any given ring, but also
for the transport of already formed instabilities from one ring to 
another. For example: if some `lump' of matter grows in one ring in the
disk, then moves by some process to a second, the amplitude of the 
Fourier components in each of those rings will be affected: one will exhibit 
a net loss in amplitude, while the other a net gain. A growth rate based 
upon amplitudes affected by such processes would no longer represent
the physical instability mechanisms present in the disk. 

In the analysis that follows, we shall use a local analysis to
identify patterns which are growing coherently over large spatial
scales, but in order to compare our results to the global analyses
of ARS and STAR, we shall utilize globally integrated quantities.

\subsubsection{SPH and PPM: A Direct Comparison of Results 
and Numerics\label{sphppm}}

Each code does well with different aspects of the evolution of 
disks.  For the example of the disks discussed here, the low noise in
the PPM calculations allows an accurate growth rate calculation,
but with our treatment of boundaries, problems develop as a simulation 
becomes nonlinear. Matter reflected from the boundaries changes the total
momentum of the system to such an extent that its center of mass 
(exhibited particularly in the position of the star) attempts to move to
infinity. Because of its ability to dynamically adapt the available 
resolution to the interesting parts of the flow and relative sensitivity
to boundaries, SPH is able to follow the nonlinear evolution much further.
These same features however, forbid simulating a disk with a low density
central hole because the steep density gradient near the inner disk edge
cannot be adequately resolved at a computationally affordable level.  
Even for disks without a hole (for which the gravitational softening at 
the inner boundary blurs the physics and allows the simulation to proceed),
the initial noise in SPH (of order $10^{-3}$) leaves very little time
for random perturbations to organize themselves into ordered global
spiral structures while remaining in the linear regime. 

Fitting growth rates to the SPH simulations requires much more
caution than is required for the PPM runs. The initial noise level is
such that only a very short time baseline is available prior to
saturation.  Typically, we observe a period during which Fourier 
components grow linearly until reaching a saturation level. This
period of linear growth lasts for about one disk orbit $T_D$ or
less for SPH and 2--3 $T_D$ for the PPM simulations.

The SPH disk simulations often reach high perturbation amplitudes close
to the star before more distant regions of the disk have become active.
To compare the two numerical methods and minimize this time scale problem,
we have simulated relatively narrow tori. Such tori have a much more 
restricted dynamic range than a disk, so that the entire system becomes 
active at once.  We use a torus with an outer to inner radius ratio of 
$R_i/R_o=5$ and a $\gamma=1$ equation of state given by eq. [\ref{eos}] 
with temperature, density and individual particle mass given by a Gaussian 
function of radius \begin{equation}\label{torus-law}
f(r) = f_0e^{ -\left({{r-r_0}\over{R_w}}\right)^2},
\end{equation}
where $r_0$ is defined at the midpoint, $r_0=(R_i+R_o)/2$, of the
torus and $R_w=(r_0-R_i)/2$, so that the torus extends about three
`standard deviations' in either direction from the highest density
point (figure \ref{torus-init}). Each simulation is then evolved
isothermally in the same way as is done with our simulations of disks.

With a $\gamma=1$ equation of state, it is difficult to find toroidal
configurations which are initially stable to axisymmetric perturbations
(i.e. $Q>1$ everywhere), except for relatively low mass tori. 
For a variety of temperature or density laws, either the high density
central region will collapse (i.e. the initial \qmin~will be less than 
unity), or the outer edge will be unbound. For our test problem, a ratio
of $M_T/M_*=0.2$  yields a minimum $Q$ of about $1.05$ near the center 
of the torus.  As before, the star mass is $M_*=0.5M_\odot$, the outer 
torus radius of $R_o$=50~AU, and thus the outer edge of the torus orbit 
period is $T_T=T_D=$500 yr.

Table \ref{cmp-tabl} summarizes the characteristics of the simulations. 
The linear and nonlinear regimes are divided by the condition that the 
amplitudes of Fourier components other than the dominant pattern (or 
patterns) reach comparable amplitudes to that dominant pattern, and total
perturbations reach $\sim$10\%.

One SPH and two PPM simulations were run with this toroidal
configuration at a resolution of $40\times150$ cells for the PPM
runs and $6998$ particles in the SPH run. One PPM simulation with
initial random noise amplitude $10^{-3}$ (comparable to the initial
noise in SPH) and one with noise of amplitude $10^{-8}$ were run. 
The $10^{-3}$ noise is input as a random density perturbation in each 
cell as
\begin{equation}
\Sigma_{ij} = \left(1 + 10^{-3}(2R-1)\right)\Sigma_{ij}
\end{equation}
where $i$ and $j$ refer to the radial and azimuthal grid indices
and $R$ is a pseudo-random number between zero and one.
The $10^{-8}$ amplitude noise is derived from truncation 
error in the initial state, as is done in the disk PPM simulations. 
Boundary conditions are identical to those used in our disk simulations.

The relatively large amplitude of the noise in the SPH simulations 
is caused by smoothing over a finite number of neighbors (see Herant
and Woosley 1994).  Increasing the number of neighbors used
in the interpolation has a small effect in decreasing the noise amplitude
but at a high computational cost. We have used a varying number of
neighbors (depending on local conditions of the run) with a distribution
centered near 15--20 neighbors per particle, a number which is standard 
for two dimensional simulations. 

The resolution of features within the torus or disk must inevitably
be less accurate in a finitely resolved system than in a physical
system.  PPM spreads shocks over at least two cells, for example, while
further loss of resolution may come from the representation of the 
gravitational potential.  In SPH, resolution is limited by
the smoothing length of the particles and the artificial viscosity 
required to adequately reproduce shocks. 

Two additional PPM and two additional SPH simulations of tori have been
run to test resolution. One PPM run has 1.5 times the resolution in each
dimension (60$\times$225--roughly doubling the number of cells) and the 
second twice the resolution (80$\times$300--quadrupling the number of cells).
The SPH simulations increase by a factor of two and a factor of four the 
number of particles in each simulation. Comparing runs of different
resolution is difficult, however, because the power spectrum of the initial
perturbations  may not be controlled to the limit required. In an attempt
to duplicate the perturbation at low and high resolution, but remain above 
the uncontrollable level imposed by the grid itself, we have input an initial
random noise amplitude of $10^{-3}$ in each 2$\times$2 block of cells in
each of the two higher resolution PPM runs.

We show the evolved configuration of each run in figure \ref{tor-cmp}.
The time at which each is shown is near the linear regime cut off
discussed below. The SPH runs are mapped onto a grid and plotted in the
same manner as the PPM runs in order to make the visual comparison as
direct as possible. In each of the runs, instability growth is dominated
by $m=2-4$ spiral patterns with the higher resolution runs tending to show
progressively less of the $m=4$ pattern and more of the $m=2$ pattern. 
The $m=3$ pattern predominates in each simulation except for the two low 
resolution PPM runs. The change in morphology in different simulations is 
probably an artifact of the resolution. As we show for the growth rates 
below, the lowest resolution simulations are apparently not converged.

In comparison, the results of Laughlin \& R\'o\.zyczka (1996) show a 
dominance of an $m=2$ component without a large presence of other 
patterns. The origin of instabilities in their systems is attributed to 
the family of vortensity instabilities with corotation exterior to the 
torus. Different initial conditions seem to be responsible for the $m=2$
rather than $m=3$ dominance. Our test simulations use a narrower torus
than theirs, with an isothermal rather than adiabatic equation of state. 
A simulation with an identical initial condition and equation of state  
compares favorably to their results.

The amplitudes and fits for growth rates for the $m=3$ spiral 
pattern at the center of the torus (at $R=30$~AU) are shown 
for each simulation in figure \ref{torm3_30}. The fit parameters are 
derived from only the portion of the curve in which the patterns are 
growing and little disruption of the large scale structure of the tori 
has begun. This disruption is characterized by an onset of fragmentation
at the inner and outer edges of the torus (SPH) or significant radial 
distortions in the torus (PPM). We also allow a short period 
($\sim 0.1T_D$) prior to the first fitted time point, for some
initial transients (e.g. the unphysical `ringed' structure in the
SPH initial state) to settle.

The pattern speeds and growth rates for the $m=1-4$ patterns
are shown in figure \ref{ppmpatgrw1-4} for each of the PPM simulations
and in figure \ref{sphpatgrw1-4} for each of the SPH simulations. The
pattern speeds for the $m\geq 2$ patterns for each of the runs agree
for both codes over the range of resolution and initial perturbation
amplitude. The growth rates from the SPH simulations differ by as much
as 50\% between runs. For the SPH simulations obtaining a constant rate 
across each ring in the torus was not possible.  For the PPM simulations,
the growth rates near the inner and outer boundaries of the tori are reduced 
due to the fact that perturbations there do not begin to grow until after 
the denser regions of the torus have been disturbed. A similar effect is 
found for the pattern speed near the inner edge.

The growth rates for the SPH runs are affected by the high amplitude of
perturbations in the initial state and the short time span over which
the fit must be derived. Longer lived initial transients caused by the 
excitation of multiple eigenmodes of the system or by small inhomogeneities 
in the initial state can cause the amplitude curves to become quite 
nonlinear in form. The PPM simulations have longer time baselines so such 
transient effects are less important. 

The growth rates for the $m=4$ pattern in the PPM simulations decrease
with increasing resolution, while the $m=2$ and 3 growth rates are less
affected. This fact and the trend towards $m=2$ and 3 spiral patterns
for higher resolution runs suggest that they may be true linearly growing 
patterns for the system. The change in character with increasing 
resolution may be due to the fact that the torus begins its life very close 
to the stability limit, \qmin~$=1.0$. Any inaccuracies in the resolution 
of the gravitational potential or the mass distribution (hence the pressure) 
will have their greatest effect in such a circumstance. The SPH simulations 
show no comparable effect, but reliable local growth rates can not be 
obtained for those simulations.

Late in each simulation the tori collapse into several condensed objects, 
but the details of the collapse vary. Not all of the spiral arms present 
during the growth of structure condense into separate objects. In many cases 
the spiral arms break up and/or merge as clumps begin to form. Figure 
\ref{tor-late} shows snapshots of each of the runs at the time at which the
spiral arms begin to collapse. Each simulation is halted at this point because 
of the influence of the boundaries on the simulation and because we did not 
properly simulate the physics important in the collapsed objects. The structures 
which develop resemble the simulations discussed by Christodoulou \& Narayan
(1993) because the tori tend to deform radially as instabilities grow. With
both codes the torus becomes so distorted radially that a line of condensations 
forms from the torus matter which has moved outwards.

We now summarize the similarities and differences between the 
results of each code.

Each code produces instabilities which grow in the tori as 
they evolve forward in time. The instabilities produced are 
multi-armed spirals structures which, at the end of each simulation,
have begun to radially distort the torus and collapse into clumps.
In both codes predominantly 2--4 armed spiral structures are produced.
The high resolution simulations each produce 2 and 3 armed structures 
while low resolution simulations (apparently incompletely converged), 
produce predominantly 3 and 4 armed structures.

The initial state of an SPH simulation begins with random noise of
amplitude $\sim 10^{-3}$ above or below an `ideal' initial value.
Near the boundaries, where particles are not distributed evenly
with respect to each other, additional differences from an ideal
initial state are present. PPM can begin with noise in the initial
state as small as machine precision for any given simulation.

The differences between one code and the other can be attributed to
several effects. First, perturbations in the initial state may
trigger more than one true eigenmode of the system which, taken
together, cause more or less observed growth in a given simulation 
with respect to another. Because the noise input for each code 
arises from such different sources, the stimulated pattern growth
may therefore initially have a much different  character.  This
growth rate variation is exhibited predominantly by the amplitude
of a given pattern `waving' above and below its true linear growth
curve and, in essence, constitutes an error estimate for a calculated
growth rate. The PPM simulations, for which the growth rates are 
calculated over longer time baselines and with a smaller initial
noise amplitude per Fourier component are not nearly as strongly
affected by such effects. We estimate errors of 10-20\% in the
growth rates due to this effect in the PPM simulations and perhaps
an additional 20-40\% in the SPH runs because of their very short
time baseline. Pattern speeds do not seem to be as strongly 
affected by these transient effects.

The adaptive nature of the resolution and high noise in SPH causes
small scale filamentary structures to become active and develop more
quickly than in our counterpart PPM simulations, which are limited
to the resolution of the fixed grid. SPH will tend toward developing
grainy and filamentary structures quickly, perhaps to a larger extent
than is physically the case.

Because the grid boundaries are far away from the main concentration
of mass in the torus, they have only a small effect until late in
any given simulation. Such is not true for the disk simulations
using the PPM code so those simulations cannot be carried out far 
into the nonlinear regime due to the growing influence of the 
boundaries at late times. The physics important for the global 
dynamical evolution of the disk ranges over a dynamic range larger 
than we are able to simulate. The state at which the PPM runs must 
be terminated (with 10-20\% perturbations) are qualitatively quite 
similar to those of the SPH runs over most of their duration. 
It may be that for the disks we discuss below, the PPM runs are 
representative of the linear regime, while the SPH simulations are 
our only representation of the late time nonlinear behavior of the 
system.

\subsubsection{Pattern Growth in Disks\label{disk-grw}}

With a clearer understanding of the numerical properties of our 
codes on a test problem, we return now to the study of disks. 
Due to the high initial noise of the SPH runs and large radial 
extent of the disks we study, saturation at small radii often occurs
well before the entire disk has become involved in the instability.
Because of this noise we do not believe growth rates calculated from these 
simulations are reliable for any Fourier component except the globally
integrated $m=1$ pattern (for which we have the behavior of the centers
of mass of star and disk), and we limit discussion of the growth rates 
in this and the following sections to the PPM simulations.

The qualitative observations of sections \ref{genobs} and \ref{tempvar}
have shown that there is rarely a single spiral pattern present in a disk.
More quantitative measurements show that growth is present in all 
Fourier components up to very high order. Such growth does not necessarily 
imply that actual spiral arms of that order are present in the simulation, 
but rather that the arms that do exist become more filamentary than pure 
sinusoids, creating power in higher order Fourier components (a Dirac
$\delta$-function will yield power at all wave numbers for example).
In order to be more definitive regarding the true morphology of each 
disk we visually examine each simulation and tabulate the dominant 
spiral patterns in Tables \ref{sph-tabl} and \ref{ppm-tabl}.

Which patterns represent linear growth in each of the systems? 
To begin to answer this question we must fit growth rates and pattern
speeds to the various spiral patterns present in each disk and
determine which patterns exhibit rates which are constant at differing
resolution, across a large portion of the system and over a large
time period. In figure \ref{m2and3hi} we show the amplitude of the 
$m=2$ and $m=3$ patterns as a function of time near the middle of the
power law portion of the disk and integrated over all radii for our 
prototype massive disk shown in figure \ref{ppm-himas}. Over long periods 
the growth is essentially linear in character. Over shorter periods it 
is punctuated by transients which can change the amplitude by up to an 
order of magnitude. The amplitude variations apparently arise as 
short-lived structures successively grow and fragment throughout the 
disk. Time dependence of pattern speeds within the disk will be discussed
in section \ref{non-lin-phenom} below.

Radius dependent growth rates and pattern speeds for the $m=1-4$ 
are shown in figure \ref{patgrwm1-4hi} for two different grid resolutions. 
The growth rates and pattern speeds are similar at both resolutions, 
suggesting that the simulations may have resolved the physical processes
important in this disk. The growth rates for the $m\geq 2$ patterns are
nearly constant with radius but the pattern speeds derived are not at all 
constant with radius; they decrease as a steep function of the
distance from the central star. 

Low mass disks show a marked absence of the dominant low order ($m=1-3$) 
spiral patterns so common among higher mass disks. Typically, the amplitudes 
and growth rates of all Fourier components are comparable.  We plot the
growth rates and pattern speeds for the same patterns ($m=1-4$) as above
for our prototype low mass disk in figure \ref{patgrwm1-4low}. We again 
find that the pattern speeds are steeply decreasing functions of the radius.
We also find that the growth rates do not exhibit the same values for 
different grid resolutions. This fact suggests that the low mass disks have 
not fully converged at the grid resolution used in our simulations. The 
systematic trend towards faster growth in the higher resolution simulation
indicates that the small scale features which dominate the morphology of
this system may be somewhat inhibited by the resolution of the gravitational
potential and the hydrodynamic quantities on the grid. Much higher resolution
simulations are required to be able to fully resolve the features important
for disks of mass less than $\sim 0.2 M_*$ than are required for more massive 
systems.

With simulations of varying stability we would ordinarily expect 
larger $Q$ values to lead to slower instability growth. Similarly,
we expect that smaller $Q$ values should imply more rapid growth of 
instability. In fact, as discussed in section \ref{tempvar}, both extremes
lead to rapid instability growth, but of different character. 

Although it begins with an extreme initial condition, the simulation
{\it pmq5} (with \qmin~$=2.0$, \mrat~$=0.8$) shows an interesting
example of the limiting behavior displayed in a highly stable disk
($Q>>1$ everywhere) with a turnover in its density profile near
the central star.  We show the $m=1$ and $m=2$ pattern amplitudes at two
locations in the disk and integrated globally in figure \ref{hiq-amp}. 
In this simulation, rather than being suppressed, the amplitude of the
instabilities begins to grow quickly in a region limited to the innermost 
portion of the disk. Further out in the disk much slower growth occurs.
The development of such instabilities in disk systems cannot be attributed
to a global, linearly growing phenomenon; its localized character and 
the different behaviors of the amplitude growth at different locations
in the system argue against that. It remains unclear to what extent 
this type of growth happens in real systems, but it seems that with
a turn-over in the density law at small radii or the less severe case 
where the density law flattens (as in our SPH simulations) can lead 
to increased local instabilities. 

It is interesting to note that Pickett \etal (1996) report similar 
behavior (which they refer to as `surge' growth) in several of
their more $Q$-stable simulations. In their work however, the initial 
mass distribution and rotation curve are somewhat different than in 
our own work. The fact that similar behavior is observed in simulations of
such different character suggests a similar mechanism may be driving the
evolution of both sets of simulations.

The lowest stability simulations also show rapid growth of spiral
instabilities. In these simulations there are no growth features 
similar to the `hump,' or sudden rise in amplitude shown in figure 
\ref{hiq-amp}. In general, the qualitative features of the growth are 
similar to those seen in figures \ref{m2and3hi} and \ref{patgrwm1-4hi}
but with as much as 50--100\% larger growth rates in the case of the
lowest stability run ({\it pmq1}).

The results of our analysis in this section show that in spite 
of its large amplitude at early times and its continued presence
for the duration of the run, our simulations do not show evidence of
a pure $m=1$ pattern. In no case is the $m=1$ growth rate or pattern
speed constant across a large portion of the disk. In contrast to 
several higher $m$ patterns, the wide variation is true of both
the growth rate as well as the pattern speed. Because of the 
variation of the growth rate and pattern speed we must conclude that a 
direct connection to the SLING mechanism is not possible. At the high
amplitude (late time) phase of evolution shown in the SPH simulations, the
$m=2$ and $3$ patterns have become dominant for disks more massive than 
\mrat~$\approx 0.2$, while at the lower amplitudes typical of our PPM runs,
$m=1$ has the largest amplitude, though the pattern itself is ordinarily
seen only as asymmetries in higher $m$ structures.

None of the disk simulations we have performed produce pattern speeds 
for any $m$ pattern which are constant across the entire disk. The 
growth rates, while ordinarily stable at a single value over the whole 
system for at least some patterns (see e.g. fig. \ref{patgrwm1-4hi}), do
not reflect the short term behavior of the system as structures 
fragment or deform over time. In this case the `linear growth modes' 
of the system, defined as the complex eigenvalue of a system of 
equations, become difficult to define or to interpret.

\subsubsection{Suggestions for the Mechanisms of Instability 
Growth\label{mechanisms}}

In each of our simulations, instabilities are generated in the 
innermost portions of the disk, eventually impacting the entire
system. Such growth occurs in spite of the fact that the inner regions 
are the most stable as measured by two of the classic stability 
indicators, namely the Toomre $Q$ criterion and the SWING $X$ parameter.
If we are to suggest a mechanism for the instability growth we 
are limited to mechanisms which can produce instabilities in what
are ostensibly highly stable regions.

We have already discussed the possibility that in some cases 
instabilities may be due to nonaxisymmetric accretion of disk 
matter onto the star or by accretion of infalling material onto the
disk, rather than to dynamical instabilities in the disk itself. In other
cases, the vortensity based instabilities of Papaloizou \& Lin 1989 
(see also Adams \& Lin 1993) may provide an answer because they can 
grow in highly `stable' regions and their growth can be local in nature.  
They discuss three classes of vortensity instabilities which can exist 
in a disk: those dependent on vortensity extrema within the disk or 
at its edge (`edge modes'), those dependent on resonances (`resonance
modes'), and those which have corotation exterior to the disk 
(dubbed `slow modes' and studied extensively by Laughlin \& R\'o\.zyczka 
1996). Because we find corotation within the disk for most times
(though at varying position), we can eliminate the last of these classes
from consideration. The remaining two, we believe, are both active at 
different times and to a greater or lesser extent in the disks we 
model. At early times, our initial condition (the softened power law
or density turn-over at small radii) implies a vortensity extremum near
the inner boundary of the disk. This condition may excite an edge mode
which over time propagates outward over the density maximum in our PPM
simulations via a resonance mode into the disk, exciting global 
instability channels such as SLING as it propagates into the main 
disk. We have not established a definite connection between the 
instabilities in our simulations and the vortensity based instabilities 
however.

We cannot definitely connect the SLING instability directly to phenomena 
present in our simulations; see  section \ref{disk-grw}. We may still
perhaps be able to make qualitative connections between phenomena predicted
to be important via linear analyses and our results. One example of such 
phenomena would be growth rates which depend upon the outer boundary 
condition imposed. Another might be a growth rate which, as a function
of disk mass, increases for disks more massive than some critical value,
as suggested by the `maximum solar nebular mass' discussed in STAR. Such
characteristics would not necessarily be limited to the $m=1$ pattern
but may also exist in $m>1$ patterns as well. 

We do see such characteristics in the variation of the growth
rates with respect to the disk/star mass ratio. For each series of PPM
simulations varying disk mass, figure \ref{disk-mratrates} shows the 
value of the globally integrated growth rates for the $m=2$ patterns.
Growth rates for other $m$ patterns appear qualitatively similar to those
shown. As one expects, growth rates of the highest mass disks are 
the largest, while instabilities in low mass disks grow much
more slowly. In the reflecting boundary runs, a distinct `turn on' mass 
is evident between $0.2<$~\mrat~$<0.4$, a value which corresponds to the
`maximum mass solar nebula' predicted by the results of STAR.
The infall series does not exhibit such a distinct onset, but rather
a continuous rise to a plateau which does not flatten out until the mass
ratio reaches \mrat~$\approx 0.5$.  

For low disk masses, the growth rates for each pattern are of order 
$\gamma_1/\Omega_D=0.15-0.2$. These rates are comparable to the rate
attributable to numerical effects. The numerical effects have their
origin primarily in the fact that mass interacting with the grid boundaries
gives an impulse to the system center of mass, which must be stable in order
to determine the amplitude of the $m=1$ spiral pattern. Higher $m$ patterns
are also affected as spiral waves reflect off the grid boundaries back into 
the simulation.

For higher mass disks, the outer boundary has a marked influence.  As ARS 
predict, details of the outer radial boundary are an important factor in the
growth pattern. The simulations with matter infalling onto the outer disk 
edge develop spiral structure with growth rates as much as 2-3 times faster 
than with a purely reflecting boundary. Simulations at two resolutions were 
run with an infall boundary to test the degree to which numerical effects of 
the boundary were affecting the growth. Both series show similar growth rates 
(fig. \ref{disk-mratrates}). 

\subsubsection{Importance of Phenomena not Included 
  in Linear Analyses\label{non-lin-phenom}}

On short time scales the pattern speeds in our disks can vary by as much
as 100\%. One example, shown in figure \ref{m2pat_tme}, is taken from the
high mass disk simulation {\it pch6}. There we show the instantaneous 
pattern speed for the $m=2$ pattern near the middle of the disk, as 
calculated by numerically differencing the pattern phase $\phi_m$, at
successive output dumps of the simulation. Such variations in time are
typical of each pattern in each disk simulation we have performed, and
appear in both local and globally integrated pattern speeds. Pattern speeds
calculated this way for the torus simulations of section \ref{sphppm} show 
much slower variations.

In the case of the $m=1$ pattern, whose global pattern speed is reflected 
in the motion of the star, we find that the star occasionally loops back upon
its own trajectory and counter-rotates with the disk for a short period.
Such a condition is not an uncommon occurrence in systems with disturbances
with different orbital (pattern) periods. In our own solar system, for 
example, the sun's motion about the solar system barycenter was retrograde
most recently in 1990, when Jupiter was on the opposite side of the sun from
the other three major planets.

The variations seem to arise because of the growth, fragmentation
and reformation processes undergone by the spiral arm structures
over the course of their evolution. Because the pattern speeds vary,
an averaged pattern speed at any location in the disk (via eq. 
[\ref{phidot}]) loses meaning and the location of the corotation
and Lindblad resonances for each pattern also vary in time. When such
variations are occurring, wave analyses, which typically assume 
stable resonances, may be of limited utility (wave analysis is of
course useful in less chaotic circumstances--see, e.g., STAR and
Laughlin, Korchagin, \& Adams 1996).

The growth of instabilities is not always suppressed as $Q$ increases, but 
the instabilities do change character; this change is due to the increasing 
importance of effects not modeled in semi-analytic treatments of disks. For 
the high \qmin~SPH runs, these effects are dominated by the nonaxisymmetric 
accretion of disk matter onto the star. As the star begins to move from the 
center of mass of the system (due to ordinary disk processes or the potential 
hump at the origin), some portion of the accretion becomes nonaxisymmetric. 
In the warmest disks, as much as 10\% or more of the disk is accreted over the 
life of the simulation. Disk matter accreting onto the star sweeps 
along some residual angular momentum which is transferred to the star
either as spin (an effect we neglect here) or as net angular 
momentum of the star about the system center of mass. In these cases,
the star may gain enough momentum to be driven further away from the
center of mass and create power in the $m=1$ pattern.

In the PPM runs with infall, the instability growth can include
a component due to the outer disk edge perturbations. These may be
due to infall itself, or to fragmentation of the disk at the boundary.
Although the linear analyses of ARS and STAR showed that the conditions 
at the outer boundary were important for the evolution of the system, 
they were unable to fully model the effects that the boundary can have
on the system (see however Ostriker, Shu, \& Adams 1992).

\subsection{Clump Formation and Characteristics \label{clump}}

Returning now to our SPH simulations, in this section we describe 
several qualitative features of clump formation and evolution in
the disks. Due to the unsteady nature of the spiral instability 
growth and the presence of multiple spiral patterns in the system,
each disk sequentially approaches and moves away from conditions in
which clump formation is likely. These conditions are most readily 
apparent in plots of the minimum $Q$ value in the disk 
and in the maximum over-density in the disk (defined as
$\Sigma(r,\phi,t)/\Sigma(r,t=0)$) with respect to time. 

The value of $Q$ is defined rigorously only for an azimuthally
symmetric disk. Nevertheless, as an indicator of the most unstable 
locations in the disk, we examine its value in nonaxisymmetric systems. 
To calculate its value locally we must first determine the epicyclic 
frequency at each point in the disk. We use the same procedure by which
SPH obtains derivatives of all other hydrodynamic quantities. By 
definition
\begin{equation}
\kappa^2 = {{1}\over{r^3}}{{d}\over{dr}} [\left(r^2\Omega\right)^2],
\end{equation}
so the value $d[(r^2\Omega)^2]/dr$, taken pairwise over each neighbor,
is weighted using the SPH kernel. The result is summed to form a local
value of the epicyclic frequency. 

Plots of maximum over-density and minimum $Q$ are shown in figure 
\ref{odqplot} for our two prototype SPH \qmin~$=1.5$ disks.  Each variable
is a global extremum. As such, the value of one could be determined
from a completely different portion of the disk than the other. However,
after only a relatively small fraction of an orbit time $T_D$, the locations 
of minimum $Q$ and maximum over-density are close, at a position between
about 10 and 30 AU.

After a few orbit periods of the inner disk regions, the over-density 
rises to about twice its initial value (of unity). A slow secular 
trend towards stronger spiral arms over the course of the run follows,
punctuated by one or more sharp, short-duration episodes of very strong 
activity in which density locally increases to 5-10 times.  Over-density 
spikes become more and more frequent as the simulations progress,
finally leading to clump formation. With the one exception \mrat~$=0.4$,
\qmin~$=1.7$ which, as noted in section \ref{tempvar}, appears to lie
on the `boundary' between clumping and non-clumping disks, simulations 
which do not eventually form clumps also do not show these large 
over-density events. We attribute the origin of the over-density 
events in our simulations to the growth of spiral instabilities into a
high amplitude nonlinear regime. In this regime spiral patterns 
present constructively interfere with each other or collide with other 
arms and orphaned arm fragments. 

The results of Adams \& Watkins (1995; hereafter AW) show that a density
enhancement within a disk will lead to collapse if the condition
\begin{equation}
{{\Sigma(r,\phi,t)}\over{\Sigma(r,0)}} > {5 \over 2} Q
\end{equation}
is met, where $Q$ is the local value (azimuth average) of the Toomre 
parameter at the location of the density enhancement. For the disks in
our study, this expression implies that an over-density factor of 3 or
higher must be present in the disk, depending on where in a disk the 
collapse event occurs. This prediction is supported by our numerical 
results, which show that disks can survive (i.e. not exhibit collapse)
for long periods with over-densities of 2-4, but collapse when 
over-density spikes of magnitude 6-10 occur.

For all disk masses, the minimum value of $Q$ rapidly falls below its
initial value to well below unity.  After the initial steep decline, a
slower decrease occurs until clumping begins and minimum $Q$ falls to
zero. The initial decline occurs most quickly in the highest mass
disks, in which instabilities of any type are most strongly felt. 
With $Q$ below unity, the disk becomes unstable not only to spiral
instabilities but also to ring formation or, in the case of isolated
patches, collapse. The collapse is slowed by the effects of rotation
within the forming clump.

We can verify that it is rotation which slows the collapse by noting that
the effects of the over-density spikes manifest themselves at only the 
20-30\% level in $Q$. We also know that the sound speed is constant in 
the proto-clump (due to our assumption that the disk evolves isothermally), 
from the definition of $Q$ we know that the rotation of an individual 
proto-clump (really the shear across the clump, measured by the local 
value of the epicyclic frequency $\kappa$) is the mechanism which inhibits 
further collapse. Only after spiral arm amplitude has reached sufficient 
levels to overcome rotation can an irreversible collapse begin. 

Clumps condense out of the spiral arms on quite short time scales
in even the least massive disks.  During and after the initial
stages of their formation, we find that the clumps show prograde 
rotation. No clumps were seen to form in any disk studied whose
initial \qmin~was greater than 1.5.  Clump formation is most common
at radii less than $\sim0.5R_D$ and usually several clumps will
form from the same disk (and even within the same spiral arm).
Less massive disks form many low mass clumps and higher mass disks
form 2-4 higher mass clumps. The mass inside the clumps is of order
1\% of the star mass at the time each simulation is ended. It is 
clear, however, that from the amount of remaining disk that no 
final mass has been determined.

The clumps form with such vigor in each of these disks because of the
strong cooling implied by the isothermal assumption. Any density
enhancements like those seen in figure \ref{odqplot} instantly lose their
pressure support and collapse rather than dispersing. With more realistic
cooling, the clumping behavior seen in our results may change. Thus our
results are most useful as an indication of the behavior of disk clumping 
and as indicator of where clumps may be most likely to form in more 
physically realistic disks. 

Figure \ref{formrad} shows a plot of the radius at which each clump 
formed for each disk in the series. Only in the case of the 
\mrat~$=0.2$ disk, in which clump formation is prolific in nearly
all regions, were any clumps formed at radii greater than 0.5~$R_D$.
With this exception, we believe the variation in the locations of clump
formation in disks of different mass in figure \ref{formrad} to be due
more to stochastic effects rather than any physical process. To test
this idea we ran a comparison series of simulations ($\times$'s), 
utilizing the Lagrangian version of the equation of state.  When such 
an assumption is made, the background noise inherent in the code 
changes character. No overall structural changes are evident in figure 
\ref{formrad}, but differences in detail are present. Also, for the disk 
with \mrat~$=0.2$, clumps were not formed at the largest radii. We believe
this lack of clumps is due in part to the radial motion of some warmer 
particles into the outer disk, causing clumping to be suppressed.

The prior results of AB92, in which clumps are seen to form at much
larger radii, correspond to a somewhat different initial
configuration. In particular, our present results use a much smaller
`core radius', $r_c$, for the density and temperature power laws. The
gravitational softening parameter for the star is correspondingly
smaller, and no initial perturbations are assumed. These differences
conspire to push collapse instabilities to larger radii in the AB92
results, since in their simulations more mass is concentrated at large
distances from the star. We believe the present conditions to be 
more realistic and thus to represent an improvement over 
the AB92 results.

\subsubsection{Initial Orbital Characteristics}

Out of the entire sample of newly formed clumps, none have an initial
eccentricity much higher than $\epsilon=0.2$, and most are
between zero and 0.1. The low mass companions now being discovered 
around nearby solar type stars show both small and large values
of eccentricity (Mayor \etal 1997; Marcy \& Butler 1996; Butler \& Marcy 
1996). Although the clumps in our simulations form only in relatively low
eccentricity orbits and are therefore dissimilar to many of those being 
discovered, considerable evolution of eccentricity can take place between 
the times corresponding to the end of our simulations and the final morphology 
of the system (see e.g., Artymowicz 1993, 1994; Goldreich \& Tremaine 1980).

\section{Conclusions\label{summary}}

By using two conceptually different hydrodynamic methods (SPH and PPM), 
we are able to simulate a broader range of problems, but gain a sobering 
insight into the limitations of these tools. It is striking that PPM 
indicates violent behavior near the inner boundary (weakly supported by SPH), 
and that SPH indicates pronounced clumping (weakly supported by PPM).
Both methods indicate that instability growth is not a steady progression 
from low to high amplitude perturbations with a single dominant pattern
present throughout. Both methods indicate a marked change in the 
character of instabilities with disk mass. Low mass disks form many
armed filamentary spiral structures while high mass disks form few armed
grand design spiral structures.

In this study of the evolution of circumstellar accretion disks, we
have found simultaneous growth of global spiral instabilities with 
multiple Fourier components. Growth of each of the components occurs 
over the course of a few orbit periods of the disk and a single component
rarely dominates the evolution of a disk. As expected, the massive disks 
are found to be the most unstable, due to self-gravitating instabilities 
within the disk. Accretion of matter onto the star itself can, in warm 
disks (i.e. those with high \qmin~values), significantly drain matter
from the disk time scales similar to the self-gravitating instabilities.
Short-term variations in the amplitude of a given component, and strong 
constructive interference behavior between different components, can 
produce `spikes' in the surface density. These spikes can eventually 
grow to such amplitude that gravitational collapse occurs resulting
in the production of one or more clumps.

Pattern growth is stimulated at early times by the rapid growth of
instabilities at small radii which eventually engulf the entire disk.
Steady spiral arm structures are not generally present. Instead, spiral
arms progressively grow, fragment and reform as time progresses. In 
cases where accretion is rapid, power can be produced in an $m=1$ spiral
pattern due to nonaxisymmetric accretion of mass and momentum onto the 
star. Understanding the dynamics of the inner region is of primary
importance for understanding the global morphology of the system.

The gross structure of low and high mass disks are markedly different
from each other. High mass disks form large, grand design spiral
arms with few arms, while low mass disks form predominantly thin,
filamentary multi-armed structures. In almost no case is the $m=1$
spiral pattern the fastest growing pattern in the disk. Typically 
a combination of $m=2-4$ patterns in high mass disks or very high
order patterns ($m\gtrsim 5$) in low mass disks dominate the 
morphology. The transition between these behaviors comes at
approximately \mrat~$=0.2-0.4$. This transition corresponds
to the `maximum solar nebula' mass discussed in STAR, above which
$m=1$ modes due to SLING are expected to grow strongly.

It is intriguing to speculate that the collapse processes seen here
are responsible for the production of brown dwarf-like companions such
as that seen by Nakajima \etal (1995) and/or of planetary companions
similar to those recently discovered around several nearby stars
(Mayor \& Queloz 1995, Marcy \& Butler 1996, Gatewood 1996). However, we
must emphasize that clump formation in self-gravitating circumstellar 
disks depends on the ability of the gas to cool efficiently. Our 
simulations here use a simple isothermal equation of state which favors 
clump formation. Additional simulations with realistic cooling functions,
including radiative transfer effects, must be done in order to clarify
this important issue.

\acknowledgments 
We wish to thank the referee, Richard Durisen, for a thorough referee
report which improved this paper substantially.  Bruce Fryxell provided
valuable insights into PPM. Greg Laughlin provided valuable discussion 
on the tori we use for our comparisons between SPH and PPM. AFN wishes 
to thanks his collaborators for patience in seeing this work through to 
its completion. This work was supported under the NASA Origins of the
Solar Systems program with grants NAGW-3406 and NAGW-2250. FCA is
supported by an NSF Young Investigator Award, NASA Grant No. NAG
5-2869, and by funds from the Physics Department at the University of
Michigan. DA is supported by NASA NAGW-2798 and NSF ASTRO 94-17346.

%% file: table-sph.tex
\singlespace
\begin{deluxetable}{lcccccc}
\tablewidth{0pt}
\tablecolumns{7}
\tablecaption{\label{sph-tabl} Disk Parameters For SPH Simulations}
\tablehead{
\colhead{Name}  & \colhead{No. of} & \colhead{\mrat} &
\colhead{\qmin}  & \colhead{End Time} & 
\colhead{Dominant\tablenotemark{a}} & \colhead{ Number} 
\\
\colhead{}  & \colhead{Particles} & \colhead{} &
\colhead{}  & \colhead{($T_D$=1)} & 
\colhead{Spiral Patterns}  & \colhead{Clumps}}

\startdata
scv0 &  7997 & .05 & 1.5 & 3.5 & $\gtrsim 12$ & 6  \nl
scv1 &  7997 & .1 & 1.5 & 1.6  & $\sim$10     & 14 \nl
scv2 &  7997 & .2 & 1.5 & 1.6  & 5-6          & 33 \nl
scv3 &  7997 & .4 & 1.5 & 1.7  & 3-4          & 7  \nl
scv4 &  7997 & .6 & 1.5 & 1.7  & 2-4          & 6  \nl
scv5 &  7997 & .8 & 1.5 & 2.4  & 1-3          & 3  \nl
scv6 &  7997 & 1. & 1.5 & 1.8  & 1-3          & 3  \nl
sqh1 &  7997 & .8 & 1.1 & 0.1  & NR           & 18 \nl
sqh2 &  7997 & .8 & 1.3 & .25  & NR           & 11 \nl
sqh3 &  7997 & .8 & 1.4 & .35  & NR           &  7 \nl
sqh4 &  7997 & .8 & 1.7 & 4.2  & 1-2          &  0 \nl
sqh5 &  7997 & .8 & 2.0 & 4.2  & 1            & 0  \nl
sqh6 &  7997 & .8 & 2.3 & 4.2  & 1            & 0  \nl
sql1 &  7997 & .4 & 1.1 & .15  & NR           & 28 \nl
sql2 &  7997 & .4 & 1.3 & 0.3  & NR           &  7 \nl
sql3 &  7997 & .4 & 1.4 & 0.4  & 4            & 1  \nl
sql4 &  7997 & .4 & 1.7 & 5.0  & 1-3          & 0  \nl
sql5 &  7997 & .4 & 2.0 & 4.2  & 1-2          & 0  \nl
sql6 &  7997 & .4 & 2.3 & 4.2  & 1            & 0  \nl
sql7 &  7997 & .4 & 2.7 & 4.2  & 1            & 0  \nl
sql8 &  7997 & .4 & 3.0 & 4.2  & 1            & 0  \nl
\enddata
\tablenotetext{a}{When only m=1 patterns are indicated, actual evolution
is apparently an accretion induced transient star/disk oscillation 
(see figure \ref{hiq-trans}) rather than a spiral arm.  NR (not 
resolved): for low stability disks, assignment of specific spiral arm
patterns loses meaning due to their rapid breakup.}

\tablecomments{Three series of runs are represented in this table. 
The first letter in each name is `s', signifying an SPH simulation.
The second, is either `c' or `q', signifying constant or varying \qmin,
and the third letter signifies that the simulation is a member of
a high (h), low (l) or varying (v) disk mass series.  Ascending
numerical order in each series refers to successive values of either
disk mass or \qmin, for each series.}
\end{deluxetable}
\doublespace

%% file: table-ppm.tex
\singlespace
\begin{deluxetable}{lcccccc}
\tablewidth{0pt}
\tablecolumns{7}
\tablecaption{\label{ppm-tabl} Disk Parameters for PPM Simulations}
\tablehead{
\colhead{Name}  & \colhead{Grid} & \colhead{\mrat} &
\colhead{\qmin}  & \colhead{End Time} & 
\colhead{Dominant\tablenotemark{a} } & \colhead{Outer} 
\\
\colhead{}  & \colhead{Res.} & \colhead{} &
\colhead{}  & \colhead{($T_D$=1)} & 
\colhead{Spiral Patterns}  & \colhead{Bndry}  }

\startdata
pcm1 &  64$\times$102 & 0.1 & 1.5 & 5.8  & NR    &  Refl.    \nl
pcm2 &  64$\times$102 & 0.2 & 1.5 & 5.0  & 2-4   &  Refl.    \nl
pcm3 &  64$\times$102 & 0.4 & 1.5 & 4.0  & 1-3   &  Refl.    \nl
pcm4 &  64$\times$102 & 0.6 & 1.5 & 3.75 & 1-3   &  Refl.    \nl
pcm5 &  64$\times$102 & 0.8 & 1.5 & 3.0  & 1-3   &  Refl.    \nl
pcm6 &  64$\times$102 & 1.0 & 1.5 & 3.0  & 1-3   &  Refl.    \nl
pch2 & 100$\times$152 & 0.2 & 1.5 & 5.0  & 2-4   &  Refl.    \nl
pch6 & 100$\times$152 & 1.0 & 1.5 & 3.6  & 1-3   &  Refl.    \nl
pqm1 &  64$\times$102 & 0.8 & 1.1 & 1.8  & 1-2   &  Refl.    \nl
pqm2 &  64$\times$102 & 0.8 & 1.3 & 2.6  & 1-2   &  Refl.    \nl
pqm3 &  64$\times$102 & 0.8 & 1.4 & 3.0  & 1-2   &  Refl.    \nl
pqm4 &  64$\times$102 & 0.8 & 1.7 & 3.0  & 1-2   &  Refl.    \nl
pqm5 &  64$\times$102 & 0.8 & 2.0 & 2.0  & 1     &  Refl.    \nl
pci2 &  64$\times$96  & 0.2 & 1.5 & 3.8  & 3-4   &  Infall   \nl
pci3 &  64$\times$96  & 0.5 & 1.5 & 2.1  & 1-3   &  Infall   \nl
pci4 &  64$\times$96  & 0.6 & 1.5 & 2.0  & 1,3   &  Infall   \nl
pci6 &  64$\times$96  & 1.0 & 1.5 & 1.6  & 1-3   &  Infall   \nl
pcl1 &  44$\times$64  & 0.1 & 1.5 & 5.6  & NR    &  Infall   \nl
pcl2 &  44$\times$64  & 0.3 & 1.5 & 4.2  & 1-3   &  Infall   \nl
pcl3 &  44$\times$64  & 0.4 & 1.5 & 4.2  & 1-3   &  Infall   \nl
pcl4 &  44$\times$64  & 0.5 & 1.5 & 2.8  & 1-2   &  Infall   \nl
pcl5 &  44$\times$64  & 0.7 & 1.5 & 2.8  & 1-2   &  Infall   \nl
pcl6 &  44$\times$64  & 1.0 & 1.5 & 2.0  & 1-2   &  Infall   \nl
\enddata
\tablenotetext{a}{NR: not resolved. For some low mass disks,
distinct spiral patterns are not possible to distinguish.}

\tablecomments{Each of these PPM runs begins with `p' to distinguish
it from SPH series. The second letter is `c' or `q' signifying a
constant or varying \qmin~value for each disk in the series.
The third letter implies a {\it l}ow, {\it m}oderate or {\it h}igh
resolution simulation. Moderate resolution infall boundary
simulations are distinquished from reflecting boundary simulations
using an `i' in place of `m'.  Numbers are successive values of disk 
mass or \qmin in each series of runs.}
\end{deluxetable}
\doublespace

%% file: table-cmp.tex
\singlespace
\begin{deluxetable}{lccccc}
\tablewidth{0pt}
\tablecolumns{6}
\tablecaption{\label{cmp-tabl} Tori and Disk Results in SPH and PPM}
\tablehead{
\colhead{Initial}   & \colhead{Hydro}      & \colhead{Linear} &
\colhead{Reason/}    & \colhead{Non-linear} & \colhead{Reason/}
\\
\colhead{Density}   & \colhead{Method} & \colhead{Regime} &
\colhead{Result}    & \colhead{Regime} & \colhead{Result}
\\
\colhead{Structure} & \colhead{}       & \colhead{}       &
\colhead{}          & \colhead{}       & \colhead{}
}

\startdata
Disk         & PPM & fails        &  inner       & not          & \nodata      \nl
(eq. \ref{denslaw})& &            & boundary     & accessible   &              \nl
             & SPH & not          & short time   & succeeds     & spiral arm   \nl
             &     & accessible   & baseline     &              & formation    \nl
             &     &              &              &              & and collapse \nl
             &     &              &              &              &              \nl
Disk w/Central & PPM & succeeds   & spiral arm   & short        & boundary  \nl
Hole           &     &            & growth       & duration     & influence \nl
(eq. \ref{denslaw-a}) &  &        &              & only         &           \nl
             & SPH & not          & short time   & \nodata      & \nodata   \nl
             &     & accessible   & baseline     &              &           \nl
             &     &              &              &              &           \nl
Torus        & PPM & succeeds     & spiral arm   & succeeds     & spiral arm \nl
(eq. \ref{torus-law})& &          & growth       &              & collapse   \nl
	     & SPH & partial      & spiral arm   & succeeds     & spiral arm \nl
             &     & success      & growth       &              & collapse   \nl
\enddata
\end{deluxetable}
\doublespace

%% file: isodiskfigs.tex
\singlespace
\begin{figure}
\plotfiddle{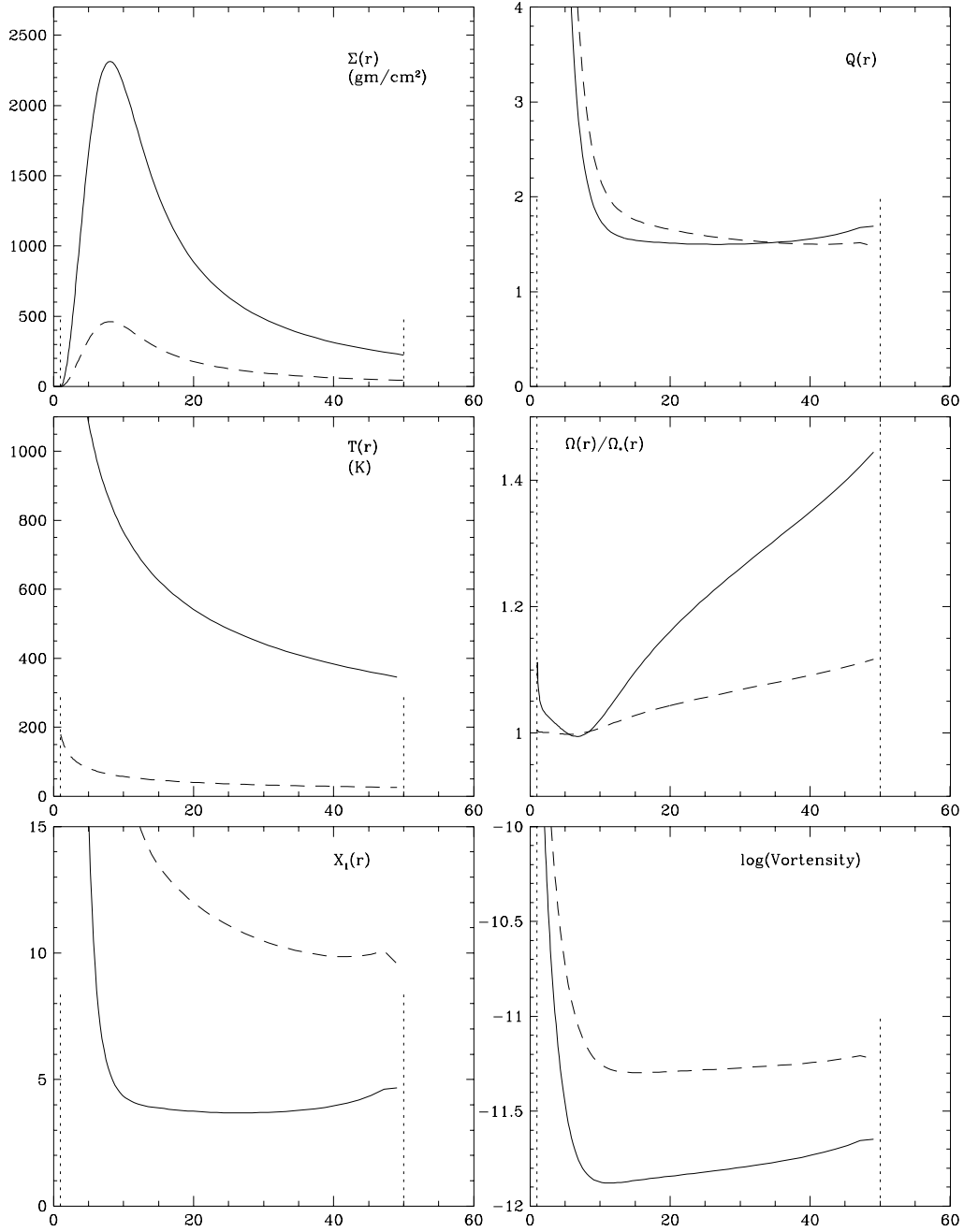}{6.75in}{0}{75}{75}{-240}{-50}
\caption[Disk initial conditions for PPM simulations]
{\label{dinit-ppm}
\footnotesize
A summary of the initial conditions for low (dashed) and high (solid)
disk mass PPM simulations (simulations {\it pch2} and {\it pch6}).
The six panels show surface density $\Sigma$, Toomre $Q$, temperature
$T$, the ratio of the rotation period at radius $r$ with the Keplerian
value. We define $\Omega_*(r)$ in the middle right panel as
$\Omega_*(r) = \sqrt{ {{GM_*}/{r^3}}}$. In the lower left
panel, we show the value of the SWING $X$ parameter for the $m=1$
pattern.  Higher order patterns ($m>1$) are smaller by a factor $1/m$
than the value shown. In the lower right panel, we show the value
of the vortensity at each radius.}
\end{figure}

\clearpage

\begin{figure}
\plotfiddle{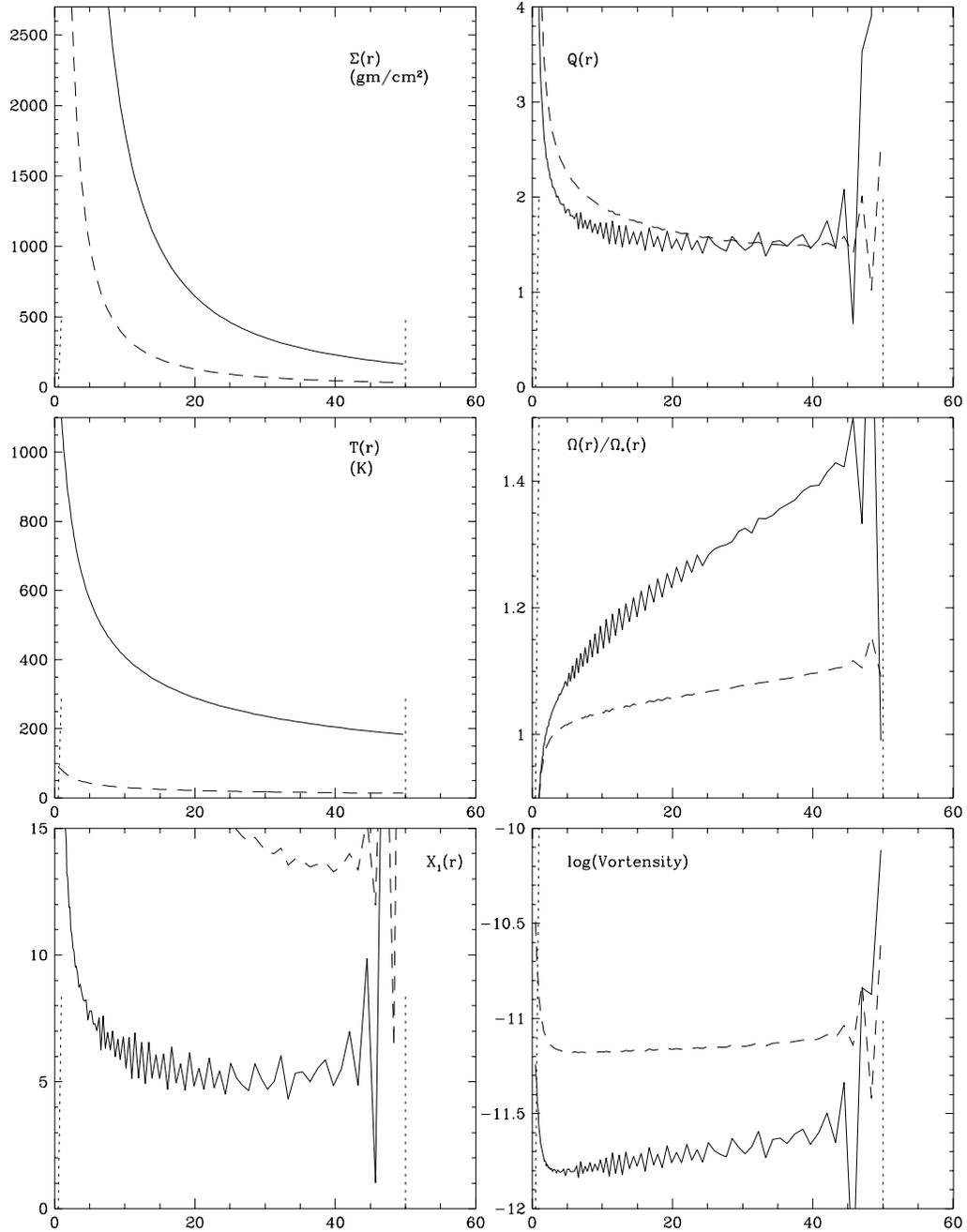}{6.75in}{0}{75}{75}{-240}{-50}
\caption[Disk initial conditions for SPH simulations]
{\label{dinit-sph}
\footnotesize
A summary of the initial conditions for low (dashed) and high (solid) 
disk mass SPH simulations (simulations {\it scv2} and {\it scv6}). This 
figure shows the same parameters as shown in figure \ref{dinit-ppm}.}
\end{figure}

\clearpage

\begin{figure}
\plotfiddle{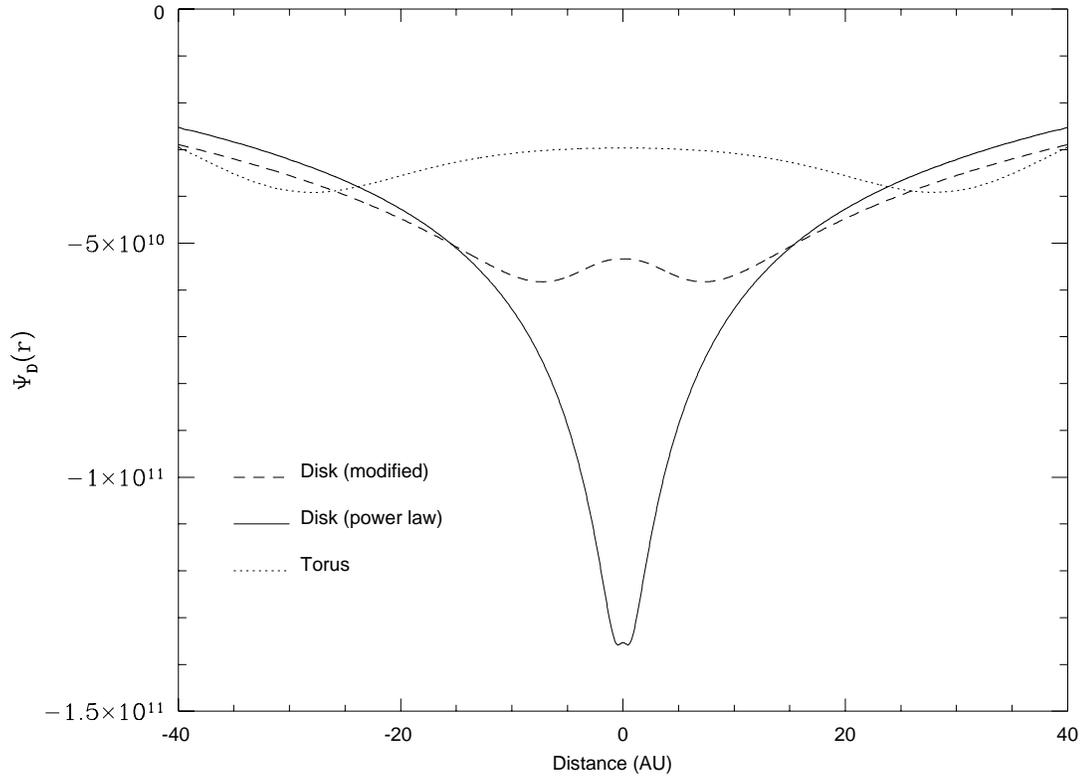}{5in}{-90}{55}{55}{-210}{400}
\caption[Gravitational potential near the star for disks and tori]
{\label{gravpot}
\footnotesize
The initial gravitational potential due to the disks and tori we 
study. We show a slice through the origin where the star is initially
located. The solid curve represents the gravitational potential
due to the pure power law as given by eq.  [\ref{denslaw}].  The
dashed curve is that due to the modified power law of eq.  
[\ref{denslaw-a}], while the dotted curve is that due to 
a torus as defined in section \ref{sphppm}. The mass in each disk
or torus is \mrat~=0.2.  Each system produces a gravitational 
potential hump at the origin which seeds the growth of $m=1$
disturbances in the disk or torus.}
\end{figure}

\clearpage

\begin{figure}
\plotfiddle{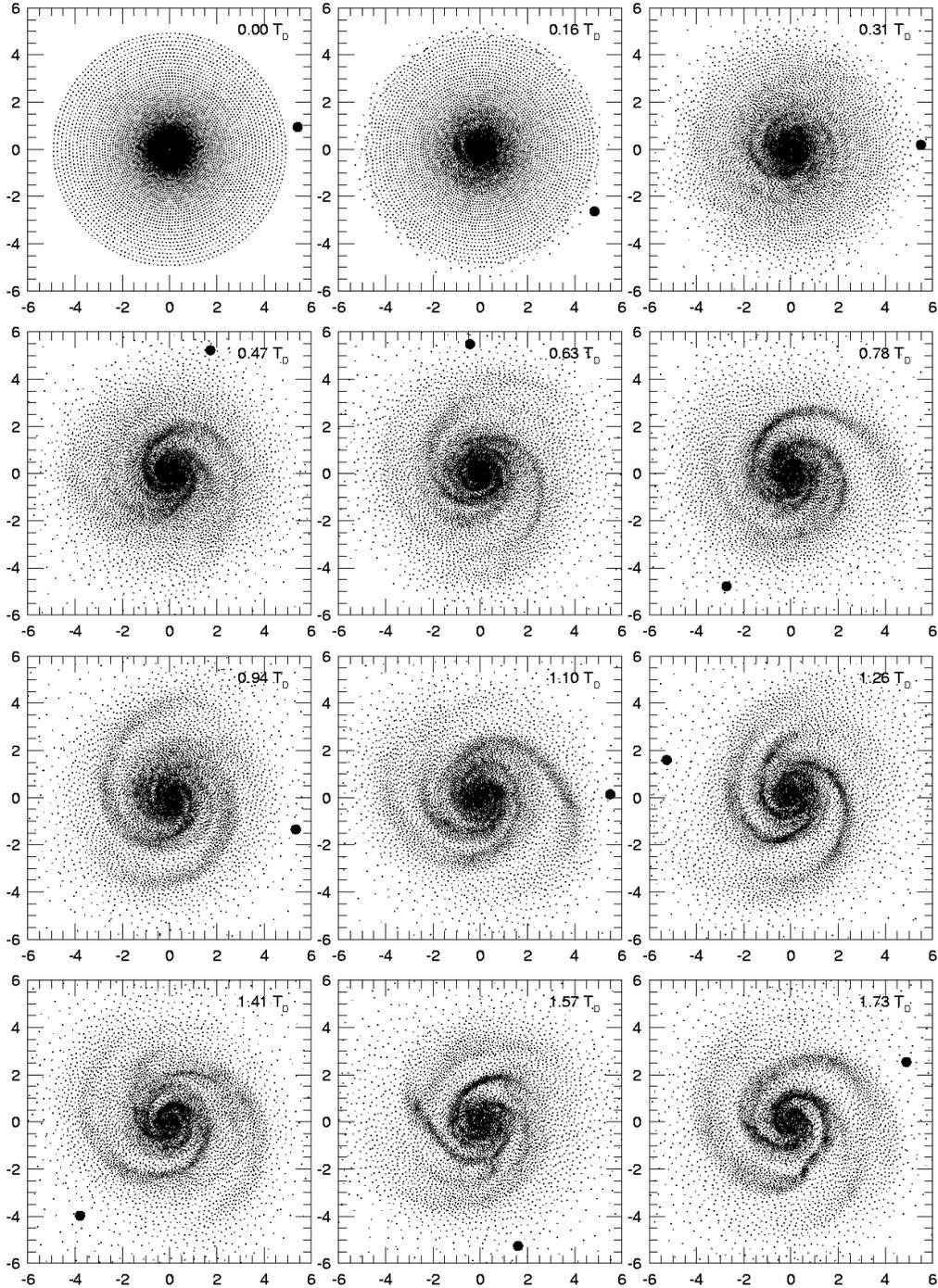}{6.70in}{0}{72}{72}{-230}{-20}
\caption[High mass disk SPH simulation]
{\label{sph-himas}
\footnotesize
A time series mosaic of SPH particle positions for a disk of mass
\mrat~$=1.0$ and \qmin~$=1.5$ (simulation {\it scv6}).
Note the strong variation of spiral structure over time. Length units
are defined as 1=10 AU and time in units of the disk orbit period $T_D$.
The large, solid dot is the angular position of the star projected
out to a distance of 55 AU, just outside the outer disk edge.
The final image in this mosaic shows the beginning stages of clump
formation as a clump begins to form in the disk at about an azimuth
angle of 5 o'clock and radius of $r=20$ AU.  A second clump which initially
formed in the other spiral arm is present but difficult to distinguish
in the image at 3 o'clock and $r\sim7$ AU.}
\end{figure}

\clearpage

\begin{figure}
\plotfiddle{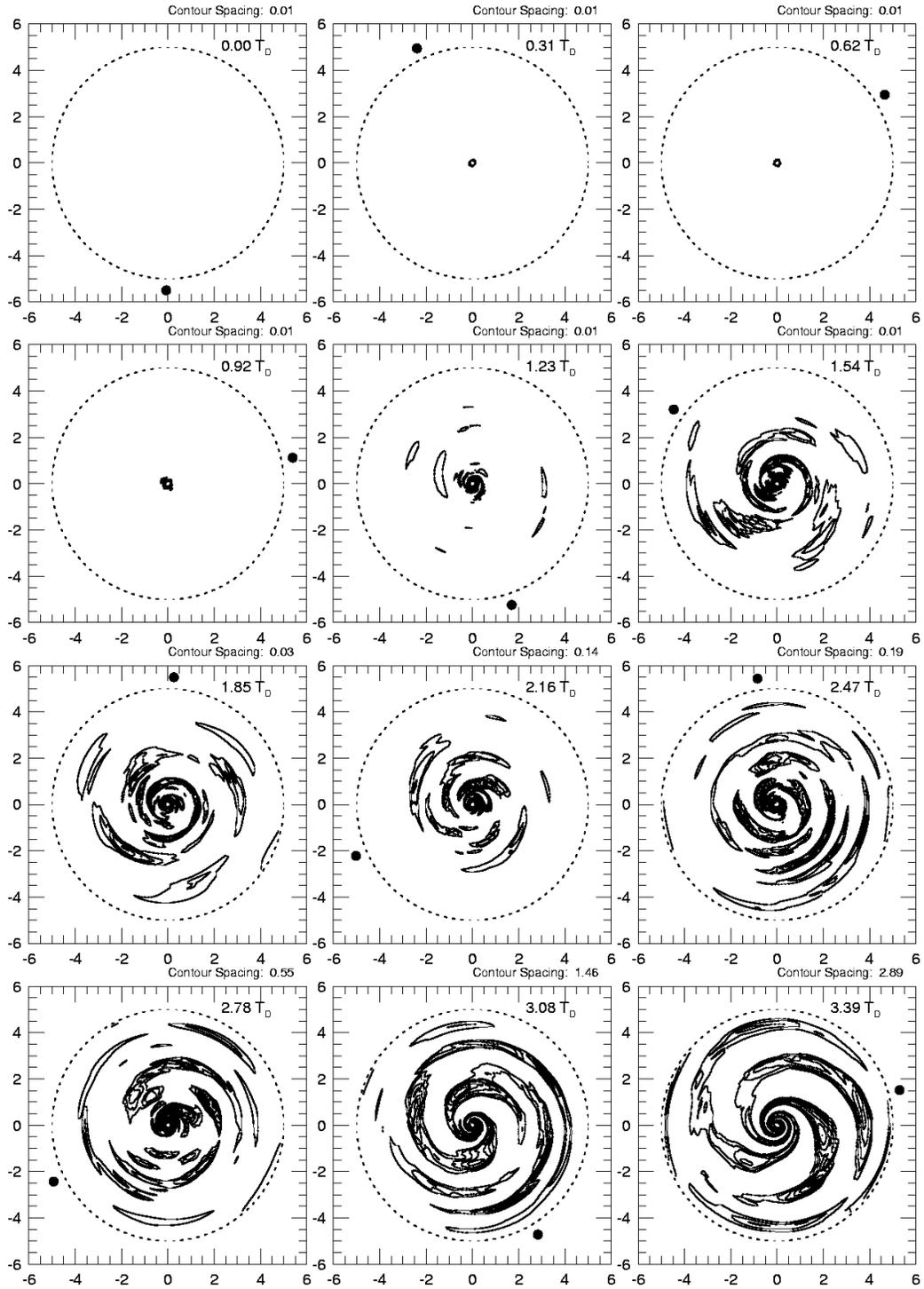}{7.05in}{0}{75}{75}{-230}{-15}
\caption[High mass disk PPM simulation]
{\label{ppm-himas}
\footnotesize
Time series of density variation in the disk for a PPM simulation
(simulation {\it pch6}). with the same initial conditions as figure
\ref{sph-himas}. Only positive density variations are plotted and the
maximum contour is derived only from deviations at radii larger than
$\sim$7 AU. Contours are linearly spaced and set to a minimum of 
.01\% variation per contour line. Larger variations are implemented
as the instability grows. The contour spacing is denoted in the
upper right corner of each frame.}
\end{figure}

\clearpage

\begin{figure}
\plotfiddle{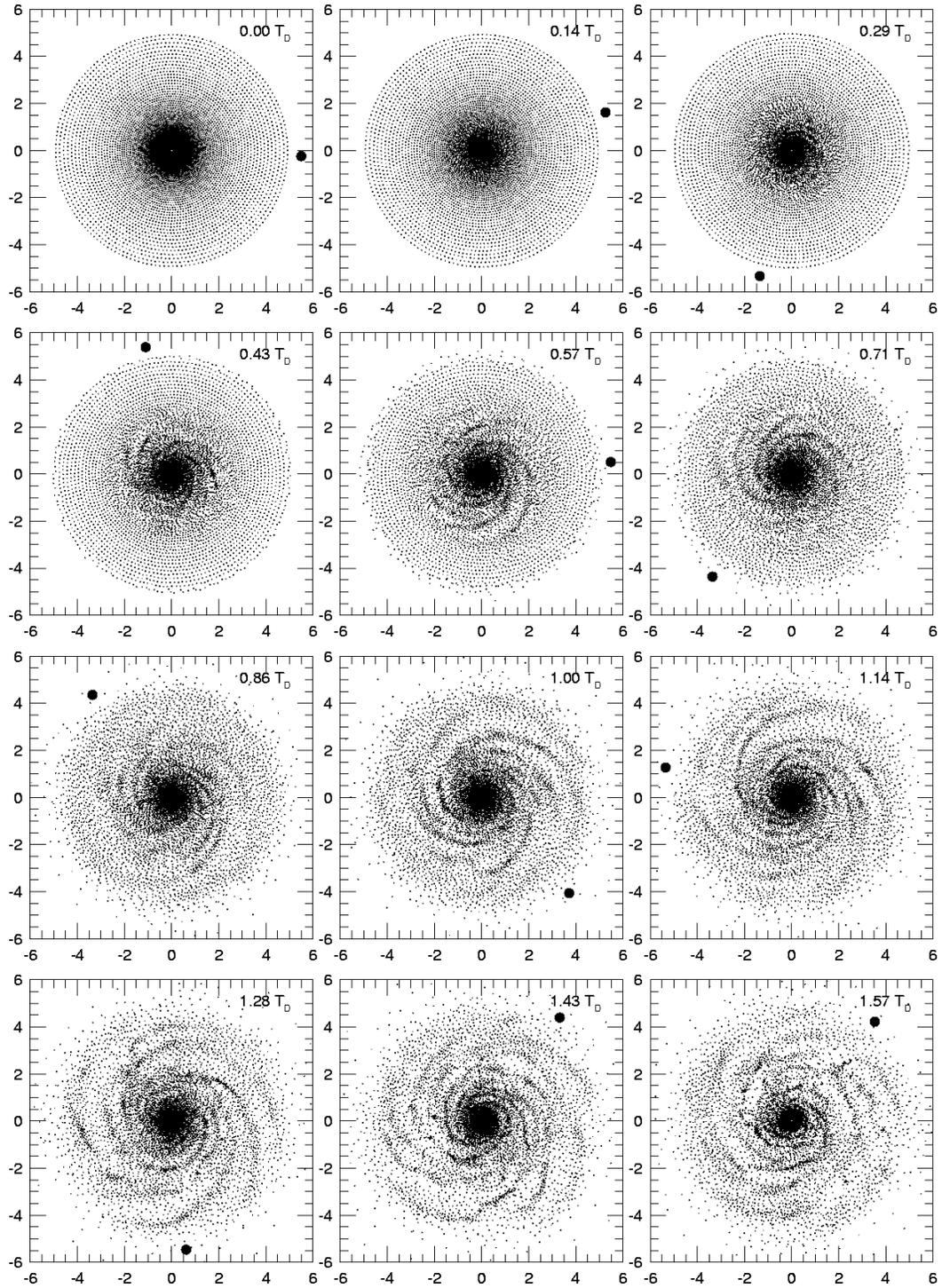}{7.25in}{0}{75}{75}{-230}{-10}
\caption[Low mass disk SPH simulation]
{\label{sph-lomas}
\footnotesize
Evolution of a disk with \mrat~$=0.2$ and with initial 
\qmin~$=1.5$ (simulation {\it scv2}).  Note the production of long
filamentary spiral arms and the production of multiple clumps at later 
times.}
\end{figure}

\clearpage

\begin{figure}
\plotfiddle{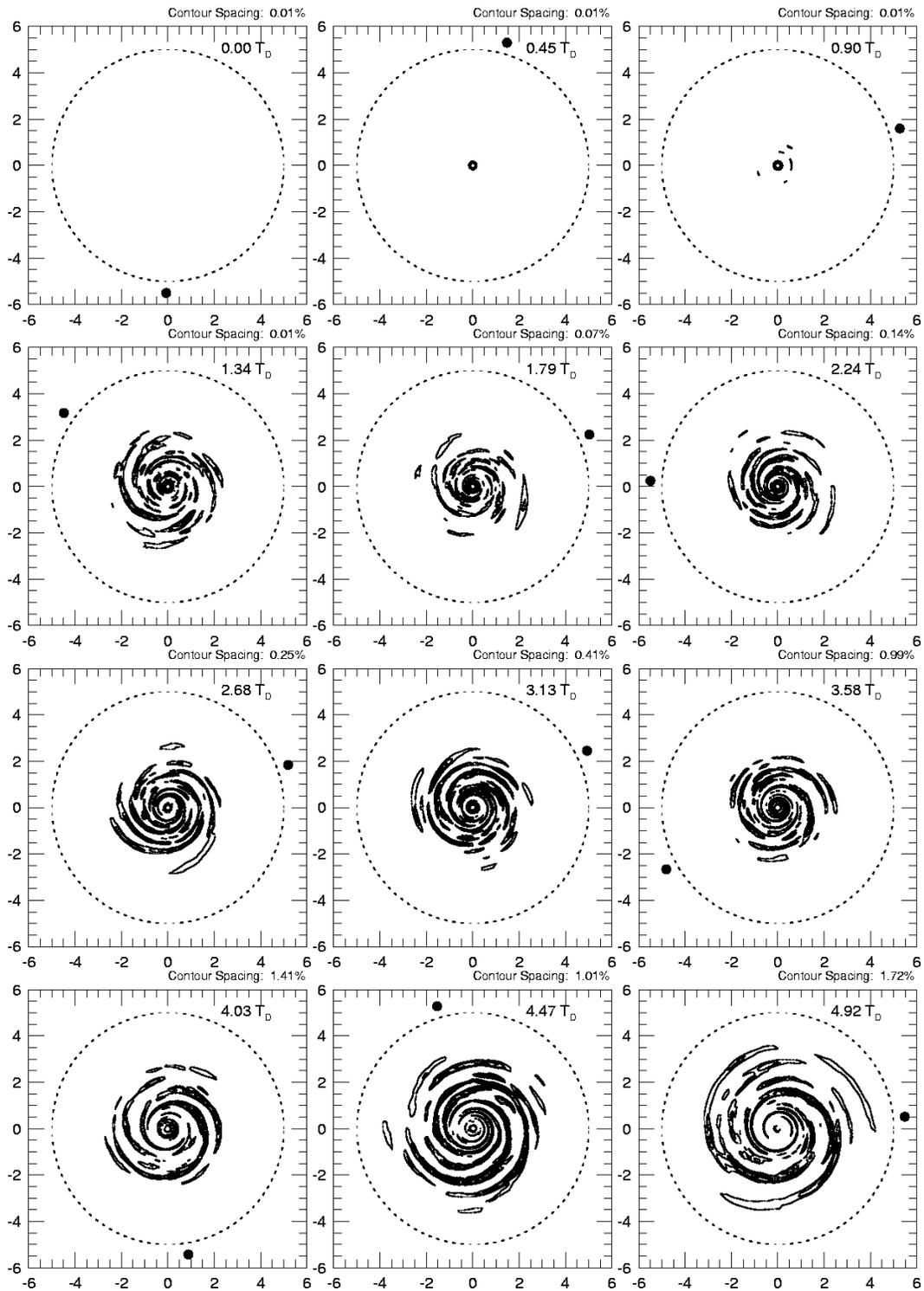}{7.25in}{0}{77}{77}{-230}{-15}
\caption[Low mass disk PPM simulation]
{\label{ppm-lomas}
\footnotesize
The same initial conditions as figure \ref{sph-lomas} with the PPM code
(simulation {\it pch2}).  A much longer evolution than figure 
\ref{sph-lomas} is possible here due to the low initial noise of PPM.}
\end{figure}

\clearpage

\begin{figure}
\plotfiddle{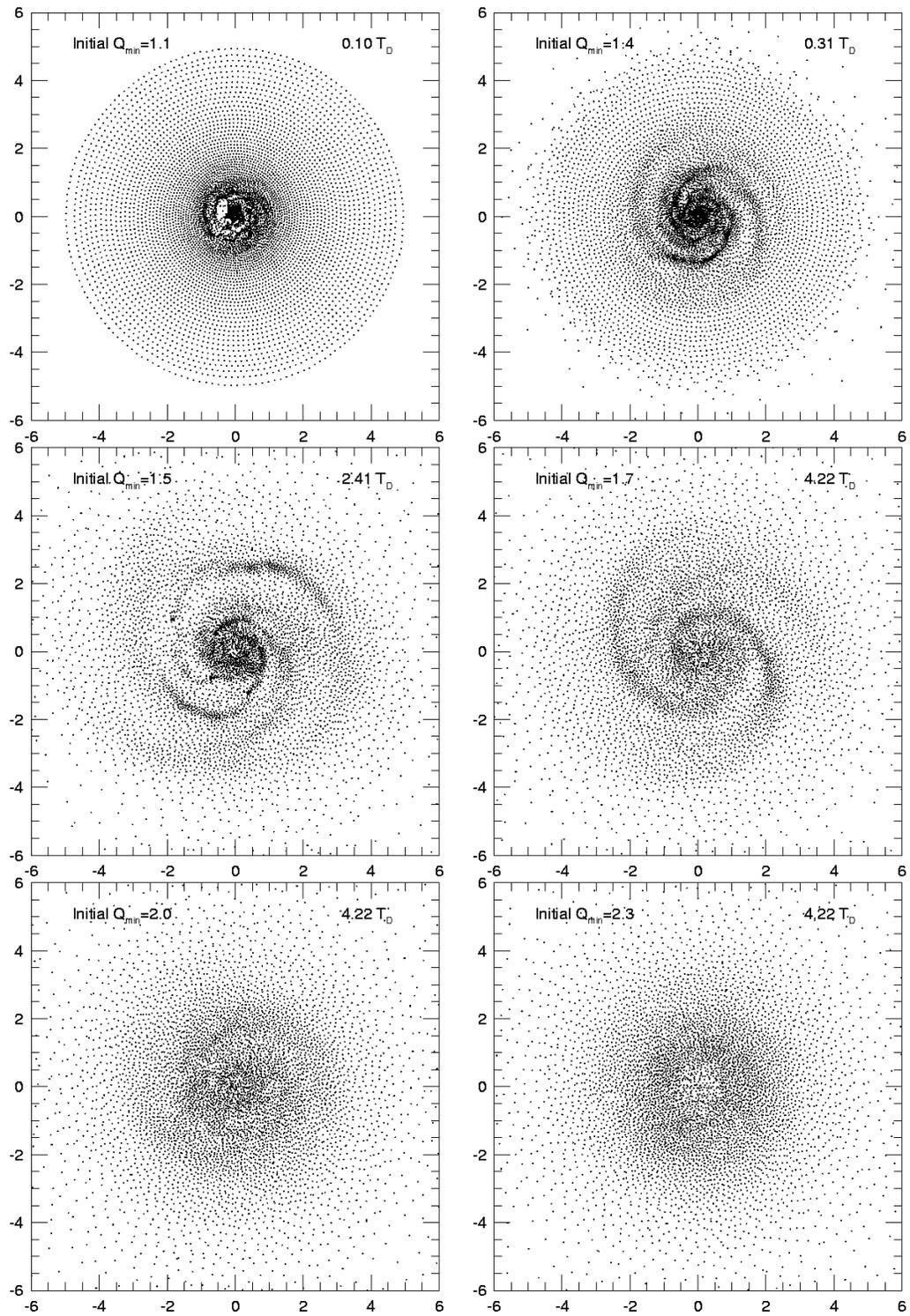}{7.25in}{0}{77}{77}{-240}{-15}
\caption[SPH simulations with identical mass disks but varying initial \qmin] 
{\label{sph-qvar}
\footnotesize
Late time snapshots of a series of disk simulations using
our SPH code. Each disk has the same disk mass of \mrat~$=0.8$ but 
varying \qmin (simulations {\it sqh1, -3,-4, -5}, and {\it -6}, as well
as {\it scv5} are shown).}
\end{figure}

\clearpage

\begin{figure}
\plotfiddle{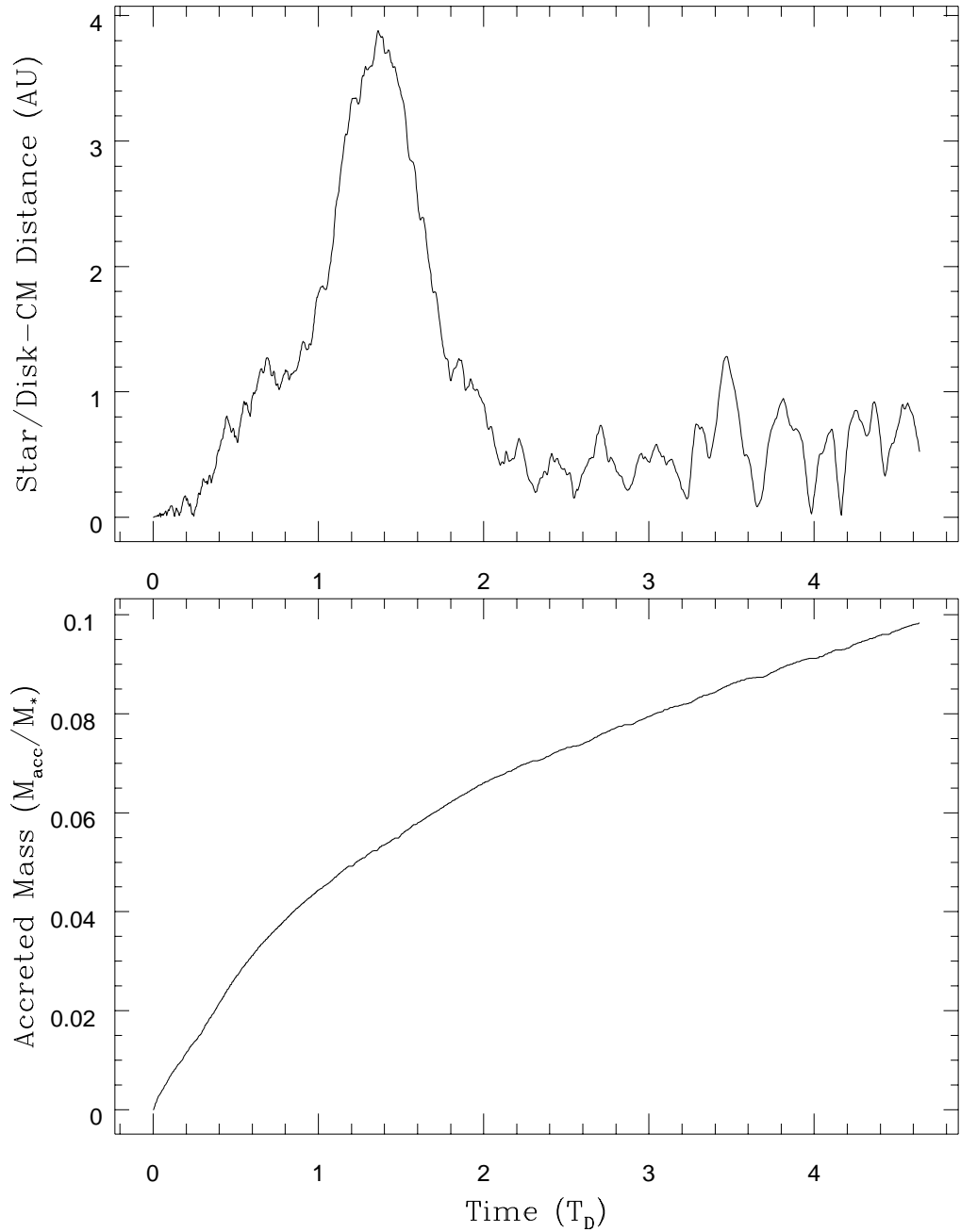}{6.5in}{0}{70}{70}{-200}{-30}
\caption[Mass accretion and the star motion for a high \qmin~SPH simulation]
{\label{hiq-trans}
\footnotesize
The distance between the star and the disk center of mass is
shown as a function of time in the top panel here, while the mass
accreted by the star is shown in the second. The simulation these
data are taken from is {\it sqh6}, which begins with \mrat~$=0.8$
and an initial minimum $Q$ value of 2.3.  With the units assumed for
our systems, the mass accretion rate is near $8\times 10^{-5} M_\odot$/yr
at its maximum.  When accretion begins to drain the inner disk
matter and the rate falls sufficiently (in this simulation, to
$\sim 3\times 10^{-5} M_\odot$/yr), the star falls to the center
of the system and returns much of its temporary increase in angular
momentum to the disk.  }
\end{figure}

\clearpage

\begin{figure}
\plotfiddle{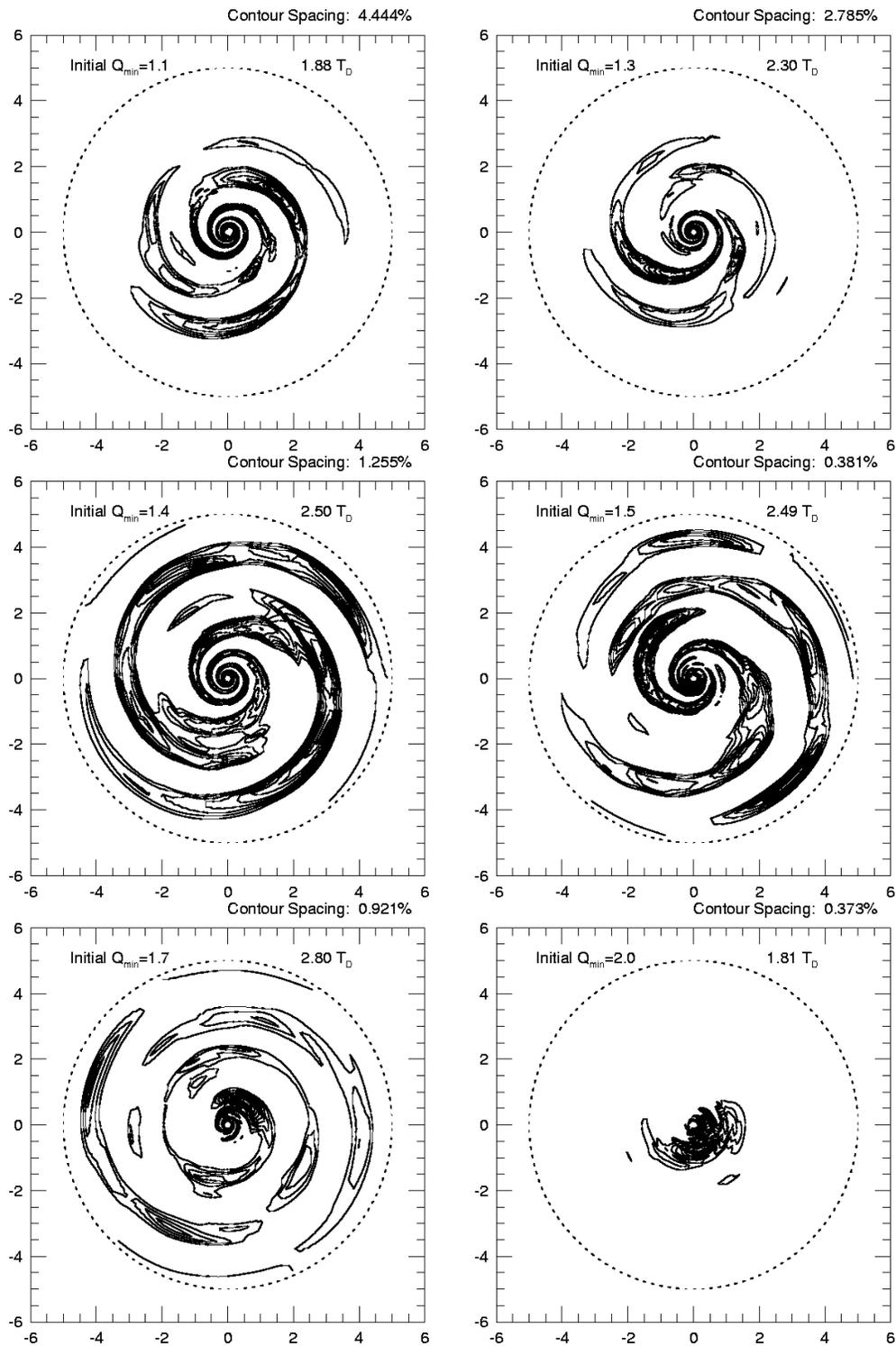}{7.25in}{0}{77}{77}{-240}{-15}
\caption[PPM simulations with identical mass disks but varying initial \qmin]
{\label{ppm-qvar}
\footnotesize
Late time snapshots of a series of disk simulations using
our PPM code. Each disk has the same disk mass of \mrat~$=0.8$ but 
varying \qmin (simulations {\it pqm1-5} as well as {\it pcm5} are
shown).}
\end{figure}

\clearpage

\begin{figure}
\plotfiddle{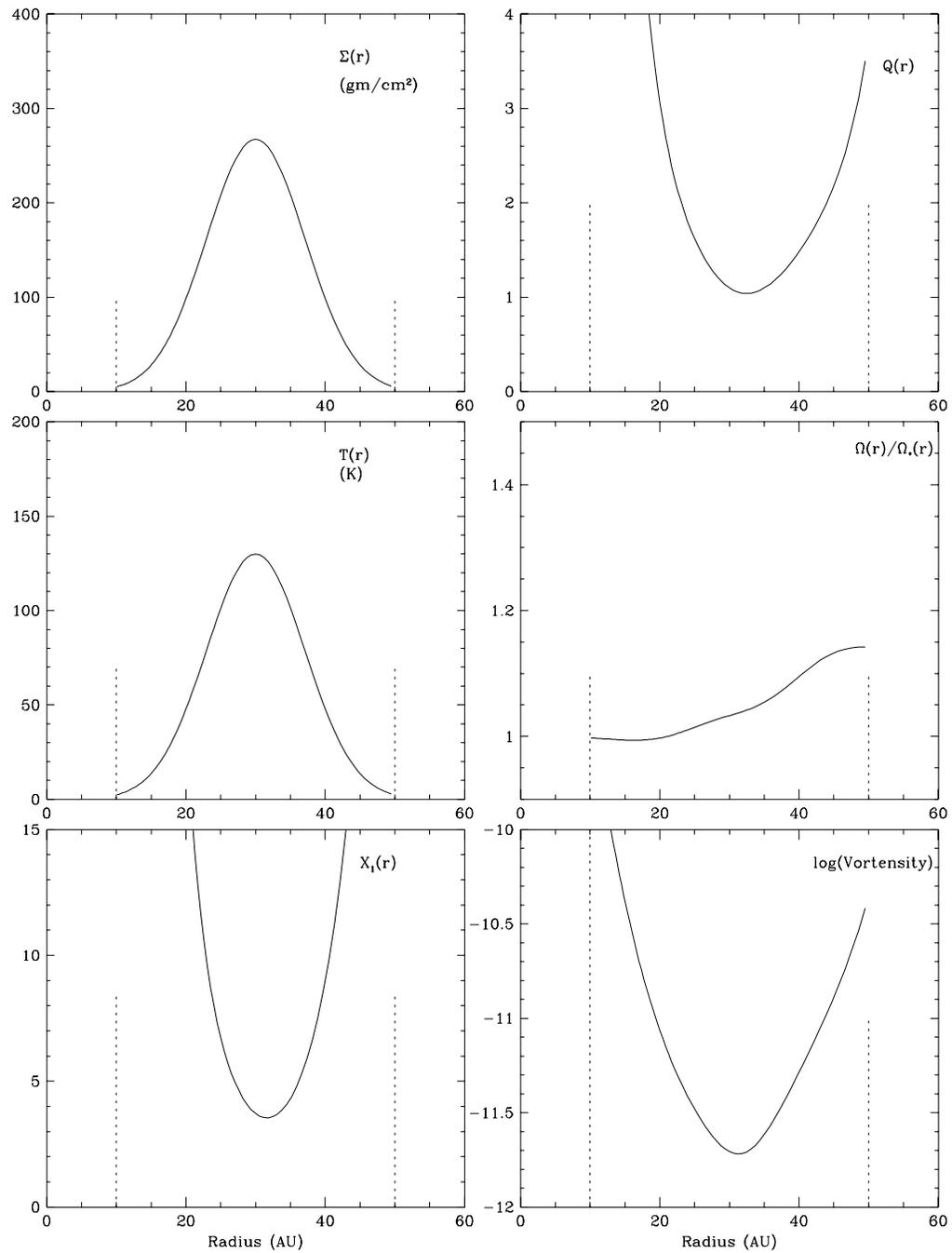}{6.75in}{0}{75}{75}{-240}{-50}
\caption[Initial conditions for Tori]
{\label{torus-init}
\footnotesize
Initial conditions for torus simulations.  Each frame contains the
same variable as in the corresponding frames in figures \ref{dinit-ppm}
and \ref{dinit-sph}.}
\end{figure}

\clearpage

\begin{figure}
\plotfiddle{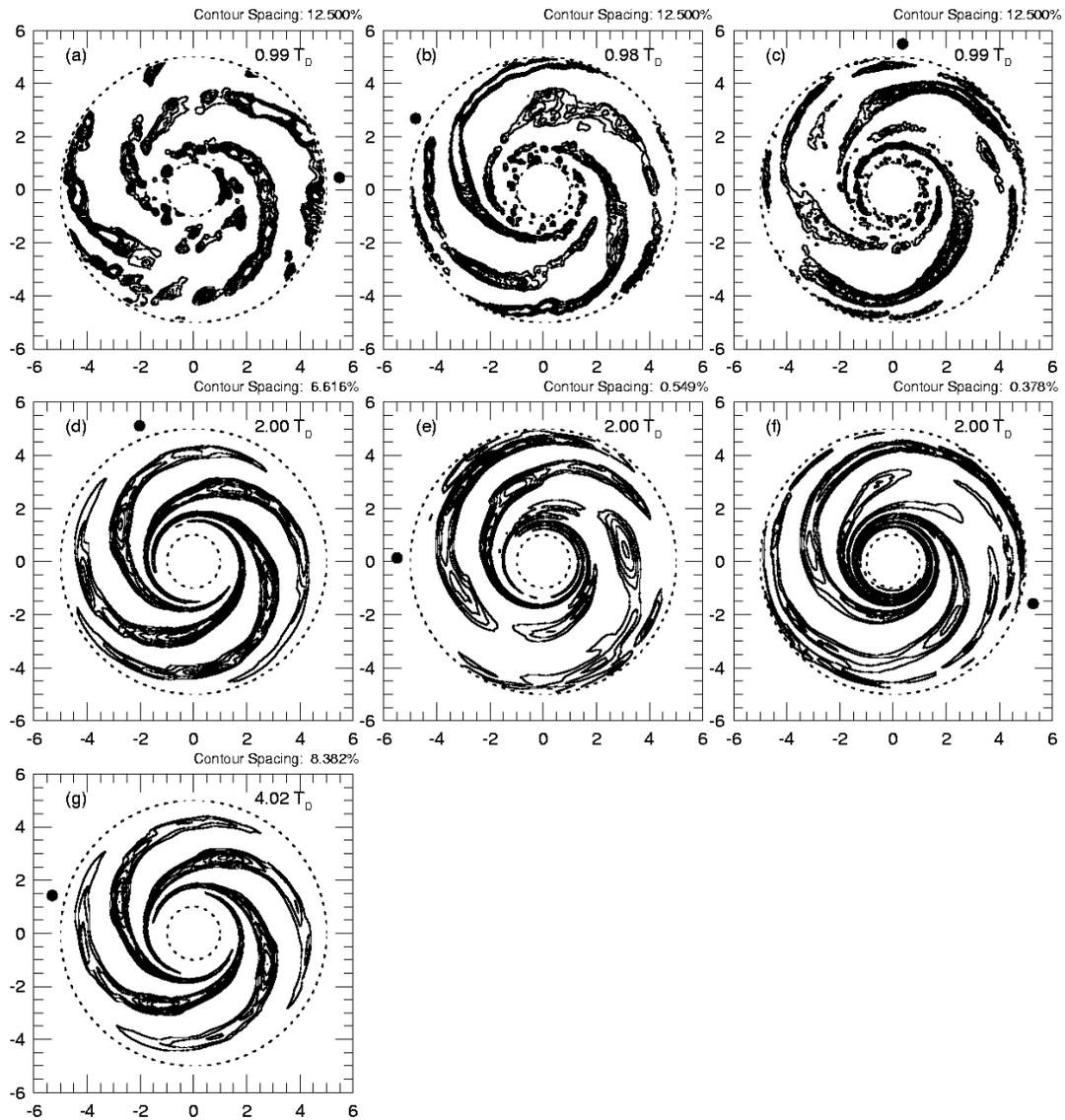}{6.65in}{0}{69}{69}{-210}{-20}
\caption[Comparison of the morphology of SPH and PPM simulations of tori]
{\label{tor-cmp}
\footnotesize
Late time snapshots of a torus with identical initial
conditions using (a-c) SPH with $\sim$7000, 14000, and 28000 particles
respectively. (d-f) PPM with 10$^{-3}$ amplitude random initial noise
at three grid resolutions: 40$\times$150, 60$\times$225 and
80$\times$300 and (g) PPM simulation with low initial noise (10$^{-8}$)
at 40$\times$150 grid resolution. For comparison purposes, the SPH runs
are mapped onto a grid identical to that used for the corresponding
PPM runs.}
\end{figure}

\clearpage

\begin{figure}
\plotfiddle{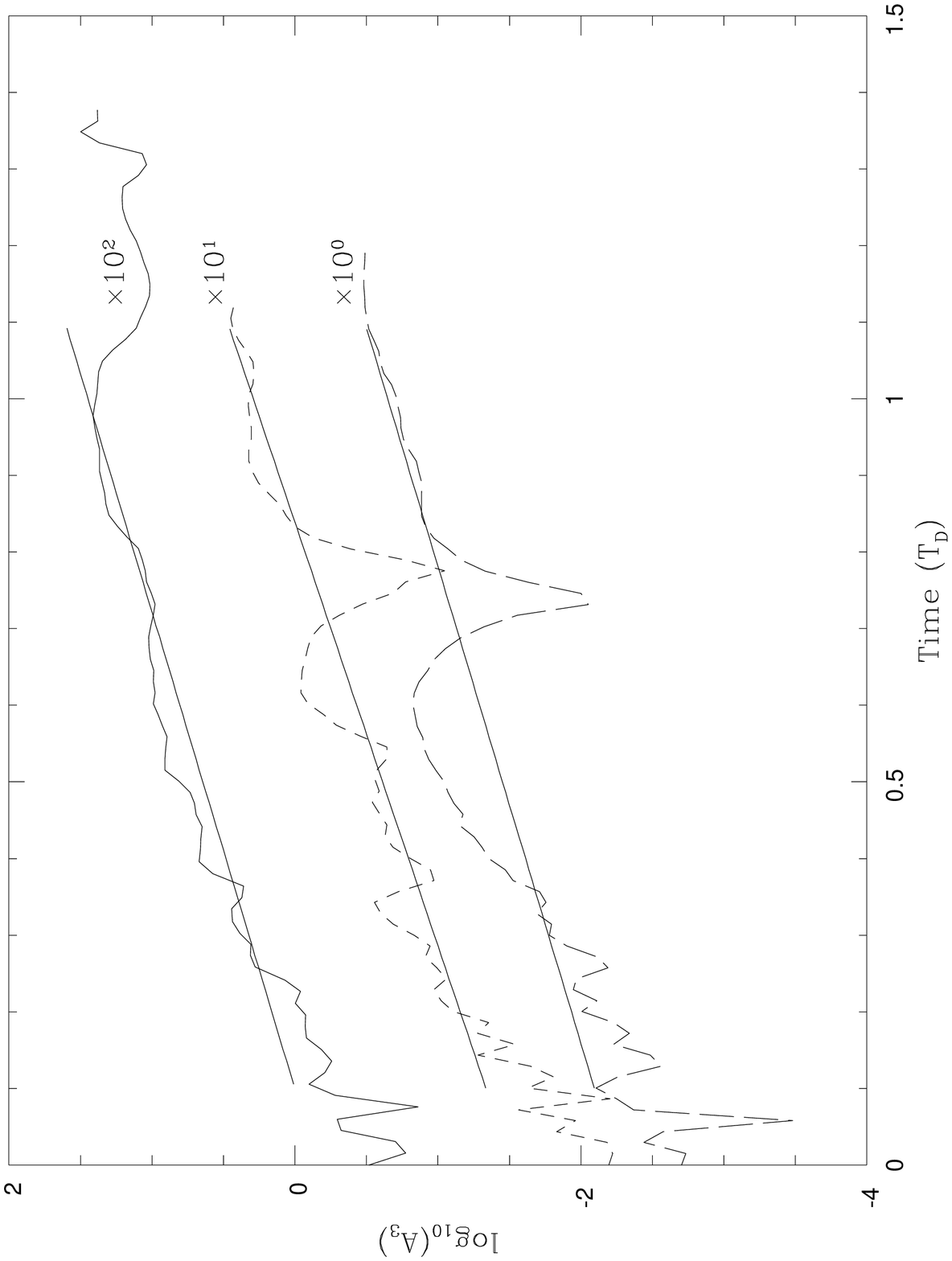}{2.99in}{-90}{42}{42}{-180}{240}
\plotfiddle{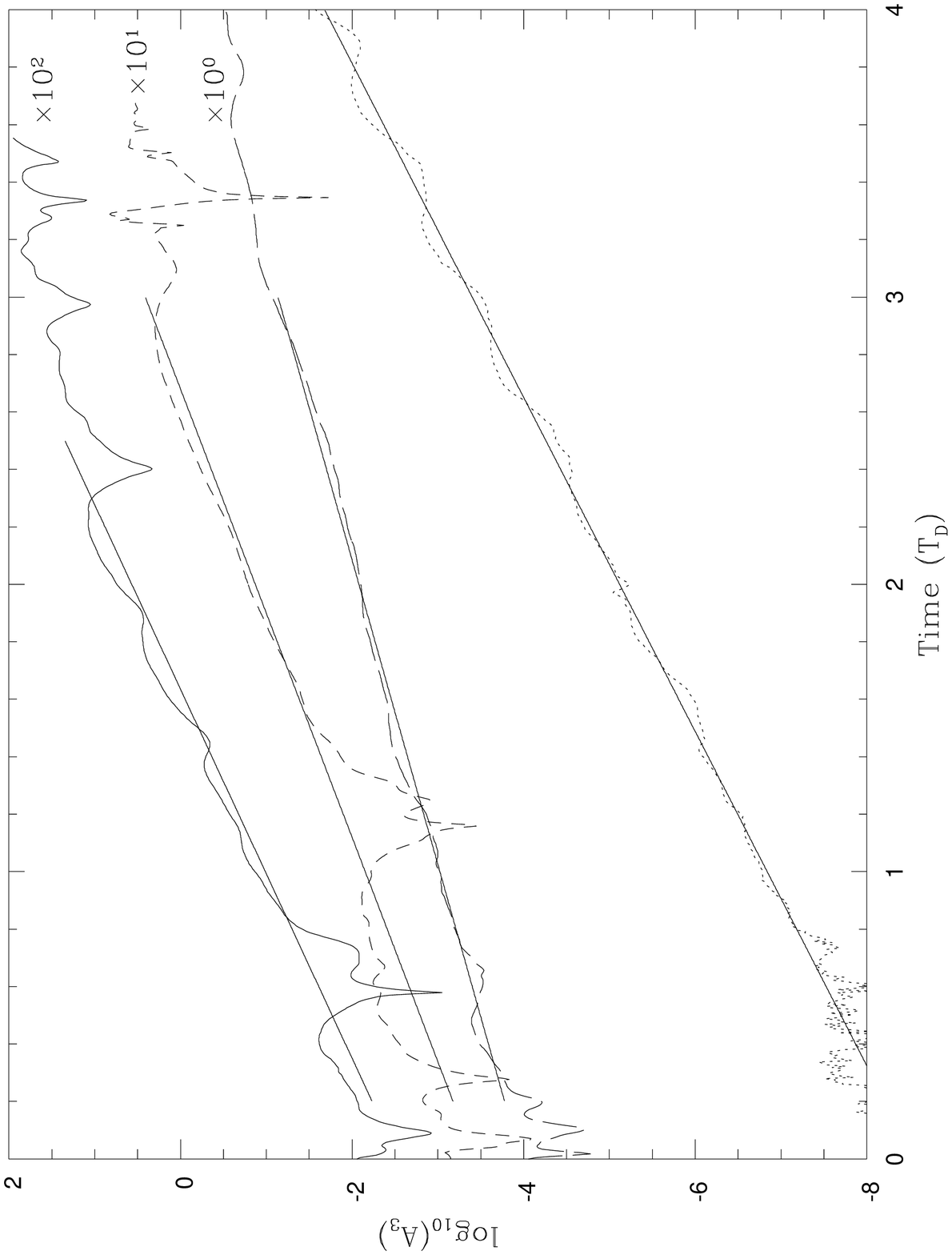}{2.99in}{-90}{42}{42}{-180}{240}
\caption[Amplitudes and best fits for the $m=3$ spiral pattern of tori]
{\label{torm3_30}
\footnotesize
Amplitudes and linear best fits for the $m=3$ pattern at the center of 
the torus ($R=$30 AU) for different resolution SPH and PPM simulations.
The top panel shows SPH simulations. The lowest resolution ($\sim 7000$ 
particles) is denoted with a solid curve while double resolution ($\sim
14000$ particles) is denoted with a short dashed curve and the highest 
resolution ($\sim 28000$ particles) is shown with a long dashed curve.  
Each of the fits are shown as solid lines. Bottom panel: PPM simulations
with the two lowest resolution runs denoted by a solid and dotted line for
the 10$^{-3}$ and 10$^{-8}$ amplitude initial noise runs respectively.
The short dashed curve represents the middle resolution and the long 
dashed line represents the highest resolution run. Solid lines denote
the best fit curves for each of the runs and displayed only for the
times for which the fit was derived. Each of the SPH runs and the PPM 
runs with 10$^{-3}$ noise are artificially multiplied by a factor of
1, 10 or 100 in order to distinguish between the different runs on the
plots.}
\end{figure}

\clearpage

\begin{figure}
\plotfiddle{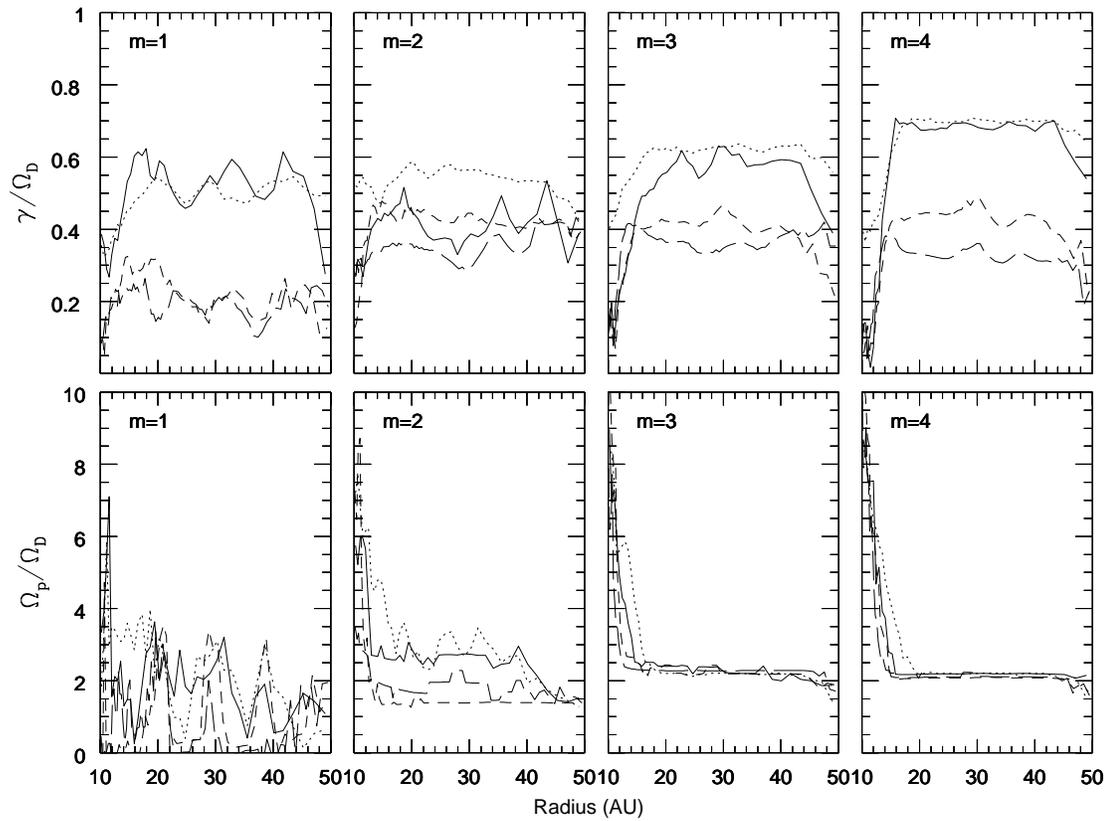}{5.5in}{-90}{58}{58}{-230}{450}
\caption[Growth rates for the $m=1-4$ spiral patterns derived from the PPM 
simulations]
{\label{ppmpatgrw1-4}
\footnotesize
Growth rates and pattern speeds for the $m=1-4$ patterns
derived from PPM simulations. The increase in the pattern
speed at the inner torus edge probably represents a boundary 
influence and we do not consider it to be significant. Each curve
uses the same representation as in figure \ref{torm3_30} to denote
low, moderate and high resolution runs. }
\end{figure}

\clearpage

\begin{figure}
\plotfiddle{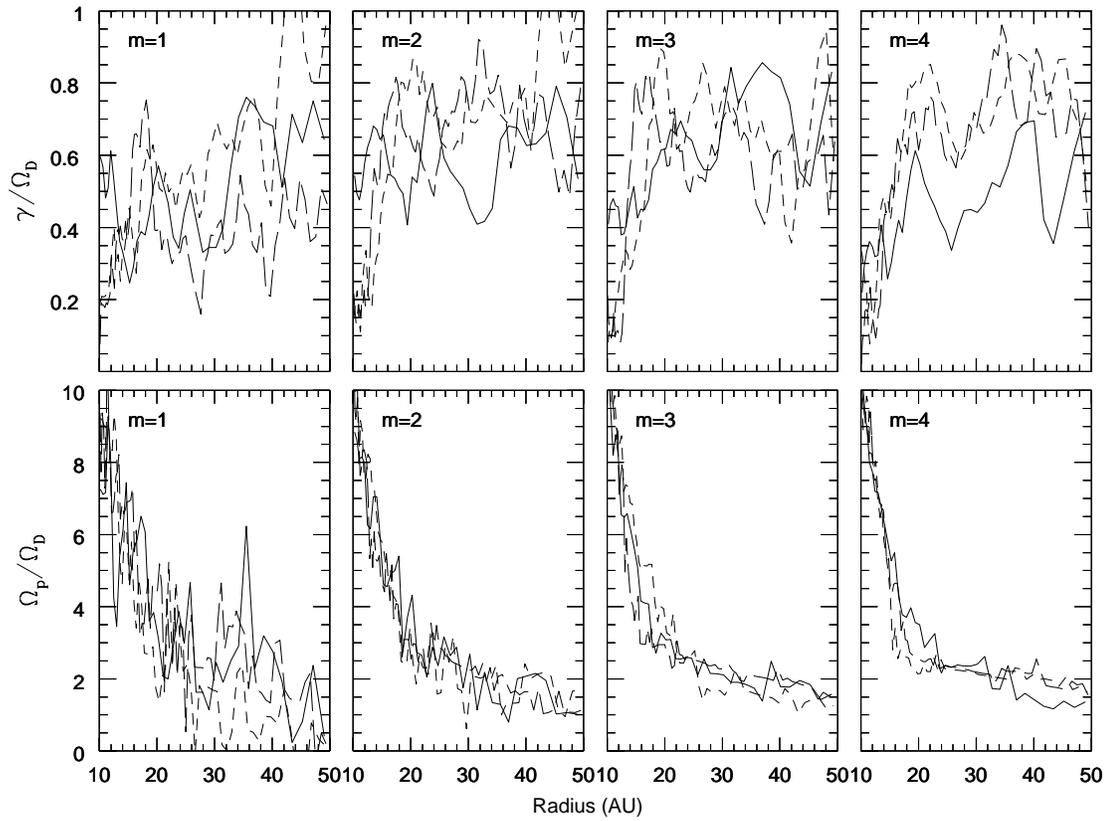}{5.5in}{-90}{58}{58}{-230}{450}
\caption[Growth rates for the $m=1-4$ spiral patterns derived from the SPH 
simulations]
{\label{sphpatgrw1-4}
\footnotesize
Growth rates and pattern speeds for the $m=1-4$ patterns derived from 
the SPH simulations. Each curve uses the same representation as in
figure \ref{torm3_30} to denote low moderate and high resolution
runs. }
\end{figure}

\clearpage

\begin{figure}
\plotfiddle{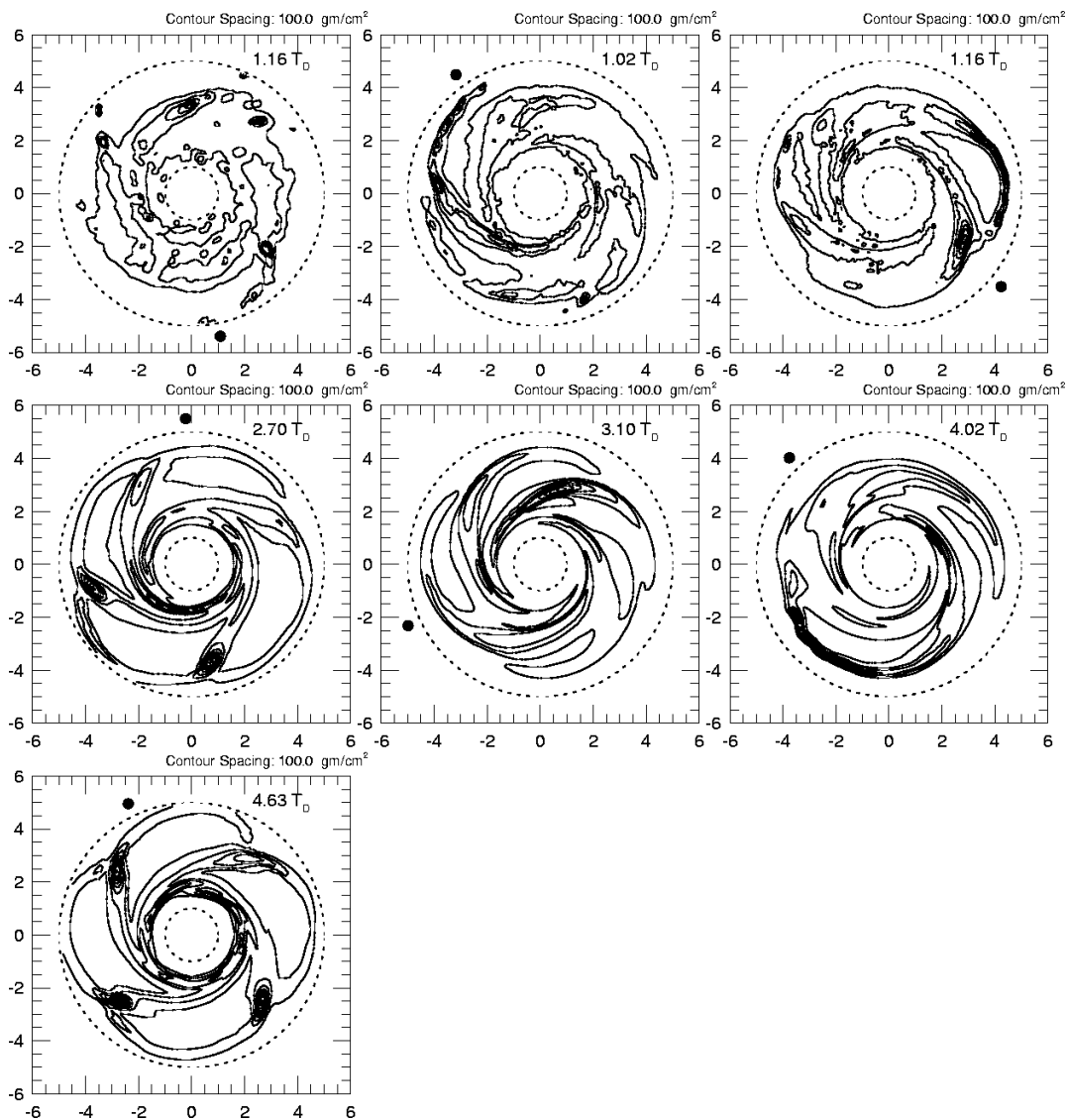}{6.52in}{0}{69}{69}{-210}{-20}
\caption[Later time comparison between SPH and PPM simulations of tori]
{\label{tor-late}
\footnotesize
Late time snap shots of the same simulations as figure \ref{tor-cmp}
above. Here we plot density rather than density variation to accentuate
collapse behavior. Contours units are gm/cm$^2$ and are linearly spaced
from 0 gm/cm$^2$ (not shown) upward with spacing between contours as
noted at the upper right of each frame. Because the collapse behavior
occurs at a somewhat different time for each of the runs, the plots
are not shown at the same time as any other plot. Rather, we show
the morphology shortly after collapse begins in each simulation,
at whatever time during the simulation that occurred. Each of the SPH
runs are mapped onto a grid identical to that used by the corresponding
PPM simulation. The dashed curves denoting the inner and outer grid
radii therefore have no meaning for these runs.}
\end{figure}

\clearpage

\begin{figure}
\plotfiddle{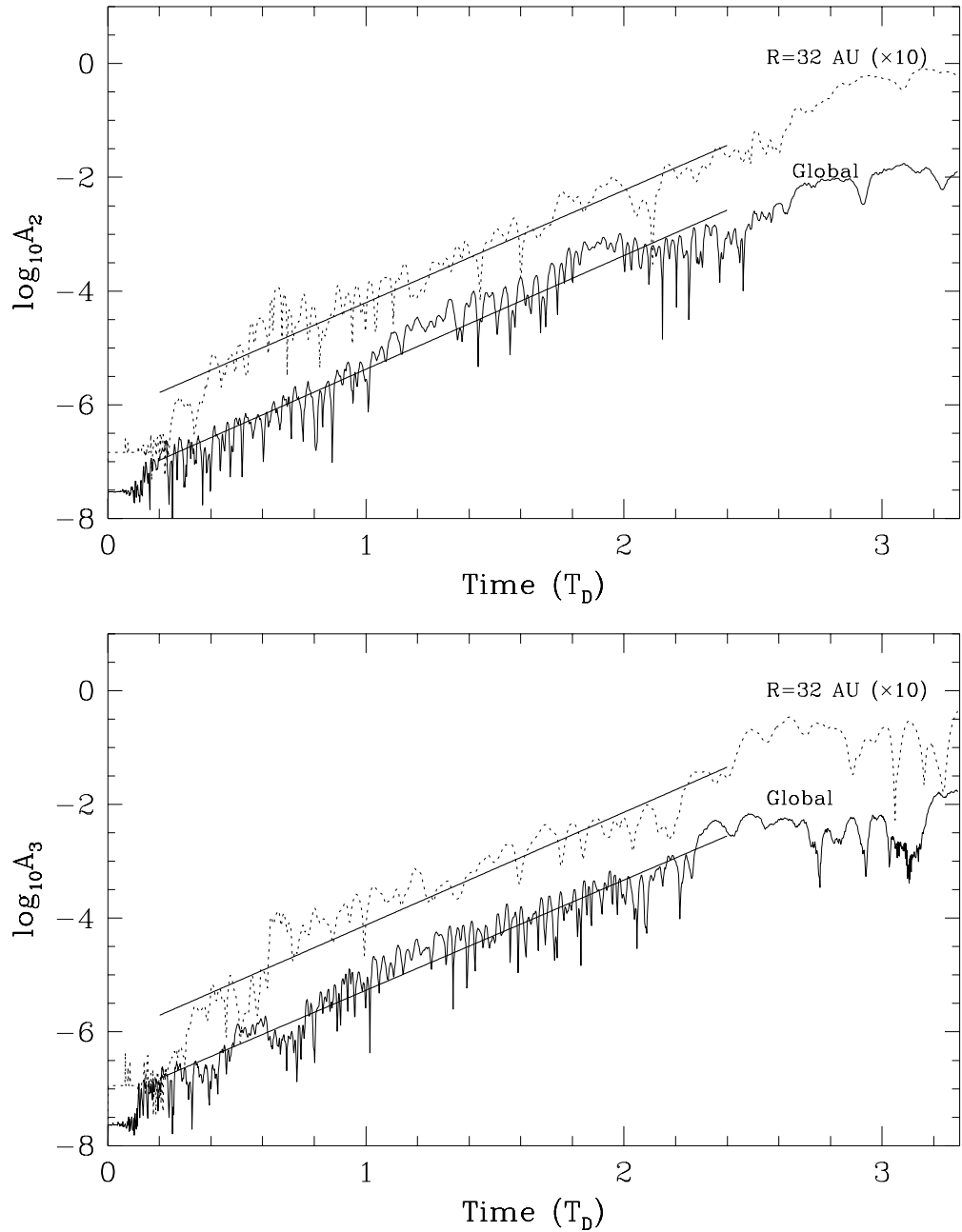}{7.05in}{0}{70}{70}{-210}{-35}
\caption[Amplitudes and fits for the $m=2$ and $m=3$ patterns of the 
simulation in figure \ref{ppm-himas}]
{\label{m2and3hi}
\footnotesize
The amplitudes and fits for the $m=2$ (top frame) and $m=3$ (bottom
frame) patterns derived from the simulation shown in figure 
\ref{ppm-himas}. The amplitude ($\times$~10) near the middle of
the power law portion of the disk as well as the globally integrated 
amplitudes for each pattern are shown. }
\end{figure}

\clearpage

\begin{figure}
\plotfiddle{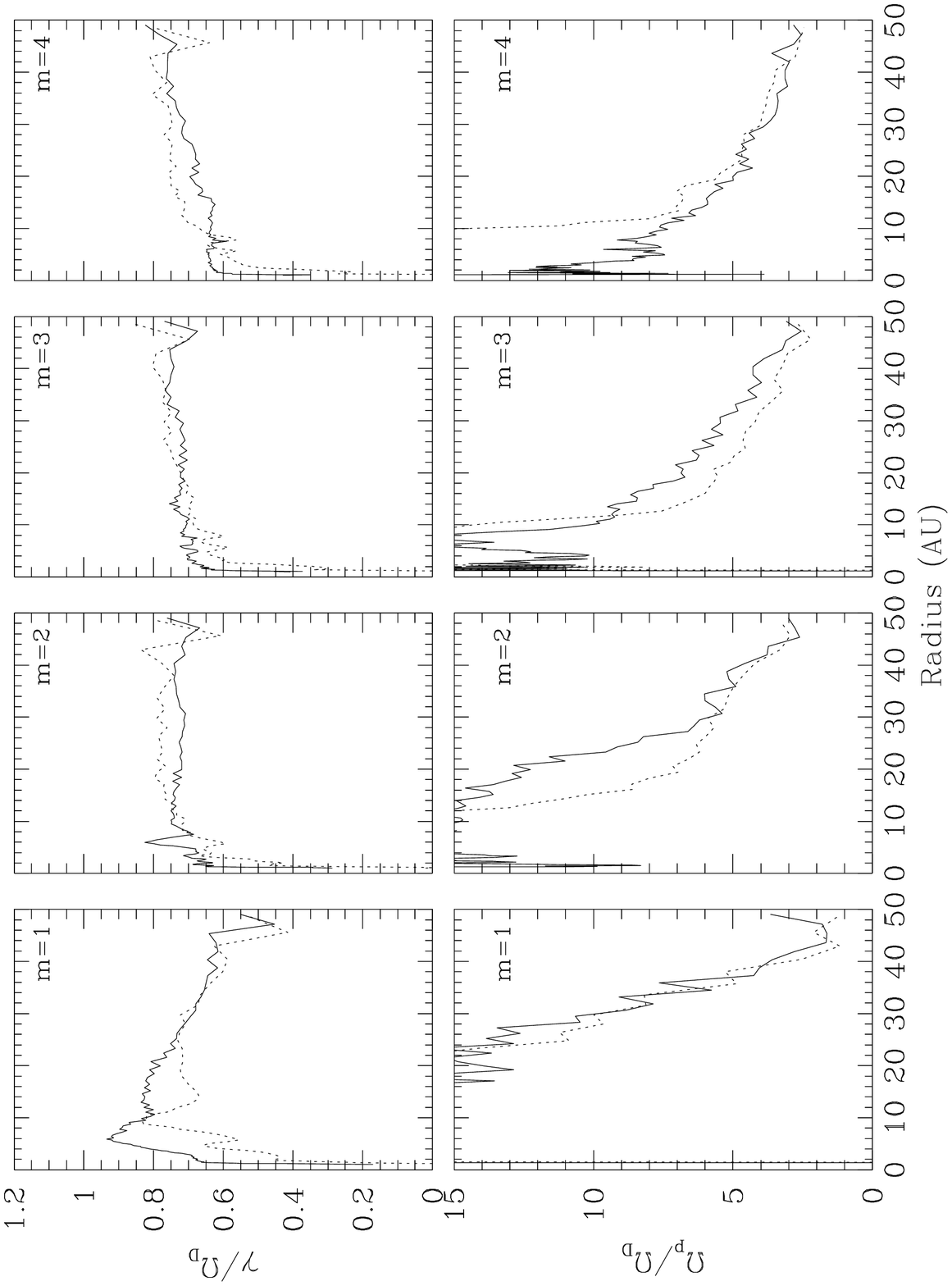}{5.5in}{-90}{56}{56}{-220}{400}
\caption[Growth rates and pattern speeds for the simulation in figure
\ref{ppm-himas} as a function of radius]
{\label{patgrwm1-4hi}
\footnotesize
The growth rates and pattern speeds for the $m=1$--4 spiral
arm patterns. The simulation from which these are derived is the same
as is shown in figure \ref{ppm-himas}. The solid lines represent the
moderate resolution simulation {\it pch6}  while the dotted lines 
represent results from the lower resolution simulation {\it pcm6}.}
\end{figure}

\clearpage

\begin{figure}
\plotfiddle{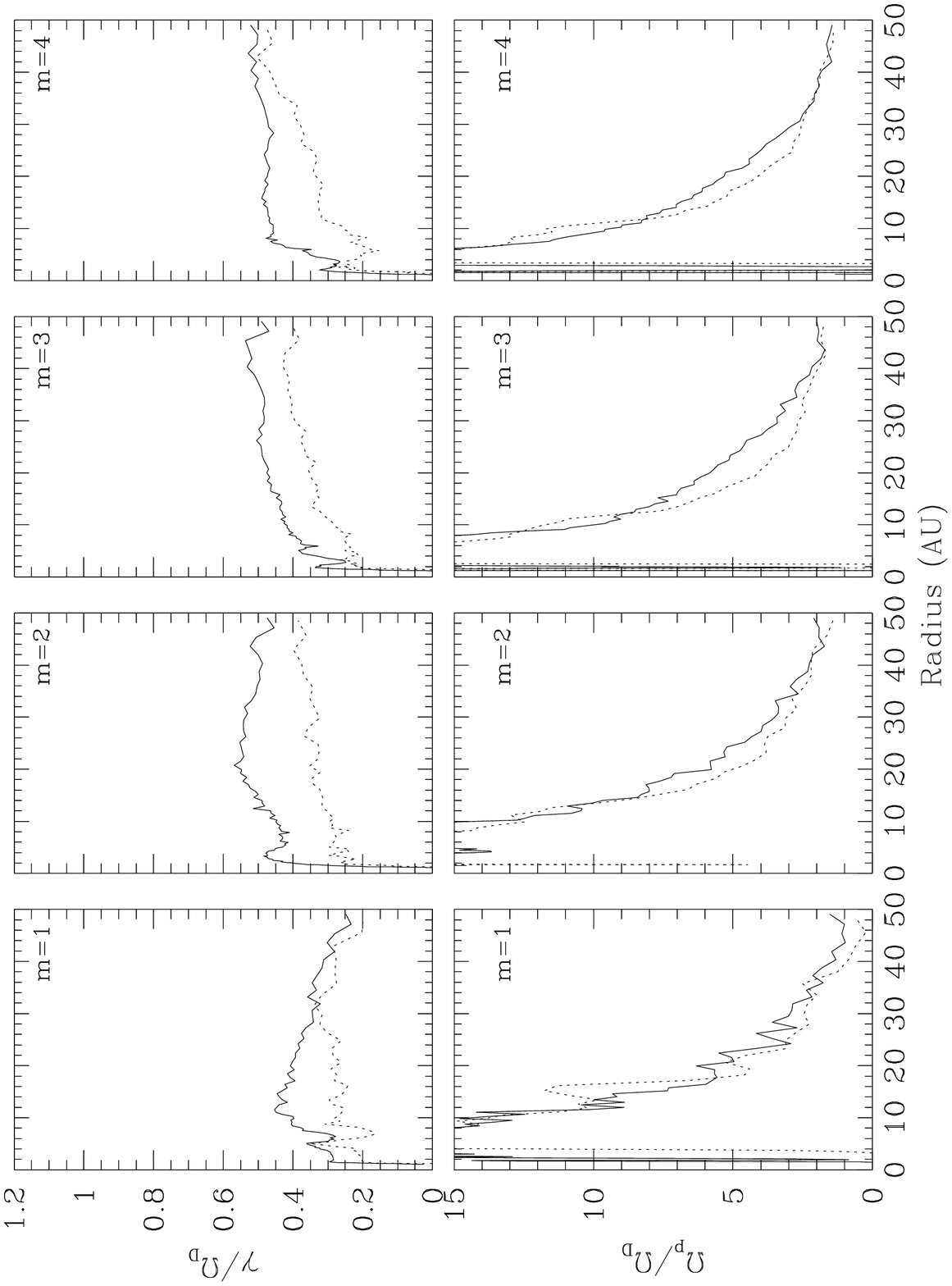}{5.5in}{-90}{56}{56}{-220}{400}
\caption[Growth rates and pattern speeds for the simulation in figure
\ref{ppm-lomas} as a function of radius]
{\label{patgrwm1-4low}
\footnotesize
The growth rates and pattern speeds for the $m=1$--4 spiral
arm patterns. The simulation from which these are derived is the same
as is shown in figure \ref{ppm-lomas}. The solid lines represent the
moderate resolution simulation {\it pch2}  while the dotted lines 
represent results from the lower resolution simulation {\it pcm2}.  }
\end{figure}

\clearpage

\begin{figure}
\plotfiddle{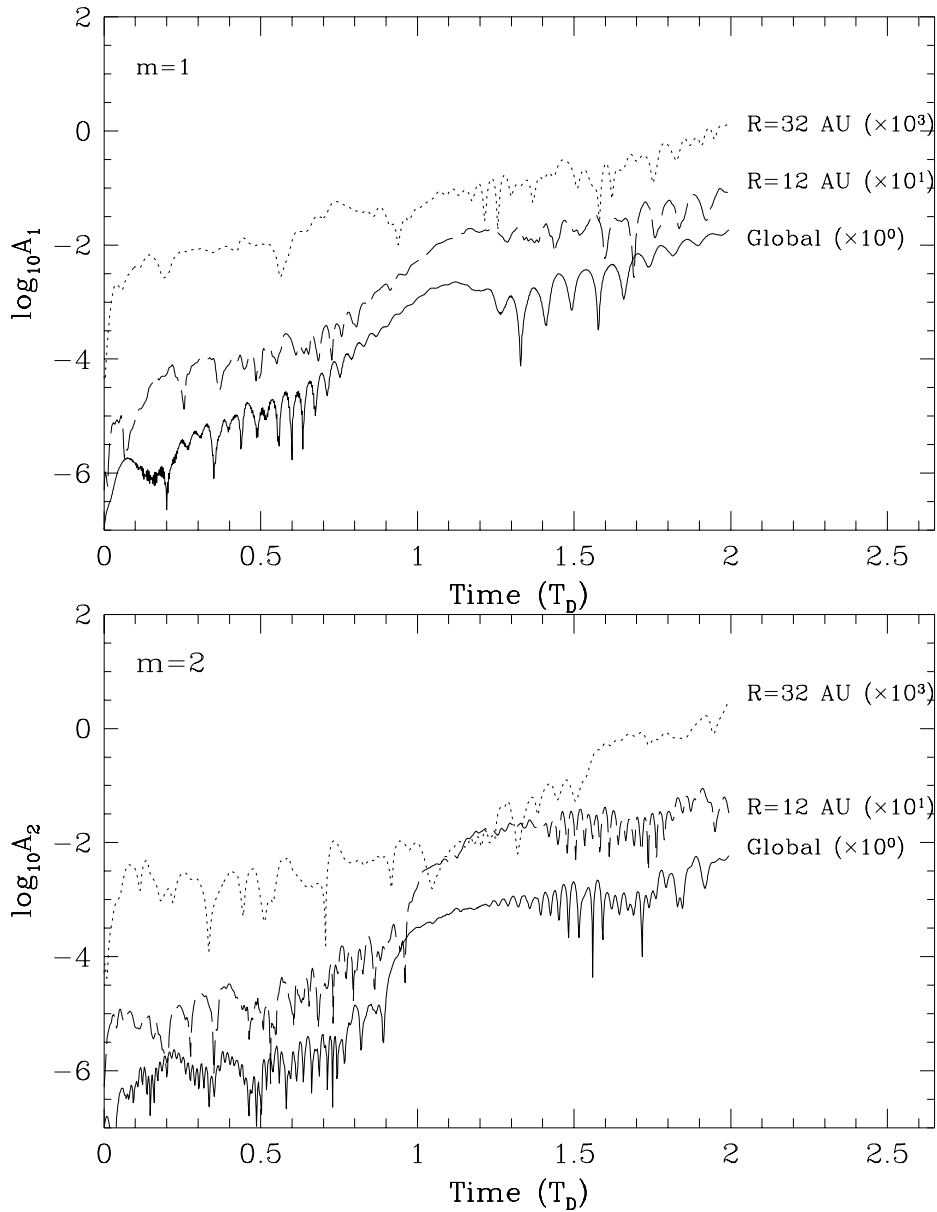}{6.0in}{0}{65}{65}{-200}{-30}
\caption[Amplitude of the $m=1$ and $m=2$ patterns at various locations
in the disk simulation {\it pqm5}]
{\label{hiq-amp}
\footnotesize
Amplitude of the $m=1$ and $m=2$ spiral patterns at various locations in
the disk simulation {\it pqm5}. The outer portion of this disk is
initially quiescent.  The amplitude of the $m=1$ pattern does begin
growing immediately, however near $T_D\sim 1$ it experiences a `hump'
in its amplitude as instability propagates towards larger radii.
The region near the density maximum ($R\sim$12 AU) experiences little
initial growth in $m>1$ patterns, but once instabilities enter that
region (cf. the lower right panel of figure \ref{ppm-qvar}) they
quickly grow to dominate the instability amplitude over the entire
system. }
\end{figure}

\clearpage

\begin{figure}
\plotfiddle{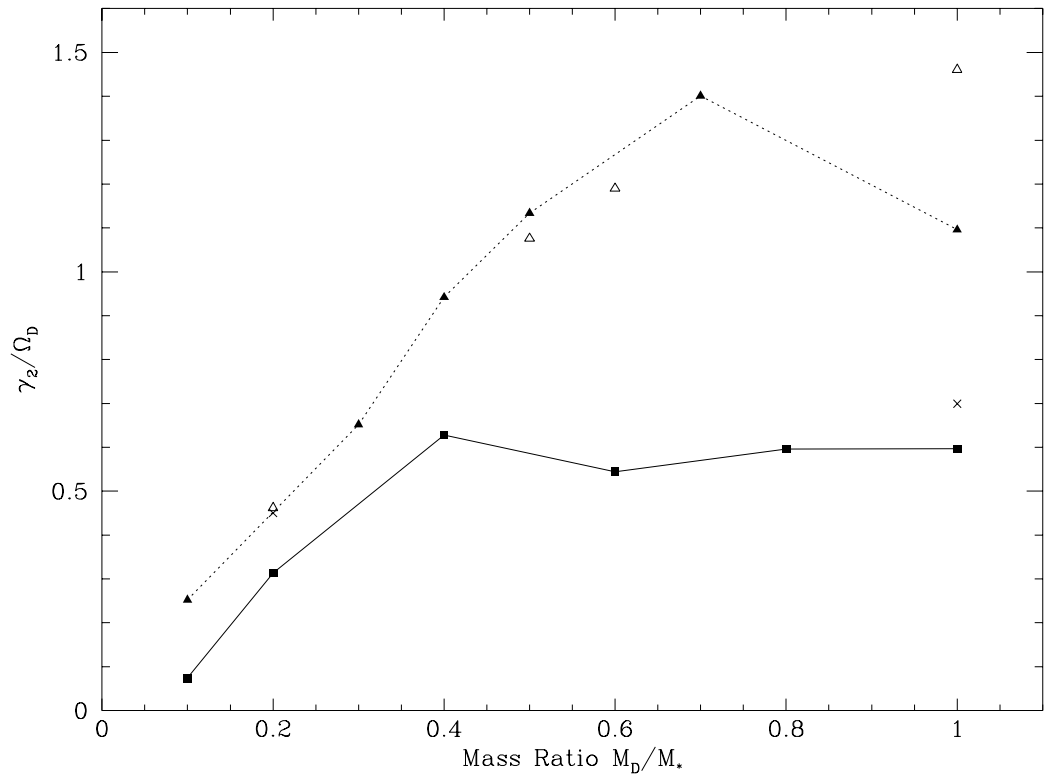}{5.5in}{-90}{55}{55}{-210}{400}
\caption[Growth rates for the $m=2$ pattern as a function of disk mass]
{\label{disk-mratrates}
\footnotesize
Growth rates for the $m=2$ pattern for PPM simulations using a reflecting
outer boundary condition at moderate resolution (solid squares) and
at higher resolution ($\times$). A second series of simulations
with an infall boundary condition are shown with solid triangles
and at higher resolution with open triangles. }
\end{figure}

\clearpage

\begin{figure}
\plotfiddle{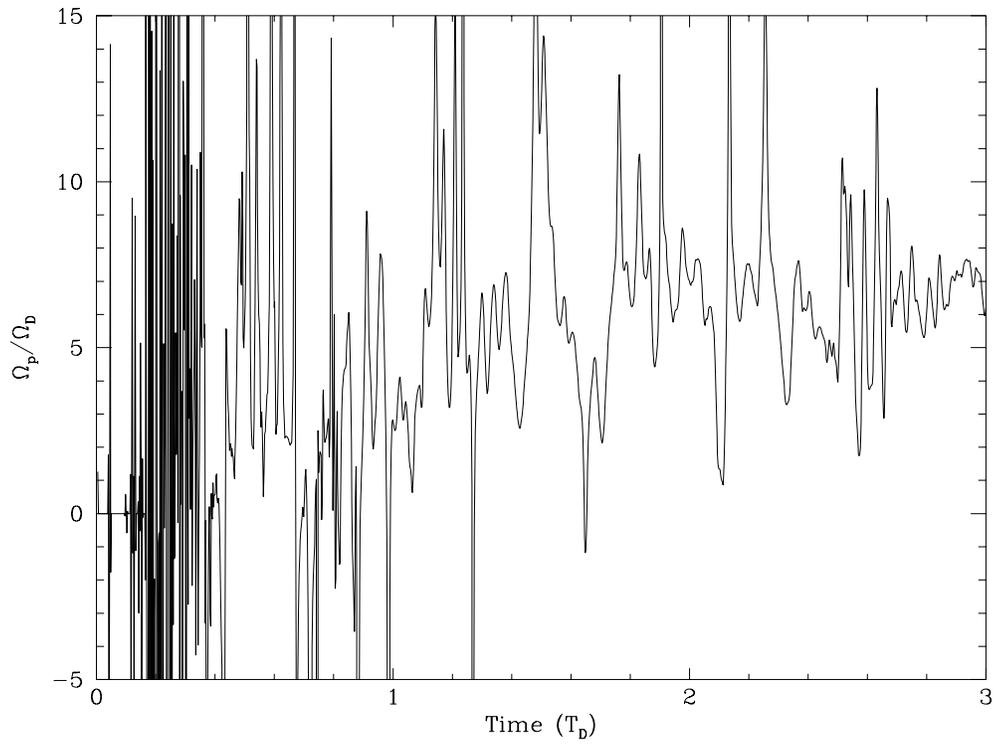}{5in}{-90}{52}{52}{-210}{400}
\caption[Pattern speed for the $m=2$ pattern as a function of time
for the disk shown in figure \ref{ppm-himas}]
{\label{m2pat_tme}
\footnotesize
Pattern speeds for the $m=2$ pattern as a function of time for the
disk shown in figure \ref{ppm-himas}.  The pattern speed is for the
pattern at a radius $R\approx$32 AU from the star, which is near 
the middle of the region where the density is a power law in form. }
\end{figure}

\clearpage

\begin{figure}
\plotfiddle{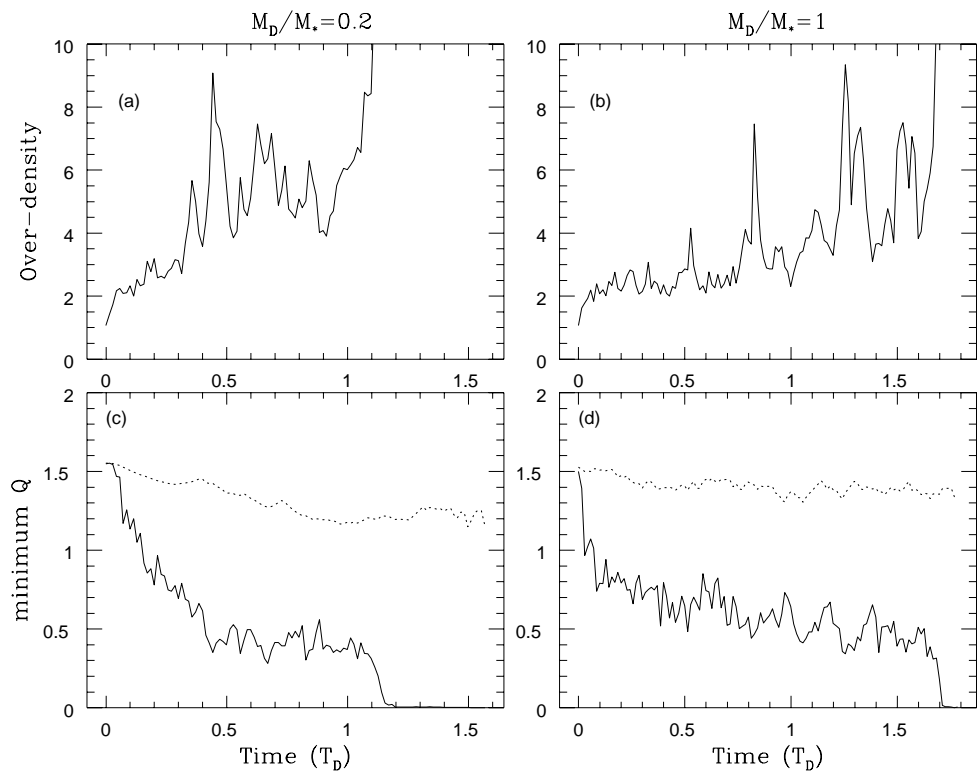}{5in}{-90}{52}{52}{-200}{400}
\caption[Over density and minimum $Q$ of low and high mass disks derived
from SPH simulations]
{\label{odqplot}
\footnotesize
Maximum over-density in SPH disks of low (a) and high (b)
disk/star mass ratio plotted vs. time (simulations {\it scv2} and
{\it scv6}).  Each disk begins with an initial \qmin~$=1.5$. 
Upon clumping the over-density assumes values 2-3 orders of 
magnitude larger than are plotted here and are omitted from these
graphs. (c) and (d) show the minimum $Q$ value for the same disks as
shown in (a) and (b) with both minimum azimuth average values 
(dotted line) and local minimum (solid line) values shown.}
\end{figure}

\clearpage

\begin{figure}
\plotfiddle{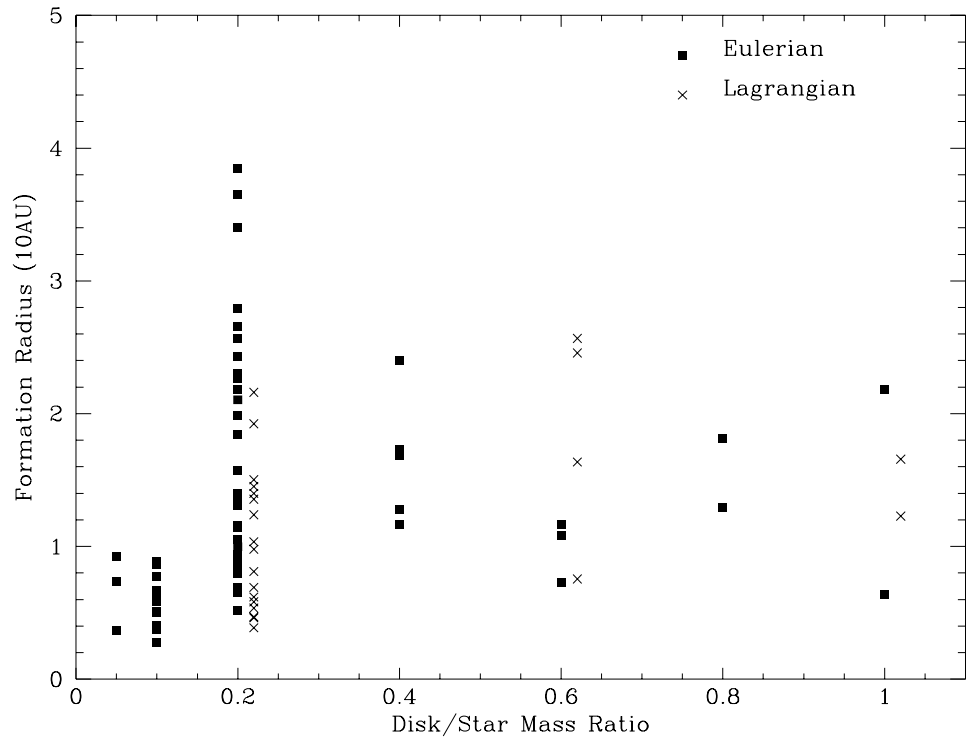}{5in}{-90}{52}{52}{-200}{400}
\caption[Radius at which clumps formed in the SPH simulations for each
of the series varying disk mass]
{\label{formrad}
\footnotesize
Formation radius (in units of 10 AU) for each clump vs. disk
mass. Each disk in the series {\it scv0-scv6} begins with an initial  
minimum $Q$ of 1.5.  Clumps form predominantly in the inner half of the 
disk, with only the \mrat~$=0.2$ disk showing clump formation over the
entire range in radius. In the simulations in which more than $\sim$10 
clumps formed an exact number becomes difficult to determine.  
Collisions between clumps and fission of a single clump into two (due 
to accretion of a large amount of angular momentum over a short 
time) make long term identification of any clump which has undergone 
a collision or fission event ambiguous and we do not include them
here. Filled triangles represent simulations evolved under an
Eulerian isothermal assumption (see section \ref{eos-sec}) while the
crosses (offset from their disk masses slightly to avoid confusion)
represent disks with the Lagrangian isothermal assumption. }
\end{figure}

\doublespace
\normalsize

%% file: interlude1.tex
\chapter{Interlude One}

The simulations described in chapter \ref{isodisk} impose a radius dependent
temperature profile on the disk which does not vary in time.  Other 
simulations (Boss 1997, Pickett \etal 1998) have implemented a similar 
prescription for both locally isothermal ($\gamma=1$) and locally adiabatic
($\gamma>1$) equations of state. In several instances the simulations done 
here or in their work have resulted in the collapse of part of the disk 
into one or more clumps, which some have interpreted as forming planetary 
or stellar companions in the limit of rather strong cooling.

Interpreting the formation of a clump as similar to the formation of a 
real physical object is premature. The dynamical character of a given
simulation can be called into question based on the same grounds.
The reason for such strong statements as these are that making the assumption
that the temperature structure of the disk is predefined is quite 
restrictive. It implies that heating and cooling are instantaneous, but
act only when the disk varies from its predefined steady state. Any 
packet of matter moving radially inward or outward heats or cools according
to the predefined temperature law, even if no other processes act upon it. 

In the following chapter, which is being prepared for publication with
W. Benz and T. Ruzmaikina, we have relaxed this temperature law assumption 
by implementing simple heating and cooling prescriptions. At each time and
at each location in the disk, we assume that the disk is locally plane 
parallel and that the gas is locally adiabatic in its $z$
coordinate. We then calculate the vertical density and temperature 
structure of the disk and (using opacity tables in the literature) determine
the temperature, $T_{eff}$, at the disk `photosphere' and cool each packet
of matter as a blackbody of that temperature. Heating is provided by $PdV$
work, viscous dissipation and shocks which are modeled through the artificial
viscosity terms incorporated into the numerical solution of the hydrodynamics. 
This prescription does not require that any gain or loss in internal energy 
be immediately restored to the system.  Packets of matter are free to
radiate their internal energy immediately, travel from one place to
another through the disk then radiate, or gain additional energy by 
further interaction with the disk.

We present a series of 2-dimensional hydrodynamic simulations of 
marginally self gravitating (\mrat=0.2) disks around protostars using 
a Smoothed Particle Hydrodynamic (SPH) code. We implement simple and
approximate prescriptions for heating via dynamical processes in 
the disk. Cooling is implemented with a similarly simple radiative
cooling prescription which does not assume that local heat dissipation
exactly balances local heat generation, however. We find that these
simulations produce less distinct spiral structure than isothermally 
evolved systems, especially in approximately the inner radial third of the
disk. These simulations also do not generally produce collapsed objects.

We synthesize spectral energy distributions (SED's) for our simulations 
and compare them to fiducial SED's derived from observed systems. The 
distribution of grains within the disk and their size distribution can 
have drastic consequences on the observed SED of a given disk. When grains
are vaporized in the midplane of a hot region of the disk, we show that 
they do not reform quickly into a size distribution similar to that from
which most opacity calculations are based. The consequences on the 
synthesized SED are dramatic. With rapid grain reformation into the original
size distribution, the synthesized spectrum of the disk does not contain 
nearly enough near infrared and optical energy to reproduce observations.
With a plausible modification of the opacity, it is possible to reproduce
observed SED's at these wavelengths.   

Known internal heating processes ($PdV$ work and shocks) are not 
responsible for generating a large fraction of the thermal energy in 
the outer part of the disk, though they produce large fraction of the
thermal energy at smaller radii. Therefore, gravitational torques, 
which are responsible for such shocks, cannot transport mass and 
angular momentum efficiently through the outer disk. Without 
external heating processes (eg. radiation from a surrounding cloud 
or from the star) or unspecified internal heating source (e.g. 
turbulence), the temperatures in the outer part of the disk are
very low (only a few degrees Kelvin), resulting in a radial 
temperature power law fit with an index of $q\sim$1.5 in the disk
midplane and an SED with only a small luminosity at long wavelengths 
($>30-50\mu$m). The temperature law derived for the disk photosphere 
is much shallower ($q\sim1.0$) due to the fact that the disk is 
optically thick over the inner half of the disk. At distances 
$\lesssim$10 AU of the central star, the disk matter becomes heated
to such an extent that it expands `upward' in the $z$-coordinate,
shading the outer part of the disk.

%% file: cooldiskchap.tex
\chapter{Heating and Cooling In Circumstellar Disks: \\
Dynamics of Circumstellar Disks~II\label{cooldisk}}

In the early stages of the formation of a star (see the review paper of 
Shu, Adams \& Lizano 1987), a cloud of gas and dust collapses and forms
a protostar with a disk surrounding it.  Later on, while the accretion
from the cloud continues, the star/disk system also begins to eject matter
into outflows whose strength varies in time. Finally, accretion and outflow
cease and over the next million or so years the star loses its disk and 
evolves onto the main sequence. A major refinement of this paradigm over 
the past decade has been to account for the formation of multiple objects 
from a single collapse. While this picture provides a good qualitative 
picture of the star formation process, many important issues remain poorly 
understood.

Once a well developed star/disk system evolves, whether as a single star 
or a multiple star system, the dynamics of the disk itself as well as its 
interaction with the star or a possible binary companion become
important in determining the system's final configuration. Depending
on the mass and temperature of a disk, one may expect spiral density
waves and viscous effects to develop and play roles of varying
importance. Each may be capable of processing matter through the disk
as well as influencing how the disk eventually decays away as the star
evolves onto the main sequence.

Until recently the primary observational evidence for circumstellar 
accretion disks has been the existence of sources with strong infrared 
excesses which extend from the near infrared to submillimeter and 
millimeter wavelengths. A number of papers (Adams, Lada \& Shu 1987, 
1988, Adams, Emerson \& Fuller 1990, Beckwith \etal 1990--hereafter 
BSCG, Osterloh \& Beckwith 1995) have successfully modeled these 
excesses assuming a geometrically thin accretion disk with or without
additional circumstellar material. Other recent observations (Roddier
\etal 1996, Close \etal 1997) have used adaptive optics to image the
disks of several young star systems. Other disk systems (so called
`proplyds') have been observed in the Orion molecular cloud (O'Dell 
\& Wen 1994, McCaughrean \& O'Dell 1996) in silhouette against
the bright cloud background or through interactions with winds from 
nearby massive stars. 

With these direct and indirect observations it has become clear that
disk systems are quite common around young stars. Many efforts to
model the dynamical processes involved in their formation (Laughlin \&
R\'o\.zyczka 1996, Bonnell \& Bate 1997) and evolution (Nelson \etal 1998,
hereafter \p1, Boss 1997, Artymowicz \& Lubow 1996, Pickett, Durisen \& 
Link 1997) have so far resulted only in a summary of what is possible 
rather than strong limits on what types of evolution are impossible.
Many gaps remain in the understanding of the physical processes 
important in different regimes and even in the configurations of
systems at various points in their history. 

Other efforts have been applied to modeling the spectral energy 
distributions (SED's) of young stellar systems. The SED's of passive 
disks (i.e. those disks which only reprocess radiation from the 
central star) have been successfully modeled in recent work by Chiang 
\& Goldreich (1997). Axisymmetric models of a disk and a mixing 
length approximation for the vertical structure (Bell \& Lin 1994,
Bell \etal 1995, Bell \etal 1997) have been used to model the most
dynamic properties of disks seen in FU~Orionis stars. Time dependent
radiative transport calculations (Simonelli, Pollack \& McKay 1997,
Chick, Pollack \& Cassen 1996) have also been incorporated into 
calculations of the structure of infalling gas and dust. They model 
the destruction of grains in material falling onto the star/disk 
system from the surrounding circumstellar cloud and find that under
many conditions grains can be partially or totally destroyed prior
to their accretion into the star/disk system: heating mechanisms in
the cloud and infalling envelope are of comparable effectiveness 
in heating the grains as in the accretion shock itself.

\p1 showed that in the limit of a disk modeled with a locally isothermal
equation of state, spiral arm formation and later collapse into clumps 
totaling at least a few percent of the disk mass was prevalent in all 
disks whose minimum initial Toomre stability was \qmin$\lesssim 1.5-1.7$.
Boss (1997) has concluded that a locally adiabatic equation of state will 
also produce spiral arm collapse as instabilities grow. Each of these
works are limited in the sense that a predefined temperature law is assumed: 
the gas is locally isothermal or locally adiabatic, but is not globally
isothermal or globally adiabatic. In this approximation, any radial motion
of gas within the disk causes the parcel of gas to heat or cool, even if no 
other processes occur to change its state. Compression and shock events are
likewise artificially managed.  Heating and cooling are instantaneous,
but only act when the state of the gas deviates from a predefined `steady
state' value. 

We present a series of numerical simulations using Smoothed Particle
Hydrodynamics (SPH) modeled under the assumption that the disk is able 
to heat or cool depending only on local conditions within the disk.
Our goal for this work is to understand the dynamical growth 
characteristics of instabilities in systems with heating and cooling
incorporated into the models and to understand which heating and cooling
mechanisms are likely to be responsible for which features in the spectral 
energy distributions of observed systems. In section \ref{coolphys},
we summarize the initial conditions adopted for the disks studied. In
section \ref{energy}, we outline heating and cooling mechanisms 
included in our study and the numerical method used to determine their
magnitude at each point and time in the disk. In section 
\ref{simulations}, we describe the results obtained from our simulations
and in sections \ref{cmparother} and \ref{coolsummary}, we compare 
our results to work in the literature and summarize their significance
in the context of the evolution of stars and star systems.

\section{Physical Assumptions}\label{coolphys}

\subsection{Initial Conditions} \label{coolinit}

The initial conditions used in this work are quite similar to those
used in \p1. We refer the reader to that work for a more complete 
discussion only summarize them here. At time zero we set equal mass
particles on a series of concentric rings extending from the innermost 
ring at a radius of $0.5$~AU to either 50 or 100 AU depending upon
the simulation (see table \ref{cool-params} below). With the number
of particles used, smoothing lengths are less than a few tenths of one
AU in the inner portion of the disk and up to $\sim$1~AU in the outer
disk. The star is modeled as a point mass free to move in response to
gravitational forces from the surrounding disk. The gravitational 
force due to the star is softened with a softening radius of $0.4$~AU
and particles whose trajectories pass through this radius are absorbed 
by the star. Magnetic fields are neglected in our simulations.

The disk mass is initially distributed according to a power law:
\begin{equation}\label{cooldenslaw}
\Sigma(r) = \Sigma_0 \left[ 1 + \left({r\over r_c}\right)^2\right]^
{-{p\over{2}}},
\end{equation}
while the temperature is given according to a similar law:
\begin{equation}\label{cooltemplaw}
T(r) = T_0\left[1 + \left({r\over r_c}\right)^2\right]^{-{q\over{2}}},
\end{equation}
where the exponents $p$ and $q$ are $3/2$ and $1/2$, respectively, and
$\Sigma_0$ and T$_0$ are determined from the disk mass and a choice 
of the minimum value of Toomre's stability parameter $Q$ over
the disk. $Q$ is defined as:
\begin{equation}
Q = {{\kappa c_s}\over{ \pi G \Sigma}},
\end{equation}
where $\kappa$ is the local epicyclic frequency and $c_s$ is the sound 
speed.  The core radius $r_c$ for the power laws is set to $r_c$=1AU.

Matter is set up on initially circular orbits assuming rotational 
equilibrium in the disk. Radial velocities are set to zero. Gravitational
and pressure forces are balanced by centrifugal forces by setting
\begin{equation}\label{coolrotlaw}
\Omega^2(r) = { {GM_*\over{r^3}} + {1\over{r}}{
       {\partial\Psi_D}\over{\partial{r}}} + {1\over{r}}{
       {{\bf\nabla}{P}}\over{\Sigma} } },
\end{equation}
where $\Psi_D$ is the gravitational potential of the disk and the
other symbols have their usual meanings. The magnitudes of the
pressure and self gravitational forces are small compared to the stellar
term, therefore the disk is nearly Keplerian in character.

\subsection{The Equation of State}

The hydrodynamic equations are solved assuming a vertically integrated 
gas pressure and a single component, ideal gas equation of state given by:
\begin{equation}\label{ideal-eos}
P=(\gamma - 1)\Sigma u
\end{equation}
where $\gamma$ is the ratio of specific heats, $P$ is the vertically
integrated pressure and $u$ is the specific internal energy of the gas.
Since we limit the motion of our particles to two dimensions, the effective
value of $\gamma$ is different from that derived from a true three
dimensional calculation (see e.g. the discussion of the equations of
motion derived for a vertically integrated torus in Goldreich, Goodman \&
Narayan 1986). For the systems we study, we have taken a simpler 
approach by assuming that only two translational degrees of freedom 
exist for each molecule. Helium is included as a monatomic ideal gas 
and metals are neglected. Coupled with the assumption that the gas is of 
solar composition, this means the effective value of $\gamma$ is no longer
the well known $\gamma=5/3$, but rather 
\begin{equation}
\gamma\approx 1.53.
\end{equation}
This value includes the contribution of hydrogen with its rotational degrees 
of freedom active but its vibrational degrees of freedom inactive. This
value will be most representative of moderate temperature regions of the
disk. In three dimensions, $\gamma\approx1.42$.

\section{Thermal Energy Generation and Dissipation}\label{energy}

In this work we relax the common practice (see e.g. \p1, Pickett \etal 
1998, Boss 1997) of predefining the temperature or adiabatic 
constant, $K$, at each location in the disk. Instead, we allow thermal
energy to be generated by internal processes and we allow the disk to
cool radiatively at a temperature solely dependent upon local conditions
at a given time. Thermal energy may be generated in one location in 
the disk but be dissipated somewhere else if matter moves there, or the
disk may heat up or cool down over time in a single location. The disk 
may therefore equilibrate to the internal energy state that the physical 
evolution of the system requires.

In our simulations, we only require that the disk be in instantaneous 
vertical thermal balance in order to determine the vertical structure.
We do not require it to be in long term vertical thermal balance. With
the latter assumption, the radiative cooling rate at each point is defined
to be equal to the local heating rate from internal processes (see e.g. 
Frank, King \& Raine 1992 section 5.4). In some cases, the assumption also 
includes energy flux radiating onto the disk from outside, so that
the radiative cooling rate includes terms due to both internally generated 
energy and passive reprocessing. Accurate quantification of the relative
contributions of each of these terms is critical because by working 
backwards from observed spectral characteristics of the disk an observer
can derive an evolutionary picture of the mass and angular momentum 
transfer through the system. For example, if radiation emitted by a disk 
comes entirely from passive reradiation, then no mass or angular momentum 
transport can occur, since such transport is due to internal dissipation 
of kinetic energy in the disk. Therefore, if the contributions due to one 
or more sources of the emitted radiation are incorrectly determined, 
an evolutionary picture derived from them will be flawed. In the following 
discussion we outline the physical basis for the heating and cooling 
processes incorporated into our simulations.

\subsection{Thermal Energy Generation}\label{energy-gen}

Thermal energy in the disk is generated in our simulations from bulk 
mechanical energy via viscous processes and shocks. In order to 
concentrate on the physics internal to the disk itself, no contributions 
to the heating from (for example) either the surrounding molecular cloud 
or from the central star are included. 

We model the energy generation using an artificial viscosity common in 
many implementations of hydrodynamic codes. Because the balance between
thermal energy generation and dissipation are important for both the 
observed character of the systems as well as their morphology and 
dynamics, we outline our implementation here. We refer the reader to 
one of the many discussions already in the literature (e.g. Benz 
1990, Monaghan 1992) for a complete treatment. 

Most hydrodynamic methods require implementation of an artificial 
viscosity to enforce stability and/or improve the treatment of shocks 
by the code. In our simulations we implement viscous pressures which 
are linear and quadratic in the velocity divergence (the so called 
`bulk' or `$\bar\alpha$' \footnote{Note that we have used the symbol
$\bar\alpha$ to denote the bulk component of artificial viscosity 
in order to distinguish it from the Shakura and Sunyaev (1973) 
turbulent viscosity parameter, $\alpha_{SS}$.} and the 
`von~Neumann-Richtmyer' or `$\beta$' viscosities) to simulate an
energy dissipation due to the presence of a viscous pressure of
a sum of particle's $j$ on particle $i$ as
\begin{equation}\label{art-visc}
{{du_i}\over{dt}} = 
      {{1}\over{2}}\sum_j m_j\Pi_{ij}
	      \left({\bf v_i} - {\bf v_j}\right)\cdot {\nabla_i W_{ij}},
\end{equation}
where ${\bf v_i}$ and ${\bf v_j}$ are the velocities of each particle,
$m_i$ is the mass of the $i$th particle and $W_{ij}$ is the value of the 
SPH kernel calculated between the two particles. The factor 1/2 in eq. 
\ref{art-visc} accounts for half of the kinetic energy dissipation being
added to each particle.  The viscous pressure, $\Pi_{ij}$, is given by 
\begin{equation}\label{Piij}
\Pi_{ij} = \cases{
	   {{-\bar\alpha c_{ij}\mu_{ij} + \beta\mu_{ij}^2}\over\Sigma_{ij}}
        &if (${\bf v_i} - {\bf v_j})\cdot({\bf r_i} - {\bf r_j}) \leq 0$;\cr
      0 &otherwise,\cr}
\end{equation}
where ${\bf r_i}$ and ${\bf r_j}$ are the positions of each particle,
$c_{ij}$ is the sound speed and $\Sigma_{ij}$ is
the mean surface density.  The velocity divergence $\mu_{ij}$ is defined by
\begin{equation}\label{diverg}
\mu_{ij} = {{h_{ij}({\bf v_i} - {\bf v_j})\cdot({\bf r_i} - {\bf r_j})}\over
           {\vert {\bf r_i} - {\bf r_j} \vert^2 + \epsilon h_{ij}^2}}
           {{ \left({{f_i + f_j}\over{2}}\right)}}
\end{equation}
where $\epsilon$ is a small value to prevent numerically infinite 
divergence as particles come very close and $h$ is the particle's 
smoothing length. 

The $f_i$ and $f_j$ terms are due to Balsara (1995) and are defined by
\begin{equation}\label{Balsara-f}
f_i =   {{ | (\nabla \cdot {\bf v_i}) | } \over 
	       { | (\nabla \cdot {\bf v_i}) | + | (\nabla \times {\bf v_i} ) |
		     + 0.0001c_i/h_i}}.
\end{equation}

Equations \ref{art-visc}--\ref{diverg} are little more than a restatement
of the standard form of artificial viscosity for SPH as discussed in 
Benz (1990). As improvements to the standard formulation, we also incorporate 
two adaptations which act to minimize the unphysically large shear 
viscosity present in the standard formulation when used in disk simulations 
(which arises because the divergence is calculated pairwise between
particles rather than as an average over some region in order to more
closely model the physical effects of shocks with the code.

First, we modify the velocity divergence from it's usual form by the 
inclusion of the factors $f_i$ and $f_j$ in eq. \ref{diverg}. This factor 
acts to reduce substantially the large, undesirable shear viscosity which 
develops in numerical simulations of disks. It is near unity when the 
flow is strongly compressive, but near zero in shear flows such as are
found in disk simulations. For the simulations we have performed we 
find that typically the reduction due to this term is a factor of 
three or better. The second improvement is due to Morris and Monaghan 
(1997). They implement a time dependence to the coefficient $\bar\alpha$
which allows it to decay within a few smoothing lengths behind a shock to
an equilibrium value much smaller than the normally utilized 
$\bar\alpha=1$, and grow to larger values in regions where strong 
compression exists and dissipation is physically appropriate. In our 
formulation, which includes both the $\bar\alpha$ and $\beta$ terms, we 
define the ratio $\bar\alpha/\beta\equiv 0.5$, but allow their 
magnitudes to vary in time and space according to the Morris \& Monaghan 
formulation. Thus, except in strongly compressing regions (shocks) 
where it is required to stabilize the flow, artificial viscosity is 
minimized.

The origin of the two artificial viscosity terms comes first from the 
fact that dissipation must be introduced to the system in order to 
reproduce the hydrodynamic quantities in shocked regions. The $\beta$
term provides a functional dependence of the viscous coefficient (usually
denoted $\nu$) itself on the velocity divergence present in the flow.
In this way, the magnitude of the dissipation becomes dependent upon
a low order approximation of the discontinuity present in a shock.
Without additional correction unphysical oscillations can still develop 
in the flow, due to the finite differencing in the numerical solution
of the hydrodynamic equations. The $\bar\alpha$ term is introduced to 
damp out such phenomena, however in many cases also provides a quite large
component of unwanted and unphysical shear viscosity in other regions of 
the flow. With this in mind we note that, while the von~Neumann-Richtmyer
term may have some approximate physical basis, its counterpart bulk term 
can only be considered a necessary nuisance.

Although we must ultimately regard it as a nuisance, since it does not 
directly model any physical process, with caution we can turn the
artificial viscosity into a useful nuisance. We have already identified
the von~Neumann-Richtmyer term as a low order representative of shock
dissipation, and we can make a similar identification of the bulk viscosity 
as a `black box' source of dissipation in the system in the same manner as
is done for the {\it ad hoc} `$\alpha_{SS}$' model of Shakura \& Sunyaev. 
As we noted in \p1, there exists a correspondence (Murray 1995, 1996) 
between the standard $\alpha_{SS}$ form of dissipation and the bulk
artificial viscosity implemented in our simulations.  In two dimensional
simulations the correspondence can be expressed for particle $j$ as
\begin{equation}\label{alpha-eq}
\alpha^j_{SS} = {{ f_j\bar\alpha_j h_j \Omega_j} \over {8c_j}}
\end{equation}
where $f_j$ is the Balsara shear reduction coefficient, $\bar\alpha_j$ is 
the bulk viscosity coefficient, $\Omega$ is the orbit frequency and
$c_j$ is the sound speed of the particle. Defined in this way, $\alpha_{SS}$
is a time and space dependent quantity, in contrast to the usual form
in which $\alpha_{SS}$ is constant everywhere or (in a few cases), varies
between an `on' and `off' state. This conversion neglects the
contribution due to the von~Neumann-Richtmyer term and so represents only
an incomplete approximation of the magnitude of the dissipation present. 
It also neglects the time dependence of the viscous coefficients noted
above. In the context of attempting to identify the source of the 
dissipation, this means that if a region experiences repeated compression 
events or shocks on short time scales, the dissipation would be accounted 
for as turbulent process rather than as a shock process. Nevertheless, 
it proves useful as an illustration of where and to what extent thermal 
energy generation processes are active.

\subsection{Thermal Energy Dissipation and the Vertical Structure 
of Accretion Disks} \label{energy-diss}

The cooling experienced by a given particle is determined first by 
calculating the approximate vertical density and temperature ($\rho, T)$
structure of the disk. Then, using these quantities we determine the 
altitude of the disk photosphere and cool the particle as a blackbody 
using the calculated photosphere temperature.

In order to calculate $\rho(z)$ and $T(z)$ without a full three dimensional
hydrodynamic calculation we make two assumptions about the disk structure.
We assume that at each location the disk has some degree of turbulence or 
convection so that it becomes very nearly adiabatic in the $z$ direction
(i.e. that $p=K\rho^\gamma$ with $K$ and $\gamma$ constant). We also assume
that it is locally plane parallel.  In this limit, Fukue \& Sakamoto (1992)
have shown that $\rho(z)$ and $T(z)$ at a known distance from the star are
determined by the solution of the second order, ordinary differential equation
\begin{equation}\label{struc-diffeq}
{{d}\over{dz}} \left({{1}\over{\rho}} {{d}\over{dz}}\left(K\rho^\gamma\right)
                +   {{GM_*z}\over{(r^2 + z^2)^{3/2}}}\right)
		=   -4\pi G\rho.
\end{equation}
A known midplane density, $\rho_{mid}$, the distance from the star, $r$,
the adiabatic constant, $K$, and the ratio of specific heats,
$\gamma$, define the conditions which completely specify the solution
in the absence of external heating of the disk surface.

In our two dimensional simulations, each SPH particle is uniquely defined 
at some time by a particular value of internal energy, surface density 
and distance from the star. These three quantities correspond to the 
three conditions $\rho_{mid}$, $K$ and $r$ which specify the structure
in $z$. The distance from the star is, of course, the same for both
the SPH and Fukue \& Sakamoto specifications. Derivation of the
quantities $\rho_{mid}$ and $K$ from the surface density and internal
energy must be done by iteration to convergence.

We supply an initial guess for $\rho_{mid}$ and $K$ and solve the 
differential equation numerically for $\rho(z)$. The $z$ coordinate
is discretized with 500 zones and the differential equation is solved 
numerically to second order accuracy. Once a tentative solution is 
reached, we integrate the density $\rho$ over $z$ using the trapezoid 
rule to derive the surface density of matter defined by the solution.
Specific internal energy is obtained by a similar integration over the 
vertical extent of the disk. The guesses of $\rho_{mid}$ and $K$ are
then revised using the downhill simplex method to converge to a self
consistent solution. Plots of the density and temperature structure
as a function of the altitude, $z$, are shown for several conditions 
typical of the disks in our simulations are shown in fig. \ref{z-strucplot}.

Implicit in this calculation is the assumption that the gas is adiabatic,
i.e. that the gas pressure and density are related by $p=K\rho^\gamma$ and 
that the heat capacity of the gas, $C_V$, (and by extension, the ratio of 
specific heats, $\gamma$) is a constant. In fact, this will not be 
the case in general because, in various temperature regimes, molecular
hydrogen will have active rotational or vibrational modes, it may dissociate
into atomic form or it may become ionized. As a matter of expediency and in 
order to retain our prescription for the structure calculation, we have 
assumed that the rotational states of hydrogen are active, but that the 
vibrational states are not. Under this assumption and including the 
contribution due to helium, the effective value for the three dimensional 
adiabatic exponent of the gas is $\gamma\approx1.42$. 

From the now known $(\rho, T)$ structure we derive the temperature of the 
disk photosphere by a numerical integration of the optical depth, $\tau$,
from $z=\infty$ to the altitude at which the optical depth becomes
$\tau=2/3$
\begin{equation}\label{tau-a}
\tau = 2/3 = \int_\infty^{z_{phot}}\rho(z)\kappa(\rho,T)dz.
\end{equation}
In optically thin regions, for which $\tau<2/3$ at the midplane, we 
assume the photosphere temperature is that of the midplane. The 
photosphere temperature is then tabulated as a function of the three
input variables radius, surface density and specific internal energy.
At each time we determine the photosphere temperature for each particle
from this table and cool the particle as a blackbody at that 
temperature. The cooling of any particular particle proceeds as 
\begin{equation}\label{dudt}
{{du_i}\over{dt}} = {{ -2\sigma_R T_{eff}^4 }\over{\Sigma_j} }
\end{equation}
where $\sigma_R$ is the Stefan-Boltzmann constant, $u_i$ and $\Sigma_i$
are the specific internal energy and surface density of particle $i$
and $T_{eff}$ is it's photospheric temperature. The factor of two 
accounts for the two surfaces of the disk. On every particle, we
enforce the condition that the temperature (both midplane and 
photosphere) never falls below the 3~K cosmic background temperature.

We use Rosseland mean opacities from tables of Pollack, McKay
\& Christofferson (1985 hereafter PMC). Opacities for packets of 
matter above the grain destruction temperature are taken from 
Alexander \& Ferguson (1994). We have chosen not to use the updated
opacity models of Pollack \etal (1994) in this work. In part this 
is due to the fact that opacity tables including both $\rho$ and $T$
variation based on this work do not exist (D. Hollenbach, personal 
communication). As the authors note however, the opacity is only a 
weak function of density (entering primarily through the change in 
vaporization temperatures of various volatiles at different densities)
and they do produce a figure comparing the new opacity with that of 
the old for a single value of the density. As we shall note in the
sections ahead, it is exactly this vaporization of grains which we
find to be an important factor in determining the character of the
SED. The functional form derived by Henning \& Stognienko (1996) to
reproduce the Pollack \etal opacity suffers from the same 
shortcoming. Hence we have chosen to implement the old version of the
opacities until such time as new tabulated values become available.
In any case, the opacities derived from the new and old works (see 
their fig. 6) are similar except in the temperature region between
200 and 450~K, where the Pollack \etal (1994) derivation exceeds
the PMC value by a factor of about three. The effect on our calculations
would be to slightly reduce the photosphere temperature in regions 
where the differences between the two versions becomes important.

\subsection{Synthesizing Observations}\label{sed-synth}

In order to connect the physical properties of our simulations of 
accretion disks to observable quantities in real systems we synthesize 
spectral energy distribution's (SED's) from our simulations. We calculate 
the SED using the derived black-body temperature of each SPH particle at
a particular time. We assume that the disk is viewed pole on and then 
determine the luminosity of each particle at each frequency as 
\begin{equation}\label{L-particle}
L_\nu^j = {{m_j}\over{\Sigma_j}} \pi B_\nu(T_{eff})
\end{equation}
and of the disk by summing the contributions of all of the particles.
In eq. \ref{L-particle}, $m_j$ is the mass of particle $j$, $\Sigma_j$ 
is it's surface density and $B_\nu$ is the Planck function. 
The area factor is given as $m_j/\Sigma_j$ in order to avoid ambiguity
in the surface area (i.e. the smoothing length and it's overlap with
other particles) defined for each particle. 

Although we neglect the luminosity of the star as a source of energy
input during the calculation, we include it in the post processed SED
calculation. We assume the star contributes to the SED as a 1~$L_\odot$ 
blackbody with temperature $T_{eff}=$4000~K, both values are 
typical of observed T~Tauri stars (see e.g. Osterloh \& Beckwith 1995,
BSCG, Adams \etal 1990). The star's contribution is included primarily
to make the visual comparison of our synthetic SED's to observed systems 
simpler and to provide a constant physically meaningful calibration
to the disk emission on the plot.  We also neglect the accretion luminosity 
of particles which are removed from the simulation due to their radial
migration inward beyond the defined accretion boundary. We expect these
two sources of luminosity to contribute primarily to the optical and UV
spectrum, while the disk will contribute primarily at longer 
wavelengths. Therefore, for our purposes, the disk luminosity will be
well separated in frequency from the stellar and accretion luminosities.

\subsection{Units: The Physical Scale of the System}

The introduction of a cooling mechanism requires an introduction of
a physical scale to the simulations. We shall assume quantities with
values typical of the early stages of protostellar evolution. The star
mass will be assumed $M_*$ = $0.5M_\odot$ and the disk radius of either
$R_D=50$~AU or $R_D=100$ AU as noted in table \ref{cool-params}. Time 
units are given in either years or the disk orbit period defined by 
\td=$2\pi\over{\sqrt{GM_*/R_D^3}}$ which, with the stellar mass and
disk radii given above is equal to about 500 or 1400 years for disk
radii of $R_D=50$ or $100$~AU respectively.

\section{The Simulations}\label{simulations}

\p1 showed that the character of disk evolution undergoes a 
marked change between disk masses of \mrat~$=0.2$ and \mrat~$=0.4$.
In this paper, we will concentrate on studies of a disk at the lower edge
of this mass boundary. In the following discussion, we present a case in 
which we simulate the evolution of a disk with a mass ratio of \mrat~$=0.2$ 
and with an assumed initial minimum Toomre stability of \qmin~$=1.5$
under varying physical assumptions. Initial parameters of our simulations
are tabulated in table \ref{cool-params}. The first column of the
table represents the name of the simulation for identification. The
second column defines the resolution (in number of particles). Initial
disk/star mass ratio and minimum $Q$ are given in columns 3 and 4, the
assumed opacity modification factor (see section \ref{cool-improve} 
below) in column 5 and the total simulation time of each simulation in
the remaining column. We examine the qualitative nature of the simulations
first, then examine in detail the structures which form and their 
characteristics.  

We have run a series of simulations under three different assumptions 
about the opacity and therefore the cooling mechanisms which dependent 
upon it. The first set of simulations proceed under the assumption that 
gas and grains exist in equilibrium everywhere in the disk and that the 
grain size distribution is well modeled by the distributions used in opacity
calculations in the literature (e.g. PMC, Alexander \& Ferguson 1994).
Vaporized material, upon entering a region cool enough for it to form 
grains, does so instantly and in such a way as to reproduce it's original 
grain size distribution, as defined by PMC. These simulations are denoted
by a leading `a' (100~AU disks) or `A' (50~AU disks) in the simulation
name in table \ref{cool-params}.

The second set of simulations relaxes the assumption that refractory
materials reform into their original size distribution quickly. Instead, 
we assume that they reform their original distribution more slowly than 
the overturn time scale for material to be processed through the disk 
midplane to high altitudes and back again so that their size distribution 
and therefore their opacities may be modified from their original form. 
These simulations are denoted with a leading `B' in table 
\ref{cool-params}.

We have also run models of disks under the same `isothermal evolution'
assumption used in \p1. These simulations are denoted with a leading
`i' or `I' in Table \ref{cool-params}. In each case, a capital letter
refers to a disk with outer edge at 50 AU while a lower case letter
refers to a disk with outer edge at 100 AU. In each case `lo', `me'
and `hi' refers to a simulation with low, medium or high resolution as
defined in column 2, and with the `B' simulations we the number 1--5
corresponds to an assumed opacity modification.

\subsection{Morphology and Spectral Energy Distributions}\label{morph}

Using the physical assumptions outlined above and cooling using the `A'
prescription we have completed a series of simulations with initial 
minimum Toomre $Q=1.5$. Snapshots of the evolution of simulation
{\it A2me} are shown in figure \ref{disk-1} and of its derived SED in
figure \ref{sed-1}. As in our previous isothermally evolved simulations 
(\p1) growth of instabilities begins in the inner regions of a disk, 
engulfing the entire system over the course of about 1~\td. Initially,
spiral structures develop in the inner disk, but are later suppressed
by the heating which occurs there.

The spiral structures which develop throughout the whole disk are 
multi-armed and change their shape and character over orbital time
scales. At times they become somewhat filamentary, but in no case
do they become as filamentary as in the isothermally evolved simulations
of \p1. At the end of simulation {\it a2me} (at \td$\approx 2.3$) a 
possible collapse of a spiral arm into a clump is present, however
at this point the cooling prescription is unable to determine the 
vertical structure of the particles near the forming clump. Additional
evolution of this simulation becomes impossible and we cannot determine
whether collapse would continue or dissipate once again into the 
background flow. Over the longer time scales available to these
simulations a substantial fraction of the matter initially located
inside $\sim$5--10 AU accretes onto the star. We will discuss this
in more detail in section \ref{rad-struc}, below. 

As was shown in \p1, SPH is unable to follow low amplitude growth
of structure in disks. As a consequence, nearly all of the evolution
is carried out in the regime in which the spiral patterns have quite large
amplitudes. Short period variation of up a factor of $\sim$3--5 occurs in
the amplitudes. However, the pattern growth has clearly saturated to the 
extent that time averages of the amplitudes must be used rather than
growth rates to characterize the system. In order to make quantitative 
comparisons between of the spiral structure of disks evolved under 
different physical assumptions we therefore examine the saturation amplitudes
of various patterns late in the simulation. We derive the time average
of a given pattern as shown in figure \ref{timeave-amp} for the one of the
largest amplitude patterns ($m=4$) produced from an isothermally evolved
simulation and the disk shown in figure \ref{disk-1}. The time averaged
amplitudes are shown as functions of radius for the same simulations
in figures \ref{radius-amp-a50} and \ref{radius-amp-iso50}.

The averages are taken over the time interval from \td$=0.5$ to \td$=1.5$
for each simulation. These limits are used in order to ensure than most
of the disk has in fact reached it's saturation amplitude (for the beginning
limit), but has not evolved long enough to form clumps (in the case of
the isothermally evolved runs). The time intervals used for both sets of runs
are identical in order to ensure that the comparison can be as close
as possible. For example, for the outer part of the disks shown in figure
\ref{timeave-amp}, it is clear than the cooled simulation reaches it's 
saturation level well after the isothermally evolved simulation forms clumps
and cannot be evolved further. In order to make a fair comparison of the 
amplitudes we must restrict the averages to the same time window.

Approximately the inner third of the cooled disk shows suppressed
pattern amplitudes relative to the isothermally evolved simulation. In \p1
we showed that this region was the region most likely to form collapsed
objects in the isothermal evolution limit. With our new series of simulations
this conclusion must be revised. No clumps form in this region in these
simulations, as they do in the isothermally evolved runs. The maximum
amplitude of each pattern is shifted to a larger radius in the cooled
simulation vs. the isothermally evolved simulation. 

The same shift of large amplitude spiral structure to greater distances 
from the star is present in disks with outer radii at 100~AU, as shown in
figure \ref{radius-amp-a100} for the cooled simulations {\it a2lo, a2me} 
and {\it a2hi}. These simulations show the same amplitude structure
as in the $R_D=$50~AU disks above. Attempts to compare these results to 
isothermally evolved simulations were unsuccessful, since such simulations
tend to develop clumps on the same physical time scale (750--1000 yr) as
with the $R_D=50$~AU isothermal runs. Evolution of the 100~AU simulations
must be terminated before the outer disk has completed even a single orbit. 

In part the smaller pattern amplitudes at small radii are a consequence of
the less efficient cooling in the present simulations. This relative 
inefficiency leads to increased temperatures in the central region 
of the disk relative to the isothermally evolved runs. Higher temperatures 
imply higher values of the Toomre $Q$ stability parameter and therefore 
smaller amplitude (or absent) spiral waves. This is true for the inner
disk but only to a much smaller degree in the outer part of the disk, where
temperatures change from their initial values by only a factor of 20-30\% 
(see section \ref{rad-struc} below). The higher temperatures and Toomre 
stability do not necessarily imply stability against all perturbations
however, since as we shall see shock activity is strongest in the inner 
disk.

The SED's synthesized from the simulation are shown in figure
\ref{sed-1}. They clearly do not reproduce the observed SED's of T-Tauri 
stars. Instead they produce a double peaked spectrum with one peak
dominated by the assumed 4000~K stellar black body contribution and the 
other just longward of 10$\mu$m ($\sim$3--400~K). Both the long wavelength 
end ($>30\mu$m) and the near infrared (defined for our purposes as 
wavelengths from $\sim 1\mu$m to 5$\mu$m) portions of the SED are poorly 
reproduced: insufficient radiation is emitted relative to other wavelength 
bands. 

The long wavelength turnover in the SED's synthesized from our simulations 
typically occurs near 30$\mu$m (10$^{13}$ Hz) rather than the 
$\sim$100--300$\mu$m (10$^{12}$ Hz) typical of observed systems. It is 
also a shallower fall-off towards long wavelengths than is the case for the 
observations. The simulations with an outer disk radius of 100~AU also
suffer from this same deficiency of radiated energy, so we are certain
that the effect is not due to modeling a disk of too small a radial 
extent. Examinations of the photosphere temperature of the particles (see
section \ref{rad-struc}, below) show that only a small fraction of the 
disk radiates at the $\sim$20--100~K temperatures required to produce
excess in the 30--300~$\mu$m wavelength regime. A test simulation identical
to {\it A2me}, but with initial \qmin=2.5 so that the initial disk
temperatures everywhere are higher, cools over the course of the first
1--3\td to resemble the conditions in simulation {\it A2me}. 

We do not believe the absence of near IR flux in our simulations is an
artifact of the relatively large (0.4 AU) truncation radius of our disk 
(recall that particles are accreted by the star inside this radius). In 
part this is because the artificial hole region will already be partially 
devoid of disk matter, due to the finite size of the boundary region between 
the inner disk edge and the star. We have verified that the hole is not 
responsible for flux deficit by running a simulation with a reduced 
accretion radius of 0.2~AU and found no significant difference in the 
derived SED.

Primarily, this system does not radiate efficiently in the near IR because 
of the effect of the large low temperature opacities and the low temperatures 
implied by our model at high altitudes above the midplane. For any conditions 
at the midplane of the disk, our vertical structure calculation produces
a region which is both cold and sufficiently dense to make the column
optically thick at high altitude. Such a condition will accurately
model real systems so long as there is sufficient vertical processing
of disk matter to retain both a vertically adiabatic structure and
a well mixed opacity source, i.e. small grains.

In several regions, our model will break down. At small distances from
the star for example, direct stellar illumination of disk material at
all altitudes will substantially alter the temperature profile throughout.
In this case, the simple vertically adiabatic assumption will break down,
perhaps leading to a more uniform vertical temperature structure, 
since the entire vertical column receives some illumination. This 
failure mode would lead to higher photosphere temperatures than are 
obtained in our model. Further, such a temperature structure may produce 
a radiative zone so that grains begin to settle to the midplane or
high altitude corona so that they are destroyed. In each of these cases, 
the disk would be able to radiate in the near IR more efficiently. However, 
none of these processes are as yet well enough understood to constrain 
the present models.

\subsection{An Attempt to Improve the Cooling Prescription}\label{cool-improve}

As is, the cooling prescription in section \ref{energy-diss} fails to 
reproduce the short wavelength spectrum (near IR) of observed circumstellar 
disks. This wavelength regime corresponds to the portion of the disk in 
which the disk midplane temperatures are warm enough to sublimate grains. 
Our cooling prescription on the other hand, assumes both that the opacity
source (grains) is evenly mixed with the gaseous disk material and that 
the grain size distribution everywhere is not substantially different from 
that of the interstellar medium distribution used to calculate the Rosseland
opacities. Is the failure of our simulations to correctly reproduce
observed disk SED's due to the failure of these physical assumptions
about the grain physics? 

To address the first assumption we note that Weidenschilling (1984) has
shown that grains smaller than $\sim$0.1--1 cm will be largely entrained
in the gas in a turbulent disk and will therefore not settle to the 
midplane. Because the smaller grains provide the largest contribution to 
the opacity, we may assume that for the purposes of our model the grains 
are well mixed. 

The second assumption (the size distribution of grains) proves much more
difficult to address. Contained within our assumption of a vertically 
adiabatic disk structure is the fact that the adiabatic condition arises
out of a convective or turbulent medium, which acts to smooth any entropy
gradients that develop. In such a case, grains entrained in the gas should
be processed through the midplane fairly frequently and, if the midplane 
temperature is hot enough, destroyed. As refractory materials are brought
to higher altitudes where temperatures are lower, they will begin to reform 
into grains. If they reform quickly (compared to their vertical motion) 
into a similar size distribution to their original distribution,
a narrow boundary region in which grains reform will delineate a 
hollowed out region within the disk as shown in figure \ref{dust-convect}a.

On the other hand, if grain reformation is slow compared to speed 
of vertical motion then the region in which the grain size is 
modified from its original distribution becomes much wider (fig.
\ref{dust-convect}b). In this case, calculations assuming one 
distribution of grains may not determine the true correct opacity,
and therefore the cooling will also be incorrectly modeled.
Using calculations based upon the coagulation models of
Weidenschilling \& Ruzmaikina (1994), we demonstrate that in fact
this scenario is the case. For a more complete description of the
code, see Spaute \etal (1991); here we shall merely summarize the 
model presented there.

The disk is divided into 20 vertical layers and particle aggregates
are accounted for as a series of 84 bins spaced logarithmically in
grain diameter with each bin $2^{1/3}$ larger than the previous 
bin. The smallest bin is assumed to contain grains of size 
1$\times 10^{-2}\mu$m. Grains smaller than 1$\times 10^{-2}\mu$m are
not accounted for and nucleation of grains from the gas phase is 
likewise neglected. Relative velocities of grains are associated
with the turbulence, settling of dust aggregates to the central
plane, and radial drift due to gas drag. The turbulent velocity
is set to $\sim260$~m/s, equivalent to $\alpha_{SS}\sim10^{-2}$ 
(assuming $v_T=\sqrt{\alpha_{SS}}c_s$).

Using the geometric cross section of each grain size, the number
density of grains of that size and the relative velocities between
grains in different size bins, we compute the number of collisions
between all possible pairs of size bins during one time step. The
result of these collisions may be coagulation, erosion or total 
destruction depending upon the relative velocities of the particles 
and their assumed strength. Collisions resulting in grain coalescence
remove aggregates from the smaller bins and change the mean mass of
aggregates in the larger bins. If a collision leads instead to 
erosion or disruption, then the fragments are distributed into 
appropriate smaller size bins. Vaporization of grains in hot
regions is modeled by lowering the grain strength so that any 
collision causes fragmentation.

The vertical density and temperature structure remain constant
throughout the calculations and are computed as outlined by the
model in section \ref{energy-diss}, with a midplane temperature 
of 1350~K and a local mass surface density of 
$\Sigma=10^3$~gm/cm$^2$ at 1~AU. In the initial state, all of the
grains are in the smallest size bin. This initial condition
is equivalent to the assumption that at some point in the
evolution of a particular column of gas all of the dust has
been destroyed and must now reform.

Fig. \ref{coagmos} shows snapshots of the grain size distribution
at one AU plotted as a function of altitude above the midplane
after a short period of evolution. Time step constraints within the
coagulation model forbid a very long time evolution of the size 
distribution, however such long term longer evolution is of limited
value because the state of the gas (it's temperature and density) 
change on these same time scales, making a grain distribution derived
from a single ($\rho, T$) configuration irrelevant physically. 

At high altitudes, grains are unable to grow to large sizes in the
time shown because of the low densities (which imply low collision 
cross sections as well) that are found there. The largest size to
which grains grow is $\sim$0.02-0.05$\mu$m. At moderate altitudes, just 
above the temperature boundary between grain destruction and reformation
occurs, larger grains ($\sim0.2-0.3\mu$m) can form over this same
time interval. At low altitudes near the midplane, grains are 
unable to grow and remain locked in the smallest size bin available.

The size distribution of grains is quite unlike that of the interstellar 
medium (ISM), as characterized by Mathis, Rumpl \& Nordsieck (1977 
hereafter MRN) or Kim, Martin \& Hendrys (1994-hereafter KMH), whose
work shows a distribution proportional to $a^{-3.5}$. Instead it is
characterized a quantity of grains in the smallest size bin, whose origin
is in the partial erosion of larger particles, and an increasing or 
near flat spectrum near the upper edge of the size distribution with
a sharp cutoff. A flat size spectrum such as this is a characteristic
feature of the collisional coagulation of grains where little destruction
takes place. The flat spectrum forms because larger grains have longer 
stopping times, encounter more grains and therefore grow faster than 
their smaller neighbors. A declining power law distribution characterizes
destructive processes and is pronounced at lower altitudes. Near the 
midplane only grains in the smallest size bin exist because the assumed
vaporization of grains (modeled in our calculation via grain destruction,
which moves grains from larger bins to smaller) is very efficient there.

These differences are important because of the common use of grain
distributions similar to those given in MRN and KMH in many opacity
calculations. Since the opacity is a function of the size distribution,
modifying the distribution from some canonical value will result in
differences calculated opacity and, for our model, a difference in
the photosphere temperature and cooling experienced by a given column 
of disk matter.

For example, PMC have shown (see their fig. 5) that a narrow size 
distribution with an average grain size much less than 1$\mu$m will produce 
Rosseland mean opacities which are reduced from their nominal values by 
a factor of $\sim 10$ at temperatures above about 150~K and an additional
factor of ten above $\sim 500$~K for very small grains. Frequency 
dependent opacities calculated for a variety of grain sizes by Miyake
\& Nakagawa (1993) and by Pollack \etal (1994) would seem to contradict
this behavior however. They find that for an ensemble of grains of a
given size, the frequency dependent opacity rises as the assumed grain
size decreases and stays roughly the same for all grain sizes less than
$\sim$1$\mu$m.

In fact the two pictures are not contradictory for two reasons. First, the
absolute scale of the opacity is dependent upon the temperature due to the 
different grain species (e.g. ice or silicates) which contribute at colder 
or warmer temperatures. Miyake and Nakagawa performed their calculations 
assuming temperatures of 150~K, while Pollack \etal (1994), consider higher 
temperature (750~K) grains and find that the frequency dependent mass
opacity is lower by a factor varying between $\sim$10 and 100 below
that at 100~K in the mid infrared (their fig. 3b). More importantly however,
changing the assumed temperature modifies the ratio of the grain size to 
the wavelength of the dominant radiation. The wavelength/grain size ratio 
is important because for different radiation temperatures the same grains 
will require an opacity calculation in quite different limits: either a 
Raleigh scattering, Mie scattering or geometric optical approximation, with
a consequent affect on the opacity.

A reduced opacity implies for our cooling prescription that the photosphere
of the disk will be found at a lower altitude and therefore an increased 
effective radiating temperature will be obtained. Using the nominal 
tabulated opacities to obtain the location and temperature of the 
photosphere will therefore underestimate the actual cooling which takes 
place. Furthermore, due to the higher effective radiating temperatures,
the SED will be modified from its previous form as more radiation is
emitted at short wavelengths.

We have investigated the effect of a modified grain opacity by 
adapting our cooling prescription to include an additional assumption.
In regions of the disk where the midplane is hot enough to vaporize
grains, we assume that the grain opacity is temporarily reduced from its 
nominal value by a constant multiplicative factor, $R$, over the entire
vertical column of disk matter above and below the midplane. In other
regions of the disk we assume the opacity remains unaffected, so that the
effective opacity is
\begin{equation}\label{mod-opac}
\kappa_{eff}(\rho,T) = \cases{ R\kappa(\rho,T) & if $T_{mid}> T_{crit}$,\cr
                              \phm{R}\kappa(\rho,T) & otherwise \cr  }
\end{equation}
where $T_{crit}$ is the grain destruction temperature and $T_{mid}$ is
the disk midplane temperature. The disk photosphere temperature and
altitude are calculated in the same manner as before, with the modified
opacity $\kappa_{eff}$ replacing $\kappa(\rho,T)$ in eq. \ref{tau-a} 
above. The disk photosphere temperature therefore will increase in 
regions where the midplane temperature is hot enough to destroy grains 
and remain unaffected elsewhere.

We have performed several simulations varying the amount by which
the opacity is modified from the PMC values over the range between
$R=0.001$ and $R=0.01$. These simulations implement the same initial 
model as simulation {\it A2me}. Simulations at our lowest resolution 
provided inadequate resolution for the inner disk so that the near 
infrared SED would often be dominated by the contribution of only a few 
particles. We show time averaged SED's derived from these simulations in 
fig. \ref{varopac-sed}. This figure shows clearly that the near and mid 
infrared emission from circumstellar disks can be `turned off' or `turned 
on' depending on the extent to which grain reprocessing in the inner 
disk affects the opacity. At either end of two extremes, too small or 
too large an opacity, the SED fails to reproduce a flat or shallow 
spectrum. With too little reduction the SED appears similar to that
shown above in figure \ref{sed-1} (whose time average is reproduced 
in the lower right panel of fig. \ref{varopac-sed}), while with too
much, it appears permanently in an `outburst' phase in which the 
1--5$\mu$m band is enhanced by as much as a factor of 5--10 over the
stellar contribution to the flux. At later times in several of the
plots shown (notably the $R=0.0025$ and the $R=0.005$) the disk again
produces an `outburst' in the same region of the spectrum as it does
in the case with $R=0.001$. In these instances, the true behavior is
episodic and will be discussed more fully in section \ref{sedtime-var}
below.

Both a frequency dependent radiative transfer code, incorporated into
the hydrodynamic calculation at each time step, and a recalculation of 
the Rosseland opacity for each $\rho$, $T$, and grain size distribution 
accessible to our simulations are beyond the scope of the present work. 
As a parameterized factor however, we can bracket the difference from the 
nominal PMC opacities to a factor between 0.001 and 0.01 times the PMC 
values in regions where the midplane temperature rises above the grain 
vaporization temperature. Below, we shall implement a `standard'
factor of 0.0075 times the tabulated PMC values in such regions. 

It is notable that the values of $R$ which are required to produce 
1--5$\mu$m flux consistent with observations are quite small. In fact,
the large modification in the opacity values are not inconsistent with
an opacity consisting only of a contribution from gas rather than from
both gas and grains. A `grain free' calculation (D. Alexander: personal
communication) of the opacity using the model of Alexander and Ferguson 
(1994) down to 1000~K produces opacities which are similar in magnitude to 
our modified values. Coupled with the near and mid infrared time variability 
present in the simulations, the interpretation we are led to is that in 
the inner portion of the accretion disk, clouds of grains in small patches 
of the disk are destroyed and reform, intermittently obscuring the hottest
parts of the disk midplane from view. Such an interpretation implies quite
naturally the existence of intermittent variability in the near and mid 
infrared spectra of star/disk systems originating from within the disk
rather than from a stellar photosphere. Skrutskie \etal (1996) observe
such variation on time scales of a few days to a few weeks in the $J$,
$H$ and $K$ bands for several young stellar systems and conclude that,
particularly in $K$ band, such variations are likely due to processes in
the accretion disk. 

\subsection{Morphology and SED's using modified opacities}\label{new-morph}

The results of a simulation with an identical initial condition, but 
with the modified cooling prescription `B' (simulation {\it B2m4}), are 
shown in figure \ref{disk-2} and the derived SED corresponding to each 
frame is shown in figure \ref{sed-2}. The gross morphology of the system 
as evolved under this modified cooling prescription is quite similar to 
that produced with the original prescription. Instabilities begin in
the inner regions of the disk and as time progresses, engulf the entire
disk, forming filamentary, multi-armed spiral structures.  A quantitative 
measurement of the system morphology as measured by its pattern 
amplitudes (fig. \ref{radius-amp-b50}) confirms the similar behavior for
these simulations. There are no significant differences in the pattern
amplitudes apparent. 

The similarities are perhaps to be expected since only the inner 
regions undergo different cooling, however if the inner regions of the
disk are truly responsible for dynamical behavior further out, the 
modifications might create a different pattern of evolution throughout 
the entire disk. Since no such differences are evident, we may conclude 
that although the instability growth begins in the inner most regions of 
the disk, its character at large radii is not strongly dependent on the 
dynamics of the inner region, at least for the two types of cooling 
assumptions we have outlined.

In both of the simulations shown in figures \ref{disk-1} and \ref{disk-2}
it appears that a substantial `hole' forms towards the middle of the disk
as time progresses. The resulting structure at first glance appears more
torus-like than disk-like. As we show in section \ref{rad-struc} however, 
the surface density only flattens out at small radii rather than evolving 
towards a true torus, in which the inner region is devoid of material.

In spite of the small differences in the system morphology, the derived
SED's show a marked difference from those produced using the original 
cooling prescription. In the present case, the SED exhibits a 
rising (toward higher frequencies) spectrum between $\sim$30--50$\mu$m
($\sim10^{13}$ Hz) and $\sim1\mu$m (a few $\times 10^{14}$ Hz), then
falls off at the highest frequencies where only the star contributes
significantly to the flux. It is also variable in time, with each
of the panels in this time mosaic of the SED's being somewhat different
from the others. The variations are concentrated in the near IR region
of the disk, for which the calculated midplane temperatures are high enough 
to invoke the modified cooling.

As before with our `A' cooling prescription, the disk does not emit 
sufficient flux at low frequencies ($<10^{13}$ Hz), since the modifications 
in section \ref{cool-improve} affect only the hottest portion of the disk. 
The temperature at the disk photosphere lies above the values required 
to produce the 30--100$\mu$m flux when grain destruction has begun to affect
the radiating temperature, or below them, where the matter contributes 
only minimally to the SED.

\subsection{Variation of the SED's with time}\label{sedtime-var}

We concluded in section \ref{cool-improve} that the time averaged SED 
in the near and mid IR is strongly dependent upon the microphysics of 
the dust grain size distribution and its effect upon the opacity, but
that our model could only bracket the magnitude of the modification
required to accurately reproduce the time averaged spectrum. The 
instantaneous emitted spectrum synthesized from our simulations is
variable in near and mid infrared wavelengths. In fig. \ref{flux-var}
we plot the emitted power, $\nu F_\nu$, at 2, 25 and 100 $\mu$m as 
well as the total luminosity of the disk as a function of time for 
each of three resolutions for our `A' simulations and our highest
resolution `B' simulation (The right hand panels will be discussed
in section \ref{therm-gen} below). With the `A' cooling model only 
small variations in time are present: no short term variations larger
than 10\% are present at any wavelength and short term variations at
longer wavelengths are smaller, less than 1\% at 100$\mu$m. At
2$\mu$m, no contribution from the disk is present; the flux is
completely dominated by the assumed constant 4000~K black body 
contribution of the star.

At all resolutions (after an initial transient) there is a slow 
systematic trend towards smaller emitted fluxes. This decay is due 
to mass depletion from the inner disk, which occurs more rapidly than 
it is replaced by matter migrating from further out. As the amount 
of mass present decreases, the inner disk becomes less dynamically 
active and consequently less energy is dissipated as heat and radiation.
A systematic difference in the disk luminosity calculated for
each simulation is apparently due to the decrease in the amount of
dissipation present at higher resolution. Variations in the flux also
decrease with increasing resolution, indicating that variation that
is present may be an overestimate.

The `B' cooling simulations show behavior identical to the `A' 
simulations in the mid and far IR, but exhibit large variations
in the near IR. At 2$\mu$m for example, the variation is about a
factor of two or more from peak to peak and the total disk luminosity
shows a similar amount of variation. Larger variations are again 
present for lower resolution simulations, indicating that the variation
shown is an upper limit.

In order to see the shorter term structure in the flux variation, 
we show the same variables as in \ref{flux-var}, but expanded to 
show a small slice in time in fig. \ref{flux-var-expand}. The time
scale of the variation in the near and medium infrared is similar to
the orbital time scales of the inner disk, which is truncated at 
0.4~AU in our simulations. At some times, only the assumed stellar 
component of the flux contributes to the flux, while at others, the
flux is dominated by the disk contribution. The variations have no well
defined periodicity. Variations occur over periods of less than a year 
and over periods of as long as ten years. Qualitatively, we can understand 
the dynamical origin the variation by noting that heating processes
such as shocks do not occur at regular intervals as the disk evolves, 
but rather occur sporadically as spiral arm structures or other 
inhomogeneities in the disk interact and dissipate orbital energy 
as heat.

The magnitude of the flux variations present in our simulations
must be considered upper limits rather than a definitive prediction
for two reasons. First, the variations are a function of the resolution
of the simulations, decreasing as resolution increases. We cannot be 
certain of the amplitude at which variations become independent of 
resolution. Second, we have not quantitatively determined the effects
of the grain vaporization, reformation and size evolution on the 
opacity. A more detailed understanding of how these variables affect
the opacity is required before spectral energy distributions of 
circumstellar disks can be self consistently incorporated into 
multi-dimensional models such as ours.

On the other hand, the time scales of the variations will be more 
reliable because such effects are dominated by the dynamical times
of the inner disk. Shorter term variations than appear in our
simulations may also occur since our our disks were truncated at a
relatively large distance from the star.

\subsection{Variation of the Dissipation with Resolution\label{resol-diss}}

In our higher resolution runs, the temperatures are lower than in
lower resolution runs and the SED's synthesized have a systematically 
smaller amount of infrared excess and total luminosity. This difference 
is due to the correspondingly lower numerical dissipation possible at 
high resolution as shown in fig \ref{alpha-fig}. Because the purely 
compressional dissipation (i.e. shocks) is better resolved at higher
resolution, we expect that as the resolution increases the shock 
dissipation term will more closely model the physical dissipation
present in shocks and purely numerical dissipation will decrease 
elsewhere in the flow. Therefore until a set of simulations converges
to a well defined amount of dissipation which is not a function of 
resolution, the dissipation which is present will represent an upper
limit on that present in a real system.

Ideally, the contribution of unidentified sources of dissipation
to the energy output of the system would be negligible. In such a
case, specification of known dissipation mechanisms and the known
passive heating mechanisms in the model assumptions would specify
the observable appearance of the system. In our models this ideal
can only be approached, rather than definitely specified. We have
previously identified the bulk viscosity term in eq.~\ref{Piij}
with the Shakura \& Sunyaev $\alpha_{SS}$, which models `black box'
viscosity. Since our resolution is finite, this black box source 
of dissipation is non-zero and we approximate it's contribution to 
the dissipation (via eq. \ref{alpha-eq}) to be of order 
$2\times 10^{-3}$ between 10 and 50~AU for the highest resolution
simulations we have run (see fig. \ref{alpha-fig}).

Although we cannot assign a physical origin to the bulk viscosity 
term, we still can constrain the magnitude of other, known sources 
of thermal energy generation by comparing their computed contribution
to those of our bulk term. In our simulations, the ratio between 
the shock and turbulent energy dissipation mechanisms varies, with 
higher resolution producing less shock dissipation relative to
lower resolution runs. At progressively higher resolution, both
the ratio and the absolute magnitude of the dissipation decrease
and we conclude that we have not fully resolved the hydrodynamics
important for energy generation in the system. Including the
contributions in our simulations from both the $\bar\alpha$ and $\beta$
viscous dissipation sources as a conservative estimate, we can 
conclude that gravitational torques produce large scale shocks 
which dissipate kinetic energy at a rate no greater than an 
equivalent $\alpha_{SS}$ dissipation of $\alpha_{SS}\sim2-5\times10^{-3}$.

\subsection{The Origin of Thermal Energy Generation}\label{therm-gen}

With the understanding that the artificial viscosity incorporated into
our simulations approximately models the underlying physical dissipation 
of kinetic energy present in the disk, we proceed to calculate the magnitude 
and origin of the thermal energy generated in different portions of the 
system. We have previously identified the bulk ($\bar\alpha$) viscosity with 
turbulence (in eq. \ref{alpha-eq}) and the von~Neumann-Richtmyer ($\beta$) 
viscosity as representative of shocks. We can therefore estimate the thermal 
energy generation present in our simulations in terms of the Shakura \& Sunyaev
viscous disk picture using eq. \ref{alpha-eq} and by quantifying the 
relative magnitude of the two dissipative terms. The goal for this section
is to understand which physical processes are responsible (and just as
importantly which are not responsible) for the luminosity produced by
observed young star/disk systems and to understand the shape of their
SED's. We attempt to minimize all other sources of parameterized or unknown
thermal energy generation mechanisms. Of particular interest will be to 
understand the origin of so called `flat' or shallow spectrum sources 
which may be representative of more massive disk systems like those in 
our study.

The azimuth averaged value of the viscous parameter, $\alpha_{SS}$, derived 
from our simulations is shown in figure \ref{alpha-fig}. Its value in
any $\delta r$ of the disk is of order a few $\times 10^{-3}$. Both the 
`A' and `B' simulations show identical viscous dissipation rates, so
only the `B' results are shown. We also plot the ratio of the dissipation 
due to the turbulent and shock artificial viscosities and find that over 
the largest portion of the disk, the magnitudes of each are within a
factor of two. The total budget of thermal energy generation from
dissipation of large scale gas motions averaged over azimuth is 
therefore within a factor of a few of that provided by an $\alpha_{SS}$ 
formulation. 

In the innermost portion of the disk, the dissipation attributed to 
shocks becomes as much as a factor of 2--3 larger than that attributed
to turbulence. Also, the value of the viscous coefficient $\bar\alpha$ 
itself is never able to relax to its lowest value in this region, which
indicates repeated strongly compressive events, and leads (by eq.
\ref{alpha-eq}) to an artificial increase in the derived Shakura \&
Sunyaev viscous coefficient, $\alpha_{SS}$, which should be attributed
instead to the shock dissipation as noted in section \ref{energy-gen}. 

Taking these two phenomena in isolation one would initially be led to
believe that shocks dominate the dissipation in the inner 5--10 AU
of the disk. It is not clear that this is the case however because
in this same region, and within only a few ten's of orbits, mass 
accretion onto the central star begins to reduce the density 
(see section \ref{rad-struc} below). Since SPH resolves the flow
using particles of finite mass, lower mass density in a given region
implies fewer particles and higher numerical dissipation and an
ambiguity in the interpretation of it's physical origin. Notably, 
higher resolution simulations produce progressively more
centrally peaked dissipations (fig. \ref{alpha-fig}), which means
that the region in which shock dissipation may be important is limited
to a smaller portion of the disk near the star. The derived value of
the turbulent dissipation at all three resolutions reaches 
$\alpha_{SS}\approx 10^{-2}$ at the inner edge, suggesting that 
this value is fairly well resolved. At early times, for which little
mass transport has yet occurred and the structures developing in the
inner disk are best resolved, the conclusion that shock dissipation
is a strong contributor to the thermal energy generation remains.

In our moderate and high resolution runs the luminosity derived from 
the simulations (figure \ref{flux-var} and \ref{flux-var-expand})
in general underestimates that from observed systems (e.g. BSCG), 
especially those thought to be younger systems for which large disk 
masses are more likely (Adams \etal 1990). Simulations at all 
resolutions underestimate the flux at long wavelengths corresponding 
to colder regions of the disk distant from the star. The luminosity
and the SED characteristics reflect the temperature profile produced
by our simulations. In section \ref{rad-struc} below we will show that
the temperatures produced by our simulations in the outer disk are 
quite low, so that very little radiation is emitted, even at long 
wavelengths.  In fact, the SED contains only very small contribution 
from the outer disk at all; it contributes chiefly in the shallow slope 
of the SED below 10$^{13}$~Hz. The effect on the integrated luminosity
is that only a few percent of emitted flux in our simulations comes from
the outer disk.

Two additional questions remain before firm conclusions about the 
physical interpretation of our simulations can be made. First, we need 
to be certain that modeling the dissipation of kinetic energy into heat
using artificial viscosities gives an reasonably accurate representation 
of the true thermal energy generation rate. Second, we need to be certain
that the assumed disk geometry does not play a large role in the details
of the synthesized SED. For example, if the assumed disk radius is 
doubled, so that the total disk surface area radiating at, say 10--30~K,
is increased by a factor of approximately four, will sufficient long 
wavelength flux be produced?

In order to investigate the first question more completely, we have
performed a simulation similar to {\it B2h3} with the time variation
of the viscous coefficients discussed in section \ref{energy-gen} 
turned off and the viscous coefficients set to $\bar\alpha=1$ and 
$\beta=2$. This simulation is denoted {\it H2h3} in table
\ref{cool-params}. Effectively, this change will increase the global 
rate of thermal energy generation because we find that the time 
dependent viscous coefficients for each particle ($\bar\alpha_i(t)$) 
fall well below the value of unity assumed in simulation {\it H2h3}. 
The time dependent fluxes are shown in the right hand panels of figure 
\ref{flux-var}. The total luminosity is increased by the viscosity 
modification but the increase comes only from short wavelengths regime
representative of the inner disk. The transient in the flux at short
wavelengths over the first 2--3 \td comes from the initially high 
density in the inner disk, its effect on dynamical activity and
therefore also thermal energy generation. With the increased heating,
a single heating event will heat a given column of matter to higher
temperatures than otherwise, leading to a correspondingly increased
flux from that column as it cools.

After the initial transient settles, the near and mid IR flux 
are increased in magnitude to that of the `B' simulation by only
$\sim 5$\%--barely large enough to be detectable on the plot. 
The long wavelength flux also increases by only $\sim 5$\% relative
to the `B' simulation. In both simulation {\it B2h3} and {\it H2h3},
little thermal energy is produced in the outer disk. More significantly,
figure \ref{sed-50-100cmp} shows that the long wavelength turnoff does 
not shift further into the far IR or submillimeter region. It remains
instead near $10^{13}$~Hz ($\sim$30$\mu$m). The insufficient long 
wavelength flux in each of the simulations leads to the conclusion 
that indeed internal heating mechanisms (due to identifiable sources 
such as shocks) incorporated into our model are an upper limit to
the heating from these mechanisms present in real systems. 

To test the second question we examined the differences in the 
long wavelength end of the SED's generated from our 50~AU 
disks (simulations {\it A2lo, A2me} and {\it A2hi}) with those 
generated from 100~AU disks (simulations {\it a2lo, a2me} and 
{\it a2hi}). Time averaged SED's for each of the simulations
care shown in fig \ref{sed-50-100cmp}. Looking specifically at
the long wavelength flux behavior, in which dust destruction
is unimportant, we find no significant differences between shape
of the SED seen in one or the other simulation.

As we will show below, the temperatures in the outer part of the disks 
are lower than are determined from observations (Adams \etal 1990 for 
example determine temperatures at the outer disk edge of order 15--25~K). 
In many parts of the disk, they are in fact also lower than those 
observed for the molecular clouds in which the disks reside (see e.g.
Walker, Adams \& Lada 1990). We can conclude from the low temperatures
and the corresponding long wavelength flux deficit that heating of 
the disk due to identifiable internal processes (e.g. large scale
shocks) in our simulations is insufficient to heat the outer disk
to the `right' temperature, i.e. temperatures warm enough to produce
SED's from our models which are similar to observed systems. This 
conclusion implies an upper limit on the amount of mass and angular
momentum transport due to gravitational torques, since such torques
are ultimately responsible for the growth and evolution of the 
spiral structures in which the shocks occur. In the inner disk, the
picture is not as clear. At early times, shocks produced by
gravitational torques there are capable of producing sufficient
thermal energy to power the near infrared SED in our simulations.
The question which remains for future consideration is whether the
surface density and temperature initial conditions assumed in our
work, especially given the immediate activity in the inner disk, 
are in fact similar to those produced by real systems.

If additional (as yet unidentified) sources of internal dissipation
exist for the part of the disk at $\gtrsim 10$~AU from the star then 
additional heating will occur there and the matter will become warmer, 
the radial temperature gradient will become shallower, and the outer 
disk will radiate at the higher temperatures required to reproduce 
observed systems. These processes might come from either phenomena
intrinsic to the disk itself, such as magnetic fields, or from three
dimensional turbulence not modeled in our calculations.

On the other hand, it may be the case that rather than being internally 
heated, the outer portion of accretion disks are externally heated, for
example by accretion of additional material onto the disk or by radiative
heating from the surrounding cloud. If in fact the mechanisms for 
heating the outer disk are external (i.e. they do not originate in an as
yet unspecified internal hydrodynamic or magneto-hydrodynamic turbulent 
dissipation), then any models which reproduce observed SED's via internal
dissipative heating alone will incorrectly model the mass and momentum 
transport which depends on such dissipation. We have not attempted to 
model such effects in this work, in part because of the radiative transport
approximations we have implemented (i.e. Rosseland mean opacities) 
preclude a reliable determination of the magnitude of such radiative
heating processes in optically thin regimes. 

\subsection{Density, Temperature and Scale Height Structure}\label{rad-struc}

Using the structure model from section \ref{energy-diss} we can derive
physical disk parameters such as the gas temperature and the disk scale
height, as well as the directly available mass density as functions of 
distance from the star.

The azimuth averaged temperature profiles late in the evolution
of the disks shown in figures \ref{disk-1} and \ref{disk-2} are
shown in figure \ref{t-struct}. Overall, the midplane temperature
shows several distinct regions which individually appear 
as power laws for suitably small radius ranges, however a single
power law for the whole range of radii provides only a poor fit
to the temperature structure. In the innermost portion ($\sim$~1 AU) 
of both sets of simulations, the radial photospheric temperature 
structure is nearly a flat function of radius. Due to the increased 
efficiency of cooling when the opacity is modified, the `B' disk 
midplane temperature profile flattens out in the inner $\sim$1~AU 
so that the temperature lies 2-300~K below that of the `A' disk. 
Between 1 and 10 AU the midplane temperature decreases roughly
according to an $r^{-1}$ power law, but further out (between
10 and 20~AU) the midplane temperature becomes much steeper as the 
disk transitions from an optically thick to optically thin regime. 
Beyond 20~AU the disk becomes optically thin to it's own radiation
so that the midplane and photosphere temperatures are the same.

The photosphere temperature follows a single power law much more 
closely over its radial range, with the exception of the inner 
few AU where the temperature again becomes flatter than in rest of the 
disk. The `A' and `B' cooling prescriptions have profiles which are 
quite similar to each other. The differences which exist are in the 
azimuthal variation of the temperature. In regions where the midplane 
is hot and the modified cooling procedure comes into play, the rms 
photosphere temperature is nearly as large as the temperature itself, 
indicating many short lived opacity holes during which hot material 
near the midplane becomes visible to the surrounding space.

We have fitted the temperature profiles to a power law whose exponent
is a free parameter and plotted the resulting exponent as a function
of time in figure \ref{t-index}. After an initial transient as the
disk becomes active the power law at the midplane of the disk is
$\propto r^{-3/2}$, while the power law at the disk photosphere is
$\propto r^{-1}$. Both of these coefficients compare quite unfavorably
to those derived from the models of BSCG, who fit observed SED's with
a model consisting of a disk with a power law temperature profile. In 
their work, exponents in the range $0.5 < q < 0.75$ were determined. In 
our simulations, the both temperature profile and the SED are available
directly. The differences are due to the lack of heating in the outer
disk discussed in section \ref{therm-gen}, this leads to insufficient
flux at long wavelengths to reproduce observations like those modeled
by BSCG.

The result of the activity and increased dissipation (i.e. larger than 
purely $\alpha_{SS}$ model dissipation) in the inner disk noted in 
section \ref{therm-gen} is that the inner disk becomes depleted of much 
of its initial complement of mass (see fig. \ref{azavesdens}) as the 
orbital energy of the gas is dissipated first as heat and then as 
radiation. This behavior is true of every simulation we study, both 
early in a simulation when activity is greatest and later when the 
density distribution in the inner disk becomes much flatter. Effectively,
the surface density profile develops a much larger core radius ($r_c$ 
in eq. \ref{cooldenslaw}) than it initially has. Initially,
the accretion rate onto the star occurs at a rate of a few
$\times 10^{-6} M_\odot/yr$. As this inner region becomes evacuated,
the mass accretion rate falls by a factor of $\sim$3. The new core
radius is approximately $r_c\approx10$~AU, but its exact value is
dependent upon the dynamics, the initial condition (i.e. the initial
core radius and assumed $r^{-3/2}$ power law) and the magnitude of
the dissipation. Due to the combined effects of the viscous and
gravitational torques a portion of the mass is driven further from the
star than in the initial configuration. We defer additional studies
of the redistribution of mass and angular momentum to a future study.

In addition to the development of the larger core radius, the density
structure exhibits another artifact of the apparently unphysical 
initial condition. The density first rises to a weak local maximum
before decreasing with radius as in the initial power law. We 
believe this maximum is little more than an artifact of our initial
condition and should not be considered as a physical manifestation
of the disk evolution. Based on each of these phenomena, we conclude
that a more physical radial density structure for circumstellar disks 
will have a shallower profile than the $r^{-3/2}$ profile assumed here.

The vertical structure model outlined in section \ref{energy-diss} can 
also be used to calculate the vertical scale height at each point in the 
disk. For the purposes of this work we define the disk scale height,
$Z_e$, at some point in the disk as the altitude above the midplane at which 
the mass density, $\rho$, decreases by a factor of $1/e$ from it's 
midplane value. We find (fig. \ref{scaleheight}) that as our simulations 
evolve, the disk becomes quite thick at small distances from the star. 
No differences are apparent between the scale height produced from the 
different cooling prescriptions. The altitude of the photosphere shows
an even more pronounced rise at small radii, extending vertically to
$Z_{phot}/R\approx0.27$ near 5~AU. Depending of the details of the grain 
distribution within a vertical column, this local maximum suggests
that the outer disk may become shaded from stellar radiation.

Further from the star the scale height $Z_e/R$ stays nearly constant 
with no flaring present except at the outermost edge. No doubt the 
vertical structure produced will be dependent upon the temperature
profile and will change when a full description of the heating
mechanisms are included. We therefore cannot regard the current
scale height results as highly significant.

\section{Comparison to other work}\label{cmparother} 

In a series of papers Bell and her collaborators (Bell \& Lin 1994,
Bell \etal 1995, 1997, Turner Bodenheimer \& Bell 1997) have 
developed an evolutionary model for accretion disks based on an
$\alpha_{SS}$ model for radial transport and a mixing length 
theory (MLT) based vertical structure model. In similar work
D'Allesio \etal (1998a, 1998b) have also developed a `1$+$1'
dimensional model. In the present work, we follow the evolution 
through a `2$+$1' dimensional model, including the evolution of
matter in both radius and azimuth, but at the expense of simpler
vertical structure and radiative transport (sec. \ref{energy-diss}). 
We investigate the dynamical evolution (and consequent thermal energy 
generation mechanisms) of the disk which can not be addressed in the 
1$+$1D work due to the input assumptions of the $\alpha_{SS}$ 
disk formulation. 

In an $\alpha_{SS}$ model, three input parameters specify the 
physics of the model. They are the accretion rate, $\dot M$,
the magnitude of the viscosity, $\alpha_{SS}$ and the radiative
flux impinging upon the disk surface. Together these three 
parameters specify the amount of active and passive heating
experienced by the disk. They also determine the temperature
and density structure throughout the disk as functions of time
and one space variable (the distance from the star). 

In deriving a vertical structure, the above works assume boundary
conditions at the $z=\infty$ boundary which include radiative 
fluxes from the star, disk and molecular cloud and the thermal
conditions of the surrounding molecular cloud. In contrast, we 
assume that the disk exists in isolation (i.e. that these external
fluxes are zero) in order to study the effect of dynamical processes
internal to the disk itself. We show that known dynamical heating 
mechanisms such as shocks or $PdV$ work do not provide enough heating
by themselves to reproduce observations. Either additional internal
heating mechanisms (e.g. dissipation via magnetic fields as in Gammie
1996 or via some other turbulent dissipation mechanism) or energy 
flux onto the disk from external sources (for example direct 
illumination of the disk by the star or the surrounding molecular 
cloud or by illumination of one part of the disk by another) must be
accounted for in order to correctly model the disk evolution. 

We find that a full multi-dimensional evolution is important for a  
description of the density structure of the systems described here
because the spiral structures typically have amplitudes, 
$\delta \Sigma/\Sigma$, of order unity. The temperature structure
of such systems does not develop such large variations in azimuth. 
An azimuth averaged temperature law will be accurate to a few percent
except in the inner region of the disk where grain destruction 
becomes important and the hot midplane intermittently becomes exposed 
to space. 

Bell \etal (1997) find that with the low disk masses implied by their
model (i.e. their input values of the mass flux through the disk, 
$\dot M$, and the magnitude of the viscosity, $\alpha_{SS}$), the
structure obtained is actually super adiabatic in $z$. Our assumption
by which the vertical structure was determined was that the structure
was vertically adiabatic and contrasts with their MLT calculation 
that shows a super adiabatic gradient. The effect that a super 
adiabatic gradient would have on our simulations would be to 
systematically reduce the temperature of the disk photosphere.
In the absence of a full three dimensional calculation of the vertical
structure, the exact nature of the gradient remains unknown. 
However, we believe that the vertically adiabatic assumption in our
simulations is justified because we model high mass systems in which
the dynamical effects present are likely create additional turbulence 
and act to additionally smooth out vertical entropy gradients.

In spite of the differences between the modeling assumptions and 
procedures utilized in our work and those noted above, the results 
produced from each are in many cases quite similar. This should perhaps
not be too unexpected since we find that shocks are not the dominant 
source of heating in the disk and our only other source of 
thermal energy generation is a viscous heating term analogous to the
standard $\alpha_{SS}$ model, extended to include limited time and
space dependence. Both methods are able to produce SED's which 
reproduce observed profiles to varying degrees. The temperature and
density profiles on which the SED's are based however are quite 
different. In our work, the inner disk provides nearly all of the 
flux and is characterized by an essentially flat surface density
profile and an $r^{-1}$ temperature law. Only at distances $>10$~AU 
does the surface density begin to fall off steeply.  At these radii,
the temperatures derived from our models are low enough not to 
contribute significantly to the flux. As is shown for the 1D case 
in Turner \etal, absorption and reradiation of infrared and microwave
photons can heat the outer disk, flatten the temperature profile 
and provide additional far infrared flux. Such a treatment in
our model would require a multi-dimensional frequency dependent 
radiative transport code, which has not yet been incorporated.

Both in our work and Bell \etal (1995), a mechanism by which the disk 
may vary it's energy output (SED) in time is explored. We consider 
the variation of the opacity due to the destruction and reformation
of grains in the inner disk, while they consider variation in the
thermal ionization state of gas within a few stellar radii of the 
star. They are able to produce very large and long term temporal 
variations typical of FU~Orionis outbursts in accretion disk SED's, 
while we find much smaller variations (factors of $\sim$2 or less) 
which occur over much shorter time scales.

The puffed up inner disk in our work is similar to the `volcano region' 
discussed by Turner, Bodenheimer \& Bell (1997) in their simulations of
the disks of FU~Orionis objects. Their models, derived from Bell \etal 
(1997), assume a much lower mass disks than our own, produce a puffed up 
region which is also much smaller in radial extent than in our work. 
Neither our model nor theirs depend upon the validity of the thin
disk assumption and both indicate that such an approximation may be
inappropriate in the inner regions of the disk.

\section{Concluding Remarks}\label{coolsummary}

In this study of the evolution of circumstellar accretion disks, we
have found that the growth of multiple armed spiral instabilities is 
suppressed relative to the growth found in \p1. In the present simulations 
spiral arms grow only weakly in the inner half of the disk. Regions more 
distant from the star are less filamentary than with an isothermal evolution.
The spiral structures also do not collapse into clumps as they did in \p1. 

The spiral structures are similar in that they do not persist as stable
structures over even a single orbit of the outer disk around the star.
The dynamics of the inner region are very important for understanding not 
only the global morphology of the system as we showed in \p1, but also its 
observational appearance as well. Transport, growth and vaporization of 
dust vertically within a column of gas can have marked observable
consequences for the spectral energy distribution of the system. A
correct model of the dynamics and spectral energy distributions of
circumstellar disks similar to those studied here must include an accurate
description of the full three dimensional spatial distribution of grains
and of their size distribution in order to accurately model the opacities 
and the radiative transfer which depends on them.

The simulations discussed here produce temperature's in the outer disk 
which are colder those produced by models of observations. The temperature
profile which results produces a power law exponent near $q=1.5$ at the 
disk midplane and $q=1.0$ at the disk photosphere, rather than the 
$\alpha_{SS}$ model prediction of $q=0.75$ or the observed
$0.5 \lesssim q \lesssim 0.75$. This steep profile produces a spectral 
energy distribution which, when compared to observed systems, emits 
insufficient flux at low frequencies ($\lesssim 10^{13}$~Hz or 30$\mu$m).
We attribute this primarily to the influence of physical phenomena present
in observed systems which are not included in our study. As modeled, our 
systems do not produce sufficient thermal heating from well defined 
internal sources such as shocks. Additional heating must be supplied from 
internal sources such as small scale hydrodynamic or magneto-hydrodynamic 
turbulence or from external heat sources such as radiative heating effects 
from the circumstellar cloud or the central star or thermal heating due to 
accretion of additional infalling matter onto the disk may provide 
additional thermal energy to the outer disk.

We find that the heating due to large scale shocks within the disk 
does not contribute a dominant portion of the internal energy 
present in the disk except perhaps in the inner few AU, where shock
dissipation provides a much larger fraction. Further, the magnitude
of shock dissipation derived from the global spiral arm structures
produced in our simulations is insufficient to power the luminosity
of observed circumstellar disks. This indicates that mass and angular 
momentum transport, which depend on such internal dissipation, may be 
over estimated in $\alpha_{SS}$ models of accretion disks which 
assume an greater than $\sim 4-5\times 10^{-3}$. This limit is derived
from fig. \ref{alpha-fig} in which the magnitude of shock dissipation 
is roughly equivalent to the black box $\alpha_{SS}$ dissipation
which is $\sim 2\times 10^{-3}$. This conclusion is limited to systems 
in which the disk is marginally self gravitating, i.e. \mrat=0.2. Spiral
arms in these systems are both filamentary and of high order symmetry
($m\ge 3$). It remains to be determined the role that gravitational
torques will play in systems with more massive disks, where the 
dominant spiral patterns are of low order symmetry ($m\le 3$). Perhaps
in this case more efficient transport is present and a flatter SED 
profile can be realized.

In both \p1 and in the present work, the assumed $r^{-3/2}$ surface
density power law quickly becomes modified in the inner part
of the disk as matter either accretes onto the star or migrates 
to somewhere further out. From these results it seems improbable that
such a steep density distribution can occur in nature but the true
nature of the density distribution as a function of radius remains
to be determined. We can suggest weak limits on the density profile 
by coupling our results with those of Laughlin \& Bodenheimer (1994).
In their study, initially toroidal initial configurations (of massive
systems) evolve towards power law-like distributions, while our power
law initial conditions evolve towards flatter profiles.

\acknowledgments
We wish to thank Phil Pinto for valuable discussions about radiative
transport, Sarah Maddison for valuable discussions about initial
conditions for SPH disks and Mike Meyer for several useful 
conversations about SED's. A conversation with Jim Stone provided 
additional insight into the role of gravitational torques. This work 
was supported under the NASA Origins of the Solar System grant 
NAG5-4380.

%% file: table-cool.tex
\singlespace
\begin{deluxetable}{lcccccc}
\tablewidth{420pt}
\tablecaption{\label{cool-params} Initial Parameters For Simulations}
\tablehead{
\colhead{Name}  & \colhead{Number of} & \colhead{\mrat } &
\colhead{\qmin}  & \colhead{Opacity} & \colhead{End Time} & \colhead{ Disk} 
\\
\colhead{}  & \colhead{Particles} & \colhead{} &
\colhead{}  & \colhead{Factor $R$} & \colhead{(T$_{\rm D}$=1)} & \colhead{Radius (AU)}}

\startdata
I2lo & \phn 16520 & .2 & 1.5 & \nodata & 1.6 & \phn 50 \nl
A2lo & \phn 16520 & .2 & 1.5 & \nodata & 6.0 & \phn 50 \nl
I2me & \phn 33399 & .2 & 1.5 & \nodata & 1.8 & \phn 50 \nl
A2me & \phn 33399 & .2 & 1.5 & \nodata & 6.0 & \phn 50 \nl
B2m1 & \phn 33399 & .2 & 1.5 &  0.001  & 6.0 & \phn 50 \nl
B2m2 & \phn 33399 & .2 & 1.5 &  0.010  & 6.0 & \phn 50 \nl
B2m3 & \phn 33399 & .2 & 1.5 &  0.050  & 6.0 & \phn 50 \nl
B2m4 & \phn 33399 & .2 & 1.5 &  0.075  & 6.0 & \phn 50 \nl
B2m5 & \phn 33399 & .2 & 1.5 &  0.025  & 6.0 & \phn 50 \nl
I2hi &     101016 & .2 & 1.5 & \nodata & 1.8 & \phn 50 \nl
A2hi &     101016 & .2 & 1.5 & \nodata & 6.0 & \phn 50 \nl
B2h3 &     101016 & .2 & 1.5 &  0.050  & 6.0 & \phn 50 \nl
H2h3 &     101016 & .2 & 1.5 &  0.050  & 6.0 & \phn 50 \nl
a2lo & \phn 16182 & .2 & 1.5 & \nodata & 3.0 &     100 \nl
a2me & \phn 33134 & .2 & 1.5 & \nodata & 2.3 &     100 \nl
a2hi &     100971 & .2 & 1.5 & \nodata & 3.0 &     100 \nl
\enddata
\end{deluxetable}
\doublespace

%% file: cooldiskfigs.tex
\singlespace

\begin{figure}
\plotfiddle{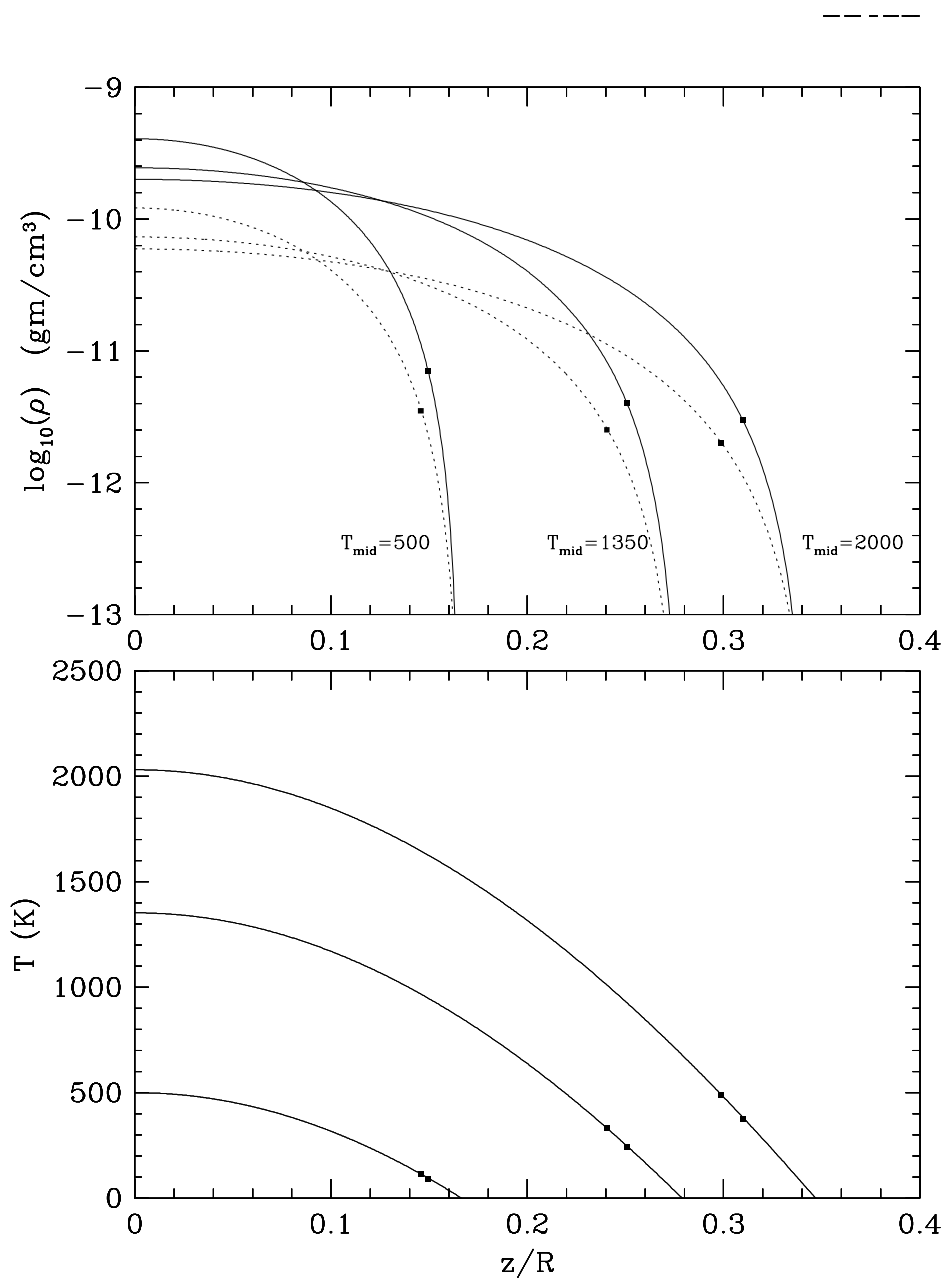}{6.40in}{0}{65}{65}{-210}{-35}
\caption[Density and temperature structure for several conditions
typical of the disks studies in our simulations]
{\label{z-strucplot}
\footnotesize
Density and temperature structure as a function of altitude above the
midplane for conditions typical of our disk simulations at a 1~AU distance
from the star. The location of the calculated disk photosphere of
the disk are marked with a solid square attached to each curve. The solid
curves on the upper frame are typical of the density derived from our
simulations of 50~AU disks (1000 g/cm$^2$), while the dotted curves represent
the density structure typical of our 100~AU disks (300 g/cm$^2$).  Each of
the three pairs of curves in the plot show the density structure for an
assumed midplane temperature of $\sim 2000$~K, $\sim 1350$~K and $\sim 500$~K
as noted. The temperatures, plotted in the lower frame, are in each case,
well below, approximately equal to and well above the grain destruction
temperature in the disk midplane. }
\end{figure}

\clearpage

\begin{figure}
\plotfiddle{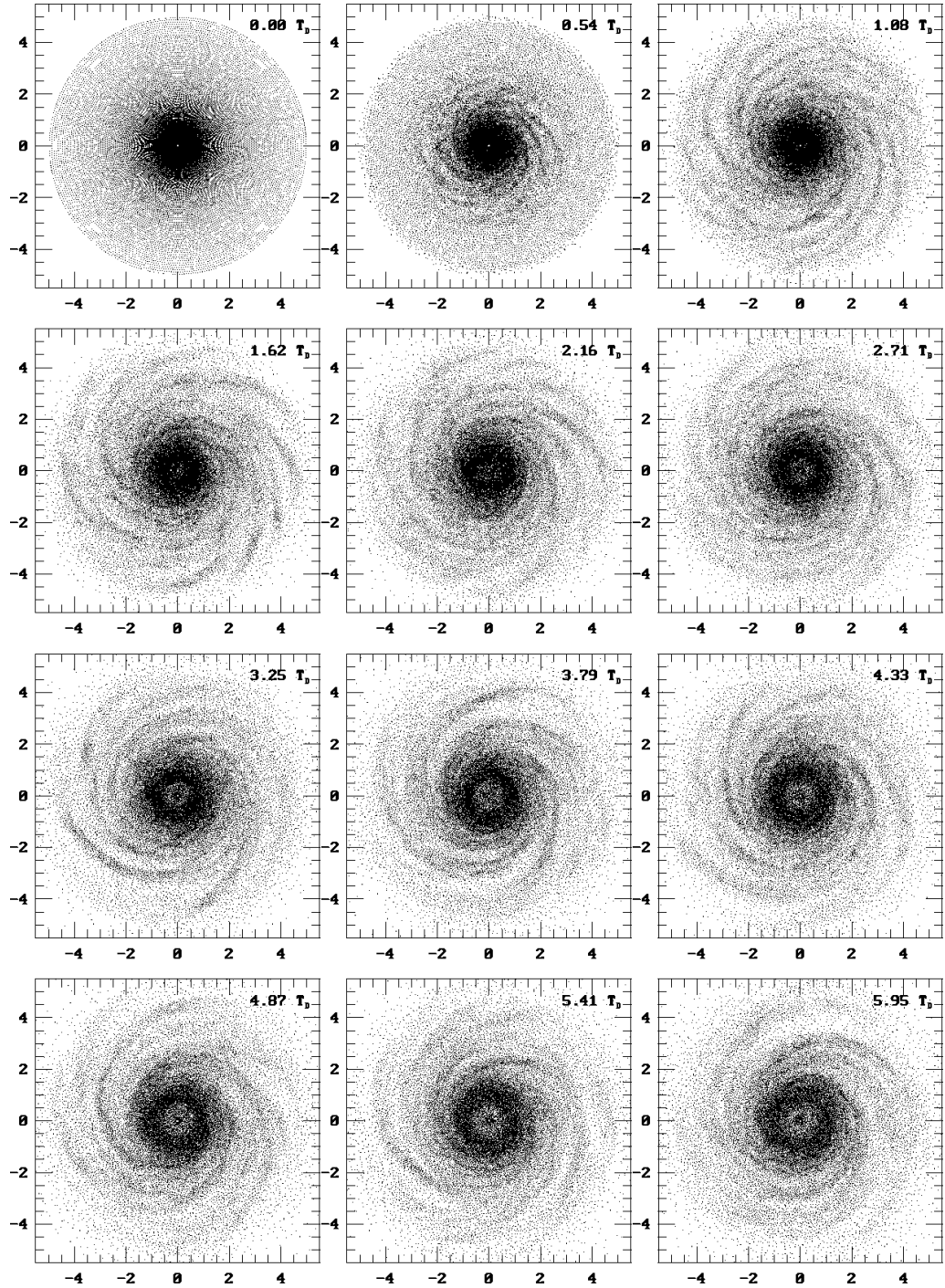}{7.20in}{0}{73}{73}{-225}{-15}
\caption[SPH simulation of a low mass disk with our `a' cooling
prescription]
{\label{disk-1}
\footnotesize
A time series of SPH particle positions for a disk of mass $M_D/M_*$=0.2
and initial minimum Q$_{min}$=1.5 (simulations {\it A2me}). Spiral structure
varies strongly over time. Length units are defined as 1=10AU and time in
units of the disk orbit period \td= $2\pi\sqrt{ R_D^3/GM_*}$. With the
assumed mass of the star of 0.5~$M_\odot$ and the radius of the disk of 
100 AU, \td$\approx$~1400~yr.}
\end{figure}

\clearpage

\begin{figure}
\plotfiddle{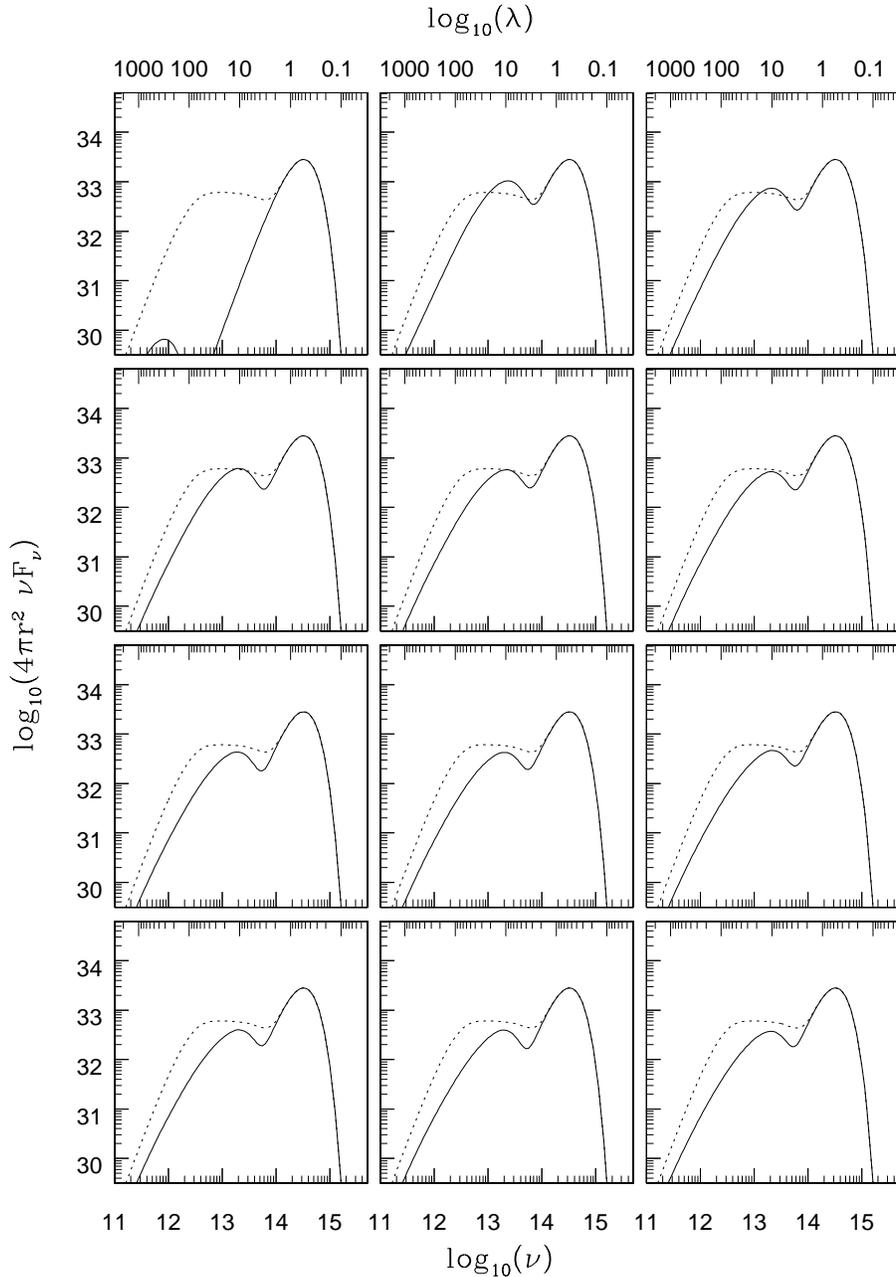}{6.53in}{0}{65}{65}{-220}{-35}
\caption[Spectral energy distribution's for the simulation shown in
figure \ref{disk-1}]
{\label{sed-1}
\footnotesize
Spectral energy distribution's for the disk shown in figure \ref{disk-1}.
Each panel corresponds to the analogous panel in figure \ref{disk-1}.  The
dotted line represents the contribution due to the central star, which is
assumed to be radiating as a $T_{eff}=$4000~K black-body with 1~$L_\odot$.
The horizontal axes of each panel are labeled in frequency (bottom tick marks)
and in wavelength (top tick marks). A `reference' SED with the same assumed
4000~K star and with a disk with the same outer radius as assumed in our
simulations, with luminosity $L_D=0.5L\odot$ and a temperature power law
exponent of $q=0.5$ is shown with a dotted line. The SED's produced clearly
do not reproduce the observed luminosity spectrum around T Tauri stars,
producing insufficient flux at both long and short wavelengths.}
\end{figure}

\clearpage

\begin{figure}
\plotfiddle{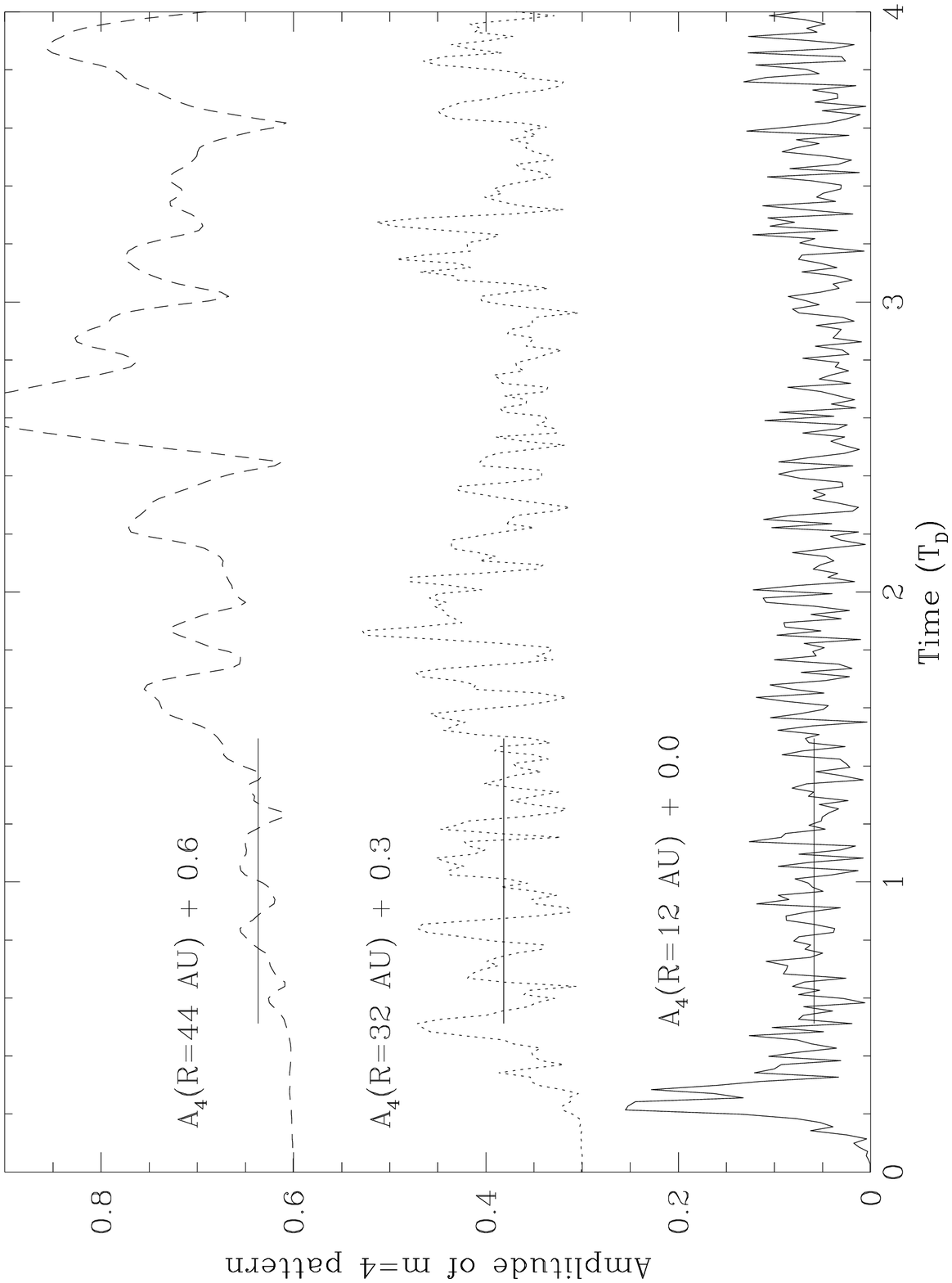}{3.02in}{-90}{44}{44}{-160}{255}
\plotfiddle{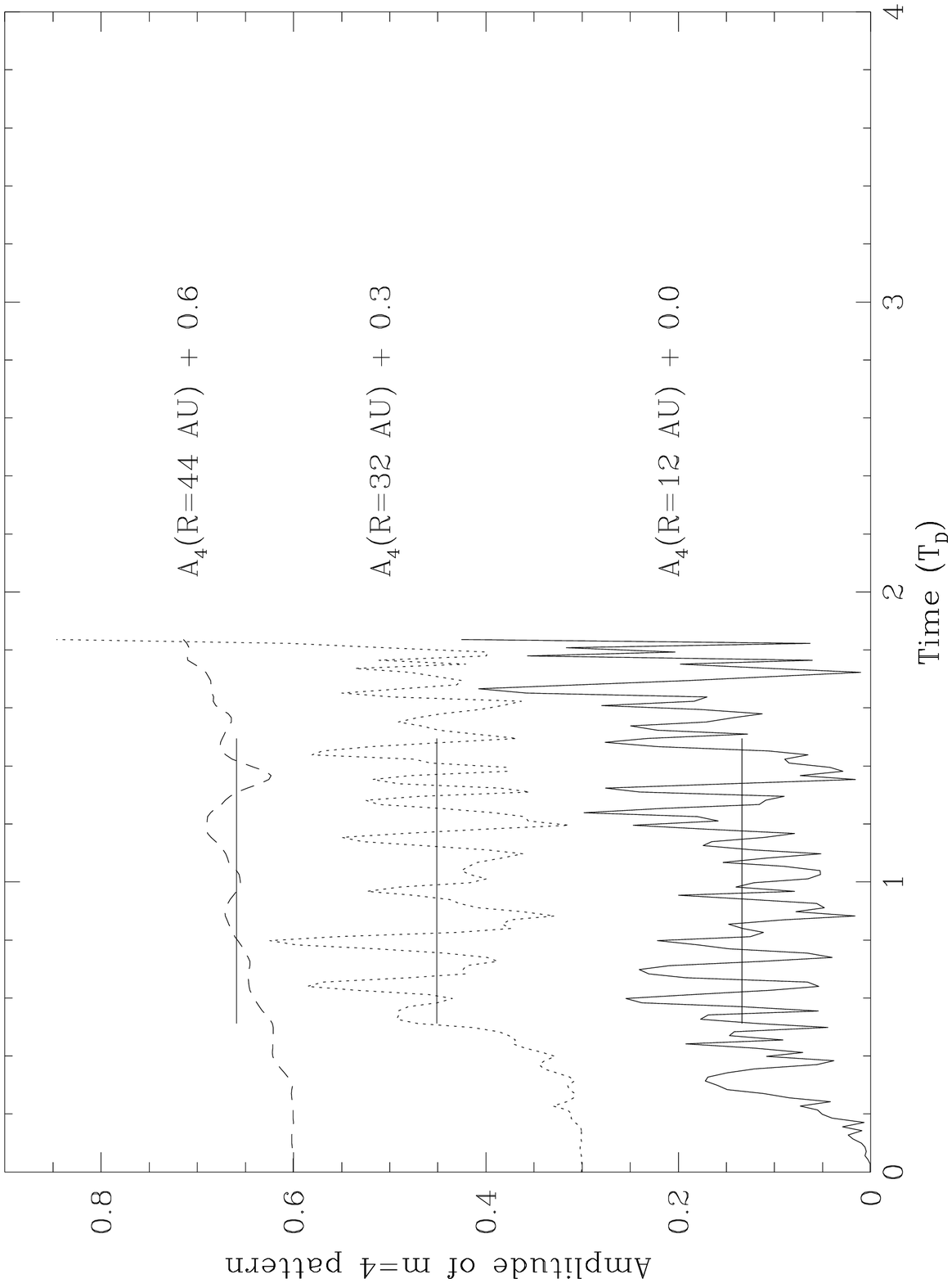}{3.02in}{-90}{44}{44}{-160}{245}
\caption[Amplitude of the $m=4$ spiral pattern at several distances from the
star from cooled and isothermally evolved simulations as a function of time.]
{\label{timeave-amp}
\footnotesize
The amplitude of the $m=4$ spiral pattern as a function of time 
at several distances from the star. The top panel show the amplitudes
derived from the simulation shown in figure \ref{disk-1}, while the bottom 
panel shows the isothermally evolved simulation {\it I2me}. For each curve, 
the amplitude of the pattern is offset from the origin by the amount noted in
order not to confuse the reader. Solid horizontal lines denote the fitted 
average pattern amplitude and extend over the time span for which the average 
was calculated. }
\end{figure}

\clearpage

\begin{figure}
\plotfiddle{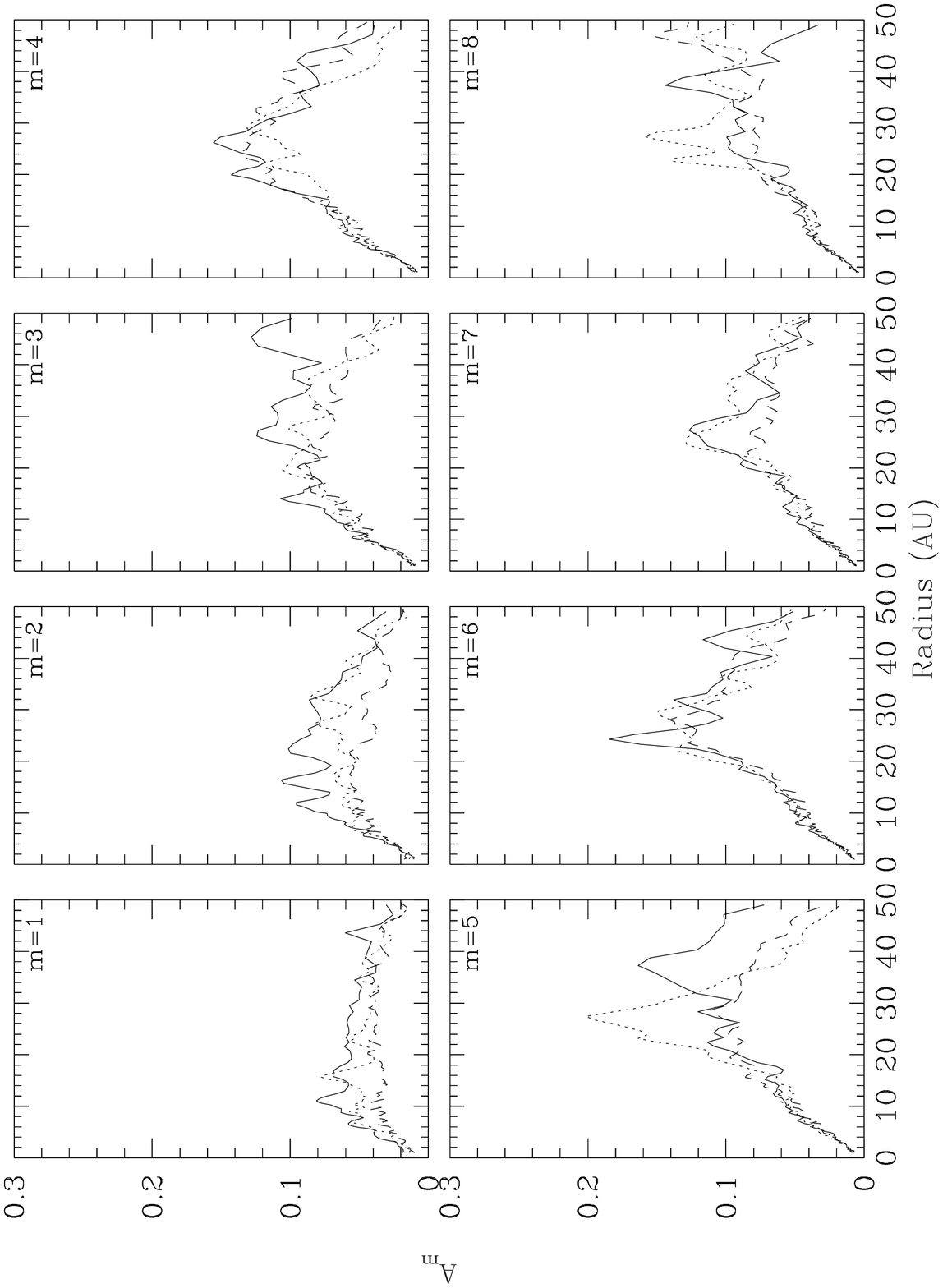}{5.5in}{-90}{56}{56}{-220}{450}
\caption[Amplitude of several spiral patterns from simulations evolved
under the `a' cooling prescription as a function of radius. ($R_D=$50~AU)]
{\label{radius-amp-a50}
\footnotesize
The time averaged amplitude of the $m=1-8$ spiral patterns as a function 
of radius for the disk shown in figure \ref{disk-1} as well as it's high
and low resolution counterparts. The solid line denotes the low resolution
run ({\it A2lo}), while the dotted and dashed lines represent the moderate
and high resolution runs, {\it A2me} and {\it A2hi}, respectively.}
\end{figure}

\clearpage

\begin{figure}
\plotfiddle{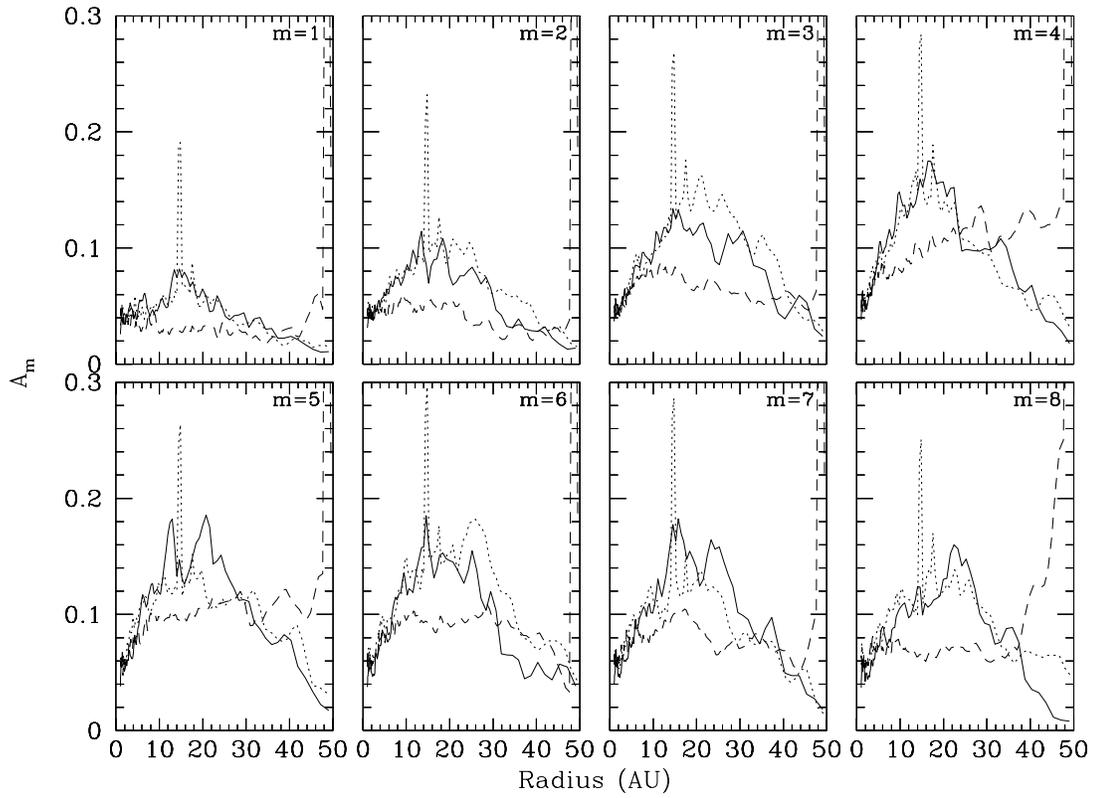}{5.5in}{-90}{56}{56}{-220}{450}
\caption[Amplitude of several spiral patterns from isothermally evolved
simulations as a function of radius. ($R_D=$50~AU)]
{\label{radius-amp-iso50}
\footnotesize
The time averaged amplitude of the $m=1-8$ spiral patterns as a function 
of radius isothermally evolved simulations with the same initial conditions
as those shown in figure \ref{radius-amp-a50}. Here again, the solid line
denotes the low resolution run ({\it I2lo}, while the dotted and dashed
lines represent the moderate and high resolution runs, {\it I2me} and
{\it I2hi}, respectively. Spike appearing in the plots for the moderate
resolution run (at $\sim$12 AU) and the high resolution run (near the
outer edge) are both artifacts of clumps which formed just prior to the 
termination of the fit. They should be disregarded in comparisons with
figure \ref{radius-amp-a50}.}
\end{figure}

\clearpage

\begin{figure}
\plotfiddle{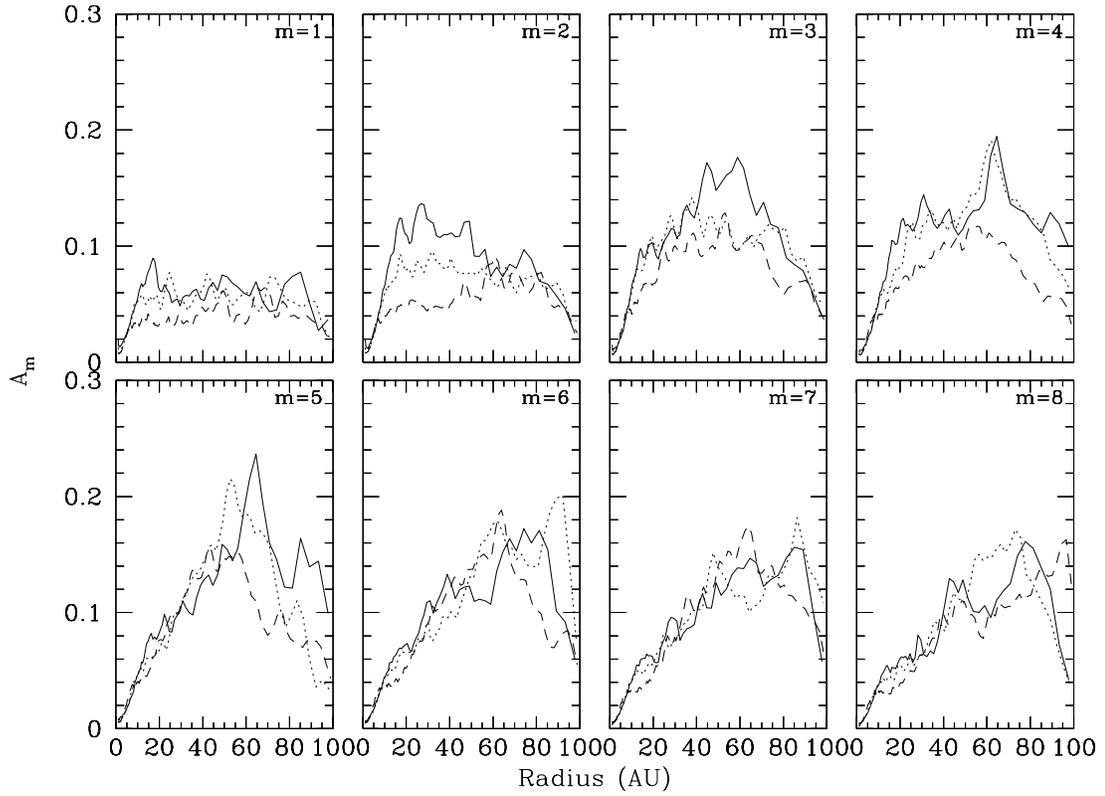}{5.5in}{-90}{56}{56}{-220}{450}
\caption[Amplitude of several spiral patterns from simulations evolved
under the `a' cooling prescription as a function of radius ($R_D=$100~AU).]
{\label{radius-amp-a100}
\footnotesize
The time averaged amplitude of the $m=1-8$ spiral patterns as a function 
of radius for the $R_D=$100~AU disks similar to those shown in figure
\ref{radius-amp-a50}. The solid line denotes the low resolution run 
({\it a2lo}, while the dotted and dashed lines represent the moderate
and high resolution runs, {\it a2me} and {\it a2hi}, respectively. The 
amplitudes which develop are quite similar to those of the 50~AU disks
shown in figure \ref{radius-amp-a50}}
\end{figure}

\clearpage

\begin{figure}
\plotfiddle{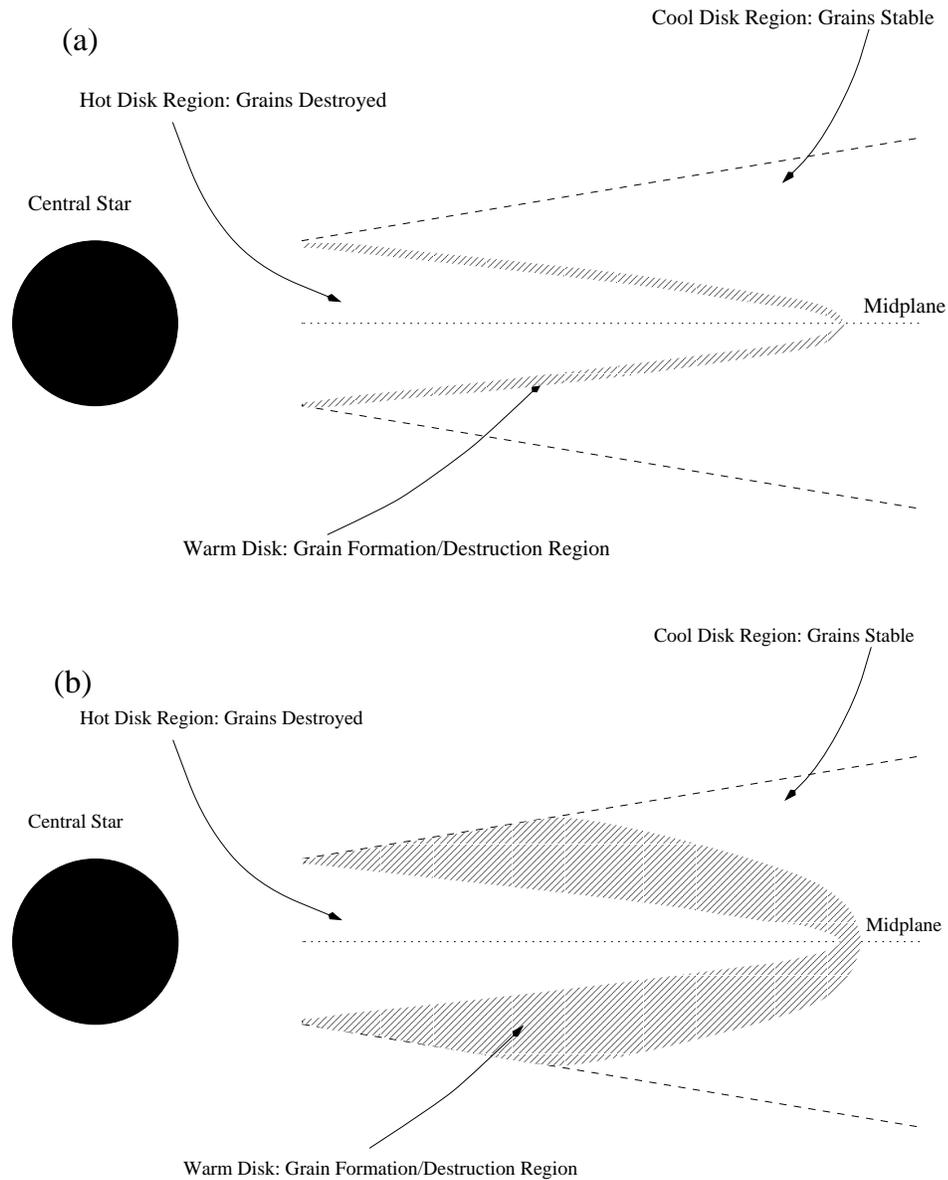}{6.61in}{0}{69}{69}{-210}{-45}
\caption[Cartoon of conditions in the inner disk where dust may be 
destroyed]
{\label{dust-convect}
\footnotesize
(a) A cartoon of the physical conditions
of cooling prescription `a' and implemented for the simulation in figures
\ref{disk-1} and \ref{sed-1}.  Under this assumption, even if the midplane
temperature lies well above the grain destruction temperature, grains 
embedded in high altitude, cool gas block radiation from the hot midplane 
matter. (b) the modified condition (cooling prescription `b') used for 
the simulation shown in figures \ref{disk-2} and \ref{sed-2} below. Under 
this modified assumption, grains are destroyed in the midplane if the
temperature is hot enough but reform only slowly at high altitudes. This
allows a particular column of gas to become less opaque so that it radiates 
at a higher effective temperature and cools more quickly.}
\end{figure}

\clearpage

\begin{figure}
\plotfiddle{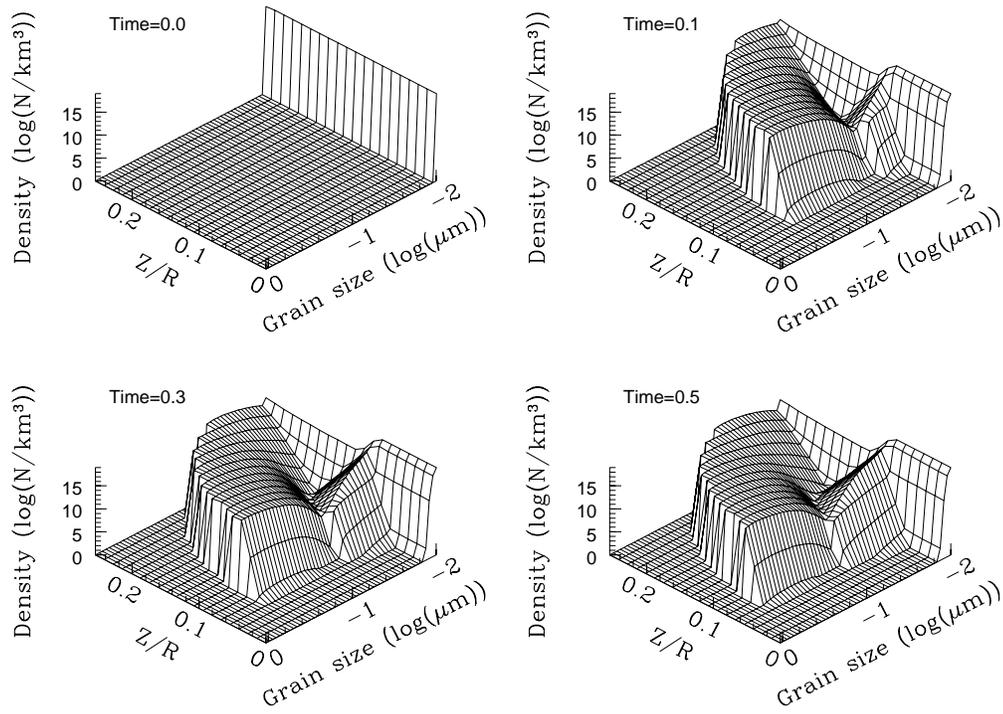}{6.0in}{-90}{52}{52}{-215}{400}
\caption[Grain size distribution at different altitudes above the midplane]
{\label{coagmos}
\footnotesize
The grain size distribution as a function of altitude above the disk
midplane. The time units in the upper left of each frame are given in
orbit periods at the assumed 1~AU distance from the star
($T=1\approx 1.44$~yr). The vertical structure is identical to that
shown in figure \ref{z-strucplot} with a surface density of 
1000~gm/cm$^2$, midplane temperature of 1350~K at a distance of 1~AU 
from the central star. In the example shown, midplane temperature 
is above the assumed grain destruction temperature ($T=1200$~K). }
\end{figure}

\clearpage

\begin{figure}
\plotfiddle{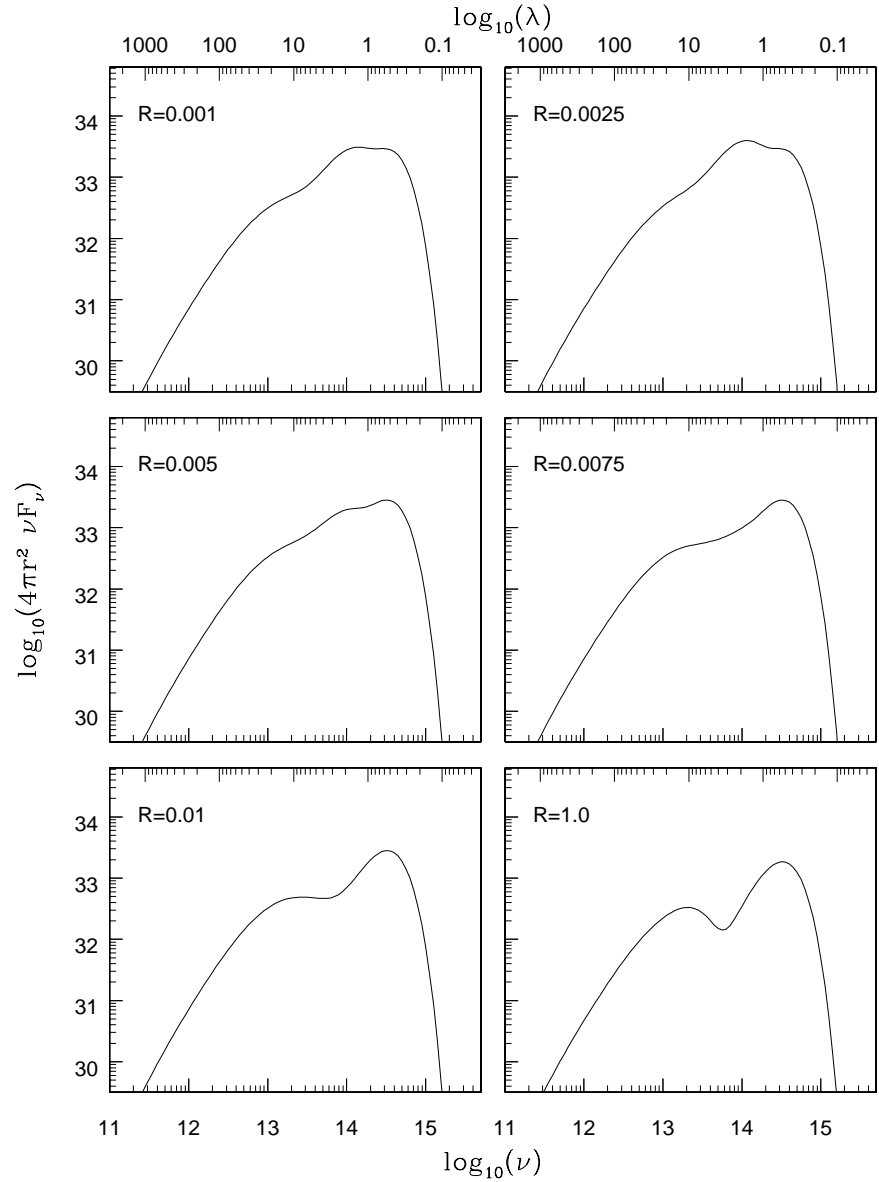}{6.0in}{0}{60}{60}{-160}{10}
\caption[Synthesized SED's of simulations with varying modifications in
grain opacity]
{\label{varopac-sed}
\footnotesize
Synthesized SED's of simulations with varying modifications in grain 
opacity. The initial conditions of these simulations are identical to 
those of simulation {\it A2me}, but are each carried out under varying
physical assumptions. To remove short period time variation, we plot
a time averaged SED over the time from \td=1 to \td=5.}
\end{figure}

\clearpage

\begin{figure}
\plotfiddle{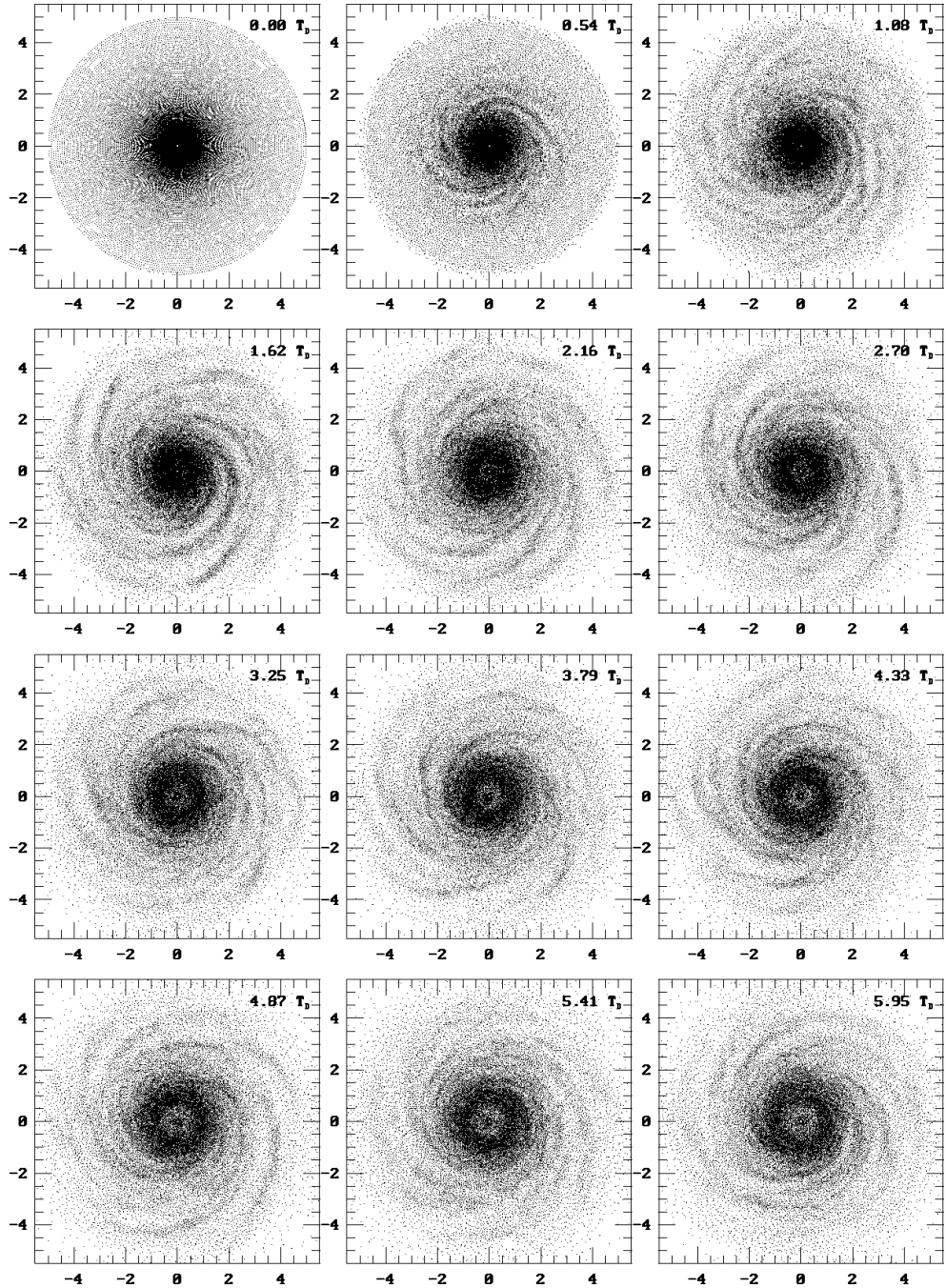}{7.05in}{0}{70}{70}{-225}{-5}
\caption[SPH simulation of a disk using our `b' cooling prescription]
{\label{disk-2}
\footnotesize
A time series mosaic of SPH particle positions for the same initial 
condition as shown in figure \ref{disk-1}. The cooling prescription 
used in this simulation has been revised to include dust destruction 
over entire vertical columns as shown in figure \ref{dust-convect}b 
with $R=0.0075$ (see text). The gross morphology of the structures
that develop is quite similar to that shown in figure \ref{disk-1}, 
even though the cooling assumptions and gas temperatures in the 
hottest (inner disk) regions are different.}
\end{figure}

\clearpage

\begin{figure}
\plotfiddle{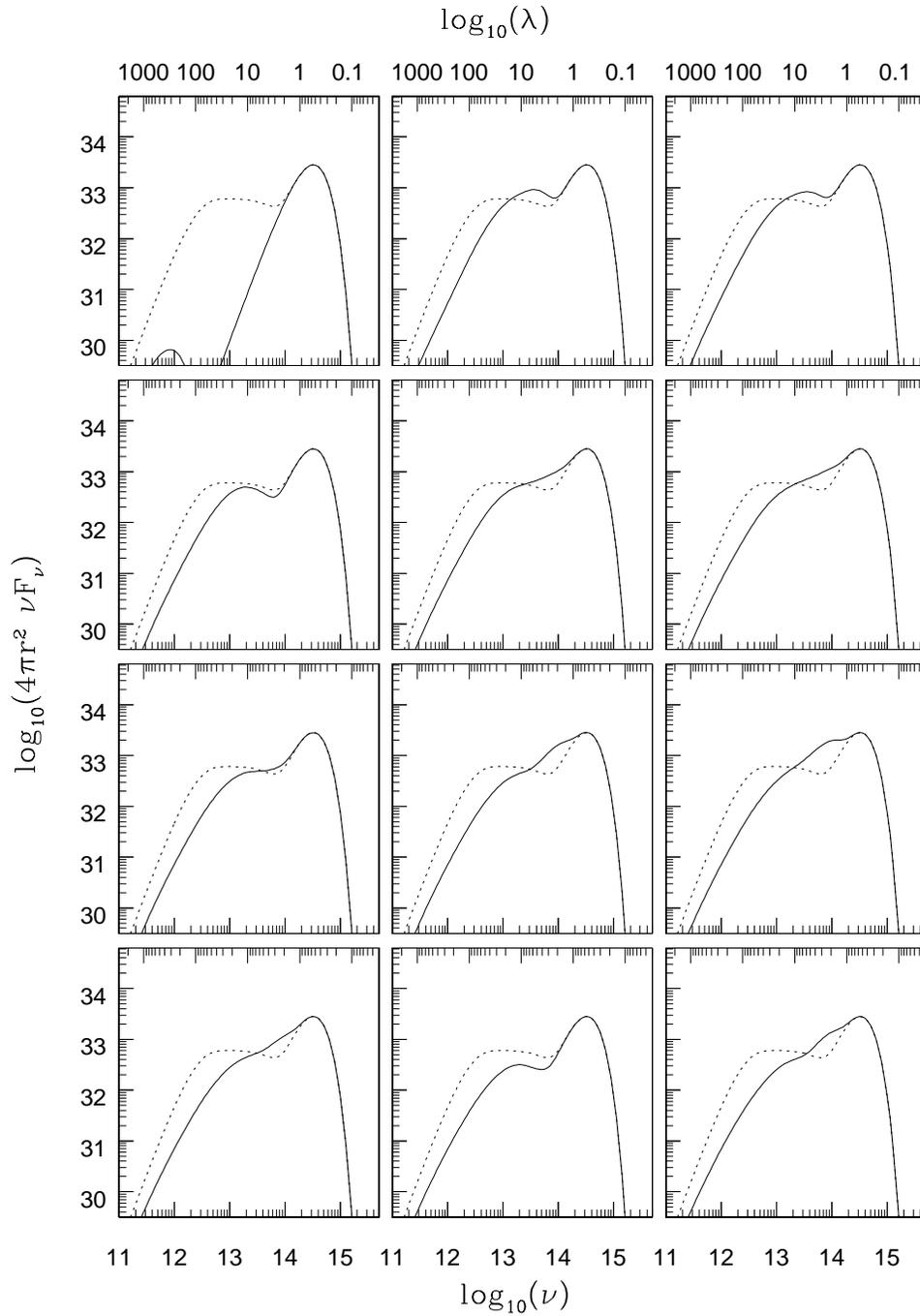}{6.53in}{0}{69}{69}{-220}{-35}
\caption[Spectral energy distribution's for the simulation shown in
figure \ref{disk-2}]
{\label{sed-2} 
\footnotesize
SED's for the simulation shown in figure \ref{disk-2}. Under the modified
cooling assumption shown here, a closer correspondence to observed systems
is found at near IR wavelengths. Substantial variations in the shape are 
seen over time scales of a few tens of years to a few hundred years. At 
some times the contribution of the star is partially masked by emission 
from the disk, while at others the star contributes nearly all of the 
short wavelength flux.}
\end{figure}

\clearpage

\begin{figure}
\plotfiddle{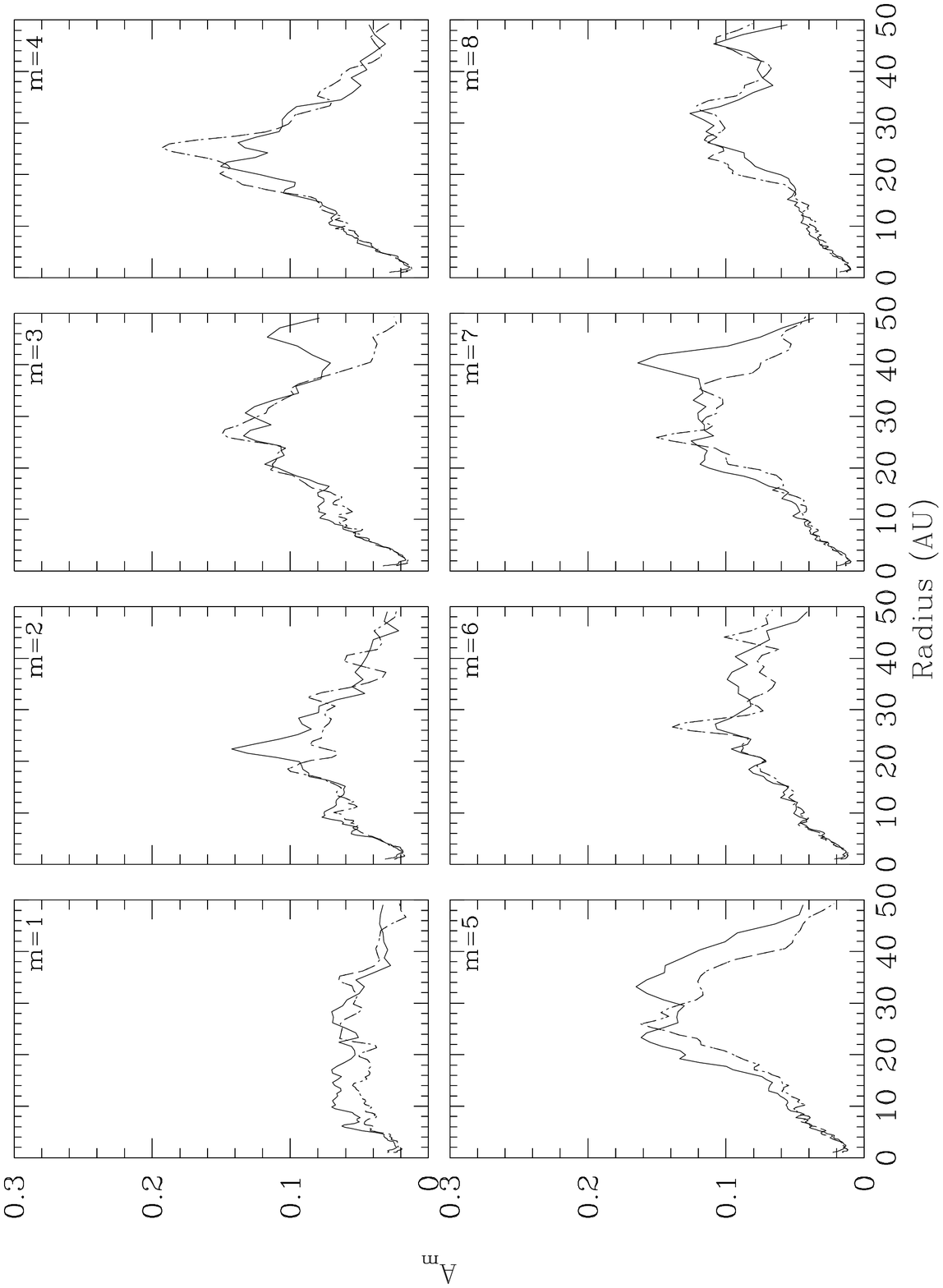}{5.5in}{-90}{56}{56}{-220}{450}
\caption[Amplitude of several spiral patterns from simulations evolved
under the `b' cooling prescription as a function of radius. ($R_D=$50~AU)]
{\label{radius-amp-b50}
\footnotesize
The time averaged amplitude of the $m=1-8$ spiral patterns as a function 
of radius for the disk shown in figure \ref{disk-2} as well as it's high
and low resolution counterparts. The solid line denotes the low resolution
run ({\it B2lo}, while the dotted and dashed lines represent the moderate
and high resolution runs, {\it B2me} and {\it B2hi}, respectively.}
\end{figure}

\clearpage

\begin{figure}
\plotfiddle{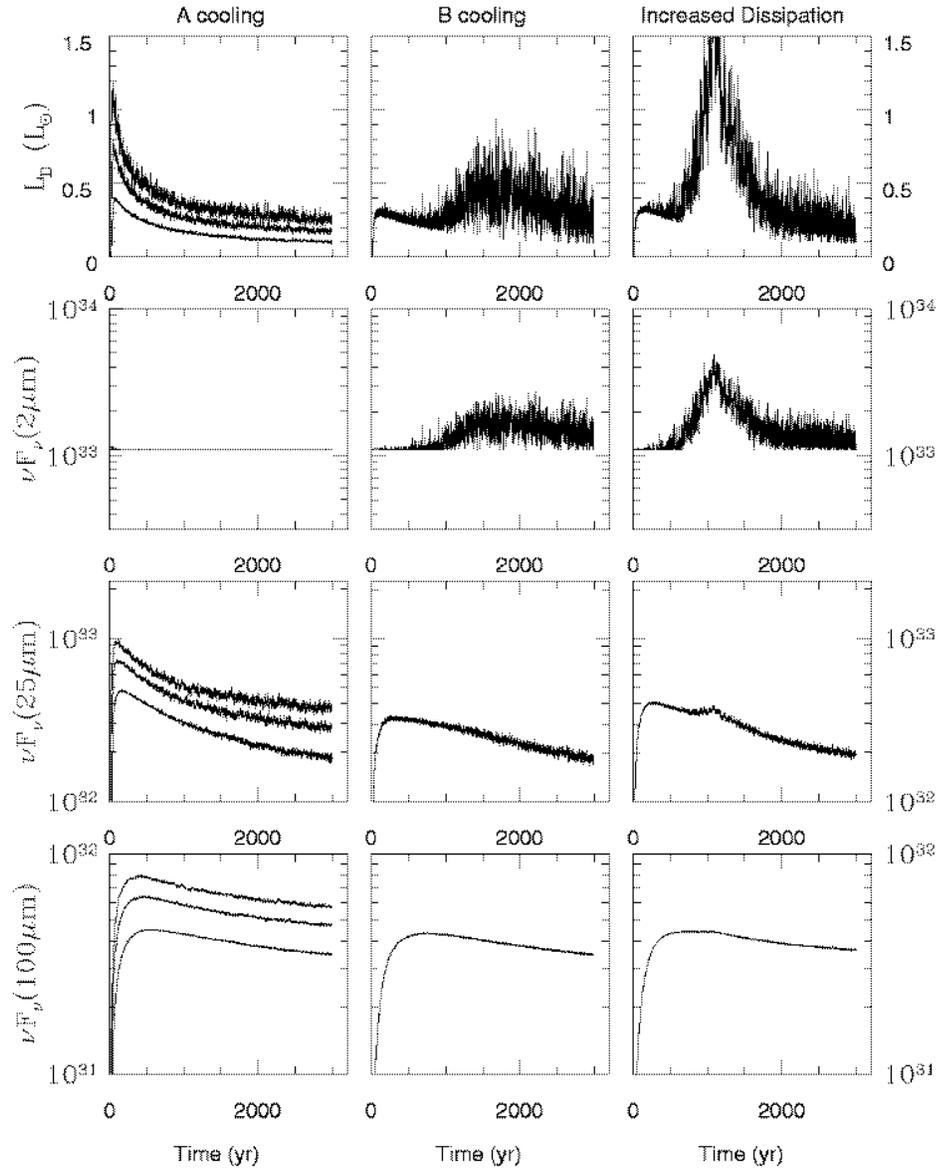}{6.75in}{0}{69}{69}{-210}{-35}
\caption[Disk flux in various wavelength bands and total luminosity
as a function of time.]
{\label{flux-var}
\footnotesize
The luminosity and emitted power at three wavelengths: 2, 25 and 100$\mu$m.
On the left are simulations {\it A2lo, A2me} and {\it A2hi} and in each panel
the top, middle and lower tracks originate from the low, middle and high 
resolution simulations respectively. The 2$\mu$m flux consists only of the
assumed constant contribution from the stellar photosphere, while the longer
wavelengths are dominated by the flux from the disk. The center panels show
only simulation {\it B2h3}. The lower resolution counterparts were suppressed
for clarity. The right panels show the results of simulation {\it H2h3}, for
which a higher effective thermal energy generation rate is present.}
\end{figure}

\clearpage

\begin{figure}
\plotfiddle{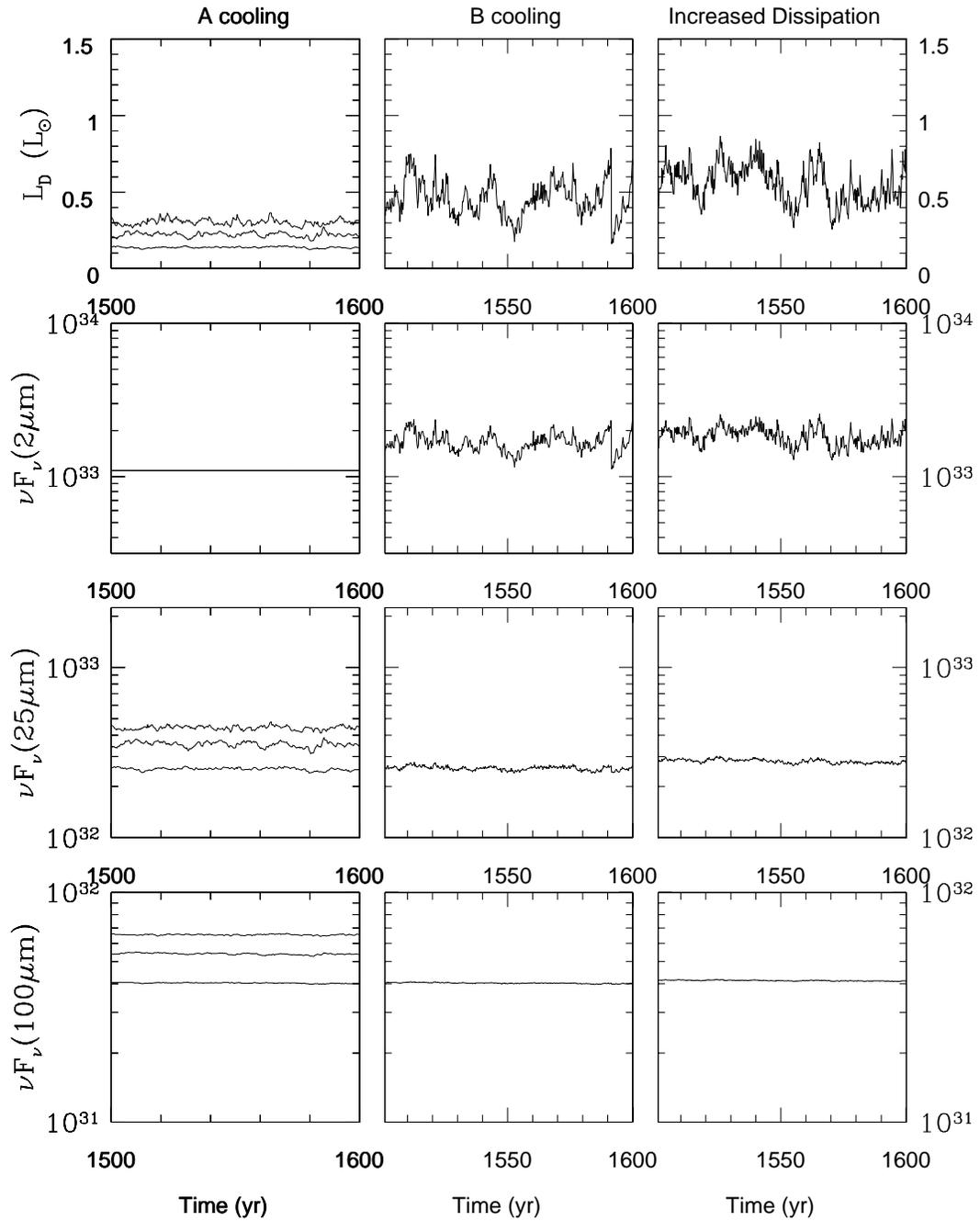}{6.75in}{0}{69}{69}{-210}{-15}
\caption[Time expanded disk flux in various wavelength bands and total 
luminosity as a function of time.]
{\label{flux-var-expand}
\footnotesize
The same as figure \ref{flux-var} but expanded in time to show the 
details of the time dependence of the flux.}
\end{figure}

\clearpage

\begin{figure}
\plotfiddle{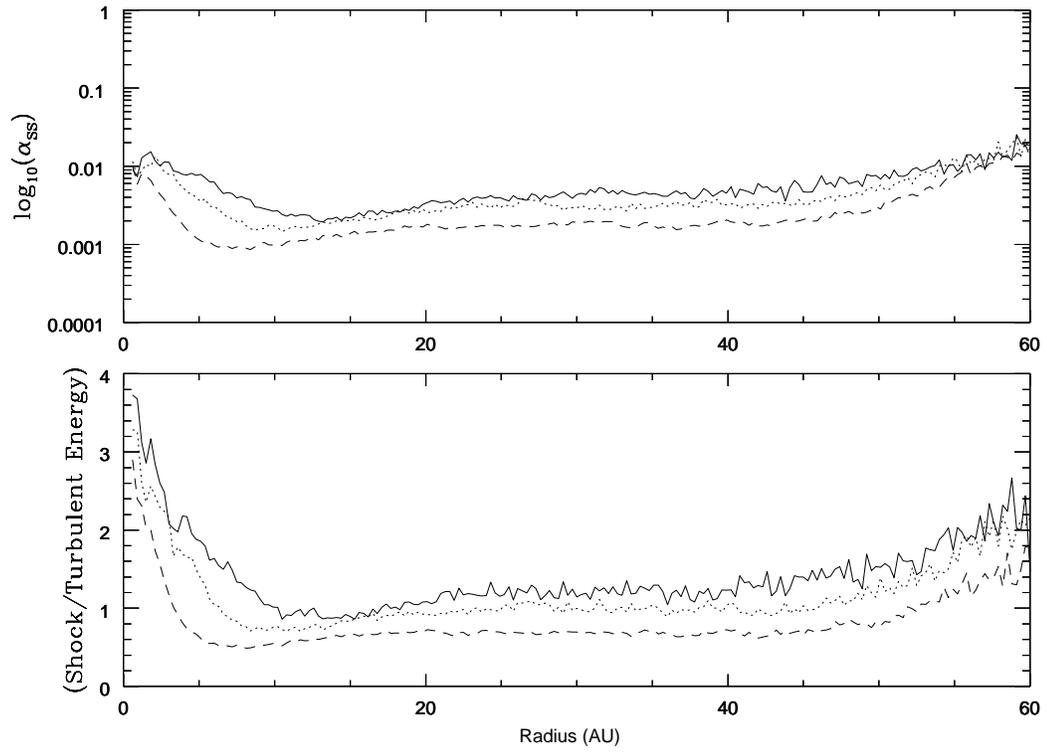}{5.5in}{-90}{53}{53}{-200}{450}
\caption[Approximate magnitude of the local dissipation in the disk
in terms of the well known `$\alpha$' parameter for viscous disks]
{\label{alpha-fig} 
\footnotesize
Top panel: The azimuth averaged value of $\alpha_{\rm SS}$ near the end
of simulations {\it A2lo} (solid), {\it A2me} (dotted) and {\it A2hi} 
(dashed). Bottom panel: The azimuth averaged value of the ratio of the 
thermal energy generation rate due to shocks and turbulence. The outer
edge of the disk has spread to $>$60~AU so that values are defined
out to the edge of edge of each panel.}
\end{figure}

\clearpage

\begin{figure}
\plotfiddle{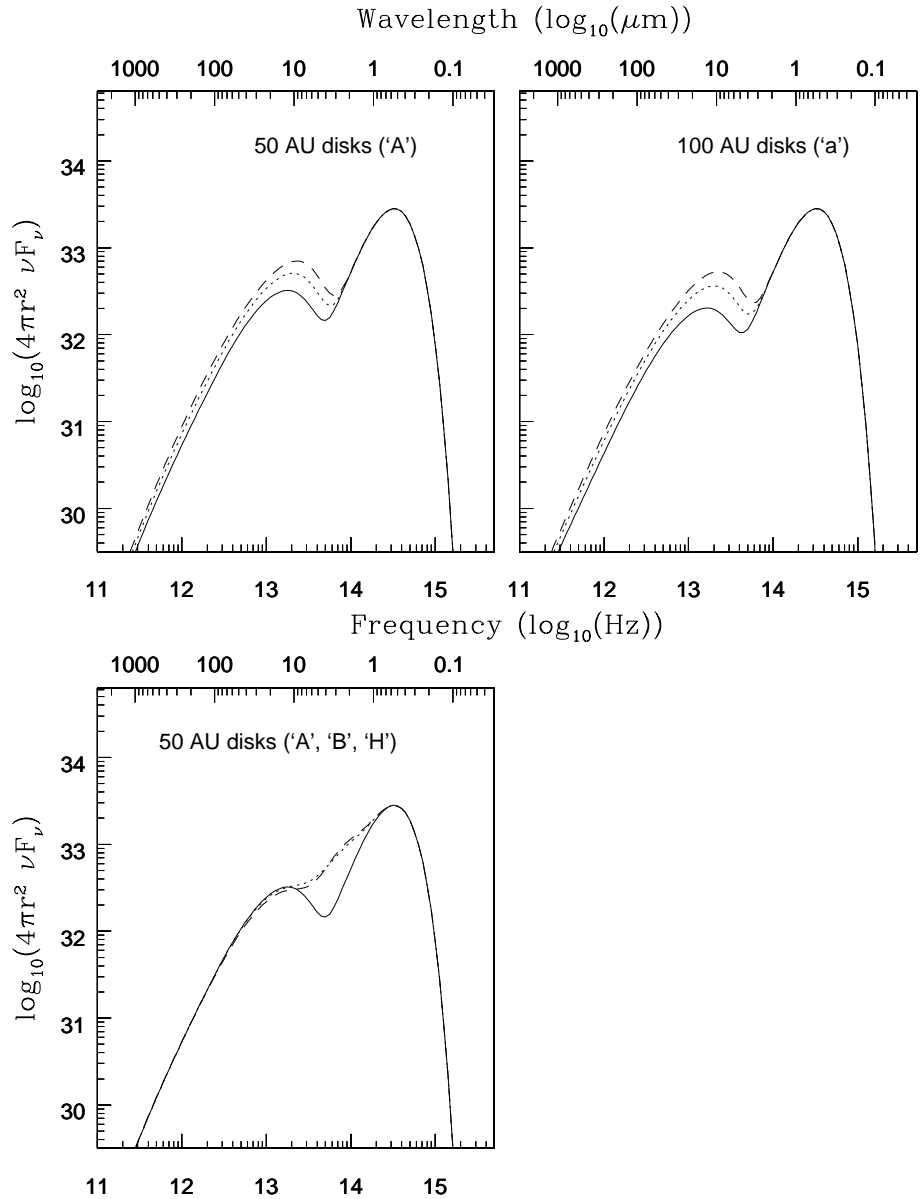}{6.4in}{0}{63}{63}{-200}{-40}
\caption[Disk flux in various wavelength bands and total luminosity
as a function of time.]
{\label{sed-50-100cmp}
\footnotesize
Time averaged spectral energy distributions of simulations of
different resolution and with assumed disk radii of 50 (top left) or
100 AU (top right). Simulations {\it A2lo, A2me} and {\it A2hi} are
designated with dashed, dotted and solid lines respectively. Similar
designations define the curves for simulations {\it a2lo, a2me} and
{\it a2hi}. Little additional long wavelength radiation is
produced in the 100 AU disks in spite of the additional surface
area. The bottom left panel shows the differences between the SED's
produced at high resolution but differing physical assumptions
(simulations {\it A2hi, B2h3} and {\it H2h3} with solid, dotted
and dashed curves respectively). The time averages are taken between
\td=1 and \td=5 except for simulation {\it H2h3}, which is taken
between \td=2.5 and \td=6 in order to reduce the effect of the
transient (see fig. \ref{flux-var} above).}
\end{figure}

\clearpage

\begin{figure}
\plotfiddle{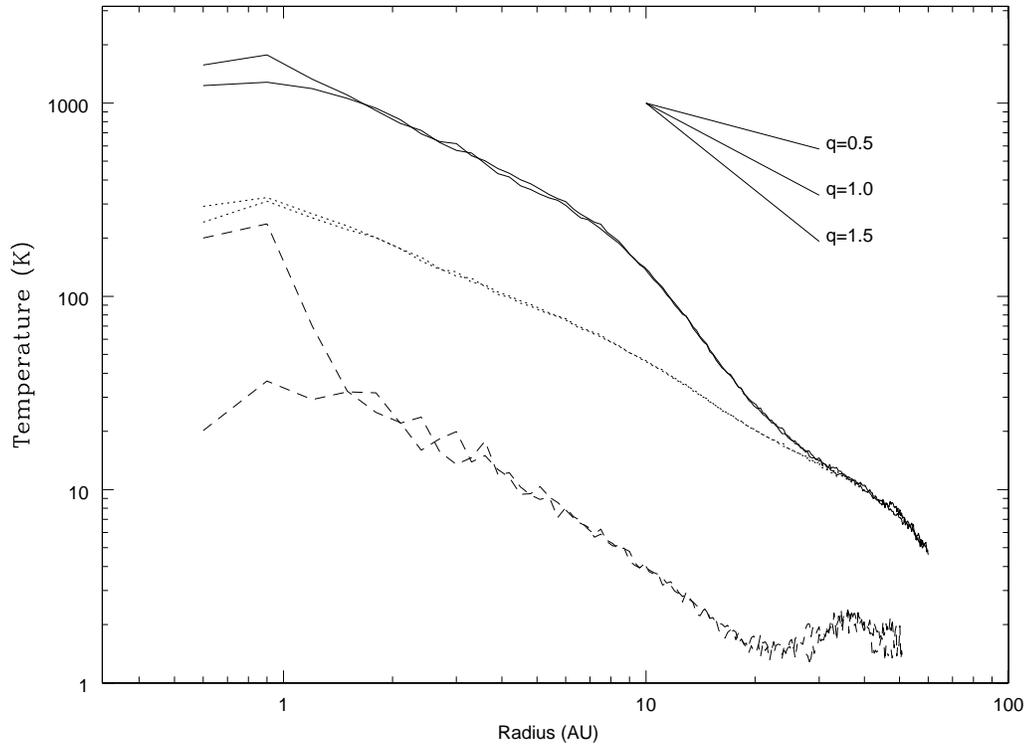}{5.5in}{-90}{53}{53}{-210}{400}
\caption[Azimuth averaged temperature structure of the disks shown in
figures \ref{disk-1} and \ref{disk-2}]
{\label{t-struct}
\footnotesize
The azimuth averaged temperature structure of the disks shown in figures 
\ref{disk-1} and \ref{disk-2}. The photosphere temperature (dotted), the
midplane temperature (solid) and the rms variation of the photosphere
temperature in each radial ring (dashed) are shown. Throughout most of
the system, the two simulations show near identical temperature structure. 
In the inner disk, the midplane temperatures for the `a' simulation differs 
from that of the `b' run by about 300~K (`a' is higher than `b'). The 
azimuth averaged photosphere temperatures are also quite similar everywhere,
however the variation in azimuth in regions where the opacities were modified 
is of the same magnitude as the temperature itself, suggesting that disk 
matter becomes transparent intermittently on time scales shorter than a 
single orbit and as local conditions dictate. The lines drawn in the upper
right of the figure represent a power law with index $q=$0.5, 1.0 and 1.5 
respectively.}
\end{figure}

\clearpage

\begin{figure}
\plotfiddle{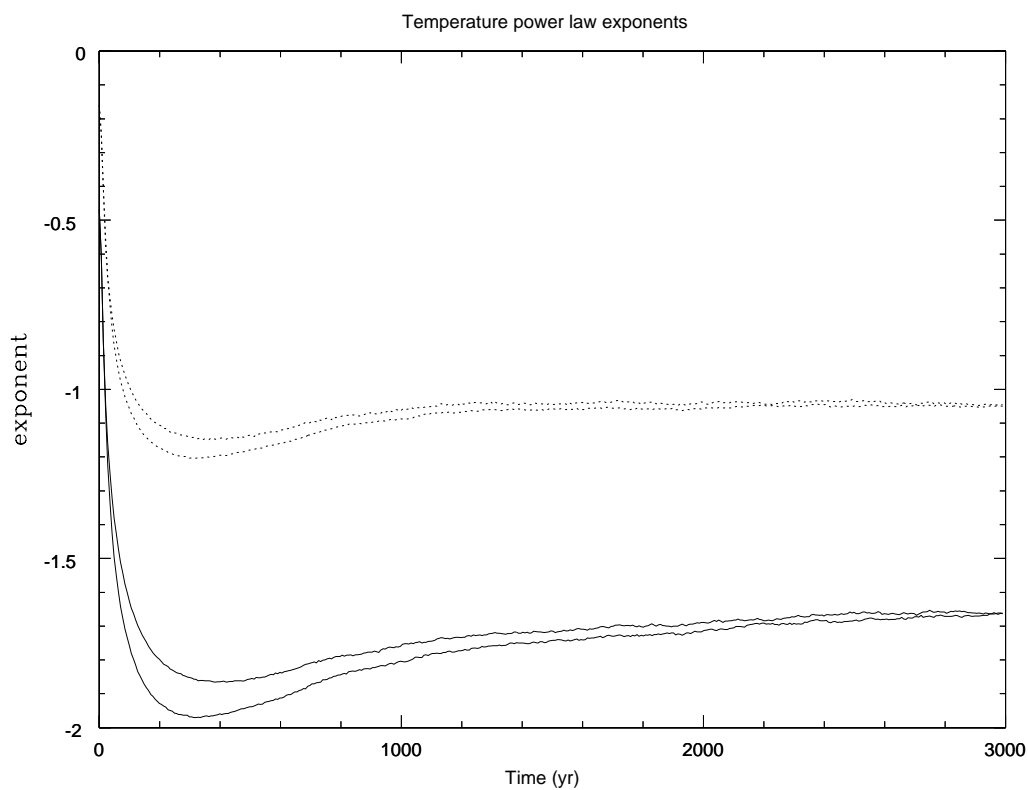}{5.5in}{-90}{53}{53}{-210}{400}
\caption[Temperature power law index at the midplane and the disk
photosphere of the simulations shown in \ref{disk-1} and \ref{disk-2}]  
{\label{t-index}
\footnotesize
The value of the temperature power law index for the simulation shown 
in figures \ref{disk-1} and \ref{disk-2} as a function of time. Indices
for both the midplane (solid) and the photosphere (dotted) of the disk
are shown. Apart from a small difference in the initial transient 
behavior the fitted exponents for each of the two simulations are 
identical. The indices for both the midplane and photosphere are far
larger than the values ($0.5 \lesssim q \lesssim 0.75$) observed in 
proto-stellar systems. We believe that the power law index is driven
to such large values by the extreme low temperatures in the outer 
portions of the disk simulation, which are lower than those derived 
from models of observations.}
\end{figure}

\clearpage

\begin{figure}
\plotfiddle{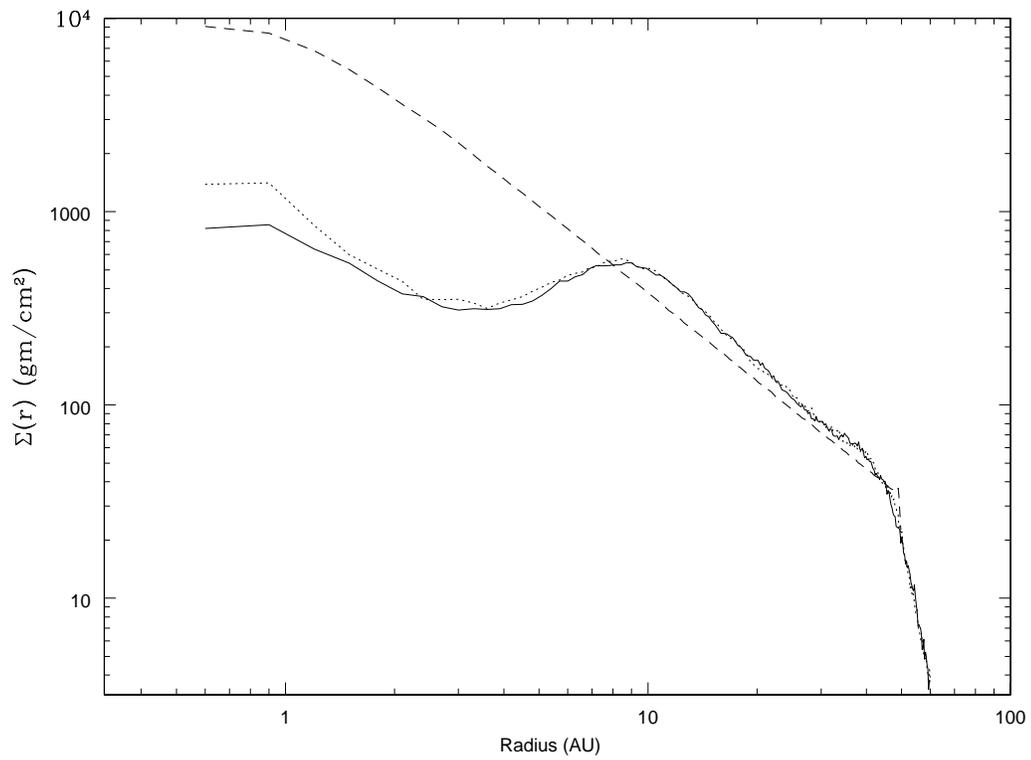}{5.5in}{-90}{53}{53}{-210}{400}
\caption[Azimuth averaged surface density of the disks.]
{\label{azavesdens}
\footnotesize
Azimuth averaged surface density of the disks in figures \ref{disk-1} and
\ref{disk-2} after evolving 4\td from their initial condition. In both the
`A' (solid line) and `B' (dotted line) simulations, the inner disk rapidly
becomes depleted of matter with respect to the initial profile, which 
increases as $r^{-3/2}$ all the way to the inner disk edge. The initial
profile is shown with a dashed line.}
\end{figure}

\clearpage

\begin{figure}
\plotfiddle{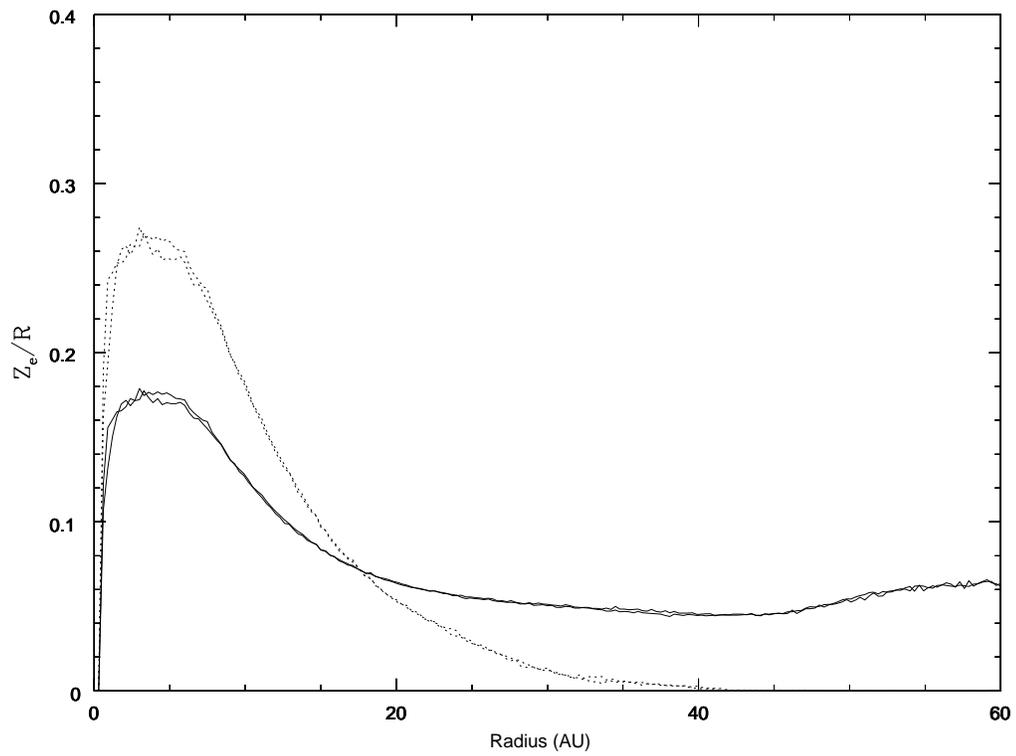}{5.5in}{-90}{53}{53}{-210}{400}
\caption[Azimuth averaged scale height of the disk]
{\label{scaleheight}
\footnotesize
Azimuth averaged scale height, $Z_e/R$ (solid), and photosphere
altitude, $Z_{phot}/R$ (dotted), of the disks in figures \ref{disk-1}
and \ref{disk-2}. }
\end{figure}

\clearpage

\doublespace

%% file: interlude2.tex
\chapter{Interlude Two}

Up until this point in my thesis we've discussed objects which are 
chronologically very young. Circumstellar gas and dust disks, almost 
by definition, can't exist for very long as steady state or equilibrium 
objects. As we've shown in chapters \ref{isodisk} and \ref{cooldisk}, 
massive disk systems can change their appearance drastically on the 
scale of only a few thousand years. For lower mass disks, where 
the effects of self gravity are not significant, the magnitude of the 
viscosity becomes the limiting factor in how long the system can exist. 
For any realistic model of what this magnitude actually is, a disk 
can only live for a few million years or, at the very outside limit,
a few tens of million years. As with any generalization though,
there are a few cases where it doesn't quite hold true. Of 
particular interest are the dusty disks around $\beta$ Pictoris and 
Vega, which are young stars quite close to the sun. 

Another example of where this generalization doesn't hold true is our 
own solar system. The planets, asteroids, comets and assorted other 
debris from the formation of the solar system have been in orbit around
the sun for more than four and a half billion years. Only a low level
of evolution of the various bodies has occured during that time, relative 
to that present during the formation of the system. There is also a tiny 
amount of matter remaining in a roughly spherical distribution (the so 
called `zodiacal cloud'), which presumably has as its source various 
volatilized materials out gassed from comets. 
In this case, the disk exists only in `fossil' form, or perhaps more 
appropriately, a circumstellar disk is an embryonic solar system. Over
the first few million years of it's history we expect a disk to evolve
into a more or less steady state, but not to go away completely with no
trace that it was ever there. It might form clumps which ultimately 
evolve into planets, brown dwarfs or low mass stellar companion.

Radial velocity measurements have proven a powerful tool for finding
planets in short period orbits around other stars. Using such measurements
a number of planets and brown dwarfs have now been discovered in orbit
around other stars. These detections have renewed interest in the
limits which can be placed on the existence of companions by current and 
future surveys. In collaboration with Roger Angel (Nelson \& Angel 1998),
I have written the paper which follows in which  we discuss some limits
which can be placed on one method for detecting such `fossil' disks around
other stars. We have developed a simple, analytic technique to relate the
detection limit obtained from a given set of data to its duration, 
precision and number of measurements. This technique, which is based
upon least squares fitting and the periodogram, delineates regions of 
mass/period parameter space accessible to radial velocity observations
of a given quality. We show that until a minimum of 15--20 measurements
have been made, it is more efficient to make more low precision measurements
than few high precision measurements.  For periods longer than the surveys
duration we derive an empirical correction to the sensitivity limit 
predicted by the analytic derivation.

We explore the effects of windowing, and also the sensitivity to periods 
longer than the total length of observations.  We show that current 
observations are not yet long or accurate enough to make unambiguous 
detection of planets with the same mass and period as Jupiter.  However, if
measurements are continued at the current best levels of accuracy (5 m/sec)
for a decade, then planets of Jovian mass and brown dwarfs will either be
detected or ruled out for orbits with periods less than $\sim$15 years.

As specific examples, we outline the performance of our technique
on large amplitude and large eccentricity radial velocity signals
recently discussed in the literature and we delineate the region explored
by the measurements of 14 single stars made over a twelve year period
by Walker et al. (1995). Had any of these stars shown motion like that
caused by the exo-planets recently detected, it would have been easily
detected. The data set interesting limits on the presence of brown
dwarfs at orbital radii of $\lesssim$5--10 AU. The most significant features
in the Walker et al. data are apparent long term velocity trends in
36 UMa and $\beta$ Vir, consistent with super planets of mass
of 2 $M_J$ in a 10 year period, or 20--30 $M_J$ in a 50 year period.
If the data are free of long term systematic errors, the probability of
just one of the 14 stars showing this signal by chance is about 15\%.

We then apply our technique to suggest an effective strategy for
new and continuing radial velocity searches. For large surveys
beginning now or proposed for the near future the factor most limiting
detections is the finite amount of telescope time allocated to the
search. Using this constraint, we suggest an observing strategy for
future large radial velocity surveys which, if implemented, will allow 
coverage of the largest range of parameter space with the smallest amount 
of observing time per star. We suggest that about 10--15 measurements
be made of each star in the first two years of the survey, then
2--3 measurements per year thereafter, provided no (or slow) variation
is observed. More frequent observations would of course be indicated
if such variations were present.

%% file: RVchap.tex
\chapter{The range of masses and periods explored by radial velocity searches 
for planetary companions\label{RVchap}}

For nearly two decades most high precision radial velocity surveys of 
nearby stars were focused on detecting radial velocity variations in
stars due to companions with mass and period of Jupiter. The signature 
would consist of changes in the relative stellar radial velocity with
a period of a decade and amplitude of a few tens of meters per second
or less, depending on orbital inclination with respect to the solar
system. The surprising recent result, triggered by the discovery
of 51~Peg~B by Mayor \& Queloz (1995), has been the finding that as 
many as 5-10\% of solar type stars have companions with mass 
$<10 M_J$ and with periods less than $\sim$3 years. No sub-stellar 
companions with periods longer than $\sim$3 years have so far been 
detected by radial velocity searches.

Are Jupiter mass companions at longer periods rare, or is it simply 
the case that current observations do not have the length or 
sensitivity to see them? Is the theoretical prediction by Boss (1995) 
correct, that Jovian planets should form preferentially at $>$4-5 AU 
separations from their primary? Our purpose in this paper is to show 
what we can learn from velocity data of a given duration and accuracy,
to help plan continued programs. 

The best measurement errors for a series of radial velocity measurements  
so far published are those of Butler et al. (1996), who observe a
magnitude $V=5$ star and quote an accuracy of 3 m/s for measurements taken
over one year. Measurements up until this work have been limited to
a lower accuracy standard of about 15 m/s.  Several other programs
(see section \ref{strategy} for a list of radial velocity search 
programs currently underway) are planning new or expanded searches with
a goal of obtaining measurements with similar accuracy. In light of
these efforts, and in expectation of their eventual success in obtaining
such accuracy, we shall use 5 m/s as a `canonical' value for the error
in many of the examples and the discussion below. Such advances in 
radial velocity calibration allow accuracy to be relatively free from
systematic error. Poissonian photon noise remains as the fundamental
limit to accuracy. In this limit, strong constraints can be placed
upon the existence of periodic radial velocity signals in a given 
set of data, given a suitable analysis technique.  

Many efforts have been made to determine whether a given set of data 
contains a signal. Most of those in common use are based upon the 
periodogram analysis techniques discussed by Scargle (1982). This 
technique is shown to be equivalent to a least squares fit for the 
signal at a given period, and he derives an exponential probability 
distribution of obtaining a false alarm from a given set of data.
Horne and Baliunas (1986 hence HB) have refined the technique by showing
that this exponential must be normalized to the total variance
of the data and derived an empirical expression for the number
of independent frequencies available to a set of data. Further
refinements (Irwin et al. 1989, Walker et al. 1995) account for
variable weighting of individual data points and correlations between
fitted parameters. 

Our work represents a different approach in which, rather than dealing 
with least squares minimization indirectly through a periodogram 
analysis, we examine the best fits to the data directly and determine
their significance.  We derive an analytic expression for the 
probability that a given best fit velocity amplitude is non-random.
We first develop analytical expressions relating sensitivity to 
planetary companions of different masses and periods, given velocity
measurements of specified accuracy, duration and number. Motion 
with periods longer than the duration of observations is detected 
with reduced sensitivity, and this reduction is explored by Monte 
Carlo methods. We illustrate our analysis technique by application 
to the published set of radial velocity data from Walker et~al. 
(1995), the longest time baseline survey so far published, with
quoted precision of 15 m/s. Limiting our analysis to the subset of
14 stars which have no known visual binary companion, we obtain 
quantitative upper limits to companions masses for orbital periods 
of a few days to periods as long as 100 years. Finally, we suggest
a strategy for efficiently implementing a search of a large number 
of stars for radial velocity signatures due to the presence of
a companion.

\section{Analysis Technique}

If a companion of mass $M_c$ exists in a circular orbit 
with inclination $i$ around its primary $M_*$, it will perturb the 
radial velocity of the star as observed from earth by: 
\begin{eqnarray} \label{vel-theo}
v(t) & = & {\left({{2\pi G}\over{PM_*^2}}\right)}^{1\over 3}M_c\sin{(i)}
                       \sin({{2\pi t}\over{P}} + \phi) \\
		      & = & K\sin({{2\pi t}\over{P}} + \phi),
\end{eqnarray}
where $K$ is the amplitude of velocity of the companion in a
circular orbit around its primary, $P$ is the period of the orbiting 
companion, $G$ is the gravitational constant and $\phi$ is an 
arbitrary phase factor. If an observer can detect the small temporal
changes in relative velocity due to a companion, then using fitting
or periodogram techniques, it becomes possible to derive a mass 
(or mass limit) for that companion.

Suppose that velocity data $v(t_i)$ have been obtained in observations 
extending over time interval $P_0$. For a given orbital period, $P$, 
we can perform a least squares fit to the data with the equation:
\begin{equation}\label{fit-eq}
v(t)= v_s\sin({{2\pi}\over{P}} t) + v_c\cos({{2\pi}\over{P}} t) + \gamma,
\end{equation}
to produce `best fit' values for the components of the motion $v_s$, 
$v_c$ and $\gamma$. At long periods and with a potential signal 
whose phase is unknown, the constant offset, $\gamma$, allows for
the possibility that a companion at a radial velocity extremum (ie.
near it's maximum or minimum) is properly modeled by the fit 
function.  For shorter periods ($P<P_0$) its inclusion or exclusion
has negligible effect so we will focus initially on this
domain.  Given fitted amplitude coefficients $v_s$ and $v_c$, a simple
trigonometric identity ($K = \sqrt{v_s^2 + v_c^2}$) 
yields the amplitude of the stellar velocity perturbation due to the 
companion.  From there we identify $K$ with the leading coefficient
in equation \ref{vel-theo} and invert to obtain a `best fit' companion 
mass:
\begin{equation}\label{mass-eq}
M_c = {{K}\over{\sin(i)}}\left({{PM_*^2}\over{2\pi G}}\right)^{1/3}.
\end{equation}
Fitting higher order harmonics would be used to refine the fit and
recover information about the orbital eccentricity of a companion.

The orbital inclination remains an unknown parameter in a set of radial
velocity data. Statistically speaking however, the average companion
mass of a set of systems randomly oriented in space which give amplitude
$K$ will be:
\begin{eqnarray}\label{mass-eq-ave}
\langle M_c\rangle={\int_0^{\pi/2} M_c\sin(i)di} 
     & = & \left({{PM_*^2}\over{2\pi G}}\right)^{1/3}
	      \int_0^{\pi/2}{{K}\over{\sin(i)}}\sin(i)di  \\
     & = & {{\pi}\over{2}}K\left({{PM_*^2}\over{2\pi G}}\right)^{1/3}.
\end{eqnarray}
Thus the average value for a companion mass is $\pi/2$ times
the directly derived $M_c\sin(i)$ value. Conversely, a companion with 
some mass will appear on average a factor of $2/\pi(\approx0.64)$ less
massive than its true value. Very large masses cannot be ruled out but 
do become increasingly improbable, with the probability that a given
mass, $M$, is exceeded being given by the formula:
\begin{equation}\label{m-exceed}
{\cal P}(M_c>M) = 1 - \sqrt{1 - \left({M_c\sin(i)}\over{M}\right)^2}
\end{equation}
For example, while values of the companion mass will be greater than twice
$M_c\sin(i)$ for 13\% of a large sample, the chance of a mass being greater 
than 10$M_c\sin(i)$ are only 0.5\%. The true companion mass will exceed
2/$\sqrt{3}(\approx 1.15)$ times the measured $M_c\sin(i)$ value in 50\%
of cases.

\subsection{Probability of a given velocity amplitude being 
exceeded by chance\label{chance-prob}}

In the absence of an unambiguous detection of a signal at some
period, we are faced with the question of whether a particularly large
fitted velocity amplitude at some period represents a real detection. 
Such spikes will occur, because the data are noisy, and the frequency 
analysis must be taken over a large number of possible periods (from a 
few days to many years). Adjacent fitted periods may have widely
different best fit velocity amplitudes even when the data have
no embedded signal.  What criterion can we apply to tell if a spectral
peak is improbably large compared to these noise spikes? More
generally, if the data for a star are analyzed in some way, what is the
probability that a given outcome would have occurred by chance? In this
section we obtain an analytical expression for the velocity amplitude 
(and hence companion mass) that will be exceeded by chance, with a
given probability and in a given frequency range.  

Suppose that in a given set of velocity measurements $v(t_i)$,
there is no real signal and that each measurement is drawn from a 
Gaussian distribution with mean zero and standard deviation $\sigma_p$.
The data can be fit to eqn. \ref{fit-eq} to produce coefficients 
of some amplitude ($v_s,v_c$). With no true signal, both $v_s$ and $v_c$
will be normally distributed about zero with standard deviation 
$\sigma_s=\sigma_c=\sigma$ and the phase of the fitted curve will be
uniformly distributed. For a set of $n_0$ measurements, taken randomly 
over a time period $P_0$, this assumption leads to the expression:
\begin{equation}\label{sigma-eq}
\sigma =  \sqrt{{2}\over{n_0}}\sigma_p.
\end{equation}
where factor $\sqrt{2/n_0}$ is derived from the least squares error 
analysis (see eg. Bevington and Robinson (1992) ch. 7) fitting a
periodic signal to random noise.

The probability of any data set with zero expectation value for $v_s$ and
$v_c$ to have any particular fit values is:
\begin{equation}
p(v_s,v_c) =  {{1}\over{2\pi\sigma_s\sigma_c}}
{e^{{-v_s^2}\over{2\sigma_s^2}}e^{{-v_c^2}\over{2\sigma_c^2}}}d{v_s}d{v_c}.
\end{equation}
which, converted to amplitude and phase gives:
\begin{equation}\label{prob-1}
p(K',\phi) = {{1}\over{2\pi\sigma^2}}{e^{{-K'^2}\over{2\sigma^2}}} 
		                          K' dK'd\phi.
\end{equation}
If we integrate this probability over all $\phi$ and from 
zero\footnote{For comparison, the integrated probability for a normal
random variable is given by:
\begin{displaymath}
{\cal P}= {{1}\over{\sqrt{2\pi}\sigma}}\int_{-K}^{K}
                        e^{{-K'^2}/{2\sigma^2}}dK'.
\end{displaymath}
} 
to some value $K$, we get the total probability, ${\cal P}$, of a fit with 
velocity amplitude $K$ or smaller:
\begin{equation}\label{conf-eq}
{\cal P} = 1 - e^{{-K^2}/{2\sigma^2}}.
\end{equation}
This probability applies to analysis of any single period. In
practice we are interested in the probability of a velocity amplitude
being exceeded by chance in a range of periods. If we assume that 
the probability of a given fit at one period is independent of 
every other period, then for $N$ periods the probability, $X$, that
no fit value exceeding a value $K_X$ will occur is the product of the
individual probabilities $X={\cal P}^N$, which to leading order gives
\begin{equation}\label{prob-eq}
X = \left(1 - e^{{-K_X^2}/{2\sigma^2}}\right)^N \approx
                         1 - Ne^{{-K_X^2}/{2\sigma^2}}.
\end{equation}
Higher order terms in the right hand equation converge to zero
as progressively higher power exponentials.  We can invert this 
equation to to derive a limit on the velocity as: 
\begin{equation}\label{v-eq}
K_X = \sqrt{-2\sigma^2\ln{\left({{1-X}\over{N}}\right)}}
\end{equation}
which expresses the velocity amplitude which will be exceeded
by any of $N$ fits to random data in a given period range
with probability $1-X$. 

The appropriate number of independent periods is related to the width 
of peaks in the frequency spectrum given by $df$=$1/P_0$. To be certain 
of sampling at a frequency that is close to the peak, we suppose 
that the sampling is made at frequency intervals $df=1/(2\pi P_0)$.
The number, $N$, of independent frequencies (or periods) in a given 
range is then given by:
\begin{equation}\label{n-eq}
N =  2\pi\left({{f_1 - f_2}\over{f_0}}\right)
\end{equation}
where $f_1$ and $f_2$ are the limiting frequencies of the range bounded
by periods $P_1$ and $P_2$ and $f_0=1/P_0$.

Finally, combining eqns \ref{sigma-eq}, \ref{v-eq} and \ref{n-eq}
we obtain an expression in terms of the accuracy $\sigma_p$, duration
$P_0$, the number of measurements $n_0$ and the probability $X$
that a velocity amplitude $K$ will be exceeded in a given frequency range:
\begin{equation}\label{vprob-eq}
K_X = 2\sigma_p\sqrt{{{1}\over{n_0}} 
        \ln{\left({{2\pi P_0\left(f_1 - f_2\right)}\over{1-X}}\right)}}.
\end{equation}
The value $K_X$ varies directly with $\sigma_p$ and varies with the 
inverse root of $n_0$, as we would expect from the central limit theorem.
Its sensitivity to the other parameters and our sampling assumptions 
depends on the details of the survey, but in general we will find that
the factor inside the natural logarithm is much greater than 1, so 
that a factor 2 change in any of the arguments produces only a small
fractional change in the value of $K_X$.

As an example, suppose that a high quality survey were made over a 
decade, with a total of $n_0=50$ observations per star and with 
rms accuracy, $\sigma_p=5$ m/s. For a false detection in a one
octave range around $P=4$ days, the velocity amplitude $K_X$ from 
eqn. \ref{vprob-eq} is 5.2 m/s. At 4 years, the amplitude is 3.9 m/s. 
If the star's mass is the same as the sun's, then from eqn. \ref{mass-eq}
we find these velocities for 1\% false detections will correspond 
to companions masses $M_c\sin(i)$ of 0.04 and 0.22 Jupiter masses
respectively. In practice, if a large number of stars are to be
sampled, say 100, and we would want a small probability of a false
detection in the sample, say 10\%, then we would want to decrease 
the probability to 10$^{-4}$ per octave per star. In this case, 
the mass limits increase to 0.047 and 0.28 $M_J$ for each range.
The small increase of only some $\sim$25\% is due to the fact 
that the argument of the natural logarithm in eqn. \ref{vprob-eq} 
is near 20 for this case.

\section{Monte Carlo Analysis\label{monte-carlo}}

Eqn \ref{vprob-eq} will fail for periods longer than the span of
observations $P_0$, under conditions in which the data collection
is periodic, or if the total number of observations $n_0$ is too 
small.  This is because the windowing may imprint its own signature 
upon the derived best fit parameters and the assumption that random 
data are fitted with random phase breaks down. We devote this section
to a Monte Carlo analysis of synthetic radial velocity data, in order
to understand the regimes in which our analysis may fail and the 
manner of its failure. In this way we can eliminate false detections 
and establish the validity of a trend in the data consistent with a
true periodicity.

For our numerical experiments we assume radial velocity data are 
gathered for either 6 or 12 years. These data are spaced 
randomly in time subject to the constraints that data be `gathered'
during the same 6 month period of each year, that they be gathered 
only during 1/2 of each 29.5 day lunar cycle and that they be gathered 
only at `night'. We run a grid of nine Monte Carlo experiments 
varying the frequency of observation over 1, 5 and 20 observations per
year and the precision for each measurement over 5, 15 and 30 m/s.

We set velocities corresponding to the time of each observation using 
a Gaussian random noise term and the input error as:
\begin{equation}\label{vrand-eq}
v_{\rm sim}(t_i) = Rv_{\rm err}(t_i)
\end{equation}
with $R$ the random noise term and $v_{\rm err}$ is the error for each
point. The value for $v_{\rm err}$ is assigned as noted above.
We use the pseudo-random number generator `{\tt ran2}' provided by
Press et al. (1992) and the rejection method to create Gaussian random
numbers. In the analysis that follows, we fit a total of 3000 data sets
for the amplitude components $v_s$, $v_c$ and $\gamma$ for each star over
period ranges from 3 days to 100 years. The boundaries of each period 
range are defined in table \ref{range-tab}.  We increase each 
successive fitted period by an amount such that the total number of 
orbital cycles over the full observation length decreases by 1/2$\pi$ 
(1 radian) as in the analysis above or the period increases by 
1/5~year, whichever gives the smaller interval. The chance of any 
particular outcome is given by the fraction of the synthetic data 
sets with that outcome.  

\subsection{Confirmation of the Analytical Results}

For the subset of experiments with assumed 5 m/s precision, figure 
\ref{mc-anal-cmp} shows the radial velocity amplitude for each fitted period 
which is exceeded in 1\% of the Monte Carlo trials, ie. there is a 99\% 
probability that a specific period analyzed will not exceed this value.
Experiments with higher or lower assumed precision produce limits
scaled upward or downward on the plot but otherwise show the same
qualitative features.  Also included are the $N=1$ limits provided 
directly by eqn. \ref{v-eq}.  In general, the Monte Carlo results 
confirm the validity of the analytical results above. The difference
between $N=1$ analytic and Monte Carlo results varies about 2\%, 
consistent with statistical fluctuations, except at the assumed windowing
periods and at periods longer than $P_0$. The analytical prediction for
the experiment with the most sparsely taken data (1 measurement per 
year for 12 years) lies some $\sim15\%$ below the Monte Carlo result
for periods less than 2 years, but agrees to $\sim2\%$ over the 
remainder of the valid period regime.  

A comparison of the limits provided by the analytical (eqn.
\ref{vprob-eq}) and Monte Carlo methods for each period range noted
in Table \ref{range-tab} are also shown in figure \ref{mc-anal-cmp}.
The assumed windowing periods are masked out of each of the Monte Carlo
limits and the results represent limits based on the remaining portion
of each range. In general, the Monte Carlo results again confirm the
validity of the analytical result to within a few percent, with the
exception of the series with only one measurement per year. In that
case, the Monte Carlo experiment produces limits which are some 50\%
or more larger than eqn. \ref{vprob-eq} predicts.  

We consider in turn in the sections below the differences between
the analytical derivation and the Monte Carlo results due to the long
period fall off in sensitivity, due to small numbers of observations,
and due to the inherent windowing in the data. 

\subsection{Loss of Sensitivity at Long Periods\label{long-per}}

The results shown in figure \ref{mc-anal-cmp} show that at periods 
longer than the 12 year window the sensitivity to a velocity signal 
drops off in very nearly power law form. In light of this
behavior, we adapt an {\it ad hoc} prescription for the velocity limit
using the eqn. \ref{vprob-eq} result at short periods and a power law
at longer periods as
\begin{equation}\label{vprob-eq-long}
\hat K_X = \cases{ K_X                  & for $P<\beta P_0$;\cr
                   K_X \left({{P}\over{\beta P_0}}\right)^\alpha
                                        & for $P>\beta P_0$; \cr }.
\end{equation}
We then fit for the free parameters $\alpha$ and $\beta$ and thereby
recover limits for periods much longer than that of the observing
window. In this equation, we assume that the value of $K_X$ used
for long periods ($P>\beta P_0$) is that defined by the last period
range prior to the onset of the fall off. This assumption ensures a
smooth joining of the two regimes.

We fit the Monte Carlo results for the constants $\alpha$ and 
$\beta$ in eqn. \ref{vprob-eq-long} for each of the experiments and plot
their values in figure \ref{mcl-bestf} for both the six and twelve year
observing windows studied. The fitted values for the 12 year window are 
typically:
$$
\alpha \approx 1.86
$$
with the turn off in sensitivity beginning between 
$$
1.4 \leq \beta \leq 1.45
$$
for the 99\% probability curve and similar values for the 99.9\%
probability curve.  A slightly steeper power law exponent 
($\alpha\approx 1.92$) is found for a 6 year window. 
If we err on the side of caution and assume that the turn-off occurs at
the {\it small} end of the range (by setting $\beta=1.3$, for both the
99\% and 99.9\% probabilities), then we provide slightly more conservative 
limits than the best possible based on our Monte Carlo analysis. Under this 
assumption, we have included in figure \ref{mc-anal-cmp} the long period 
fits for the velocity limits placed upon the data by eqn. \ref{vprob-eq-long},
and the shorter period limits for 11 period ranges less than $1.3\times$12 
years. 

\subsection{Limits of Sparse Data\label{sparse}}

When a data set contains only a few measurements, a least squares 
analysis will depend strongly upon the measured value and placement 
in time of each measurement. How many data are needed to assure that
the random data/random phase assumption is reliable and we are able 
reproduce the results of equations \ref{v-eq} (with $N$ set to unity)
or \ref{vprob-eq} (for octave period ranges)? Is there a difference in 
the number of measurements that must be made if we assume a strategy 
of taking, say, one or two measurements per year over a long period 
or taking several measurements per year but over a much shorter
baseline?

Taking the first strategy, we assume the data are gathered over a 6
year span with an error in each measurement of 5 m/s. If a star is
observed with a frequency of one observation per year, we find 
(figure \ref{vlim-sparse6}) that the eqn. \ref{v-eq} limits with
$N=1$ underestimate the Monte Carlo results by more than a factor of 
two for periods shorter than 1 year, and by a smaller margin at all 
periods. The same experiment with a 12 year span shown in figure 
\ref{mc-anal-cmp} shows a much smaller ($\sim$15\%) difference. 
Increasing to three observations per year for 6 years the analytic
equation underestimates the limits by $\sim$10\%, while 5 measurements 
per year duplicates the analytic results to 5\% or better.

For octave sized ranges, the analytical and Monte Carlo results converge
somewhat more slowly. Figure \ref{mc-anal-cmp} shows that a single
measurement per year over 12 years is sufficient only to provide limits
a factor of two higher than would be predicted analytically for periods 
less than 2 years. When data are gathered at the higher rates shown 
(5 and 20 obs/yr), the agreement is excellent. An experiment with 
two measurements per year (not shown), for a total of 24 measurements, 
is sufficient to recover the analytical form to $\sim$10\% in all 
period bins. With the six year baseline shown in figure 
\ref{vlim-sparse6}, agreement at the $\sim$20\% level is reached if
three measurements per year (18 total) are taken.

Taking the second strategy, we assume data are gathered over a
two year window. We do not believe we can rely upon octave range
limits for such a short data gathering period because of the large 
effects of windowing, which we discuss below. The long period fall
off is similarly affected. We therefore limit our discussion for these
experiments to limits for individual periods, shorter than about one 
year. With a two year data window and a total of 6 measurements (three
measurements per year), we again find (figure \ref{vlim-sparse2}) that 
the Monte Carlo limits exceed those of eqn. \ref{v-eq} with $N=1$ by 
more that a factor of two.  Increasing to 6 measurements per year (12 
total), we lose only 15\% of the maximum sensitivity for $N=1$, while 
12 measurements per year (24 total) recovers the analytic results with 
only a $\sim$5-7\% difference.

In order to obtain limits which retain the benefits of a given 
precision to within 15\% at any single period, we find that at
least $\sim$12 or more observations of a star must be made. This 
number of observations produces limits a factor of two larger than 
predicted over octave ranges.  To reduce the difference
to $\sim$5\% for a single period and 15\% over octave ranges
requires at least 18-20 measurements. Barring windowing effects, 
these minimum requirements do not seem to depend strongly upon the
time span over which the data were gathered, but only upon their 
accuracy and number.

\subsection{Windowing\label{windowing} }

Sensitivity loss of a factor of two or more is present in `blind spots' 
for any single period near the assumed lunar and annual windowing periods
for every experiment performed. There are also double period counterparts
and beat periods between the lunar and annual data windows, though lower
sensitivity loss is evident there. Day/night windowing effects are not 
visible in the limits due to their extreme short periodicities. When the 
data are sparse and the data are gathered over a short period $P_0$, the 
effects are especially pronounced.  Figure \ref{vlim-sparse2}) shows
that for a period $P_0$ of two years, the lunar windowing effects are
observable not only at the lunar orbital period, but also at the double, 
triple and quadruple period aliases. Additionally, fitting for the long
period turn-off becomes of little use because the turn-off occurs at 
a period with lower sensitivity than can be modeled analytically. 

Based on these results, we suggest that the limits which can be placed 
on signals at periods corresponding to a lunar or annual windowing period 
cannot be reduced beyond a factor of two greater than that given by eqn.
\ref{v-eq} with $N$ set to unity at a windowing period or a
factor of $\sim$3/2 at one of its double or beat period counterparts.

\section{Comparison to Periodogram Techniques\label{compare}}

To obtain definitive probability that a signal that been detected at
some period is nonrandom, nothing less than a full Monte Carlo
analysis is adequate. For a large survey which is continually updated
as more data are gathered, such analysis is unfeasible because
of the considerable commitment of computational facilities to perform
a statistically meaningful analysis. Even for the computers of today,
a sample of 500 stars might prove unmanageably burdensome. To reduce
the effort required per star, either periodogram or fitting techniques
such as ours may offer a lower cost alternative.  We will now make a
comparison of our technique to periodogram techniques in common 
use.

Each technique is based upon a $\chi^2$ analysis of the data.
Indeed, for equally weighted data least squares analysis and 
periodogram analysis have been shown (Scargle 1982) to be equivalent.
The main difference lies in the fact that on the one hand,
a periodogram utilizes a normalized measure of the power of the signal
at some period while our technique relies directly on the value of 
the best fit velocity amplitude. Additionally, with the present 
analysis, we allow the data to be fit with unequal weights, though 
the amplitude limits derived are based upon only upon equally 
weighted data.

Let us examine the least squares fitting procedure and, for 
purposes of illustration, limit ourselves to the case of fitting 
for only the coefficients $v_s$ and $v_c$ in eqn. \ref{fit-eq}.
In this case, the best fit coefficients derived from the $\chi^2$
minimization at some frequency $\omega=2\pi/P$ for a set of 
$n_0$ velocity measurements, $v_i$, are: 
\begin{equation}
v_s =  C_{ss}\sum_{i=1}^{n_0}{ {v(t_i)\sin \omega t_i }
                     \over{\sigma_i^2}}
           + C_{sc}\sum_{i=1}^{n_0}{{v(t_i)\cos \omega t_i }
                     \over{\sigma_i^2}}
\end{equation}
and
\begin{equation}
v_c =  C_{cs}\sum_{i=1}^{n_0}{{v(t_i)\sin \omega t_i }
                             \over{\sigma_i^2}}
           + C_{cc}\sum_{i=1}^{n_0}{{v(t_i)\cos \omega t_i }
                             \over{\sigma_i^2}}
\end{equation}
where the subscripted $C$ terms are the four components of the 
covariance matrix used to derive the fit (see for example Press et al. 
1992 ch. 15.4 for a discussion). When these terms are combined to
form the velocity amplitude $K$ as $K=\sqrt{v_s^2 + v_c^2}$ and data
are translated in phase by a value $\tau=\tan^{-1}(v_c/v_s)$ (derived 
by setting $C_{cs}=C_{sc}=0$) then, as was shown by Lomb (1976),
the square of the best fit velocity amplitude, $K^2$, becomes the 
unnormalized power of the periodogram at that period. With the
identification of $K^2$ with the periodogram power, we note that 
false alarm probabilities are given in each case is given as an 
exponential of $K^2$, with a normalization given by the variance,
$\sigma^2$, of the data.

The use of the velocity amplitude rather than a normalized measure
of its square represents an improvement to existing techniques for 
several reasons. First, a physically meaningful limiting velocity amplitude
(or equivalently, a companion mass $\times\sin(i)$) is explicitly a part 
of the definition of the probability. A potential weakness of this method 
is that because it utilizes amplitude as a figure of merit rather than 
power, its dynamic range is more compressed on a given plot. A single 
dominant peak will not stand out to nearly the extent that occurs in a 
periodogram. In spite of this somewhat minor defect, we submit that
a best fit amplitude is a far more useful quantity to an observer than 
is the power.

In sections \ref{sparse}-\ref{windowing} we have outlined the
regimes for which our analysis is valid and the manner in which it
fails for sparse or windowed data and for very long periods. Because 
of the similar origins of our analysis and periodograms we expect that 
similar failure modes also apply to periodograms. Hence probabilities
derived from sparse data ($n_0\lesssim 10-15$) and at `windowed'
periods such as the annual cycle using a periodogram will yield erroneous
results. Extensions to standard periodogram techniques (Irwin et al. 1989
and Walker et al. 1995) which explicitly account for unequal statistical
weights and correlations between fit parameters may provide more accurate
limits than our eqn. \ref{vprob-eq} in such regimes. 

Our extensions to long periods explicitly provide limits on
the amplitude of the signal (and therefore $M\sin(i)$) possible at
any given period at least 10 times as long as the data window. The 
limits account for the fact that a long period signal may in fact be
near an extremum during the time over which most or all of the data 
were gathered.

Both techniques may be used to determine the probability of
a signal being nonrandom for a single period, for a period range
or over all independent periods. The Scargle (1982) and HB false 
alarm probability generates the probabilities, in the ideal case, by
requiring a Monte Carlo analysis to specify the number of independent 
periods, $N$. Their analysis to determine $N$ is limited to sampling 
frequencies below the Nyquist limit however. With unevenly sampled data, 
it is well known that higher frequencies are accessible without aliasing. 
How far above the Nyquist limit a signal can be detected and how many
additional independent frequencies (if any) are required remains 
unknown. 

We also require a specification of the number of independent 
periods, however our analysis uses a definition of the number
of independent periods (not equivalent to the HB definition) based
only upon the width of a spectral peak. We make no distinction between
potentially aliased spectral peaks at high frequencies and those
found at lower frequencies. The excellent correspondence between
our analytical formalism and our Monte Carlo analysis for each period
range shows that the definition of $N$ made in eqn. \ref{n-eq} is 
reasonable. The functional dependence of the amplitude limit on $N$ is 
quite weak, going only as $\sqrt{\ln{N}}$. When $N$ is large, as is the 
case for the shortest period bins, our definition will yield slightly
more conservative (higher) limits than the comparable HB limits, while for 
longer periods when $N\lesssim 100$, our limits may be somewhat lower. 

\section{Application to Real Data\label{realdat}}

In this section we apply our analysis technique to data for 
two stars obtained by Mayor and his collaborators at the Geneva 
Observatory, data obtained by Marcy et al. (1997) for the star 51 
Pegasi, and to the data obtained by Walker et al. (1995) in their 
12 year search for extra-solar planets. The Walker et al. radial
velocity data are for a set of 21 stars with data taken over a 12 
year period from 1980-1992. The data used in our analysis were 
originally archived at the Astronomical Data Center 
(URL http://hypatia.gsfc.nasa.gov/adc.html) by Walker et al. upon
publication of their work.  We limit our analysis to the subset of 
14 stars for which no visual binary companion is known, shown in 
Table \ref{star-tab}.  For these stars, no other periodic radial velocity
signatures which could obscure a planetary signature are present, and 
no significant periodicities attributable to planetary companions 
were found by the Walker et al. search. 

Using equations \ref{mass-eq} and \ref{vprob-eq} we can derive for 
any period (or period range) of interest the limit below which random
data is fit with probability $X$ to be: 
\begin{equation}\label{m-eq}
M_c = {{{\hat K_X}\over{\sin(i)}}
	      \left({{PM_*^2}\over{2\pi G}}\right)^{1/3}} 
\end{equation}
where we assume the orbital period, $P$, is at the midpoint of some
range of periods shorter than $\beta{P_0}$ or that $P>\beta{P_0}$.
This mass limit depends upon both the velocity amplitude limit $K_X$, which 
changes slowly, and also the period for which we fit the data. The minimum 
mass detectable by a set of measurements increases only as the cube root 
of the period at short periods, but for $P>\beta{P_0}$, this dependence 
becomes much steeper, increasing faster than $P^2$.

Once we have mass or velocity amplitude limits for a set of data,
we can define quantities $M_{99}$ and $M_{999}$ via eqn. \ref{m-eq} as 
the mass exceeded by chance by a fit at the 1\% and .1\% level of 
probability in each period range. For a sample of radial velocity 
measurements of say 10 stars observed for 10 years divided up into 10
period ranges, these values are of interest because if there are no 
true periodicities in the data, we would expect to find by chance one
apparent planet with best fit mass $M > M_{99}$, but to find a mass
greater than $M_{999}$ with only $\sim$10\% probability.

\subsection{Determining the Measurement Uncertainty}

In order to obtain a value of $\sigma_p$ for use in eqn. \ref{vprob-eq}
or \ref{m-eq}, we assume that no strong periodic signals are present 
in the data and that each datum is drawn from the same statistical 
distribution. Then we may use the rms scatter of all the data for 
a star as an empirical measure of the error, $\sigma_p$, for each
measurement of that star. For data in which no clear signal is 
observable, this measurement of the error will give a more reliable
estimate of the true value than from internal estimates.  In Table
\ref{star-tab}, we show both $\sigma_p$ as derived directly from the
data as well as the average of the internal errors for each star 
($\sigma_i$) quoted by Walker et al.

\subsection{Detecting Large Amplitude and Large Eccentricity Signals}

In this section we show that our analytic technique is capable of
detecting large amplitude signals and signals with high eccentricity.
As an initial test we obtained the data for the original discovery 
of the companion to 51 Pegasi, taken by Mayor and his collaborators
at the Geneva Observatory. These data consist of the original 35 
measurements as published by Mayor and Queloz (1995) as well as
their observations of the star since that time. The total number
of observations used in our analysis was 89 radial velocity 
measurements made over 2.4 years with internal errors of 15 m/s.
The value for $\sigma_p$ was obtained from the rms scatter of all
of the velocity measurements and was $\sigma_p = 44.5$ m/s. An 
independent set of measurements (Marcy et al. 1997) was also used
to compare the technique using higher precision data.  A total of 116
measurements are characterized by an rms scatter of $\sigma_p$=40.6~m/s
and were gathered over a total time span of 325 days with internal 
errors of 5 m/s.

We show the results of these two tests in figure \ref{51peg-vel}.  
For each set of measurements, we detect a clear peak in the best 
fit velocity amplitude at a period $P=4.23$ days, as expected. 
For the Mayor and Queloz data, we also detect a number of side
lobes peaks which represent the radial velocity signal `beating'
against other periodicities in the data such as the 29.5 day 
lunar cycle (the data were taken predominantly during the same 
half of each lunar cycle). The Marcy et~al. data show one 99\%
significant peak just shortward of one year. We consider this
peak to be an artifact of the short time baseline of their data
(less than one year) and do not consider it very significant. 

A second 99\% significant period is detected at 23.84 days in the
Mayor and Queloz data. We have re-analyzed the residuals  of the
measurements (with the 4.23 day periodicity removed) and found
that the peak remains and so cannot be attributed to an alias
of 4.23 day periodicity.  Is it an artifact of the rotational
period of the star itself? We note that the period is roughly
in a 2:3 ratio to the observed 37 day rotation period for 51 Peg 
(Baliunas, Sokoloff \& Soon 1996). We suspect that with the incomplete phase 
coverage for periods near 24 and 37 days, the stellar rotation 
period may be aliased to the observed 23.84 day period. Only 
complete phase coverage may be able to determine the origin of 
this signal.

The independent radial velocity observations of Marcy et al.
do not show a similar periodicity and the precision of their
measurements is only half the best fit amplitude from the Geneva 
data. In their work more than 3/4 of the data were gathered in
less than two rotation periods of the star. In such a case, it 
is unclear whether a rotation signature would be observable in
Doppler spectroscopy data. 

As a second test, we obtained another set of radial velocity data 
from Mayor. In this case the data were obtained under the condition
that the identity of the data and whether they contained a signal
not be disclosed until the conclusion of the test. These data consisted 
of 45 measurements taken with the CORAVEL spectrometer with internal 
errors of 300 m/s. The data were gathered over $\sim$15 years and 
the value for $\sigma_p$ obtained from the rms scatter in the velocity
measurements and was $\sigma_p = 800$ m/s.

The best fit velocities and the corresponding 99 and 99.9\%
probability limits for the fits are shown in figure \ref{mayanon}.
In this case, we were unable to detect a significant periodicity
in the data except a possible long term signal near 15 yr.  The data
show no obvious periodicity in the velocities for orbits of $\sim$15 yr,
but do show that several measurements are some 3--5 $\sigma$ away from 
the mean of any other velocity measurements of the star. These data
were gathered within 6 days of each other in the fall of 1996.

To test the effect of these data we deleted them from the sample and
reapplied our analysis. In this case, the rms scatter was reduced 
to $\sigma_p = 480 $ m/s and $n_0$ to 40 while the duration of the
measurements $P_0$ remained the same. Figure \ref{mayansub} shows
the results of this reanalysis.  In this case a peak in the best
fits for a circular orbit, well above the 99.9\% probability curve,
becomes visible at 275 days.  With the detection of this peak, we
concluded our test and obtained the identity of the star from which
the observations were taken. The star from which the data were 
obtained was HD~110833, for which Mayor et al. (1996) published an
orbital solution with a best fit companion mass $M\sin(i)=17~M_J$,
a period of $P=270$ days and an orbital eccentricity $e=0.7$ using
data from both the CORAVEL and ELODIE spectrometers. The difference
in the period derived from our analysis and the Mayor et~al. fit 
is due to the inconsistency between the high eccentricity of the 
companion and assumption made in our analysis that the orbit is 
circular.

In discussions with M. Mayor and D. Queloz, several issues were brought 
forward. First, the data from the fall 1996 run may have been faulty due to 
a combination of several factors, including a slight misfocusing of the 
telescope or a temperature instability in the spectrometer itself. However,
while the measurements in question are unusually distant from the mean of 
the other measurements for that star, they were gathered during a single 
periastron passage of the companion and therefore may not contribute as 
strongly to the detection of the periodicity as they would otherwise. A 
fit for the set of orbital elements would then yield a higher eccentricity 
than is truly the case. 

This case may therefore expose a degeneracy between results using data derived 
from a companion which is truly in a highly eccentric orbit and data 
with possible systematic biases. Our technique is based upon only the lowest 
order Fourier components of the signal (i.e. a circular orbit) and does not 
account for eccentric motion.  The fact that so many data (5 of 45 measurements) 
were so far from the mean and that they were from the same observing run 
suggests that removing the data from that run from our analysis is 
justified. Omitting them, we are able to recover a strong periodicity near
the best fit period for the orbital solution.

With the results from this section we can be confident that our analytic
technique is capable of detecting signals with large amplitude and/or
large eccentricity.

\subsection{Masses from Best Fit Velocities of the Walker et al. 
Sample, and Analysis of Significance}

We have shown that our analytic technique is capable of detecting 
large amplitude periodicities in radial velocity data. In this
section we move to lower amplitude signals and upper limits to
companion signatures. For each star in the Walker et al. sample,
best fit velocities for periods between 3 days and 100 years were 
determined by the least squares method using eqn. \ref{fit-eq}. 
Statistical weights for each datum were taken to be 1/$\sigma_i^2$,
where $\sigma_i$ is the quoted internal error for each point
as given by Walker et al. Plotted in figure \ref{datamc-bf} are the
corresponding companion masses $M_c\sin(i)$ obtained from eqn. 
\ref{mass-eq} using the stellar masses from table \ref{star-tab}. In 
order to determine the significance of any particular best fit value 
we compare the best fit mass for some period to the limit provided by 
our analytical analysis and to Monte Carlo experiments similar 
to those described in section \ref{monte-carlo} for synthetic
data. Each of these limits are shown in figure \ref{datamc-bf}.

While as expected from the results of the original analysis of 
Walker et al., there are no clear cut companion signatures, in 
several cases the data produce statistically significant fits. We 
tabulate each of these periods in table \ref{signif-tab}. Are any 
of these signatures due to the existence of a companion? Stellar 
processes such as pulsation, rotation or magnetic cycles can affect
the measured radial velocity for a star and in many cases it is 
quite possible to fit such signals with orbital solutions. Early 
in this century for example (see eg. Jacobsen 1925, 1929), the 
radial velocity variations of Cepheid variable stars were fit 
with Keplerian orbits. Although today no one would attribute
Cepheid radial velocity variations to a companion, the principle
that processes intrinsic to the star must be eliminated from 
consideration remains if we are to be certain that a given 
radial velocity detection is definitely due to a companion.

Many of the signals in table \ref{signif-tab} do in fact correlate
with known periodicities due to stellar rotation or magnetic cycles
in the star. For example, we find a $>$99.9\% significant $\sim$10 yr 
period in $\epsilon$ Eri and two short period signals (at 11.9 and
52.5 days) with $>$99\% probability. Walker et al. establish that the
10 year and 52 day peaks are aliases of each other and McMillan et~al.
1996 have definitely connected this periodicity to a stellar magnetic
cycle.  Gray and Baliunas (1995) have observed an 11.1 day periodicity 
in the Ca H\&K S-index with an extensive data set. They comment that
subsets of their data taken during different observing seasons produce 
peaks varying in period from 11 to 20 days. We conclude that we
are seeing a comparable effect in the Walker et al. radial velocity 
data and are in fact detecting the rotational signature of the star.  

Further work by the same group (Gray et al. 1996) on the star 
$\beta$~Com provides evidence of a magnetic cycle. However their 
measurements have sufficient duration only to have observed a minimum, 
and a period is not known. Figure \ref{datamc-bf} shows that for 
$\beta$~Com the probability that the best fit velocity amplitude
is not random exceeds 99\% for periods near 10 years. Assuming
a 10 year period, we calculate that the best fit radial velocity 
curve went through its minimum in 1988/1989, which is coincident with 
the photometric and Ca H\&K minimum observed by Gray et al. (1996).
Significantly, the radial velocity minimum is not coincident with
the velocity span minimum derived from their line bisector analysis.

The star $\eta$ Cep shows $>$99\% significant periodicities at 
$P=$164 days and $\sim$10 yrs. Walker et al. have speculated that
the 164 day periodicity was due to stellar rotation. No periodicities 
are detected in line asymmetry to $\sim$19 m/s by Gray (1994) with
measurements spanning four years. However we find the best fit radial
velocity amplitude at each of these periods is only 16 and 13 m/s 
respectively. If a direct correlation between a radial velocity 
measurement and a line bisector measurement exists, such signals would
be below his detection limit. By analogy to $\epsilon$ Eri, we
speculate that the 10 year periodicity in $\eta$ Cep might be 
linked to a magnetic cycle, but we cannot be certain of its origin.

Other marginal periodicities appear in the data for HR 8832 and 
$\theta$ UMa. Again by analogy to other stars, in this
case $\epsilon$ Eri and $\eta$ Cep, we might speculate that these
shorter periodicities are due to stellar rotation, however no 
certainty can be attached to their origin. 

We also find that two stars in the subset (36~UMa and $\beta$~Vir)
show best fit minimum masses which, for fitted periods longer than 
$\sim$12 years, rise above the curve for which the best fits are 
random with 99\% and 99.9\% probability. Walker et al. find similar 
trends in these stars but make no firm conclusions based upon their
analysis.  Are these signals indications of long period companions, 
or are they also due to stellar effects? We show the data for these 
stars in figure \ref{star-rv}, both raw and binned by year. While the 
raw data show no obvious signals, the binned data, particularly for 
36~UMa, show some indication of a partially complete sinusoid. We 
note that in both cases, the curvature in the velocity trends is of 
the same sign and the portion of the sinusoid is similar, which
suggests a long term calibration error.  However, the binned data for 
all stars taken together shows no such trend so a systematic 
explanation seems less likely.

\subsection{A Check by Monte Carlo Analysis}

Each of the stars in the Walker et al. sample average 3-5 measurements 
per year over the 12 year period and, according to the results of section 
\ref{sparse}, this number should be sufficient for application of 
our analytic apparatus. We note however, that implicit in our analytical
derivation of mass limits is the assumption that the measurements be at
least somewhat regularly distributed. In the case of the Walker et al.
data, this is not always the case. The data are irregular on both short
time scales (ie. 3 night runs consisting of 1--3 velocity measurements
per star per run) and longer time scales, for which more data may be
loosely clustered on several year time scales due to changes in 
observing procedures etc. Because of these irregular sampling patterns
fits may be less tightly constrained than a more evenly spaced data 
set, and the limits provided by our analytic apparatus may become 
misleading.

Since the Walker et al. data are rather irregular, we examine the effect
on our analysis technique by performing a Monte Carlo experiment and 
comparing the result to our analytic formalism. We create synthetic data
sets using a constant value of the error, $\sigma_p$, equal to the rms
scatter of the observed velocity measurements for each star. This value
is input into eqn. \ref{vrand-eq} to derive individual simulated velocity
measurements. We use the observation times given by the data itself. 
We fit the measurements and derive best fit velocity amplitudes (and
corresponding $M_c\sin(i)$ values) for periods between 3 days and 100 years
Each synthetic datum used in the fit is weighted with the internal error 
in that point (quoted by Walker et al.) as $1/\sigma_i(t_i)^2$.

The results of these experiments are shown as the solid histograms
in figure \ref{datamc-bf}. In general, the agreement between
the analytic limits and the Monte Carlo experiments is good. However
a small systematic trend towards larger limits for the Monte Carlo
experiments is found. Typically the difference is 10\% or less,
however, in the most extreme case ($\theta$~UMa), the limits produced 
are about 20-30\% higher than with the analytic method. This star has
the shortest time baseline of any in the sample as well as one of the
largest degrees of clumping of any star in the sample, as measured
by the ratio of the number of data to the number of runs. A test in 
which we replace in eqn. \ref{vprob-eq-long} the number of data, $n_0$ 
with the number of runs recovers the Monte Carlo results for this
star quite well. The star $\epsilon$~Eri, for which the data are 
the most highly clumped of any star in the sample, also produces 
analytic limits lower than reproduced with the Monte Carlo experiment. 
In this case, replacing $n_0$ with the number of runs produces limits
much larger than the Monte Carlo result, so we cannot recommend such
a procedure for general use.

In the case of one star (HR 8832), two measurements were made with
a 3-1/2 year separation from any other measurement for that star.
This case provides an interesting test in the limit of very 
irregularly spaced data.  We find that the limits derived from the
Monte Carlo experiment are quite similar to the analytic result
except in the range between about 8 and 12 years, where limits
some 20\% larger than those derived via eqn. \ref{m-eq} are found. 
The longer period fall-off characteristics are unaffected by the
irregularity.

The sensitivity fall off at long periods for each star is similar to 
that produced in the synthetic data. We show the derived fit values for 
the power law exponent and the long period turn off in figure 
\ref{datal-bf} for the sample of 14 stars. The sharp upturn in limiting 
velocity at long periods produces a power law exponent which is 
best fit with values near $\alpha=1.86$, while the turn-off period, $\beta$,
is best fit with values near $\beta=1.45$, but with a larger scatter 
than is present in the synthetic data. Because the scatter is by its
nature rather unpredictable from star to star, we retain the low 
$\beta=1.3$ value found for synthetic data when determining limits
via eqn. \ref{vprob-eq-long} or \ref{m-eq}. 

The limits provided by our analytic expression produce upper bounds
which are ordinarily $\lesssim$10\% different than those produced via 
Monte Carlo experiments. In the most irregularly spaced data, 
a difference of up to 20-30\% can occasionally be produced. In several
cases, the difference results in possibly spurious `detections' of 
marginal signals by the analytic technique where the Monte Carlo limits
do not show that the periodicities are significant. In some of these
cases we are able to attribute the detections to physical processes
discussed in the literature. In no case do the analytic limits exceed 
the 99.9\% level of probability where the Monte Carlo result did not 
also show at least a 99\% probability. Despite this level of difference,
our conclusions about the significance or lack of it for any periods 
and companion masses for each star in the Walker et al. data remain 
unchanged. We are confident that this method can be relied upon to
obtain probabilities that a given set of data contains a periodic
signal.

\subsection{Sensitivity to Short Period Planets\label{limits}}

For the star with the lowest companion mass limits
($\epsilon$ Eri), we have also plotted in figure \ref{datamc-bf} 
several recent planet detections (Mayor and Queloz 1995, Marcy 
and Butler 1996, Butler and Marcy 1996, Latham et al. 1989, 
Gatewood 1996, Noyes et al. 1997) and Jupiter. Extra solar
planets with combinations of period and mass like those shown
would have been readily detected by Walker's radial velocity
measurements. These stars do not have such companions.  For most
of the stars in the sample, the data are too noisy to have 
reliably detected a radial velocity signature such as would
be predicted for the companion to Lalande 21185, announced by
Gatewood (1996) but which remains unconfirmed.  Planets such as
Jupiter, which would appear at $P=12$ years with a typical value
of $M_c\sin(i)$=0.64$M_J$ would not have been reliably detected.
The best fit values exceed this period/mass combination in 40\%
of the sample. 

The analysis of Walker et al. sets upper limits to the mass 
of companions in their sample ($\times \sin(i)$) of $\leq 1 M_J$ 
and $\leq 3 M_J$ in periods of less than 1 year and 15 years 
respectively. In general, our analysis provides limits which are 
somewhat lower than theirs in both long and short period orbits.
For one year periods, we can limit companion $M\sin(i)$ values
to $\leq 0.7 M_J$ for all but three stars in our subset and
$\leq 1.0 M_J$ for the rest. In 15 year orbits, our analysis
limits possible $M\sin(i)$ values to $\leq 1.5 M_J$ for every star
except $\theta$ UMa, for which only 6 years of data were gathered.
For this star, the limit is $\leq 4.0 M_J$.  We have also extended
range over which companion signatures are constrained to shorter
periods than were analyzed in Walker et al. The limits for these
extreme short period orbits ($P<40$ days) correspond to companion
masses ($\times \sin(i)$) below 0.4~$M_J\sin(i)$. 

Under either our own analysis or the original analysis of Walker 
et al., the companion mass limits derived from the data essentially
eliminate brown dwarfs and large Jovian planets with periods
$\lesssim 15$ years, barring very unfortunate inclinations. 
Given the detections of significant periodicities by either our  
analytic treatment or Monte Carlo experiments, we find more signals 
present than can be attributed to purely random data. In some cases,
such detections may be due to physical mechanisms other than a 
companion, and we have compared these to known periodicities due
to stellar rotation or magnetic cycles, where they have been 
identified in the literature. In no case do the limits eliminate 
the possibility of gas giants such as exist in our solar system
or low mass brown dwarfs, especially in the period/radius range
$\geq$12~yr/5 AU where theory predicts such companions.  

\section{Strategies for Large Radial Velocity Surveys\label{strategy}}

There are currently six active groups with programs for radial 
velocity searches at the $<$10-20 m/s level. Three groups began
searches at this precision in 1987-88 (Cochran \& Hatzes 1994
(Texas), McMillan et~al. 1994 (Arizona), Marcy \& Butler 1992
(Lick)), while one (Duquennoy \& Mayor 1991 (Geneva)) have used
lower precision measurements with the CORAVEL spectrometer 
($\sim300$ m/s) to investigate stellar binary companions and
have recently built a new spectrometer (ELODIE) to allow 
$\lesssim$15 m/s precision measurements to be made.  The latest 
high precision searches (K\"urstner et~al. 1994 (ESO), Brown 
et~al. 1994 (CfA)) began in 1992 and 1995 with quoted precision
of $\sim$4-7 m/s and $\sim$10 m/s respectively.  Another group
(Walker et~al. 1995 (UBC)) concluded a 12 year search in 1992.
Two others (Mazeh et al. 1996 (CfA), Murdoch et~al. 1993 
(Mt John NZ)) obtain precision of $\sim$500 m/s and $\sim$60 m/s 
respectively.  

The recent discoveries of sub-stellar mass companions around other
stars have stirred new interest in very large radial velocity 
surveys.  The Geneva group for example, intends to expand their
search to $\sim$500 stars in the northern hemisphere (ELODIE)
and another $\sim$800 in the southern hemisphere (CORALIE), and
other groups have similar expansions underway. In order to observe
as many stars as possible with a finite telescope allocation, 
such large surveys must necessarily aim toward the most efficient
use of the available observing time. The goal of such large surveys
might be properly stated as ``What fraction of stars have a companion
(or a system of companions) and what is the distribution of the 
masses, periods and eccentricities of those companions?''. 

In order to answer this question three criteria must be met. First,
an observer must first detect a variation in the radial velocities 
measured for a star about which a prospective companion orbits. 
Second, the observer must determine the origin of such variations
by fitting a Keplerian orbit and by making additional photometric
or spectroscopic observations to constrain effects due to the 
stellar photosphere. Finally the observer must determine the extent
to which the survey is complete: what fraction of stars which were
observed may have companion signatures which went undetected over
the course of the survey? Based on the analysis in this paper, we
can suggest strategies for the most efficient methods of detecting
radial velocity signatures and which also provide meaningful upper
limits on the amplitudes of undetected signatures.

Let us suppose that the random error for each measurement is 
dominated by photon noise, ie. that $\sigma_p\propto 1/\sqrt{t}$,
where $t$ is the length of a single observation. This should be
the case provided detector read noise is not significant. It 
follows from eqn. \ref{vprob-eq} that the limiting amplitude,
$K$, is proportional to $1/\sqrt{t_n}$, ie. $K$ depends on the
total integration time devoted to a star, $t_n$, independent
of number of observations making up that time. In other words,
as long as eqn. \ref{vprob-eq} holds and the total integration
time is the same, making many lower precision measurements is
equivalent to making fewer high precision measurements. Since
constraining additional orbit parameters such as eccentricity
is at its most simplistic level an exercise in detecting higher
order Fourier components of the signal, this equivalence holds
for orbit determinations as well as for detection of a periodic
signal. We caution, however that with lower precision data,
larger amplitude systematic errors may go undetected. With 15 m/s
precision for example, the signal of Jupiter could be completely
obscured by a hidden systematic error of amplitude $\sim$10~m/s. 

From section \ref{monte-carlo}, to insure than eqn. \ref{vprob-eq}
holds, the number of observations, $n_0$, must be at least $\sim$12 
and preferably as high as 20 in order to constrain octave period 
ranges.  The limits are degraded most severely for periods less 
than 1--2 years. Longer periods limits are nearly identical to the
analytic prediction even for very sparse data (see figures
\ref{mc-anal-cmp} and \ref{vlim-sparse6}). This is because 
with only a few observations of each star, a companion signature 
could still slip through undetected if by some unfortunate 
coincidence its radial velocity ``zero crossings'' corresponded 
to the times at which the star was observed. For $P\lesssim 1$ year,
there are very many independent periods, so that the possibility 
of any one of them coincidentally undergoing such a zero crossing
event is very high. For $P\gtrsim 1$ year, where there are relatively
few independent periods, such a condition becomes much more unlikely.

The cost in observing time to obtain useful limits if there are 
few observations is great. When the data are sparse and eqn.
\ref{vprob-eq} breaks down, for example with a total of either 
$n_0=6$ or $n_0=12$ observations and the same total integration
time, our Monte Carlo simulations show that the limiting amplitude
is in fact twice as big for any single period, and ten times as 
big for octave period ranges. Because 12 much lower precision
measurements would identically constrain short period signals as 6
high precision measurements, the increase in sensitivity translates
to a reduction factor of 4 or 100 in the amount of observing time
required to identically constrain the existence of companion 
for any single period or over octave ranges in orbits of
$\lesssim 2$ years. 

As an example of a strategy which addresses this concern, suppose 
a survey is to observe 500 stars and is to last at least 12 years.
Let us also suppose that 1/4 of the use of a telescope is dedicated 
to the radial velocity measurements, yielding about 400 hours of
integration per year to be divided among the stars in the survey. 
The total number of observations to obtain 12 for each star is 6000.
A good ``quick look'' could be obtained after the first two years if
each observations takes 2$\times$400/6000 hours = 8 minutes.

Butler et~al. (1996) report that precision of $\sim$3~m/s can be 
obtained in a 10 minute exposure of a magnitude $V=5$ star on a 
3 meter class telescope. However, in a large survey most stars will 
be dimmer than $V=5$, with a practical limiting magnitude between
$V=7$ and $V=8$, depending on the size of the survey. If the average
star is of magnitude $V=7$, a measurement with 3 m/s precision
would nominally require about 60 minutes.  With such long duration
measurements each star in the program would average less than one 
observation per year and would make a large, high precision survey 
unfeasible. In order to complete a large survey at $\leq$5 m/s
precision a large allocation of time on an 8--10 m class telescope
would be required. If instead we allow reduced precision measurements
of $\pm$10~m/s, using our assumption that the precision is
proportional to $1/\sqrt{t}$, a single measurement would require only 
about 6 minutes on a 3~meter telescope, which would be feasible for 
a large survey.

For this to be practical without poor observing efficiency, the 
time from the end of an observation to the beginning of the next
on a new star must be short, ideally a minute or less.  During this
time, the telescope must be slewed to the new star, while the 
CCD with the exposed spectrum is read out.  Automatic slewing 
and acquisition would make this quite practical.  Also, for a
typical spectroscopic CCD with around 3 million pixels, the 
required read rate of 100 kpixel/sec should be readily achievable
at negligible read noise with current devices. A 8 minute cycle
time with 6 minutes data acquisition would thus be a reasonable
target, and yield 72 minutes of integration for each star, spread
through the first two observing seasons.

With this strategy, an observing program should be able to sustain
6 measurements of every star every year that the
program is continued. If after two years, variations are detected
in some stars, additional measurements of those stars would
be possible if constant velocity stars were observed only 1--3 
times per year. This compromise has the advantage that limits on
companion masses are well constrained by such density of points
and orbital solutions, should a star's velocity later be observed
to vary, would also be well constrained. A second advantage is that
after the first two years, strong limits on the existence of a 
companion signal are available for short periods and these limits
extend to longer periods incrementally as long as the program is
maintained.  In contrast, a high precision/sparse observation 
strategy with say 1-2 measurements per year, will strongly limit
short period signatures only after 6 or more years of the program
has passed.

As a second example of observing strategy, we consider a search 
for a Jupiter mass planet with the same 12 year period as
Jupiter, around a star with the same mass as the sun, what 
accuracy measurements are needed over what period to ensure only
1\% probability of a false detection for any given star? For sets
of observations spanning 6 and 12 years figure \ref{mlim-anl} 
shows the limiting mass above which a companion would be detected
with 99\% probability in a given set of data at any single period.
A 1~$M_J$ companion in a 12 year orbit around a solar twin will 
have a best fit mass of $\sim 0.64 M_J$ assuming a random 
inclination. We require via eqn. \ref{m-eq}, that data be taken
with precision $\pm$5 m/s for 12 years with a single observation
per year in order to detect such a companion. Increasing to 5 or
20 observations per year, only 15 or 30 m/s are required, 
respectively, with the requirement that no hidden systematic errors
are also present in the data. For an identical number of 
measurements per year, a 6 year baseline requires more than 6 
times the precision in each measurement to similarly constrain 
a long period companion, or 36 times the observing time each 
year. Clearly the cost of impatience is very high.

\acknowledgments
We would like to thank the referee, Paul Butler, for helpful comments
and criticisms in his referee's report. Alan Irwin also provided 
helpful criticism of this manuscript. We are grateful to several
people for providing the radial velocity data used in this work.
Among them are Gordon Walker and Alan Irwin for providing a portion of
their radial velocity data prior to publication, Michel Mayor for
sharing the data for 51~Peg and HD~110833 and Geoff Marcy for sharing
an electronic copy for a second, independent data set for 51 Peg.  
M. Mayor provided commentary on the results of our analysis and 
a discussion with Didier Queloz outlined reasons for which the data
for HD~110833 may have contained systematic errors in several 
measurements. W. Benz provided impetus for the section on sensitivity
to large signals and Bob McMillan, Adam Burrows, Paul Harding and Heather
Morrison provided other helpful discussion. This work was partially
supported under NASA Grants NAGW-3406 and NAS7-1260.

%% file: range-tab.tex
\singlespace
\begin{deluxetable}{c}
\tablecolumns{1}
\tablewidth{1in}
\tablecaption{\label{range-tab} Period Ranges}
\tablehead{
}
\startdata
3-6d  \nl
6-12d  \nl
12-24d  \nl
24-48d  \nl
48-96d  \nl
96d-0.5yr  \nl
0.5-1yr  \nl
1-2 yr  \nl
2-4 yr  \nl
4-8 yr  \nl
8-12 yr  \nl
$>$12 yr  \nl
\enddata
\end{deluxetable}
\doublespace

%% file: startab.tex
\singlespace
\begin{deluxetable}{rrrcccccc}
\tablecolumns{8}
\tablewidth{0pt}
\tablecaption{\label{star-tab} The Subset of 14 Stars Included in our Analysis}
\tablehead{
\colhead{HR}  & \colhead{HD} & \colhead{Name} & \colhead{M/M$_\odot$}
& \colhead{$\sigma_i$ } & \colhead{ $\sigma_p$ } 
& \colhead{No. }     & \colhead{No. }  & \colhead{Duration } \\
\colhead{}    & \colhead{}   & \colhead{}     & \colhead{}
& \colhead{(m/s)}       & \colhead{(m/s)}
& \colhead{Obs.} & \colhead{Runs} & \colhead{ (yr)}
}

\startdata
 509 &  10700 & $\tau$ Cet       & 0.87 & 13 & 17 & 68 & 39 & 11.7 \nl
 937 &  19373 & $\iota$ Per      & 1.15 & 15 & 18 & 46 & 29 & 10.8 \nl
 996 &  20630 & $\kappa^1$ Cet   & 0.98 & 13 & 20 & 34 & 22 & 10.0 \nl
1084 &  22049 & $\epsilon$ Eri   & 0.82 & 14 & 16 & 65 & 34 & 11.1 \nl
1325 &  26965 & o$^2$ Eri        & 0.84 & 14 & 19 & 42 & 28 & 11.0 \nl
3775 &  82328 & $\theta$ UMa     & 1.45 & 24 & 21 & 43 & 23 & \phn6.0 \nl
4112 &  90839 & 36 UMa           & 1.08 & 16 & 21 & 56 & 36 & 10.7 \nl
4540 & 102870 & $\beta$ Vir      & 1.22 & 14 & 26 & 74 & 48 & 11.7 \nl
4983 & 114710 & $\beta$ Com      & 1.09 & 16 & 18 & 57 & 40 & 11.4 \nl
5019 & 115617 & 61 Vir           & 0.98 & 13 & 18 & 53 & 35 & 11.4 \nl
7462 & 185144 & $\sigma$ Dra     & 0.85 & 13 & 19 & 56 & 37 & 11.5 \nl
7602 & 188512 & $\beta$ Aql      & 1.30 & 12 & 14 & 59 & 39 & 11.4 \nl
7957 & 198149 & $\eta$ Cep       & 1.36 & 12 & 19 & 58 & 39 & 11.2 \nl
8832 & 219134 & \nodata          & 0.79 & 11 & 15 & 32 & 23 & 10.6 \nl
\enddata
\end{deluxetable}
\doublespace

%% file: signif-tab.tex
\singlespace
\begin{deluxetable}{rrcccc}
\tablecolumns{6}
\tablewidth{0pt}
\tablecaption{\label{signif-tab} Detection of Significant 
Periodicities}
\tablehead{
\colhead{Period range}  & \colhead{Star} & \colhead{Period\tablenotemark{a}} &
\colhead{Companion\tablenotemark{b}}  &
               \multicolumn{2}{c}{Probability of Chance} \\
\colhead{}  & \colhead{} & \colhead{} & 
\colhead{Mass (M$_{\rm J}$)}  &
	       \multicolumn{2}{c}{Detection in Period Range} \\
\colhead{}   &  \colhead{}   &  \colhead{}   & 
\colhead{}   & \colhead{Analytic}  & \colhead{Monte Carlo} 
}
\startdata
3-6d      &  \nodata        &          &                 &          &          \nl
6-12d     &  $\epsilon$ Eri &    11.9d &   0.14          & $<$1\%   & \nodata  \nl
12-24d    &  \nodata        &          &                 &          &          \nl
24-48d    &  \nodata        &          &                 &          &          \nl
48-96d    &  $\epsilon$ Eri &    52.5d &   0.24          & $<$1\%   & \nodata  \nl
96d-0.5yr &  $\eta$ Cep     &     164d &   0.54          & $<$0.1\% & $<$1\%   \nl
          &  HR 8832        &     165d &   0.35          & $<$1\%   & \nodata  \nl
          &  $\theta$ UMa   &     179d &   0.63          & $<$1\%   & \nodata  \nl
0.5-1yr   &  \nodata        &          &                 &          &          \nl
1-2 yr    &  \nodata        &          &                 &          &          \nl
2-4 yr    &  \nodata        &          &                 &          &          \nl
4-8 yr    &  $\epsilon$ Eri &  $>$7 yr &   0.7\phn       & $<$1\%   & $<$1\%   \nl
8-12 yr   &  $\epsilon$ Eri &    10 yr &   0.95          & $<$0.1\% & $<$1\%   \nl
          &  $\eta$ Cep     &    10 yr &   1.2\phn       & $<$1\%   & $<$1\%   \nl
          &  $\beta$ Com    &    10 yr &   1.05          & $<$1\%   & \nodata  \nl
          &  36 UMa         &    10 yr &   1.1\phn       & $<$0.1\% & $<$0.1\% \nl
$>$12 yr  &  36 UMa         &    15 yr &   2.0\phn       & $<$0.1\% & $<$0.1\% \nl
          &                 &    25 yr &   5.3\phn       & $<$0.1\% & $<$0.1\% \nl
          &                 &    50 yr &  24\phd\phn\phn & $<$0.1\% & $<$0.1\% \nl
          &  $\beta$ Vir    &    15 yr &   1.9\phn       & $<$0.1\% & $<$0.1\% \nl
          &                 &    25 yr &   5.0\phn       & $<$0.1\% & $<$1\%   \nl
          &                 &    50 yr &  23\phd\phn\phn & $<$1\%   & $<$1\%   \nl
\enddata
\tablenotetext{a}{We exclude `significant' periods which coincide with
known annual or lunar windowing periods}
\tablenotetext{b}{Best fit mass assuming that the periodicity is
actually due to a companion.}
\end{deluxetable}
\doublespace

%% file: RVfigs.tex
\singlespace
\begin{figure}
\epsscale{.9}
\plotfiddle{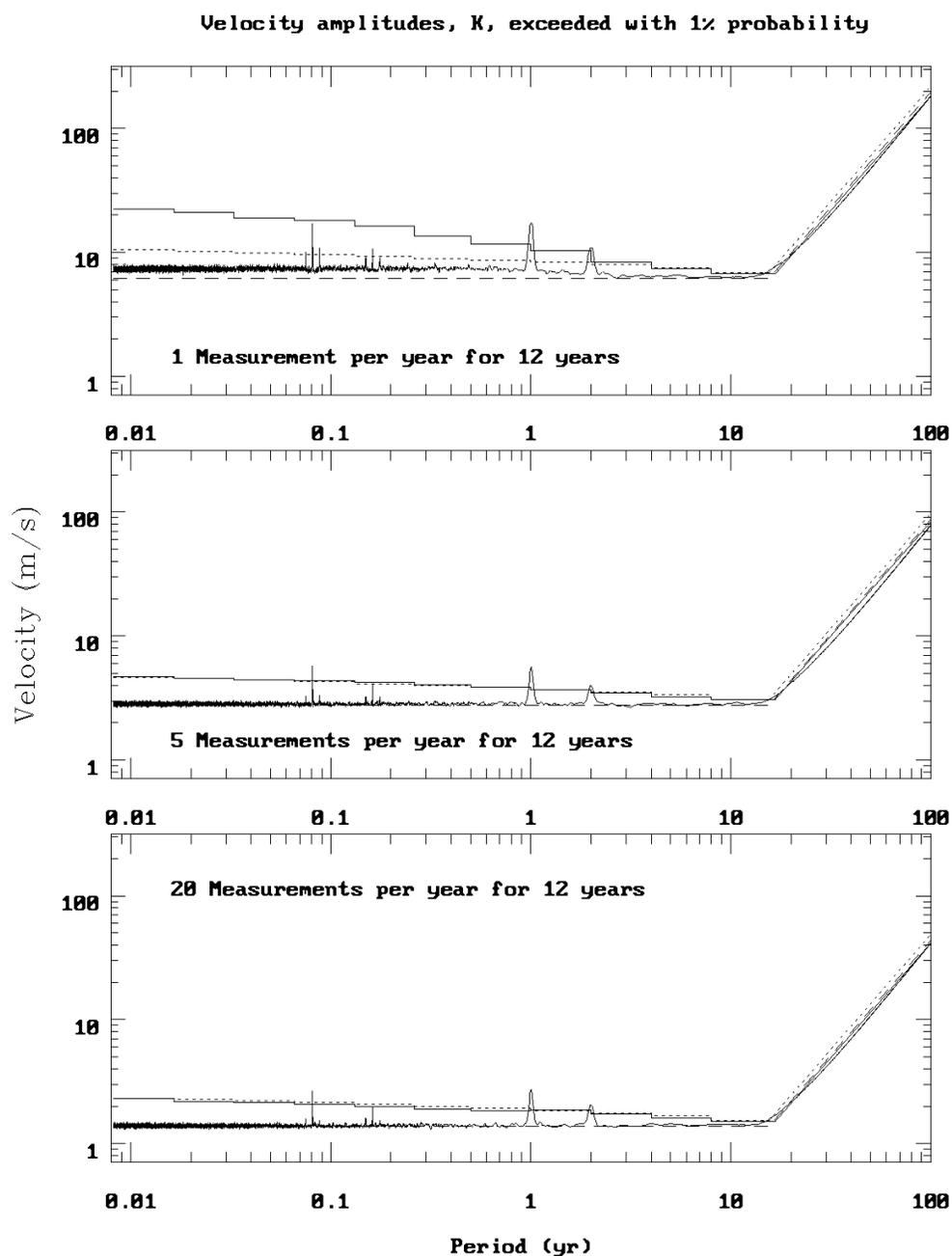}{6.70in}{0}{67}{67}{-210}{-20}
\caption[Results of Monte Carlo simulations of fitted velocity
amplitudes]
{\label{mc-anal-cmp}
\footnotesize
Monte Carlo (solid line) and analytic results (dashed line) for 
the velocity amplitude which, for a given fitted period, is 
exceeded by chance in 1\% of analyses of simulated random data. A
second solid line (histogram) shows the Monte Carlo results for
the amplitude which is exceeded at any period within a range of
approximately one octave, while the dotted histogram line shows
the result from our analytic expression, eqn \ref{vprob-eq}.
For each of the experiments in the bottom two frames (5 and 20 
obs./yr), the analytic and Monte Carlo results are indistinguishable.
Assumed windowing at the lunar and annual periods as well as their 
double period counterparts are excluded from the Monte Carlo limits 
in their respective ranges.}
\end{figure}

\clearpage

\begin{figure}
\plotone{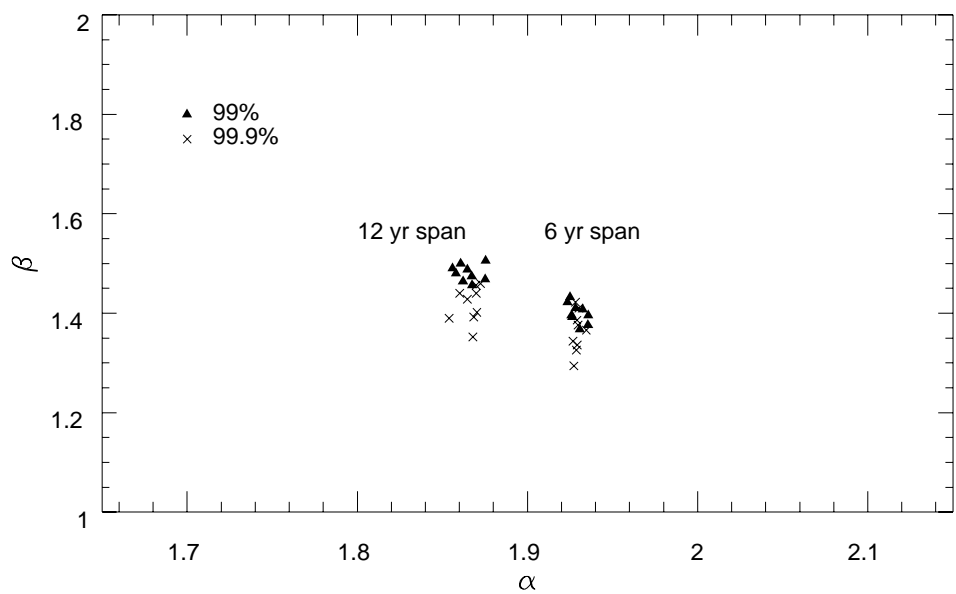}
\caption[Best fits for the long period sensitivity fall off parameters]
{\label{mcl-bestf}
\footnotesize
Best fit values for the long period sensitivity fall
off parameters $\alpha$ and $\beta$ for the nine Monte Carlo
experiments run for both 6 and 12 year baselines.}
\end{figure}

\clearpage

\begin{figure}
\plotfiddle{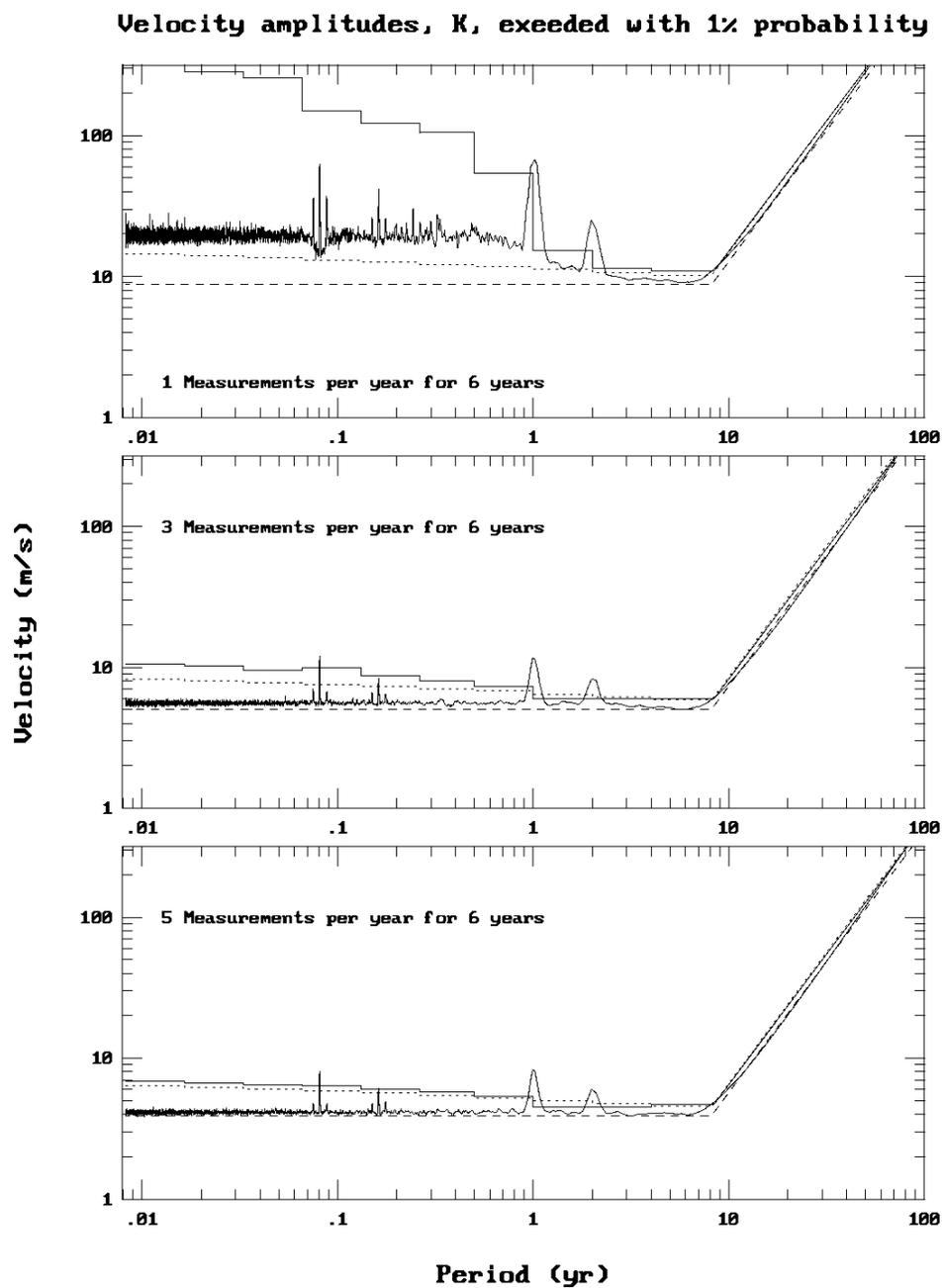}{6.79in}{0}{67}{67}{-210}{-20}
\caption[Results of sparse data Monte Carlo simulations with
a 6 year time baseline]
{\label{vlim-sparse6}
\footnotesize
Velocity limits given by Monte Carlo simulations (solid), 
single period ($N$=1) analytic limits (dashed). The histograms
represent the Monte Carlo (solid) and analytic (dotted) limits
for octave sized ranges. Each of the histogram limits ignore the
periods affected by the assumed lunar and annual windowing of 
the data. A 6 year window is assumed with 1 (top), 3 (middle) and
5 (lower) observations per year.}
\end{figure}

\clearpage

\begin{figure}
\plotfiddle{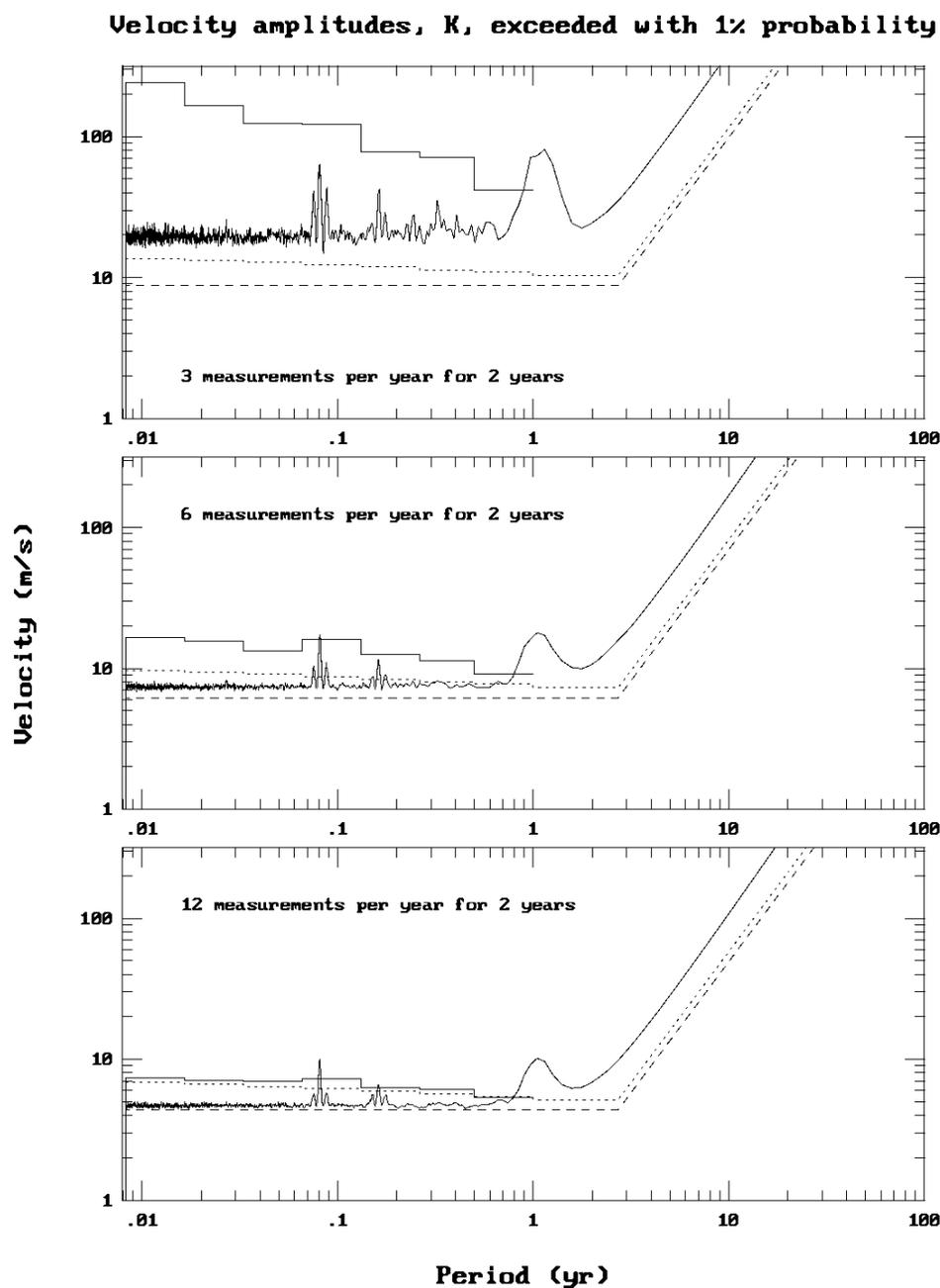}{6.79in}{0}{67}{67}{-210}{-20}
\caption[Results of sparse data Monte Carlo simulations with
a 2 year time baseline]
{\label{vlim-sparse2}
\footnotesize
Velocity limits given by Monte Carlo simulations (solid), 
single period ($N$=1) analytic limits (dashed). A 2 year window is
assumed with 3 (top), 6 (middle) and 12 (lower) observations per 
year. As before, histograms represent the Monte Carlo (solid) and
analytic (dotted) limits for octave sized ranges and each of the
histogram limits ignore the periods affected by the assumed lunar
and annual windowing of the data. In this plot however, for periods 
longer than 1 year the Monte Carlo histogram limits are suppressed.}
\end{figure}

\clearpage

\begin{figure}
\plotfiddle{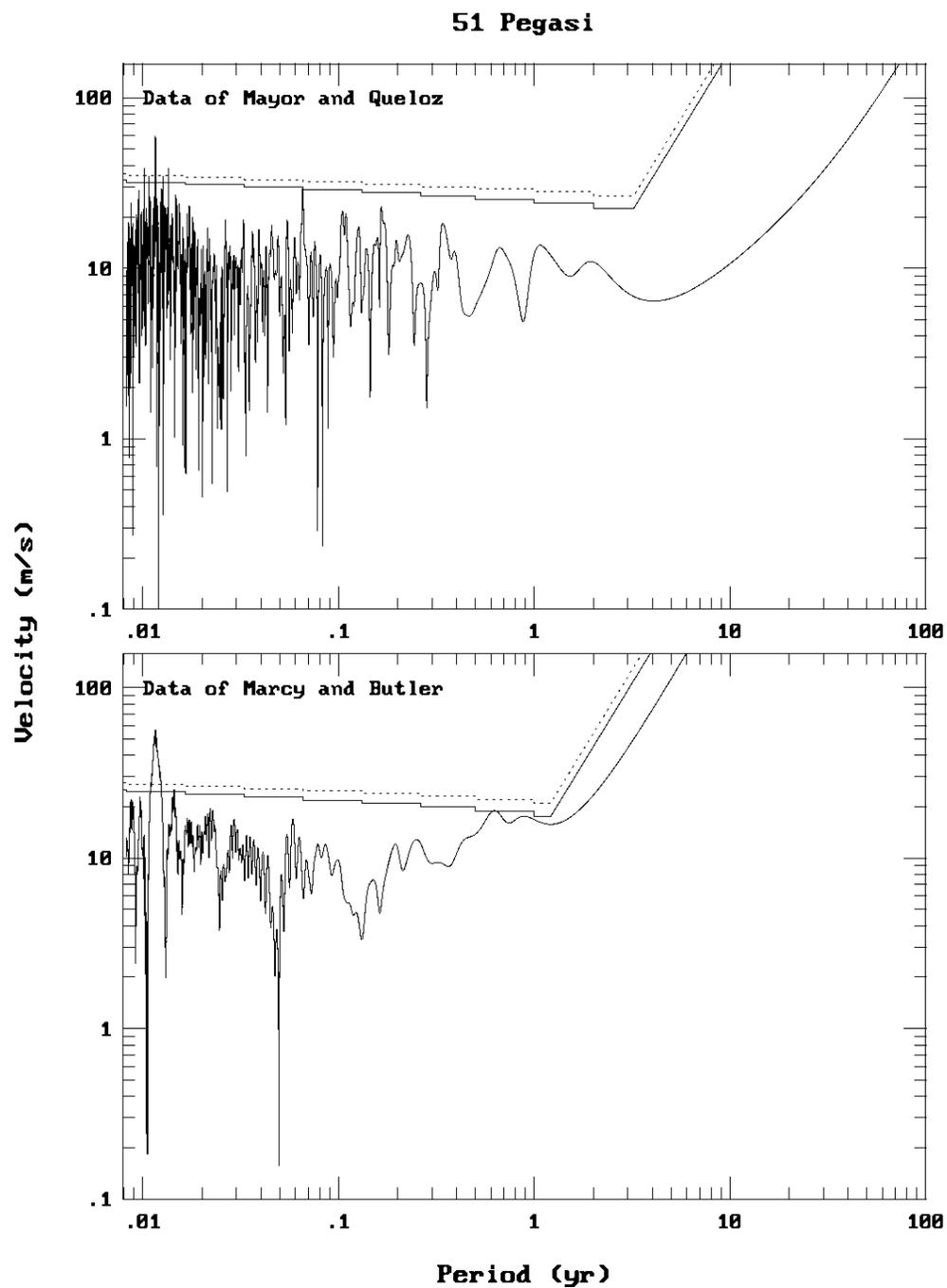}{6.79in}{0}{70}{70}{-230}{-20}
\caption[Best fit reflex velocities for the star 51 Pegasi derived from
two radial velocity programs]
{\label{51peg-vel}
\footnotesize
Best fit velocities for the star 51 Pegasi. The solid and dotted
histograms denote respectively, the limits below which random data 
would occur with 99 and 99.9\% probability. }
\end{figure}

\clearpage

\begin{figure}
\plotone{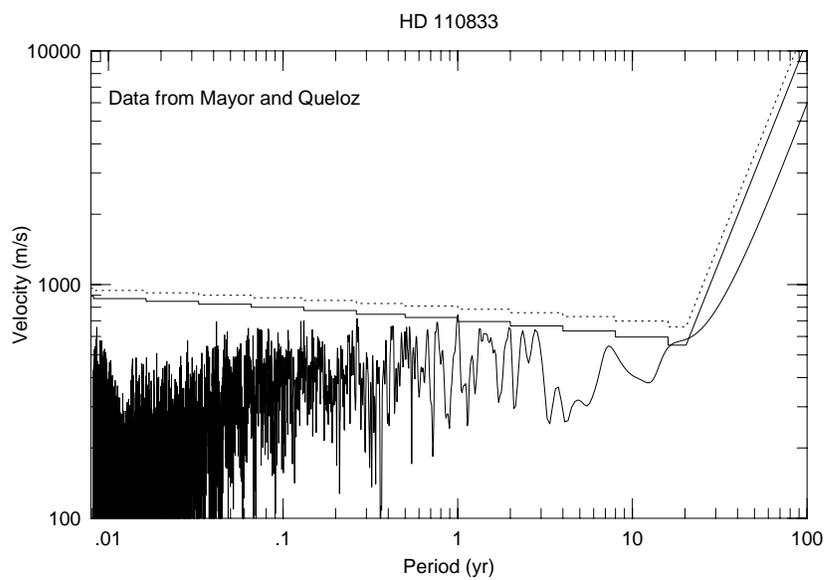}
\caption[Best fit reflex velocities  for the star HD 110833 derived
from the data of Mayor and Queloz]
{\label{mayanon}
\footnotesize
Best fit velocities for the data derived from star HD 110833.
A 15 year periodicity is apparent in the best fits with $>$99\%
probability that it is nonrandom.}
\end{figure}

\clearpage

\begin{figure}
\plotone{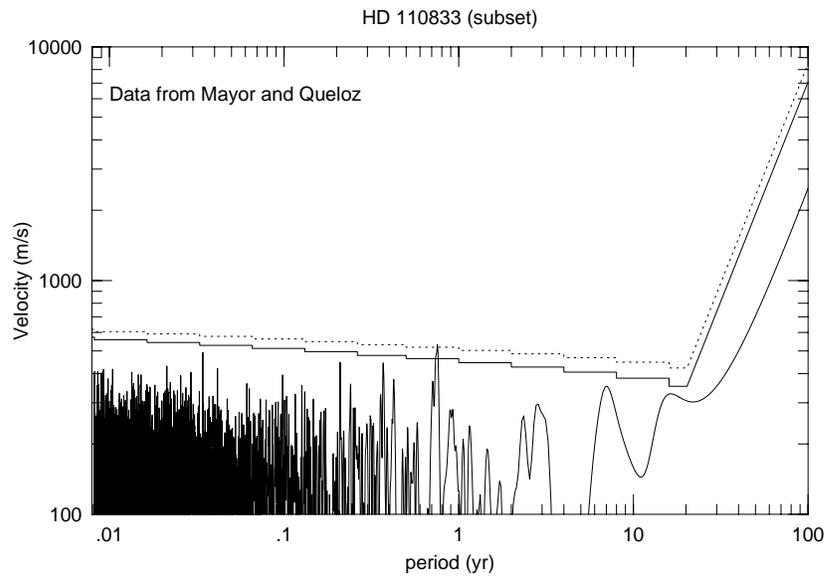}
\caption[Best fit reflex velocities for the subset of data from
HD 110833 with 5 measurements removed]
{\label{mayansub}
\footnotesize
Best fit velocities for the the subset of the data from HD~110833 
with 5 measurements removed. In this case, a periodicity is
present at 275 days with probabability $>$99.9\% that it is 
non-random. }
\end{figure}

\clearpage

\begin{figure}
\epsscale{0.8}
\plotfiddle{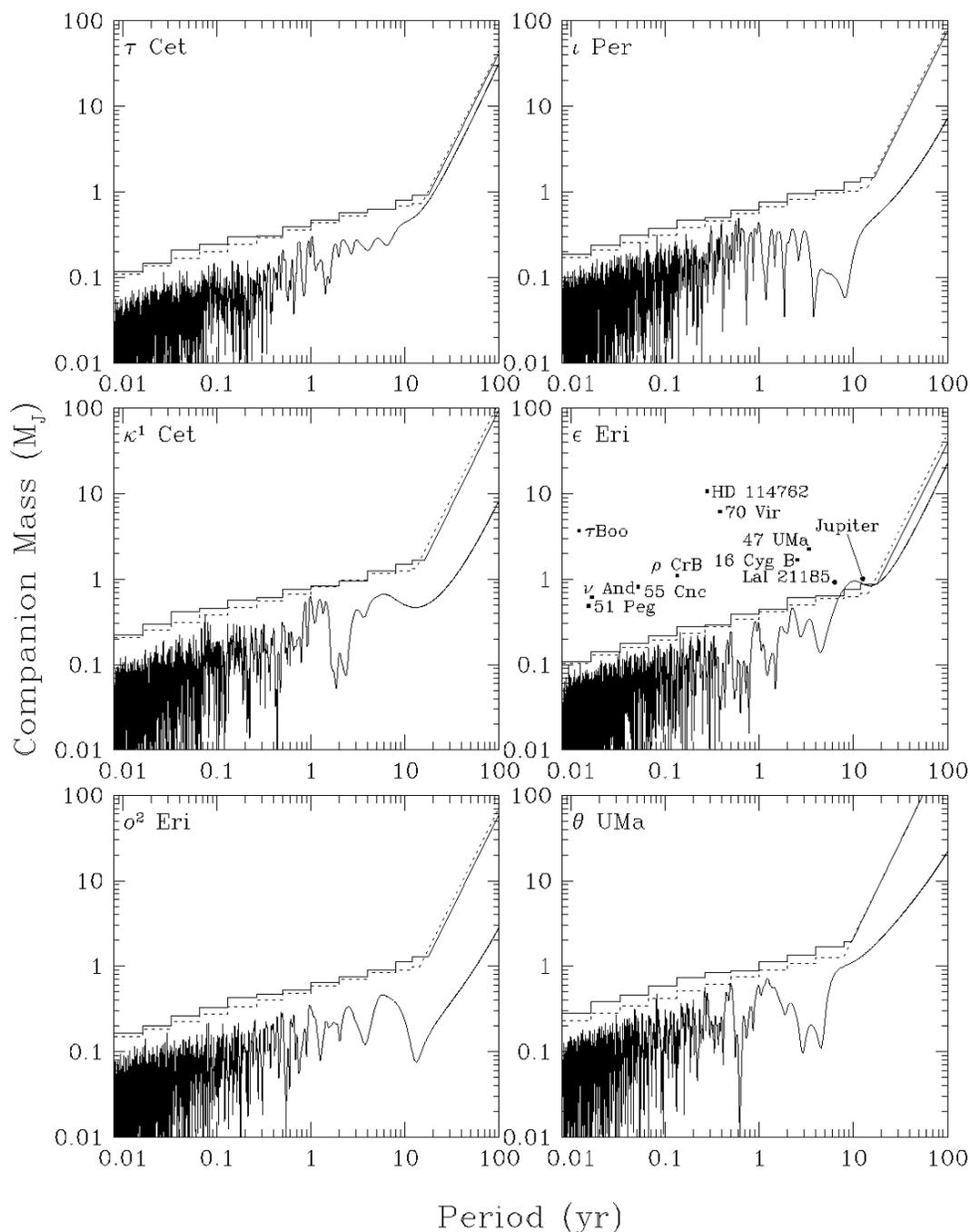}{6.85in}{0}{70}{70}{-210}{-20}
\caption[Best fit companion masses and limits for a subset of 14 stars 
studied by Walker \etal]
{\label{datamc-bf}
\footnotesize
Best fit companion masses ($\times \sin(i)$) for each star over a 
range of periods corresponding to the most sensitive range of the 
data. The histograms represent the 99\% mass limits for each
specified period range as given by our analytic formulation (dotted 
histogram) and based on a Monte Carlo experiment (solid histogram) 
consisting of 3000 simulated data sets.  Also plotted (squares) 
are the measured $M_c\sin(i)$ values of recent planet detections 
(see text). Jupiter and the astrometrically detected (but unconfirmed)
companion to Lalande 21185, are shown omitting a $\sin(i)$ 
correction (circles).}
\end{figure}

\clearpage

\begin{figure}
\figurenum{\ref{datamc-bf}--continued}
\plotfiddle{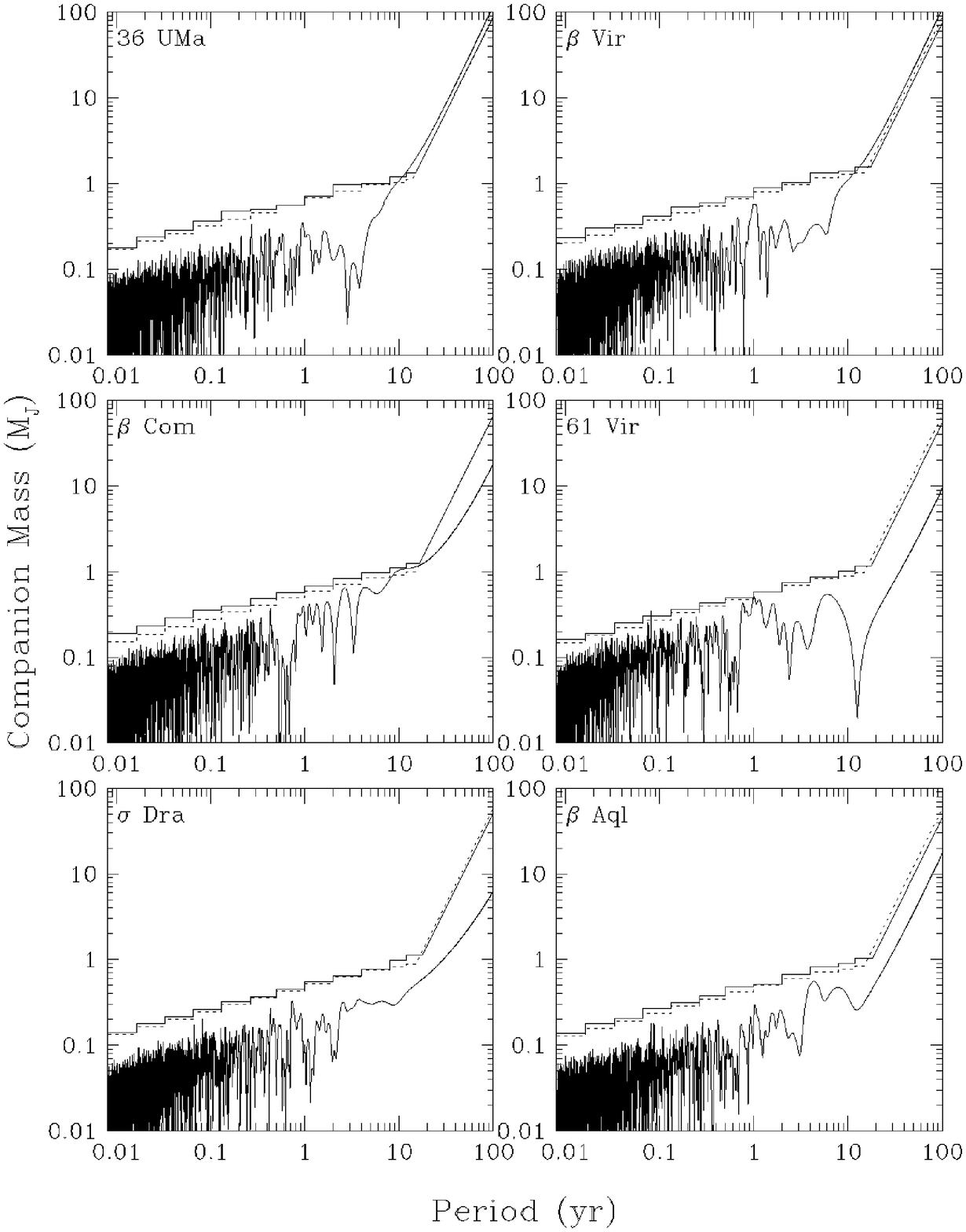}{7.2in}{0}{71}{71}{-220}{0}
\caption{}
\end{figure}

\clearpage

\begin{figure}
\figurenum{\ref{datamc-bf}--continued}
\plotfiddle{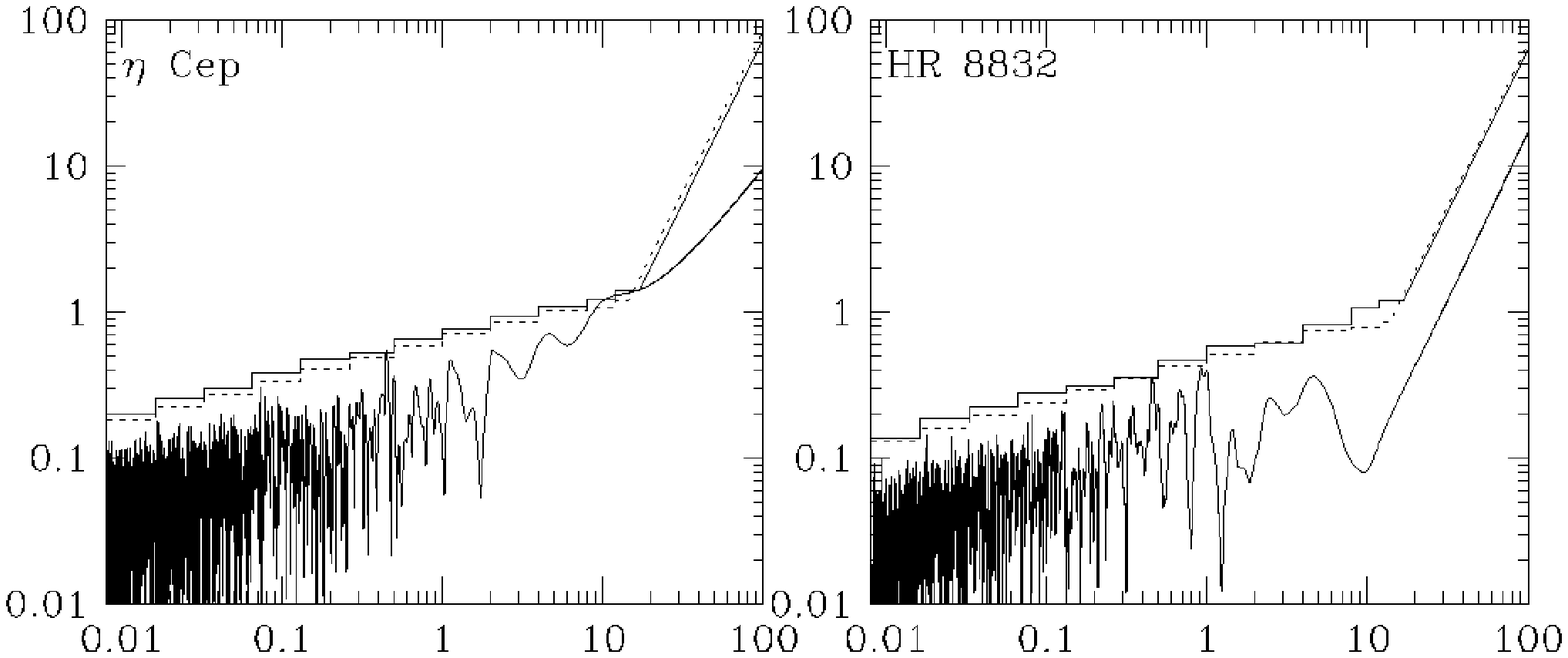}{7.2in}{0}{67}{67}{-220}{100}
\caption{}
\end{figure}

\clearpage

\begin{figure}
\plotone{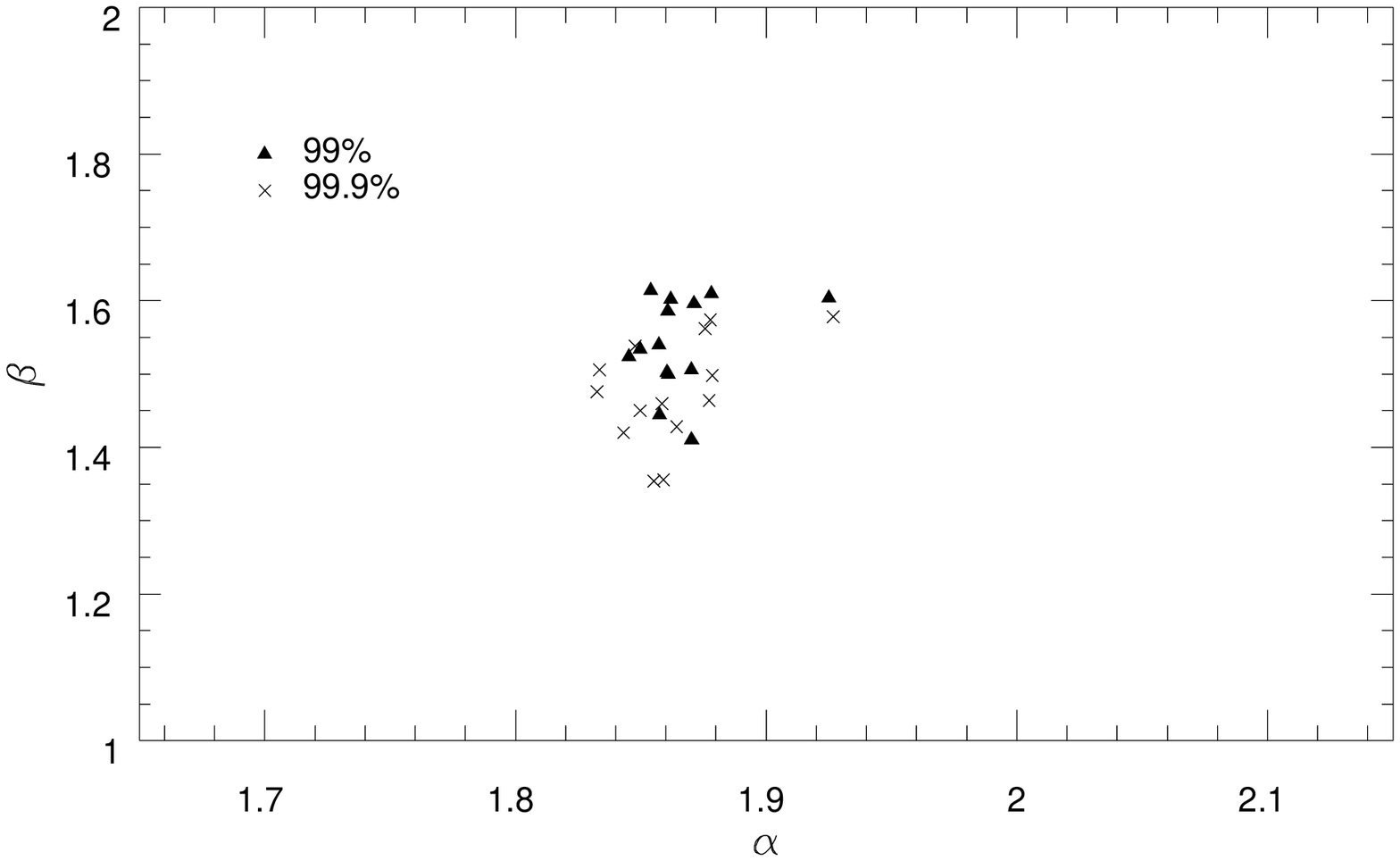}
\caption[Best fits for the long period sensitivity fall off parameters derived
from the Walker \etal sample]
{\label{datal-bf}
\footnotesize
Best fit values for the long period sensitivity fall off parameters
$\alpha$ and $\beta$ derived from the data. Values for both 99\% 
(triangles) and 99.9\% ($\times$'s) fall-off are shown. The two
points lying to the right of the main group originate from the star
$\theta$ UMa and are consistent with the Monte Carlo results for
a 6 year data span.}
\end{figure}

\clearpage

\begin{figure}
\plotone{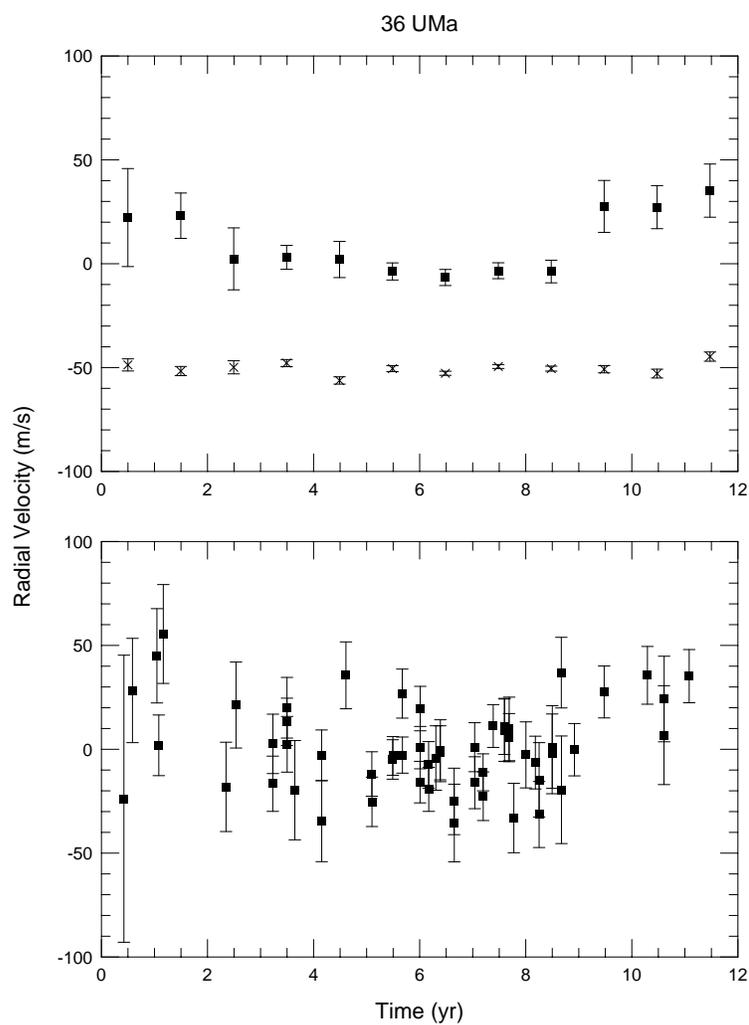}
\caption[Radial velocities measurements for the stars 36 UMa and $\beta$ Vir
from the Walker \etal sample]
{\label{star-rv}
\footnotesize
Radial velocity data for the stars (a) 36 UMa and (b) $\beta$
Vir, for which long term trends are observed using both our analysis
technique and the Walker et al. analysis. Both the published velocity 
data and the weighted mean of the velocities for each year are shown.
The binned data for the sample of 14 stars taken together is shown offset 
by $-50$ m/s. No trend similar to that found in either 36 UMa or $\beta$
Vir is present, indicating that a systematic error common to all stars
is unlikely.  Errors are taken from Walker et al. 1995, while the errors 
in the binned points are derived from the central limit theorem 
1/$\sqrt{n}$ improvement in the error of a multiply sampled mean.}
\end{figure}

\clearpage

\begin{figure}
\plotone{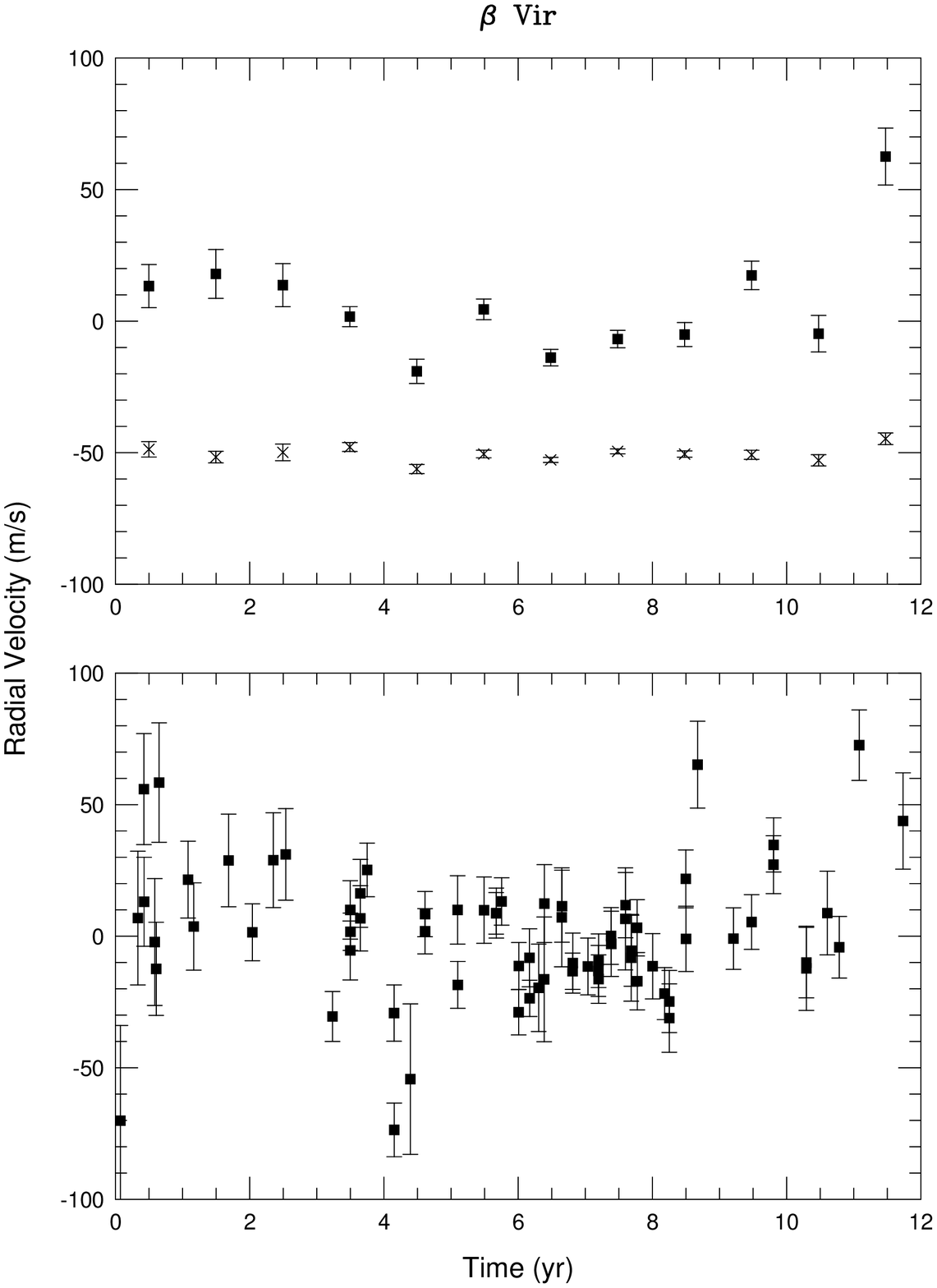}
\figurenum{\ref{star-rv}b}
\caption{}
\end{figure}

\clearpage

\begin{figure}
\plotone{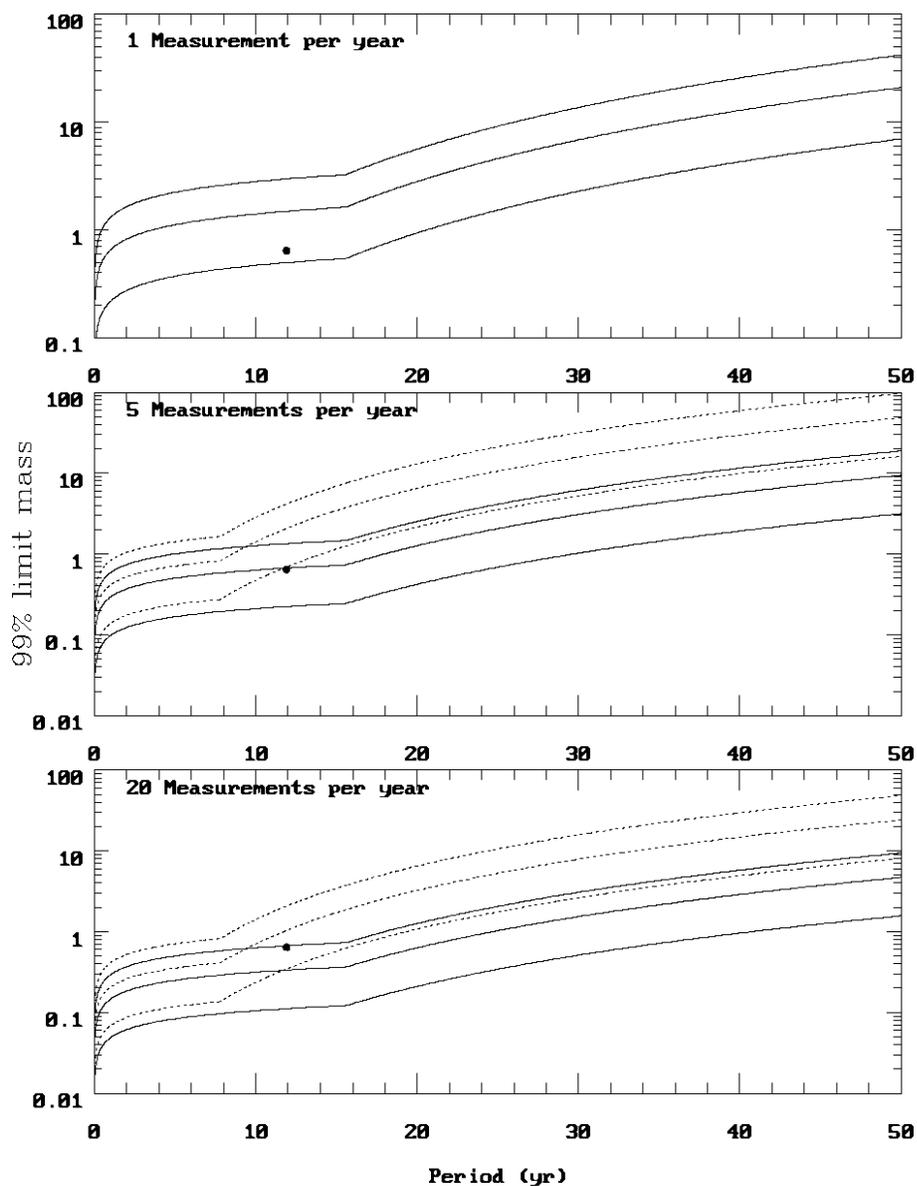}
\caption[Companion mass limits possible for data of a given quality] 
{\label{mlim-anl}
\footnotesize
Limits above which data of a given quality and duration 
constrain the mass of companions at the 99\% level for any single 
period. Solid lines represent a 12 year span of data taken with 5, 15 
and 30 m/s precision while dotted lines represent a comparable 6 year
span.  Limits for the 6 year baseline with one only measurement per 
year are omitted here because they do not correspond to results from 
Monte Carlo experiments (see section \ref{sparse}).}
\end{figure}

\doublespace

%% file: conclusions.tex
\chapter{Concluding remarks and where we ought to go from 
here\label{conclusions}}

Over the past few hundred years, we've developed an extremely detailed
understanding of the mechanics and characteristics of objects in our
own solar system. Over the past few decades we've started to unravel
a few of the problems associated with it's formation and origin.
The kinds of questions now under study for our own system are not of
the type `what objects are in the solar system and how do they move?' but
rather `what is the internal structure of this or that object, how did 
it get that way and what is it likely to do in the future?'. 

The two thrusts of star formation research: first, to acquire a more 
complete inventory of the types of solar systems which exist and second, 
to more completely understand how those systems came to be the way they 
are now, have each been addressed in this thesis. Our focus has been 
primarily on understanding the physical processes important during the
formation of solar type stars, but we've also focused on the observable
signatures of such systems and limits which can be placed on their 
detection by observers here in our solar system. Working with Willy
Benz, Dave Arnett, Fred Adams and Tamara Ruzmaikina, I have numerically
simulated the evolution of massive circumstellar disks and analyzed the 
results for their dynamical and observable spectral characteristics. With 
Roger Angel, I have outlined the detection limits available to radial
velocity searches for low mass (ie. planet or brown dwarf) companions. 
We then used these limits to suggest an efficient strategy for new or
continuing large radial velocity surveys.

Many of the problems which remain require a considerable allocation 
of resources in numerical computation or telescope instrumentation, so
it is important that we address the questions which most efficiently
provide answers to our remaining questions. On the theoretical side 
for example, many of the uncertainties which remain require that no
symmetries be assumed which reduce a problem from three to two, or from
two to one spatial dimensions. A complete treatment of all three spatial
dimensions is required. On the observational side, developing an 
inventory and understanding of the kinds of systems which exist requires
both very high spatial resolution of the sky and very high contrast 
resolution, in order to detect very faint objects (planets and 
brown dwarfs) near very bright objects (stars).

The past decade or two has seen a great deal of work attempting to 
understand the origin of various morphologies of stellar systems. Not 
much attention has yet been paid to what the systems simulated would 
actually look like to an observer attempting to detect such objects
in some wavelength band. It is not too much of an exaggeration to say
that the current state of affairs is one in which we can now make
stellar systems of nearly any morphological type (single, binary, 
multiple etc.) through either cloud collapse or disk evolution. 
However, several questions about the formation morphology of disk 
systems and the formation of planetary systems do remain. Also, the 
field is nearly wide open in terms of understanding the observational
characteristics of simulations of the newly formed systems of all
morphologies.

Several of the areas requiring further development which build on the 
work presented here are (1) to characterize the full inventory of low 
mass objects which exist in the solar neighborhood and in the galaxy
as a whole and similarly, to understand the characteristics of forming
systems. Just as importantly, we also require reliable limits on what 
types of objects and systems are not detected and limits on our 
sensitivity to detect various classes of objects (2) to understand in
detail the initial formation processes of circumstellar disks, from 
small scales outward, (3) to understand the transport and evolution of 
ice and dust within the nebula in order to better understand both 
radiative transfer processes (and the consequent observational 
characteristics) and the formation of low mass companion objects through
agglomeration and (4) to understand the early stages of growth and 
migration of Jovian companions within disks.

\section{Taking Inventory: What is Possible Using Various Detection 
Techniques} 

In chapter \ref{RVchap}, we discussed limits on the detection of low 
mass companions to nearby stars via a single technique: radial velocity 
measurements. There are also a variety of other possible techniques
which are in various stages of study or implementation, including
micro-lensing, astrometry, occultation and direct imaging.   

Of the many different techniques that are now or soon will be available,
which provides the best chance of detecting companions of other stars, or
of detecting low mass objects not orbiting a star? What kind of objects
should we expect to detect with one technique or another and what kinds
of systematic biases exist for each? How can we know how complete the
sample of detected objects is and what kinds of objects are likely to
be missed? A study which compares the sensitivities of the various
techniques and their systematic `blind spots' would be of great use
in establishing the true distribution of low mass objects.

Some of these other techniques have in fact begun to bear fruit, with a 
variety of low mass objects floating in free space (Hillenbrand 1997, 
Reid \etal 1998, Luhman \etal 1998), bound to stellar primaries (Nakajima 
\etal 1996) or even bound to each other (Basri \& Martin 1998). Others, such as 
micro-lensing (Peale 1997) and occultation surveys (Borucki \etal 1998),
are not yet underway. 

Of all of the techniques mentioned, only direct imaging is capable of detecting
free floating low mass objects. As marvelous as it is to have a picture of 
the object you're looking for, it is also true that direct detection suffers 
from some systematic biases. Among them are the fact that a substellar mass 
companion will be orders of magnitude dimmer than it's primary. Detection will 
therefore require quite high angular resolution on the sky in order to make 
sure that the point spread function of the star on the focal plane of the
telescope is small enough to enable the detection of the slight additional
flux from the companion.

Other techniques also have biases. For example, it is well known that 
radial velocity techniques are biased towards detecting companions close
to their primary because of the larger reflex velocities imposed on the
motion of the star by a close companion. For a similar reason, astrometry
is biased toward large companions at large distances from their primary,
since a planet in a distant orbit has a larger `lever arm' to perturb 
the motion of the primary. Both have a systematic bias against detecting
companions in large orbits due to the attention span (and in some cases
life span) of the observer and/or the apparatus used to make the measurements.

Micro-lensing and occultation searches also suffer from systematic biases.
In the case of these two methods, the biases arise because of the large
area of sky which must be searched at a very high rate of observation.
For example, occultations of a star by a companion will occur in only a 
small fraction of stars which have companions (of order a few percent)
and will last for only a few hours. Micro-lensing events will undoubtedly
prove to be quite limited for detecting individual objects since an
event can only be observed once. On the other hand, it may prove to be
of great use in establishing larger statistical distributions of
objects than can be obtained by any other means.

Given these data on the possibilities of various detection techniques,
what is the best strategy for getting a large enough sample of 
low mass companions of every type (planets, brown dwarfs or low mass
stars) around stars near the sun, floating freely in space and in the
galaxy as a whole? 

\section{Disk Formation}

Circumstellar disks extend over a very large range of radii: from a few
$R_{\sun}$ to several hundred AU. They form from the inside out, as 
higher and higher angular momentum material falls into the system. As we
found in the work discussed in chapters \ref{isodisk} and \ref{cooldisk}, 
they are notoriously difficult to model due to the vastly disparate 
time scales involved in their evolution. Because of this problem, numerical
simulations of star formation have generally fallen into two general
categories: simulations of collapse and simulations of already formed, narrow
tori or disks. Only a few collapse calculations have been done which also 
follow the evolution until a disk forms (eg. Laughlin and Bodenheimer 
1994; Boss 1993, 1996). Each of these works resolves the innermost regions
on a scale of several AU, which implies that the spatially small, inner
disk regions are not important for the evolution of the gross structure
of the system. They sacrifice the resolution of the small scale features 
of the forming disk in favor of understanding the large scale morphology 
of the system.

We have shown (chapters \ref{isodisk}, \ref{cooldisk}) that the inner regions 
of already formed disks are in fact very dynamically active and do affect the 
gross structure of the system. If a spatially large disk ($>50$~AU) is to 
form we must understand how the dynamic inner region can sustain itself
while more distant regions form. The small scale features of the collapse
from the star outward must be resolved. I suggest a series of numerical
simulations modeling the formation of circumstellar disks from this 
perspective. 

Stahler et al. (1994) has extended the work of Cassen and Moosman 
(1981) to show that as the infall proceeds and the infalling material no 
longer intersects the stellar surface, three distinct regions form. Each
region expands radially as $t^3$. Innermost is a Keplerian disk, while in 
the outer regions the disk is characterized by infalling matter with 
comparable radial and azimuthal velocities. In between lies a transition 
region (modeled in their work as a discontinuity) in which the matter 
loses much of its radial velocity and moves into the Keplerian inner 
region. 

I propose that numerical hydrodynamic techniques like the PPM code discussed 
in chapter \ref{isodisk} be used to simulate the early evolution of the systems
like those studied by Stahler et al. Such a study should investigate several 
questions not possible to address in their work. In particular, they study 
only axisymmetric and inviscid systems; each fluid element conserves $j_z$,
the $z$-component of specific angular momentum. Transport through the disk
is provided by accretion of additional low angular momentum material onto 
the surface of the disk from above and below. They also posit a massive 
`fly-wheel' ring in their transition region, which is modeled only as a 
discontinuity. Such assumptions are unsatisfactory because the forming disk
will develop shocks and turbulence and therefore become dynamically 
unstable. Shock dissipation and the eventual formation of non-axisymmetric
structures will play a crucial role in determining the structure and 
subsequent evolution of this region and of the disk as a whole. 

In order to model the dynamics of the forming disk it is likely that 
three dimensional simulations will be required, at high computational cost.
A useful first step can be made if we assume that the collapse originates 
from a cloud in solid body rotation (see Stahler \etal figure 2). Then at
any given instant the highest angular momentum material falling onto the disk
will come from the equatorial regions of the collapsing cloud.  With this
assumption, material falling onto the disk from above and below can be 
neglected and the system can be modeled in only two dimensions ($r$ and
$\phi$) but still preserve many of the important physical phenomena.  

Such a study is a natural outgrowth of this thesis because the dynamically 
important effects we have described require only small modifications of
existing code to address. For example, a number of the simulations performed 
in chapter \ref{isodisk} assumed an infall onto the outer edge of the disk, 
so implementation of the infalling cloud matter will be a trivial adaptation
of already existing code. Also, because the collapse and disk growth are self
similar, modeling the disk formation for the relatively short time scale allowed
by a multi-dimensional hydrodynamic scheme will give a good picture of the 
disk even at much later times.

This study will provide a wealth of information on the properties of a 
previously unexplored region of parameter space. Several important questions
which become accessible with this work are: What is the initial mass
distribution in the disk? Is it a power law like that given by the results 
of Stahler et al. and others, or is it flatter/steeper? To what extent can 
it be fit to a power law at all? What is the form of the transition region 
and how does it vary in time? What is the character of the boundary layer 
between the disk and star? 

Ultimately, when high resolution three dimensional simulations become
computationally feasible, another set of questions can be addressed.
For example, what is the origin of the bipolar jets observed to be
coming from forming star/disk systems? Undoubtedly magnetic field
effects are important in this same region. Some work has already been 
done (Stone \etal 1996, Shu \etal 1994a, 1994b, Najita \& Shu 1994,
Ostriker \& Shu 1995, Ouyed \& Pudritz 1997a,b) to try and understand 
what their effects are on the system, but more work is required. An interesting
question that bears on both the star/disk boundary layer and the effect of 
magnetic fields is the question (proposed by Shu \etal) of how and whether
mass is accreted onto the star through magnetic flux tubes at the interface. 

\section{The Transport and Evolution of Dust in the Disk: Effects on Energy
Transport\label{dust-energy}} 

We have shown (chapter \ref{cooldisk}) that a correct model of the spectral 
energy distributions (SED's) of observed systems requires a detailed 
understanding of the processing of dust within the disk. In the regions of 
the disk within a few AU from the star, gas may be heated to temperatures 
above the destruction temperature of dust grains contained in the nebula. 
We have shown that if dust which is destroyed in the hot midplane of
the disk reforms quickly as it is processed to high altitudes, the SED's
synthesized from our numerical simulations produce insufficient flux in 
the near IR ($\sim$1--5 $\mu$m) to reproduce observed systems. On the 
other hand, if dust reforms on a longer time scale, comparable or longer than
the convective overturn time scale so that most of the refractory material
is found in the gas phase or in a grain size distribution strongly 
modified from its original distribution, then we may be able reproduce 
the SED's of observed systems in the near infrared. 

The assumptions underlying our models are that the disk is locally
plane parallel and vertically adiabatic at each location in the disk.
No true three dimensional calculation of the vertical structure,
no self consistent modeling of the transport of dust from low to high
altitudes is done and no evolution of the grain size distribution is
incorporated into the hydrodynamic calculation. For a complete picture of
the evolution all of these effects must be included in the calculation. 
Work to study the evolution of turbulence in disks (Cabot 1996, Balbus, 
Hawley \& Stone 1996) for some conditions. However they have not studied
its effect on the grains swept along with the gas. 

Rather than studying its origin I propose a study of the observational
consequences of turbulence. More specifically I propose a study of the 
processing of disk gas and dust in the vertical coordinate of the disk. 
Cabot (1996) has shown that full three dimensional simulations may not be 
required to obtain some information from such a study because the large 
Keplerian shear quickly removes azimuthal variations. A two dimensional 
study in $r$ and $z$ may therefore provide many of the physical properties
of the disk necessary to model the evolution of the dust and ice present.

This work will require that several species of dust and ice be evolved
separately through each simulation. The `PROMETHEUS' code, which utilizes
the PPM hydrodynamic method and which is used in chapter \ref{isodisk},
can be easily adapted to this task. In its original implementation, it is
used to study the evolution of nuclear species inside supernovae (see eg.
Bazan \& Arnett 1997). Much of the machinery for multi-species evolution
has already been thoroughly tested. Remaining tasks involve adaptation
of the code specifically to the problem of ice and dust vaporization and
reformation reactions and the specification of the initial state of the 
system.

The specific simulations I propose should concentrate on modeling the 
vertical structure of the inner disk where it is warm enough that ice and 
dust can be vaporized. This region extends outwards from the stellar 
surface to perhaps 1--2 AU from the star. The outcome of these simulations
will lead to a better understanding of the influence of solid material 
on the transport of energy through the disk and its dissipation into 
space. It is possible that these simulations of the inner disk may also
provide insight into the mechanisms involved in forming a bipolar outflow
from the system.

Grain and planetesimal growth processes have been studies for many years,
both on small scales and larger scales (Safronov 1969, Weidenschilling 
1980, Greenberg \etal 1978, Wetherill \& Cox 1984, 1985, Greenberg 
\etal 1991, Spaute \etal 1992, Tanaka \& Ida 1996). The studies outlined
in section \ref{dust-energy} will certainly be able to incorporate many
of the results of this work (for example the reaction network physics
required to couple grains of different species and size to each other).
Most of the work done so far has been in a celestial mechanical framework,
with the gas modeled as a `black box' turbulent fluid with some characteristic 
velocity. Grains are more or less entrained in the fluid, depending upon 
their size and collide with each other, break apart or stick at relative 
velocities defined by the input turbulence assumption. 

The calculations I suggest may be able to extend the previous work by relaxing 
two of the most important of these assumptions. First, the mixing in the 
previous calculations is assumed to be rather uniform in space: any migration 
of different sized particles from one place to another is assumed to be mixed 
instantaneously into the surrounding medium. If instead the grains do not
mix well, so that some regions retain grains of whose size distribution 
is quite different than nearby regions, drastically different grain 
evolution may result. Secondly, with a hydrodynamic code, the turbulence 
may be resolved to some extent (depending upon the available computing 
power), so that relative velocities of grains becomes much better 
constrained and the dependent collisional properties become more narrowly 
defined. A solution to these questions will lead to a better understanding
of not only the radiative transport processes in disks (through grain 
absorption and scattering of light), but also of planet formation, through
grain growth to larger and larger sizes.

\section{Migration of Jovian Planets}

The classical models for Jovian planet
formation (ie. that gaseous disk matter collapses onto a 10--15 $M_{\earth}$ 
rock/ice core), predict that Jovian planets will form at distances of 
$\sim$5 AU or more from the central star (Boss 1995) due to the condensation
of ices at that distance. The detection of low mass companions in orbits
very close to their primary (as small as 0.05~AU) has forced theorists to
postulate that a companion can form far from its primary, then migrate to
its present location. The migration models proposed (see Takeuchi \etal 1996 
and references therein), have shown that gravitational torques upon the 
companion by a circumstellar disk are sufficient to change its orbit 
drastically on a time scale of only a few thousand years. Migration is 
so efficient that under normal assumptions about the character of a disk,
it is far too {\it easy} for a companion to continue its migration and 
simply fall into the star. 

The migration models proposed have been limited in the sense that 
they are linear analyses and are one dimensional. They omit effects
such as accretion onto the companion, do not take full account of shock
dissipation of waves excited in the disk and assume a zero eccentricity
orbit.

In (so far) unpublished calculations, I have performed a set of numerical 
simulations incorporating an already formed companion into a two 
dimensional ($r, \phi$) disk. These calculations show that an already
formed Jovian mass companion can form a wide gap in the disk in only
a few $\times 10^{2}$ years. In so doing, it moves inward to nearly
half its original orbit radius and then proceeds for the remainder of
the simulation on a slow, secular inward evolution. Time step 
constraints forbid us from following the evolution of the system for
the long periods ($\gtrsim 10^4$yr) necessary to follow the orbital 
evolution to it's conclusion. 

The simulations I have performed, although providing interesting results
in themselves, begin with an initial condition which is quite artificial. 
A $\sim$1--2 $M_J$ companion would certainly have formed a gap or otherwise
modified the mass distribution in the disk. Instead of concentrating
on the evolution of the disk once a companion has grown to appreciable
size, I suggest simulations modeling the evolution of a low mass core 
($\sim$10--30 $M_{\earth}$) as it begins to accrete large amounts of gas
from the disk and perhaps carve out a gap.

Three dimensional calculations of Bondi-Hoyle accretion have been done
(Ruffert 1997 and references therein) and have shown that accretion 
proceeds at a rate not unlike the Bondi-Hoyle rate even in flows with
transverse velocity gradients. No calculations have been performed in
which the accretor lies in a medium undergoing Keplerian shear however.
I suggest extending the previous calculations to the problem of an 
accreting proto-planet. As a first step, these calculations will 
implement initial conditions including a transverse Keplerian shear
but excluding density variations in the $r$ or $z$ directions.
This condition is equivalent to simulating the earliest stages of growth, 
when the proto-planet is only able to affect regions of the disk close to 
the midplane.

The outcome of these simulations will be to determine the mass and angular
momentum accretion rates of the proto-planet as it begins to grow quickly. 
Comparison of these rates to the rate at which gaps form by gravitational
torque processes will provide insight into the eventual final state
of the system as a whole. Depending on the outcome of this first round of
simulations, it may become of interest to include density gradients in 
two dimensions. In this case, the question of how large the accretor must be
before it first begins to deplete mass from an entire vertical column of 
the disk. For an accretor of that size, accretion may be reduced due to the
loss of material accreting onto the poles of the proto-planet and a gap 
may form, slowing additional migration through the disk.